\documentclass[11pt,a4paper,twoside,openright]{report}

\usepackage[italian,english]{babel} 
\usepackage{graphicx} 
\usepackage{tabularx} 
\usepackage{algorithm} 
\usepackage{algorithmic} 
\usepackage{amsmath} 
\usepackage{amssymb} 
\usepackage[scriptsize]{caption} 
\usepackage{stdclsdv} 
\usepackage{tocloft} 
\usepackage{varioref} 
\usepackage{xspace} 
\usepackage[usenames,dvipsnames,svgnames,table]{xcolor} 
\usepackage{fancyhdr} 
\usepackage{subcaption} 
\usepackage{rotating} 
\usepackage[bookmarks=true,colorlinks=true,linkcolor={blue},citecolor={blue},urlcolor={blue}]{hyperref}

\cftsetpnumwidth{3em} 

\def \nibf#1{\noindent\textbf{#1}}

\newcommand{\stay}{stay} 
\newcommand{\gs}{study group\xspace}

\newcommand{\GS}{Study Group\xspace}
\newcommand{\se}{study group extraction\xspace}

\newcommand{\SE}{Study Group Extraction\xspace}
\newcommand{\pa}{propensity analysis\xspace}

\newcommand{\PA}{Propensity Analysis\xspace}
\newcommand{\oa}{outcome analysis\xspace}

\newcommand{\OA}{Outcome Analysis\xspace}
\newcommand{\db}{Mimic II Clinical Database\xspace}
\newcommand{\los}{length of stay\xspace}
\newcommand{\Los}{Length of stay\xspace}
\newcommand{\LOS}{Length of Stay\xspace}
\newcommand{\mort}{mortality\xspace}
\newcommand{\Mort}{Mortality\xspace}
\newcommand{\gettingdiuretics}{the administration of diuretics\xspace}

\newcommand{\GettingDiuretics}{The Administration of Diuretics\xspace}
\newcommand{\illness}{health condition\xspace}
\newcommand{\Illness}{Health condition\xspace}

\newcommand{\DiureticsDecision}{DiureticsDecision\xspace}
\newcommand{\diureticsDecision}{diureticsDecision\xspace}
\newcommand{\ModelA}{\textsc{ModelA}}
\newcommand{\ModelB}{\textsc{ModelB\xspace}}
\newcommand{\ModelC}{\textsc{ModelC\xspace}}
\newcommand{\gettingDiureticsvar}{diuretics decision variable\xspace}

\newcommand{\pvalue}{p-value\xspace}
\newcommand{\Pvalue}{P-value\xspace}

\newcommand{\SAPSTZ}{SAPS-$T_0$\xspace}
\newcommand{\SAPSONE}{SAPS-$T_1$\xspace}
\newcommand{\SOFATZ}{SOFA-$T_0$\xspace}
\newcommand{\SOFAONE}{SOFA-$T_1$\xspace}
\newcommand{\ELIX}{Elixhauser Score\xspace}
\newcommand{\Elix}{\ELIX}
\newcommand{\SAPSD}{SAPS-${t0}$\_diureticsDecision\xspace}
\newcommand{\betacoeff}{beta coefficient\xspace}
\newcommand{\betaCoeff}{Beta coefficient\xspace}

\newcommand{\ModelCLessSick}{\textsc{ModelC.LessSick}\xspace}
\newcommand{\ModelCSick}{\textsc{ModelC.Sicker}\xspace}
\newcommand{\ModelCSicker}{\ModelCSick}
\newcommand{\group}{quintile}
\newcommand{\Group}{Quintile}
\newcommand{\groups}{quintiles}
\newcommand{\Groups}{Quintiles}

\begin{document}

\pagestyle{empty} 
\normalfont
\thispagestyle{empty}
\label{title}

\vspace*{-1.5cm} \bfseries{

\begin{center}
  \large
  UNIVERSIT\'A DEGLI STUDI DI MILANO BICOCCA\\
  \normalsize
  Corso di Laurea Magistrale in Informatica\\
  FACOLT\'A DI SCIENZE MATEMATICHE FISICHE E NATURALI\\
  \begin{figure}[htbp]
    \begin{center}
      \includegraphics[scale=0.55]{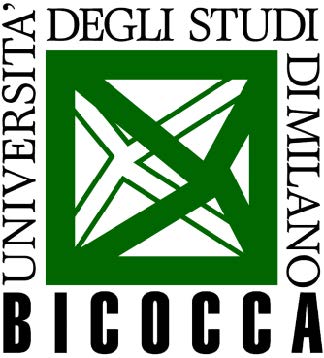}
    \end{center}
  \end{figure}
  \vspace*{0.3cm}
  \LARGE
  \textbf{An Observational Study:\\
   The Effect of Diuretics Administration on Outcomes of Mortality and Mean Duration of I.C.U. Stay}\\
  \vspace*{.75truecm}
  \large
  Evolutionary Design and Optimization Group\\
  Computer Science and Artificial Intelligence Laboratory\\
  Massachusetts Institute of Technology
\end{center}

\vspace*{3.0cm} \large

\begin{flushleft}
  Supervisor: Prof. Giancarlo Mauri\\
  Supervisor: Prof. Leonardo Vanneschi\\
  Supervisor: Prof. Una-May O'Reilly
\end{flushleft}

\vspace*{1.0cm}

\begin{flushright}
  Master Degree Thesis by:\\
  Daniele Ramazzotti, 725339
\end{flushright}

\vspace*{0.5cm}

\begin{center}
  Academic Session 2011-2012  
\end{center} \clearpage

}

\clearpage \mbox{} \clearpage 
\vspace{17cm}
\label{dedication}

\begin{flushright}
\itshape{To My Parents,\\My Grandparents\\and My Friends}
\end{flushright}

\clearpage \mbox{} \clearpage 
\pagestyle{fancy} 
\fancyhead[LE,RO]{\thepage} 
\normalfont
\pagenumbering{Roman} 
\setcounter{page}{1}
\chapter*{Acknowledgments}
\label{ack}
\addcontentsline{toc}{chapter}{Acknowledgments}

\vspace{-0.2in}
The prospect of carrying out my master thesis at the Massachusetts Institute of Technology (MIT) was proposed me for the first time a year ago by Professor Vanneschi. That possibility was amazing and I immediately decided to accept. My experience at the MIT Computer Science and Artificial Intelligence Laboratory (CSAIL) began on October 9, 2011 and I stayed in Boston until February 13, 2012 to  then return for another five weeks between May and June.

During the 5 months I spent in Boston, I learned an incredible amount of things that allowed me to grow a lot both in professional and personal terms. I can say now that it was an amazing experience that has exceeded even the big expectations I had before leaving.

First of all I would like to thank Professor Leonardo Vanneschi and Professor Una-May O'Reilly for giving me this great opportunity.

I first met Professor Vanneschi during the Soft Computing course. It was really a great experience to learn from him and use the Soft Computing techniques during my stay at the MIT. I would like to thank him the most for offering me the chance to work on my master thesis at the MIT.

I first met professor O'Reilly when I arrived at the CSAIL and during my stay there she taught me a brand new way of working, different from how I was used to. Working with her was incredible.

I would like to thank all the CSAIL team and in particular Dr. James McDermott for helping me to set up and perform the work during the first 4 months of my stay at the CSAIL and Dr. Kalyan Veeramachaneni for helping me in completing the thesis.

I would like to thank Leo Celi M.D. and John Danziger M.D. for their  support regarding all the medical issues and the definition of the problem.

I would like to thank my family for always supporting me in my decision to go to Boston and for helping me during all the 5 months.

This experience was incredible. Thank you all for making this possible.

\clearpage \mbox{} \clearpage 
\chapter*{Abstract}
\label{abs}
\addcontentsline{toc}{chapter}{Abstract}

\vspace{-0.2in}
This thesis conducts an observational study into whether diuretics should be administered to ICU patients with sepsis when length of stay in the ICU and 30-day post-hospital mortality are considered. 
The central contribution of the thesis is a stepwise, reusable software-based approach for examining the outcome of treatment vs no-treatment decisions with observational data. The thesis implements, demonstrates and draws findings via three steps: 
\begin{description}
\item[Step 1.]Form a study group and prepare modeling variables.
\item[Step 2.]Model the propensity of the study group with respect to the ad-ministration of diuretics with a propensity score function and create groups of patients balanced in this propensity.  
\item[Step 3.]Statistically model each outcome with study variables to decide whether the administration of diuretics has a significant impact.
\end{description} 
Additionally, the thesis presents a preliminary machine learning based method using Genetic Programming to predict mortality and length of stay in ICU outcomes for the study group. 

The thesis finds, for its study group, in three of five propensity balanced quintiles, a statistically significant longer \los when diuretics are administered. For a less sick subset of patients (SAPS ICU admission score $< 17$) the administration of diuretics has a significant negative effect on mortality.

\clearpage \mbox{} \clearpage 
\listoffigures 
\addcontentsline{toc}{chapter}{List of Figures}
\clearpage 
\listoftables 
\addcontentsline{toc}{chapter}{List of Tables}
\clearpage 
\listofalgorithms 
\addcontentsline{toc}{chapter}{List of Algorithms}
\clearpage 
\tableofcontents 
\clearpage \mbox{} \clearpage 

\pagenumbering{arabic} 
\setcounter{page}{1}
\chapter{Problem Statement}
\label{cpt:problemStatement}

\section{Introduction}
\label{cpt:problemStatementIntro}
Diuretics are drugs which promote urination. They are often used to bring fluids levels down to normal after  intravenous (IV) fluids have been intensively infused. They could be harmful in some circumstances but there are no randomized clinical trials to date which provide evidence for the benefit or harm of these drugs in general. This thesis conducts an observational study into this question.

Generally, all ICU\footnote{An Intensive Care Unit (ICU) is a highly specialized department of a hospital that provides intensive-care medicine, concerned with the diagnosis and management of life threatening conditions requiring sophisticated organ support and invasive monitoring.} patients with sepsis\footnote{Sepsis is a potentially deadly medical condition that is characterized by a whole-body inflammatory state and the presence of a known or suspected infection. For a precise description of sepsis see Appendix~\ref{apx:medicalBackground}.} are infused with high levels of fluids as soon as they enter the ICU in order to improve their low blood pressure and treat their medical condition. This practice is called fluids resuscitation. When a patient is recovering and still has high fluid levels, clinicians face a decision on whether to prescribe diuretics, which will bring about a reduction of a patient's fluids levels to normal, or to let fluids levels decrease naturally. This is a grey area of clinical medicine: different doctors, even when presented with similar patients, can choose either to prescribe diuretics or not.

The long term and broad aim of the developed analysis is to provide clinicians with quantitatively reasoned decision support when they face the choice of treating or not treating. The central contributions include  clearly delineated steps describing the analysis see Section~\ref{cpt:steps}, and a set of software tools. The software tools are re-usable. They support the creation of a new study group by providing software that extracts, intersects and filters \db records. Additional software supports developing co-variate statistically balanced quintiles of the study group with respect to propensity to receive diuretics. It can be used to develop any propensity score function whether there are treated and untreated patients. Another software module supports outcome modeling with logistic and linear regression accompanied by p-value derivation. The final component of software is slightly modified genetic programming-based machine learning code which executes symbolic regression for classification and regression plus code calling a library function that performs clustering, a form of unsupervised learning. Specifically, in this thesis, the aim is to retrospectively examine the data of ICU patients with sepsis and, while controlling for the propensity for \gettingdiuretics and considering the possibly \textbf{confounding factor} of illness, to determine whether \gettingdiuretics has a significant effect on 30 day mortality outcome post-ICU or mean duration of ICU \stay.

\section{\GS}
\label{cpt:group}
The analysis attempts to address a specific group of patients, in particular adult patients with a large amount of fluids in their bodies. Therefore, the analysis will be conducted on patients over 18 years old\footnote{Neonatal sepsis is not subject of study in this work.} and with a sepsis diagnosis\footnote{It is difficult to define sepsis. In this work, as described in Appendix~\ref{apx:medicalBackground}, have been used the definition described in\cite{SEPSIS-EPIDEMIOLOGY}.}. From this group, CMO\footnote{Comfort Measures Only refers to medical treatment of a dying person where the natural dying process is permitted to occur while assuring maximum comfort.} patients have been filtered out because their outcome with respect to mortality is distinctive. Patients who had been taking diuretics before entering the ICU have also been eliminated because of the compliating nature of this on a decision for \gettingdiuretics (or continuing to take them).  Finally, patients who had multiple admissions, both in ICU and in the hospital have been eliminated.\footnote{First ICU visit data is a good potential alternative filter in a follow up study group creation.}.

\subsection{Variables of the Outcome Study and/or Propensity Model}
\label{cpt:variables}
A certain number of variables have been used as variables in the study to describe the condition of a patient during his/her stay in the ICU. Now a brief medical description of those factors is presented. Much of the text is directly quoted because of the need for medical precision.

\begin{itemize}
\item SAPS II score: this point score is based upon a severity of disease classification system\cite{SAPS-II}. It is calculated from 12 routine physiological measurements during the first 24 hours, information about previous health status and some information obtained at hospital admission. 
\item SOFA score: this point score is based upon a scoring system to determine the extent of a person's organ function or rate of failure\cite{SOFA}. The score is based on six different scores, one each for the respiratory, cardiovascular, hepatic, coagulation, renal and neurological systems.
\item Elixhauser score: this score integrates a list of 30 comorbidities relying on the ICD-9-CM coding manual. The comorbidities were not simplified as an index because each comorbidity affected outcomes (length of hospital stay, hospital changes, and mortality) differently among different patients groups. The comorbidities identified by the Elixhauser comorbidity measure are significantly associated with in-hospital mortality and include both acute and chronic conditions. Walraven et al.\cite{Elixhauser} has derived and validated an Elixhauser comorbidity index that summarizes disease burden and can discriminate for in-hospital mortality.
\item Creatinine: this is a break-down product of creatine phosphate in muscle. It is usually produced at a fairly constant rate by the body (depending on muscle mass). In our study, this factor is included because it can be indicative of kidney disease.
\item (Administration of) Vasopressors: Vasopressors indicates whether the patient was administered any sort of vaso-suppressor. Vasopressors are drugs that constrict the blood vessels and thereby elevate blood pressure. Usually, before a patients is considered able to leave the ICU, vasopressors are suspended. The administration of any vaso-suppressor have been included as a factors because vaso-suppressors are indicative of \illness.
\item Mechanical ventilation: this boolean variable indicates whether or not the patient was mechanically ventilated. Mechanical ventilation assists or replaces spontaneous breathing. Ventilation may involve a machine called a ventilator or the breathing may be assisted by a physician, respiratory therapist or other suitable person compressing a bag or set of bellows.
\item Arterial blood pressure: this quantity is the pressure exerted by circulating blood upon the walls of blood vessels and is one of the principal vital signs. The blood pressure in the circulation is principally due to the pumping action of the heart. In the study, the value for arterial blood pressure refers to systolic\footnote{During each heartbeat, blood pressure varies between a maximum (systolic) and a minimum (diastolic) pressure. Systolic blood pressure is a measure of blood pressure while the heart is beating, while diastolic pressure is a measure of blood pressure while the heart is relaxed.}.
\item Mean arterial blood pressure:  This quantity is defined as arterial pressure during a single cardiac cycle. Is is calculated as $\frac{2 \cdot diastolic + systolic}{3}$.
\end{itemize}

\section{Analysis Steps}
\label{cpt:steps}
There are 4 steps in the developed statistical analysis.

\begin{description}
\item [\textbf{Step~1:}] Form a \gs and prepare modeling variables. In this step, subsets of patient records are extracted from the \db and the subsets are fused then filtered according to clinician input. The variables of the \gs are next prepared for subsequent analysis. This step is described in Chapter~\ref{cpt:datasetExtraction}.
\item [\textbf{Step~2:}] Model the propensity of the \gs with respect to \gettingdiuretics with a propensity score function and create groups of patients balanced in this propensity. This step is described in Chapter~\ref{cpt:propensityAnalysis}.
\item [\textbf{Step~3:}] Statistically model outcome with study variables to decide whether \gettingdiuretics has a significant impact on mortality and \los in ICU while considering \illness and propensity of \gettingdiuretics. This step is described in Chapter~\ref{cpt:outcomeAnalysis}.
\item [\textbf{Step~4:}] Design a preliminary machine learning based method using Genetic Programming to predict mortality and length of stay in ICU outcomes for the \gs.  This step is described in Chapter~\ref{cpt:GPML}.
\end{description}

An overview of this process is shown in Figure~\vref{fig:p-def}. Finally, Chapter~\ref{cpt:conclusions} summarizes the analysis and its findings and lists possible future work.

\begin{figure}
\centering
\rotatebox{90}{
\stepcounter{figure}
\begin{minipage}{\textheight}
\centering
\includegraphics[width=\textheight]{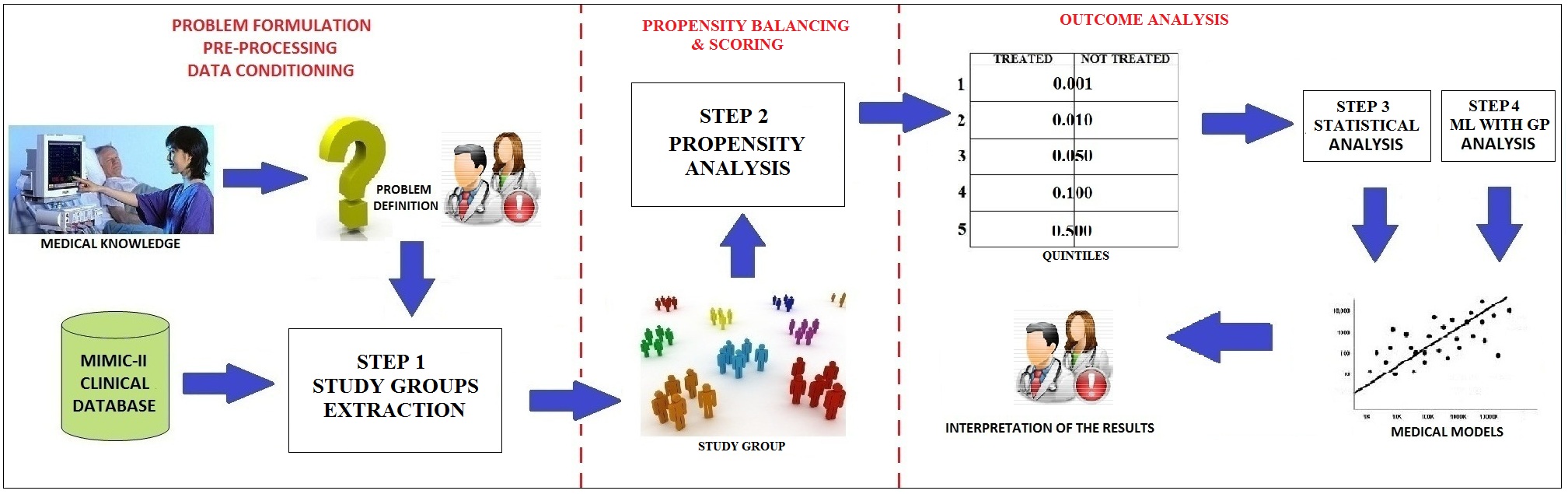}
\addtocounter{figure}{-1}
\captionof{figure}[Problem Definition Overview]{The workflow used to define and solve the problem. The definition of the problem and in particular of the \GS required a lot of work in collaboration with medical experts. The process was subject to a progressive refinement at the end of which the dataset extraction was performed.}
\label{fig:p-def}
\end{minipage}
}
\end{figure}

\clearpage \mbox{} \clearpage 
\chapter{\SE}
\label{cpt:datasetExtraction}

The aim of this Chapter is to show Step~1 of the analysis in which the \gs is identified by means of extraction, intersection and filtering of records from the \db and the variables are conditioned for subsequent propensity balancing and outcome analysis modeling.

In Figure~\vref{fig:p-whole} the software modules supporting Step~1 are shown.
\begin{figure}
\centering
\rotatebox{90}{
\stepcounter{figure}
\begin{minipage}{\textheight}
\centering
\includegraphics[width=\textheight]{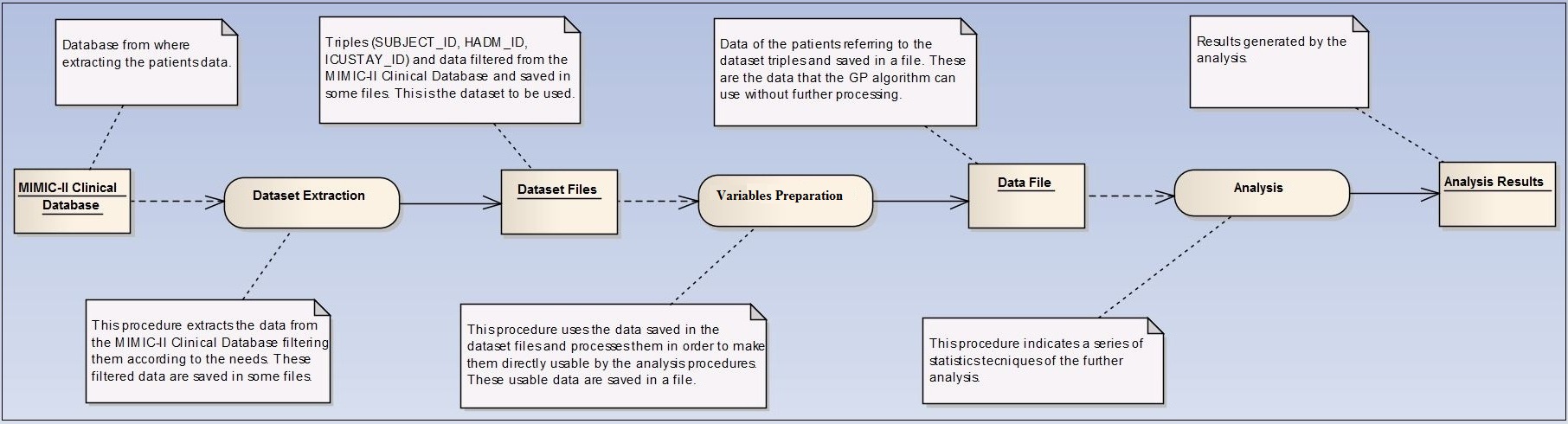}
\addtocounter{figure}{-1}
\captionof{figure}[Process Overview Diagram]{The whole process of analysis in details. After the extraction, the variables preparation was performed to save all the needed data.}
\label{fig:p-whole}
\end{minipage}
}
\end{figure}
The output after dataset extraction from the \db is a series of flat files. In the dataset preprocessing module, the flat files provided by the dataset extraction were then merged into a flat file containing all the data.  A detailed description of the contributed software is provided in Appendix~\ref{apx:software}.

The chapter proceeds in the following manner: Section~\ref{cpt:db} starts with a description of the \db. All the contents of this Section are drawn from \cite{MIMIC-GUIDE} and a deeper description of the database can be found there. A description of how the \gs was formed via a series of extractions, intersections and filters then follows in Section~\ref{cpt:extraction}. Descriptive statistics on the \gs is also provided. Section~\ref{cpt:var-preparation} explains how each of the variables for the modeling steps were prepared. Timeline oriented variables were worthy of explicit attention.  It provides a complete list of every variable prepared for modeling and a breakdown of how many patients \gettingdiuretics vs those not were distributed for the 
variable. 

\section{\db}
\label{cpt:db}
The \db (Multiparameter Intelligent Monitoring in Intensive Care) records data from all ICU patients in the Beth Israel Deaconess Medical Center. It is notable for three factors: it is publicly and freely available; it encompasses a diverse and very large population of ICU patients; and it contains high temporal resolution data including lab results, electronic documentation, and bedside monitor trends and waveforms. The database can support a diverse range of analytic studies spanning epidemiology, clinical decision-rule improvement, and electronic tool development.

The process and the sources of the data collection for the \db is shown in Figure~\vref{fig:mimic-collection}.
\begin{figure}[htbp]
\centering
\includegraphics[width=\textwidth]{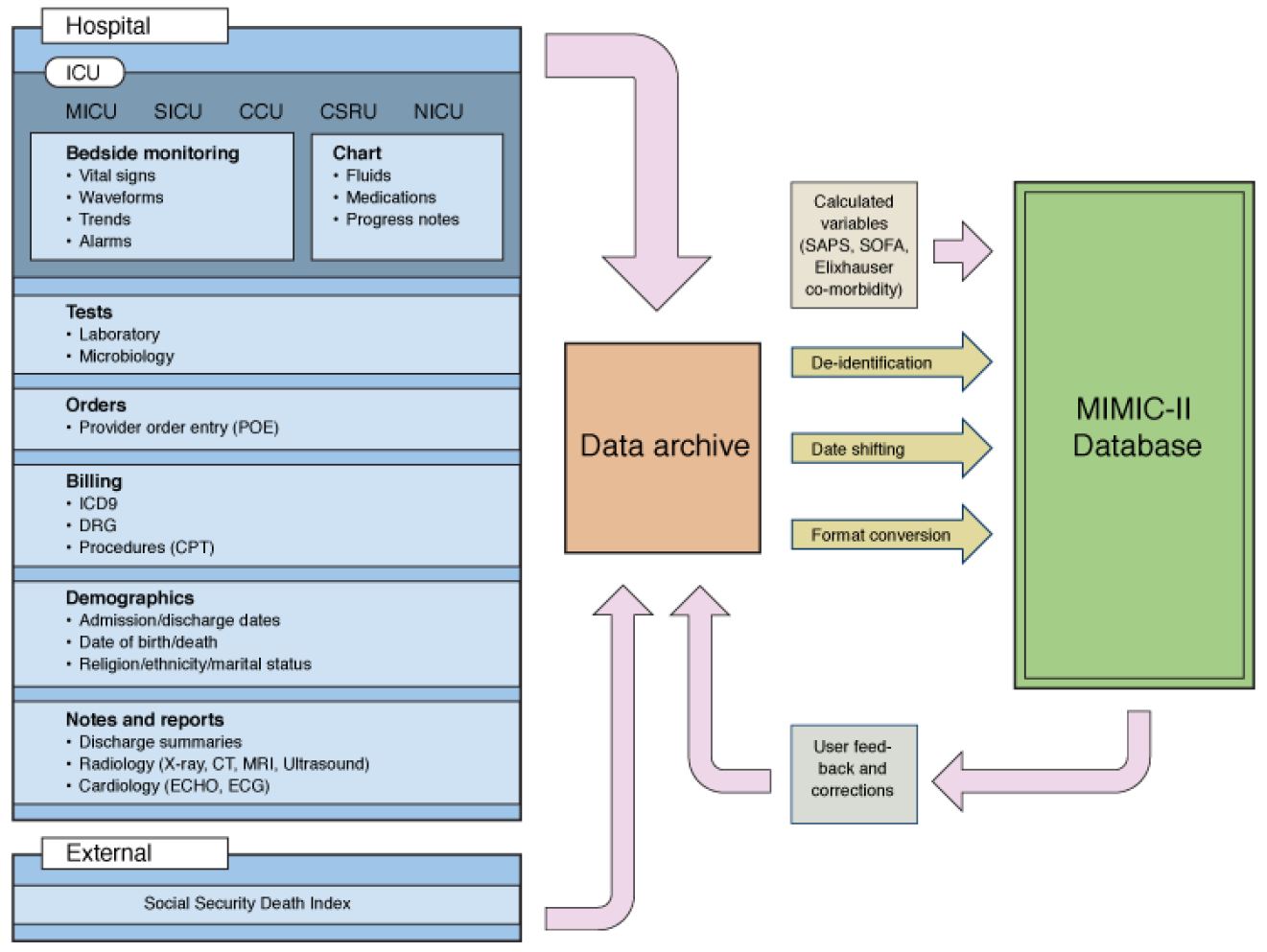}
\caption[\db Data Collection]{Schematic of data collection and database construction. Source data consists of: bedside monitor waveforms and trends, the ICU clinical databases, the hospital archives and the Social Security Death Index. These data are assembled in a protected and encrypted database which is then de-identified to provide one relational database plus associated flat file bedside waveforms and trends.}
\label{fig:mimic-collection}
\end{figure}
The data is collected dating from 2001 from Boston's Beth Israel Deaconess Medical Center (BIDMC). Any patient who was admitted to the ICU on more than one occasion may be represented by multiple patient visits. The adult ICUs (for patients aged 15 years and over) include medical (MICU), surgical (SICU), coronary (CCU), and cardiac surgery (CSRU) care units. Data were also collected from the neonatal ICU (NICU).

Clinical data are recorded far less frequently than bedside monitor data and come from a variety of databases. These include the laboratory results, pharmacy provider order entry (POE records, admission and death records, demographic details, discharge summaries, ICD-9 codes, procedure codes, microbiology and lab tests, imaging and ECG reports and the ICU central database (which includes some subset of the bedside monitor trends, drip rates, free text nursing notes and nurse-verified down-sampled trends, amongst other information).

\subsection{Definition of Patient Record}
\label{cpt:patient-record}
Since a patient may have been admitted several times during the period in which our data were collected, it is important to understand exactly how to identify patients and his/her stay(s).

There are essentially four identifiers for data associated with any given patient:
\begin{itemize}
\item {\bf SUBJECT\_ID:} to identify the patient. It is an integer number identifying a particular patient. This can be thought of as a substitute for a unique medical record number. In the flat file data posted on PhysioNet, the number representing the SUBJECT\_ID is left padded with zeros to five digits and preceded by the letter s. In the relational database, the SUBJECT\_ID has no preceding letter or leading zeros.
\item {\bf HADM\_ID:} to identify the admission in the hospital. It is an integer number identifying a particular admission to the hospital. Each patient may have many HADM\_IDs associated with his/her unique SUBJECT\_ID.
\item {\bf ICUSTAY\_ID:} to identify the admission in the ICU. It is an integer number identifying an ICU stay. An ICU stay, refers to the period of time when the patient is cared for continuosly in an Intensive Care Unit. Each patient may have one or more ICU stays associated. An ICU stay is considered to be continuous if any set of ICU events (such as bed transfers or changes in type of service) belonging to one SUBJECT\_ID which are fewer than 24 hours apart. Longer breaks in the patient's stay automatically cause a new ICUSTAY\_ID to be assigned.
\end{itemize}
Figure~\vref{fig:mimic-record} illustrates the possible data available for a given individual, identified by a ICUSTAY\_ID. Time progresses from left to right, and the different types of data collected are shown vertically. Each subject can have multiple hospital admissions, identified with HADM\_IDs. Each hosptial admission can contain multiple ICU stays, identified with ICUSTAY\_IDs. Laboratory and microbiology tests are performed throughout a hospital stay and can therefore take place outside the ICU stay. Vital sign validation, medications, fluid balances and nursing notes are only performed in the ICU and are not available during the remainder of the hospital stay. Date of death is recorded in-hospital and has also been obtained from social security records for out-of-hospital mortality.
\begin{figure}
\centering
\rotatebox{90}{
\stepcounter{figure}
\begin{minipage}{\textheight}
\centering
\includegraphics[width=0.75\textheight]{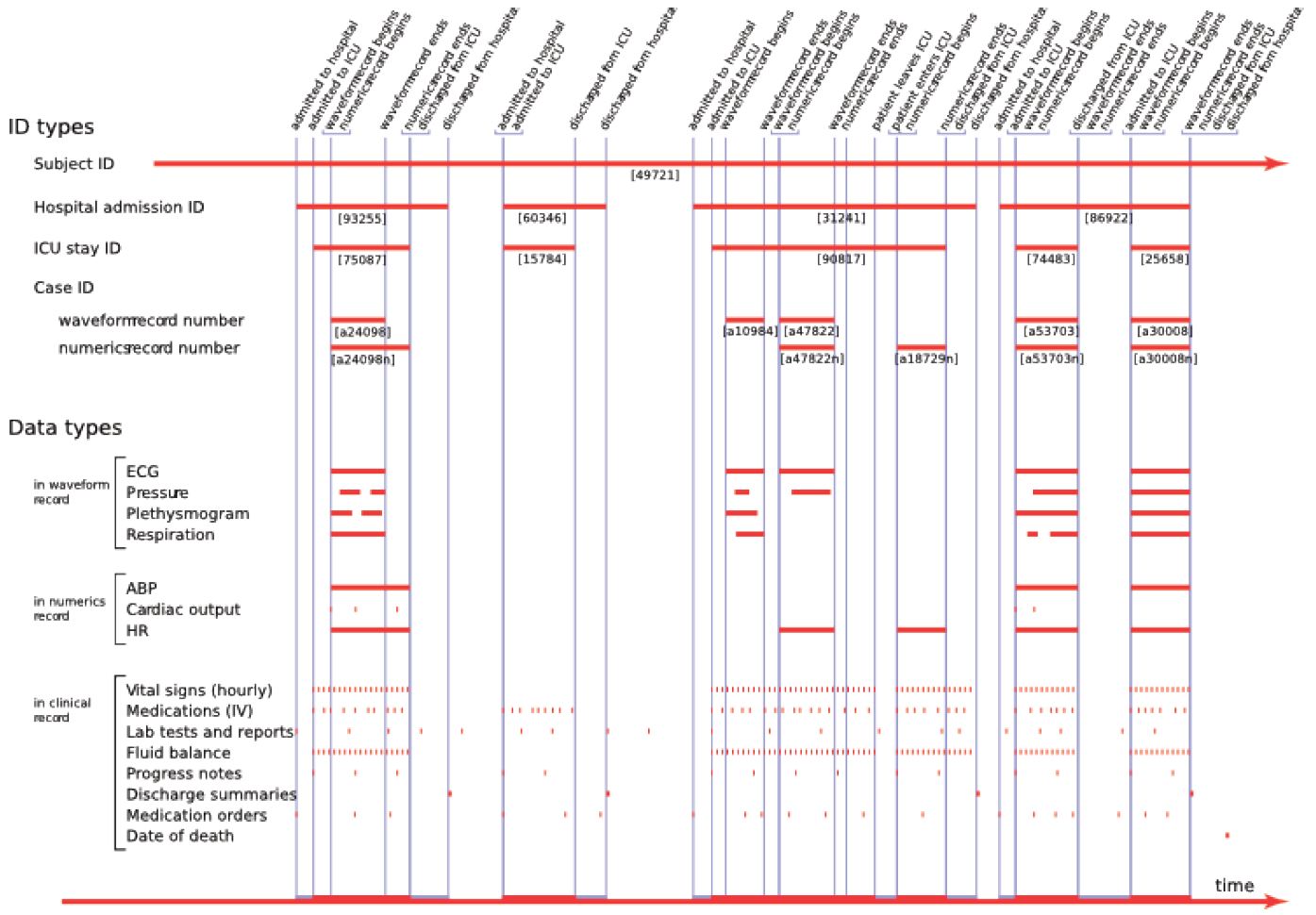}
\addtocounter{figure}{-1}
\captionof{figure}[\db Record]{Schematic of a patient record. Note that the patient may experience several hospital admissions and ICU stays, for which differing amounts of data are available.}
\label{fig:mimic-record}
\end{minipage}
}
\end{figure}
The above illustrates an ideal case where the timestamps associated with the data fall within the hospital and/or ICU stay. Unfortunately, real-world issues can complicate matters allowing data to be recorded outside of a patient stay. For example, a patient could be physically present in the ICU and connected to monitors before their admission has been entered into the system. This results in a waveform recording which starts before the subject's ICU admission. Furthermore, missing/mistaken data can mean that ICU stays exist where there is no matching hospital admission record.

Note that a patient may move between ICUs during any given admission. If the move is longer than 24 hours, it is defined to be a new ICU stay. Note also that the amount of data varies during and between ICU stays and that data are often missing.

The \db is a relational database.

\section{Dataset Extraction}
\label{cpt:extraction}
The \gs was extracted with a series of extractions, intersections and filters.

\subsection{Extractions, Intersections and Filtering}
\label{cpt:extraction-steps}
Step 1 consists of a series of queries to the \db. These queries extract the data to 21 files and generate record sets. Of the 21 extracted files, the first 3 contain the triples (SUBJECT\_ID, HADM\_ID, ICUSTAY\_ID). These triples, which identify respectively the patients, the admission to the hospital and the admission in the ICU, identify uniquely the data for each feature extracted in the remaining files.

An other file is generated which contains the patient {\it discharge summaries}. They are reports in a English text-like format where the clinical history of the patients before and during the stay in the hospital is described.

Finally, the last 17 extracted files contain the variables of interest for each patient.

In the \db there are 39,919 ICUSTAY\_ID records. The impact on the filter is shown for each step. The values referred as {\it original db} report how the single step impacts on the \db. The values referred as {\it previous subset} report how the single step impacts on the subset after the previous step. The values referred as {\it all conditions} report how the interactions between all the steps applied till that point impact on previous step. The sequence of filtering steps follow:
\begin{enumerate}
\item Extract patients with all the 3 IDs (SUBJECT\_ID, HADM\_ID, ICUSTAY\_I(d) available and different from null:
\begin{itemize}
\item 36,708/39,919 â 91.95\% (original db)
\end{itemize}
\item Extract patients with only one admission in the hospital and in the ICU:
\begin{itemize}
\item 26,027/39,919 â 65.19\% (original db)
\end{itemize}
\item Intersect 1 and 2:
\begin{itemize}
\item 26,027/36,708 â 70.90\% (previous subset)
\item 26,027/39,919 â 65.19\% (all conditions)
\end{itemize}
\item Extract patients with at least 1 full day of data in the ICU:
\begin{itemize}
\item 29,462/39,919 â 73.80\% (original db)
\end{itemize}
\item Intersect 3 and 4:
\begin{itemize}
\item 18,770/26,027 â 72.11\% (previous subset)
\item 18,770/39,919 â 47.02\% (all conditions)
\end{itemize}
\item Extract patients with age equal or over 18. The aim of doing this is to remove neonates:
\begin{itemize}
\item 31,859/39,919 â 79.80\% (original db)
\end{itemize}
\item Intersect 5 and 6:
\begin{itemize}
\item 15,176/18,770 â 80.85\% (previous subset)
\item 15,176/39,919 â 38.01\% (all conditions)
\end{itemize}
\item Extract patients with sepsis according to \cite{SEPSIS-EPIDEMIOLOGY}:
\begin{itemize}
\item 3,818/39,919 â 9.56\% (original db)
\end{itemize}
\item Intersect 7 and 8:
\begin{itemize}
\item 1,644/15,176 â 10.83\% (previous subset)
\item 1,644/39,919 â 4.11\% (all conditions)
\end{itemize}
\item Extract patients not CMO (comfort measure only):
\begin{itemize}
\item 39,651/39,919 â 99.32\% (original db)
\end{itemize}
\item Intersect 9 and 10:
\begin{itemize}
\item 1,644/1,644 â 100\% (previous subset)
\item 1,644/39,919 â 4.11\% (all conditions)
\end{itemize}
\item Extract patients with a discharge summary available:
\begin{itemize}
\item 31,475/39,919 â 78.84\% (original db)
\end{itemize}
\item Intersect 11 and 12:
\begin{itemize}
\item 1,638/1,644 â 99.63\% (previous subset)
\item 1,638/39,919 â 4.10\% (all conditions)
\end{itemize}
\item Extract patients who are diuretics naive. See note in~\ref{cpt:naive}:
\begin{itemize}
\item 30,646/39,919 â 76.77\% (original db)
\end{itemize}
\item Intersect 13 and 14:
\begin{itemize}
\item 1,606/1,638 â 98.04\% (previous subset)
\item 1,606/39,919 â 4.02\% (all conditions)
\end{itemize}
\item Filter 15 for missing data:
\begin{itemize}
\item 1,606/1,638 â 94.76\% (previous subset)
\item 1,522/39,919 â 3.81\% (all conditions)
\end{itemize}
\end{enumerate}
For a detailed description of the results of this step see Appendix~\ref{apx:dataset}.

\subsubsection{Diuretics Naive Status}
\label{cpt:naive}
The study was intended to consider patients who had not been administered diuretics before entering the ICU. This is intended to avoid having data which would be conditioned before the interval of study. Unfortunately this information is not directly available from the \db. It needs to be parsed out of the discharge summary. The discharge summary is saved in the database in an English text-like format. It consists of a summary of what has happened to the patient before and during his admittance at the hospital. The document is hand written by clinicians so it does not have a well defined structure. The needed pieces of information were extracted using a complicated parser which searched the English text for the names of a list of diuretics decided by the doctors. Those patients who had not been administered diuretics before entering the ICU were considered {\it diuretics naive}.

\subsection{Filters}
\label{cpt:filter}
Step 2 takes as an input the files provided by the SQL queries and the list of diuretics naive patients. The aim of this step is to combine the results of the SQL queries to the one of the diuretics naive procedure. Moreover, the procedure discards the patients who have missing data for any variable needed for the analysis.

The output of this phase is a series of files where all the available data are saved. A file per variable is then created.
\begin{itemize}
\item Results 1,522/39,919 patients (3.81\%)
\item To 189/1,522 patients (12.41\%) diuretics have been given, $D^+$
\item To 1,333/1,522 patients (87.59\%) diuretics have not been given, $D^-$
\end{itemize}
A summary of all the steps of the dataset extraction are shown in Table~\vref{tab:extraction-steps}.
\begin{table}
\centering
\begin{tabular}{|c|c|c|c|}
\hline {\bf Step} & Type & Effects & \%\\
\hline 1 & Extract(A) & 36,708 & 91.95\%\\
\hline 2 & Extract(B) & 26,027 & 65.19\%\\
\hline 3 & A $\cap$ B $\to$ C & 26,027 & 65.19\%\\
\hline 4 & Extract(D) & 29,462 & 73.80\%\\
\hline 5 & C $\cap$ D $\to$ E & 18,770 & 47.02\%\\
\hline 6 & Extract(F) & 31,859 & 79.80\%\\
\hline 7 & E $\cap$ F $\to$ G & 15,176 & 38.01\%\\
\hline 8 & Extract(H) & 3,818 & 9.56\%\\
\hline 9 & G $\cap$ H $\to$ I & 1,644 & 4,11\%\\
\hline 10 & Extract(L) & 39,651 & 99.32\%\\
\hline 11 & I $\cap$ L $\to$ M & 1,644 & 4.11\%\\
\hline 12 & Extract(N) & 31,475 & 78.84\%\\
\hline 13 & M $\cap$ N $\to$ O & 1,638 & 4.10\%\\
\hline 14 & Extract(P) & 30,646 & 76.77\%\\
\hline 15 & O $\cap$ P $\to$ Q & 1,606 & 4.02\%\\
\hline 16 & Filter(Q) $\to$ R & \colorbox{yellow}{1,522} & \colorbox{yellow}{3.81\%}\\
\hline
\end{tabular}
\caption[Steps of the Dataset Extraction]{Summary of the 16 steps of the dataset extraction. Next of each step, is shown how it impacts on the \db. The number of patients in the final subset is 1,522.}
\label{tab:extraction-steps}
\end{table}
The final number of patients considered for the study is 1,522\label{FINAL-NUMBER-OF-PATIENTS}.

\section{Variables Preparation}
\label{cpt:var-preparation}
Critical care medical data is arguably the most valuable clinical data supporting medical informatics. This is because the ICU is the crucible of a hospital. It accepts the most acutely ill of patients, it uses pervasive monitoring, and intensivists encounter frequent medical episodes for which they must make rapid interventions.

ICU medical doctors, specialists known as intensivists, are key members of any knowledge mining team which consults a data resource such as the \db. Data engineers, with expertise in modeling and machine learning, request a lot of information from a team's intensivists. A key request they make is the variables that should be selected for predictive or explanatory models which will be mined from the data.

For example, in a study on the necessity of so called {\it diuretics} (drugs) for diuresis (fluid shedding), after fluid resuscitation in the ICU, it must be determined what intravenous diuretics data should be included as model variables.

Indicator selection calls upon intensivists' theoretical and clinical knowledge and their ICU experience. The selection process is intensely deliberative and uncertain. The intensivists recognize how the events of each patient's ICU stay are unique and how the care administered is both subjective and informed by medical knowledge. They are uncertain as to whether some of the collected data variables confound the outcome that is to be explained. As well, they are aware that among intensivists there exists a propensity to treat similar patients differently. They find it very challenging to choose variables that essentially pinpoint some time point or variable of a complex human health process. For example, while including an feature expressing the amount of a diuretic is an obvious decision, how the amount is described is open to debate.

The uncertainty of the decisions imply that the feature selection process is challenged to be systematic and unbiased and impacts the quality and accuracy of modeling in an obviously critical way. In addition to the uncertainty, from a broader perspective, the selection of variables prior to their experimentation in model regression is problematic. It forces a model design decision that is premature because the appropriate information is unavailable.

In this work, the definition of the problem and descriptive variables to be included in the model was carried out through interviews and meetings with a group of doctors, experts in the field. In particular, from an initial definition of the problem due to the intuition and experience of the medical experts, through this process a more formal definition has been gained of all variables needed to define a comprehensive model. This process went hand in hand with an increasing acquisition of medical knowledge needed to define, from a statistical point of view, an effective model that describes the problem. Another fundamental endeavor has been made in understanding the \db and the related problems that occurred in the preprocessing step.

\subsection{Timelines}
\label{cpt:timelines}
Some of the variables have timelines, this means that for those variables there are values at different times. It is difficult to choose good times where to take the values avoiding forward-looking variables. In the study the clinical data of the patients are available during the whole ICU stay, but the doctors, while making their decision whether giving or not diuretics, are considering only the values till the current day. In this sense, values available after the diuretics decision points are considered forward-looking variables, in fact their values may be caused by the diuretics decision itself.

To avoid such problems, for those variables the extraction have been performed at the following times:
\begin{itemize}
\item {\bf T1} {\it Diuretics Decision Timepoint}: This is when the decision to give diuretics was 'theoretically' or actually taken. There are two possibilities:
\begin{itemize}
\item for a patient who got diuretics, the actual time of the first dose;
\item for the patients who didn't get diuretics, the {\bf T1} timepoint is day 4. The decision to use day~4 was reached by examining the patients who got diuretics in the first week of their ICU stay. Among this group the first day they were administered diuretics was examined. From this, 'first administration day', data, the median was extracted (day~4) as the timepoint.
\end{itemize}
\item {\bf T2} {\it Max Fluids Ratio Timepoint: Day of highest fluids ratio}: Ideally this timepoint would be determined individually for each member of the study group. It would express variables at the time when a patient has his/her highest fluids ratio. However, this would have required us to look forward in the data which would make calculation of the timepoint in reality impossible. Instead, day~3 in the ICU T2 has been selected. It was selected by examining the day of highest fluids ratio for all patients in the \gs and choosing the median. The fluids ratio is calculated as
\begin{equation}
\frac{inputs(t-1)+inputs(t)}{outputs(t-1)+outputs(t)}
\end{equation}
\item {\bf T3} {\it 2nd Highest Fluids Ratio Timepoint: Day of 2nd highest fluids ratio}: The timepoint of second highest fluids ratio has been added because the day of highest fluids ratio is typically very close to the first day of ICU admission when fluids are infused. It may reflect recent infusion more than a delay in shedding fluid.  

Ideally this timepoint would be determined individually for each member of the study group. It would express variables at the time when a patient has his/her second highest ratio of fluids in the body. However, this would have required us to look forward in the data which would make calculation of the timepoint in reality impossible. Instead, T3 have been selected to be day~4. This day was selected by examining the day of the second highest fluids ratio for all patients in the \gs and choosing the median.
\end{itemize}

Figure~\vref{fig:timepoints} provides a visual summary of the considered times and of the ICU timeline in general. In Appendix~\ref{apx:dataset} are available more details on how these times were chosen.
\begin{figure}[htbp]
\centering
\includegraphics[width=\textwidth]{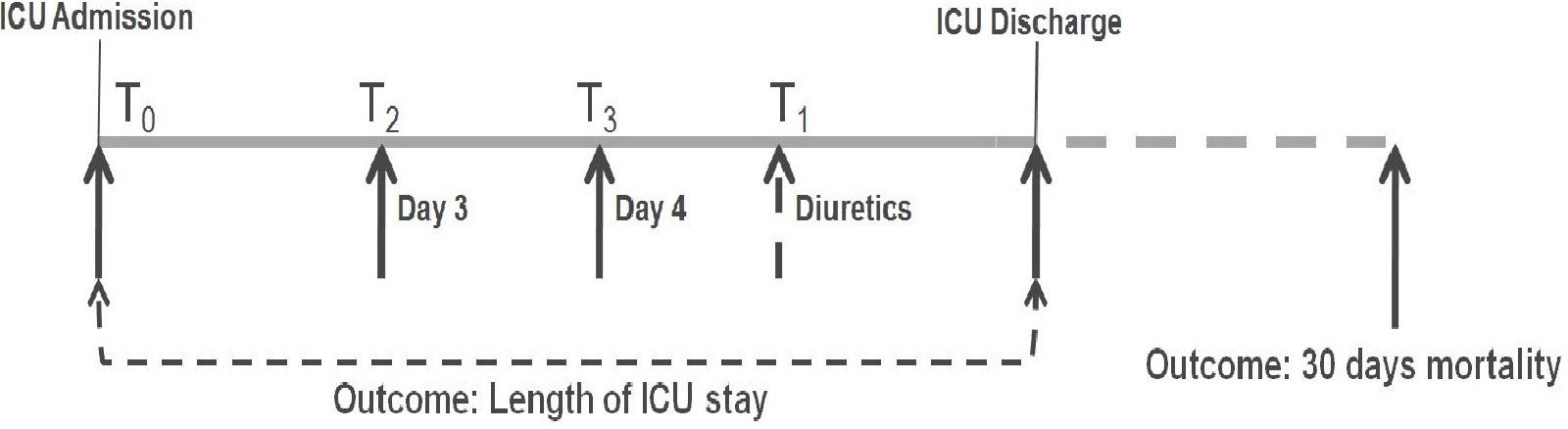}
\caption[Timepoints of Interest]{The timepoints where the values for timeline variables were acquired. All the values refer to the ICU stay. T0 is day 1, T2 and T3 are days 3 and 4. T1, the diuretics decision time, is day 4 for the patients who didn't get diuretics and it is the actual day when the drug was first administered in the patients who got diuretics. The choice of these timespoint allows us to avoid forward-looking variables.}
\label{fig:timepoints}
\end{figure}

\subsection{List of Variables}
\label{cpt:var-list}
The flat files produced by the previous modules are the input of a series of procedures which elaborate and save them in a format directly usable as an input for the further analysis. As has been said, the inputs of these modules are a series of files provided by the previous phase, then on those files are computed the time points as discussed in~\ref{cpt:timelines}.

In the \db the values are saved in an irregular sampling rate, each values is recorded when available tipically a few times each days. As for this study the values were needed on a daily basis, the original irregular sample rate has been downsampled to fit a regular daily rate of values. This has been done by computating the median value of the available values each day. The median value also made the system robust to outlayers.

Then the list of variables are generated. In the following list is reported the type of the feature, and where needed a brief description:
\begin{enumerate}
\item Diuretics in the ICU:
\begin{description}
\item[$x_1$] Binary: -1 not given, +1 given;
\end{description}
\item Age when admitted in the ICU:
\begin{description}
\item[$x_2$] Numeric;
\end{description}
\item Gender:
\begin{description}
\item[$x_3$] Binary: -1 male, +1 female;
\end{description}
\item Race (white vs not white):
\begin{description}
\item[$x_4$] Binary: -1 not white, +1 white;
\end{description}
\item Saps:
\begin{description}
\item[$x_5$] Average from day 1 to day T1;
\item[$x_6$] Mean of values during the first day;
\item[$x_7$] Mean of values during day T1;
\item[$x_8$] Mean of values during day T2;
\item[$x_9$] Mean of values during day T3;
\end{description}
\item Sofa:
\begin{description}
\item[$x_{10}$] Average from day 1 to day T1;
\item[$x_{11}$] Mean of values during the first day;
\item[$x_{12}$] Mean of values during day T1;
\item[$x_{13}$] Mean of values during day T2;
\item[$x_{14}$] Mean of values during day T3;
\end{description}
\item Elixhauser overall\footnote{The sum of all the parameters of the Elixhauser score, as explained in Chapter~\ref{cpt:variables}.}.
\begin{description}
\item[$x_{15}$] Numeric;
\end{description}
\item Elixhauser binary: 9 of the 30 fields composing the Elixhauser score are selected: congestive heart failure, cardiac arrhythmias, valvular disease, hypertension, diabetes uncomplicated, diabetes complicated, renal failure, liver disease and obesity.
\begin{description}
\item[$x_{16} \to x_{24}$] Binary: -1 not present, +1 present;
\end{description}
\item Creatinine:
\begin{description}
\item[$x_{25}$] Average from day 1 to day T1;
\item[$x_{26}$] Mean of values during the first;
\item[$x_{27}$] Mean of values during day T1;
\item[$x_{28}$] Mean of values during day T2;
\item[$x_{29}$] Mean of values during day T3;
\end{description}
\item Fluids inputs in liters:
\begin{description}
\item[$x_{30}$] Average of sums from day 1 to day T1;
\item[$x_{31}$] Sum of values during the first day;
\item[$x_{32}$] Sum of values during day T1;
\item[$x_{33}$] Sum of values during day T2;
\item[$x_{34}$] Sum of values during day T3;
\end{description}
\item Fluids outputs in liters:
\begin{description}
\item[$x_{35}$] Average of sums from day 1 to day T1;
\item[$x_{36}$] Sum of values during the first day;
\item[$x_{37}$] Sum of values during day T1;
\item[$x_{38}$] Sum of values during day T2;
\item[$x_{39}$] Sum of values during day T3;
\end{description}
\item Fluids balance in liters (fluids inputs - fluids outputs):
\begin{description}
\item[$x_{40}$] Average of sums from day 1 to day T1;
\item[$x_{41}$] Sum of values during the first day;
\item[$x_{42}$] Sum of values during day T1;
\item[$x_{43}$] Sum of values during day T2;
\item[$x_{44}$] Sum of values during day T3;
\end{description}
\item Use of vasopressors in the ICU:
\begin{description}
\item[$x_{45}$] Binary: -1 not given, +1 given;
\end{description}
\item Mechanical ventilation in the ICU:
\begin{description}
\item[$x_{46}$] Binary: -1 not happened, +1 happened;
\end{description}
\item Arterial blood pressure:
\begin{description}
\item[$x_{47}$] Average from day 1 to day T1;
\item[$x_{48}$] Mean of values during the first day;
\item[$x_{49}$] Mean of values during day T1;
\item[$x_{50}$] Mean of values during day T2;
\item[$x_{51}$] Mean of values during day T3;
\end{description}
\item Mean arterial blood pressure:
\begin{description}
\item[$x_{52}$] Average from day 1 to day T1;
\item[$x_{53}$] Mean of values during the first day;
\item[$x_{54}$] Mean of values during day T1;
\item[$x_{55}$] Mean of values during day T2;
\item[$x_{56}$] Mean of values during day T3;
\end{description}
\item Mortality within 30 days:
\begin{description}
\item[$x_{57}$] Binary: -1 alive, +1 dead;
\end{description}
\item Length of stay in the ICU after the first dose of diuretics:
\begin{description}
\item[$x_{58}$] Numeric, in days;
\end{description}
\end{enumerate}
For the medical meaning of some of those variables, see Chapter~\ref{cpt:variables}. Table~\vref{tab:features-bin-summary} shows a summary for the values for the binary variables.

\begin{table}
\centering
\centering
\begin{tabular}{|l|c||c|c|c|c|}
\hline \multicolumn{2}{|c|}{\bf Variable} & \multicolumn{2}{|c|}{$D^+$ (189)} & \multicolumn{2}{|c|}{$D^-$ (1333)}\\
\hline Name & {$i$} & \# Positive & \% & \# Positive & \%\\
\hline DIU & 1 & 189 & 100\% & - & -\\
\hline GEN & 3 & 81 & 42\% & 566 & 42\%\\
\hline RAC & 4 & 1 & 0.5\% & 29 & 2\%\\
\hline EL1 (CHF) & 16 & 91 & \colorbox{SkyBlue}{48\%} & 418 & \colorbox{SkyBlue}{31\%}\\
\hline EL2 (Cardia arrythmia) & 17 & 70 & \colorbox{Grey}{37\%} & 335 & \colorbox{Gray}{25\%}\\
\hline EL3 & 18 & 26 & 13\% & 110 & 8\%\\
\hline EL4 & 19 & 51 & 26\% & 366 & 27\%\\
\hline EL5 & 20 & 47 & 24\% & 258 & 19\%\\
\hline EL6 & 21 & 9 & 4\% & 77 & 5\%\\
\hline EL7 & 22 & 10 & 5\% & 115 & 8\%\\
\hline EL8 & 23 & 16 & 8\% & 125 & 9\%\\
\hline EL9 & 24 & 5 & 2\% & 18 & 1\%\\
\hline VAS (vasosuppressor) & 45 & 163 & \colorbox{Melon}{86\%} & 865 & \colorbox{Melon}{64\%}\\
\hline VEN (ventilation) & 46 & 177 & \colorbox{Melon}{93\%} & 904 & \colorbox{Melon}{67\%}\\
\hline MOR & 57 & 63 & 32\% & 504 & 37\%\\
\hline
\end{tabular}
\caption[Descriptive Statistics for Unbalanced, Binary Variables in the \gs]{Descriptive statistics for unbalanced, binary variables in the \gs. The abbreviation of the variable, its $x_i$ subscript, the number of patients positive for the variable within the patients who got diuretics and within the patients who did not get diuretics are reported. The 4 variables where there is a noticeable, major difference between the diuretics treated and non-treated patients have been highlighted however the \gs is not balanced in terms of covariates.}
\label{tab:features-bin-summary}
\end{table}
Tables~\vref{tab:features-summary}, ~\vref{tab:features-summary2} and ~\vref{tab:features-summary3} show a summary of the values for the not binary variables, for the variables with timeline which refer to the clinical condition of the patient and for the variables with timeline which refer to the fluids measurements.
\begin{table}
\centering
\begin{tabular}{|c|c||c|c|c|c|c|c|}
\hline \multicolumn{2}{|c|}{{\bf  Variable}} & \multicolumn{2}{|c|}{$D^+ \cup D^-$ (1522)} & \multicolumn{2}{c|}{$D^+$ (189)} & \multicolumn{2}{c|}{$D^-$ (1333)}\\
\hline Name & $i$ & Mean(SD) & Me & Mean(SD) & Me & Mean(SD) & Me\\
\hline AGE & 2 & 66.1(16.9) & 68.3 & 66.2(15.4) & 68.9 & 66.1(17.1) & 68.2\\
\hline ELI & 15 & 3(1.6) & 3 & 3.1(1.4) & 3 & 2.9(1.6) & 3\\
\hline LOS & 58 & \colorbox{SkyBlue}{7.4(11.8)} & \colorbox{SkyBlue}{3} & \colorbox{Melon}{15.1(14)} & \colorbox{Melon}{11} & \colorbox{SkyBlue}{6.3(11)} & \colorbox{SkyBlue}{3}\\
\hline
\end{tabular}
\caption[Descriptive Statistics for Unbalanced, Non-Binary Variables]{Descriptive statistics for unbalanced discrete variables in \gs. The abbreviation of the variable, its $x_i$ subscript, the mean, standard deviation and median of the variable within the \gs, patients who got diuretics and within the patients who did not get diuretics are reported. There is a noticeable difference between the length of stay of diuretics treated and non-treated patients which has been highlighted however the \gs is not balanced in terms of covariates.}
\label{tab:features-summary}
\end{table}
\begin{table}
\centering
\begin{tabular}{|c|c||c|c|c|c|c|c|}
\hline \multicolumn{2}{|c|}{{\bf Variable}} & \multicolumn{2}{|c|}{$D^+ \cup D^-$ (1522)} & \multicolumn{2}{c|}{$D^+$ (189)} & \multicolumn{2}{c|}{$D^-$ (1333)}\\
\hline Name & $i$ & Mean(SD) & Me & Mean(SD) & Me & Mean(SD) & Me\\
\hline SA1 & 5 & 15.8(4.6) & 15.4 & 16.2(3.4) & 16 & 15.7(4.8) & 15.3\\
\hline SA2 & 6 & 17.1(5.4) & 17 & 17.9(4.9) & 18 & 17(5.5) & 17\\
\hline SA3 & 7 & 15(5.2) & 15 & 15.7(4.3) & 16 & 14.9(5.3) & 15\\
\hline SA4 & 8 & 15.3(5.2) & 15 & 16.3(4.6) & 16 & 15.2(5.2) & 15\\
\hline SA5 & 9 & 17.1(5.4) & 17 & 17.9(4.9) & 18 & 17(5.5) & 17\\
\hline
\hline SO1 & 10 & 8.3(4.2) & 7.6 & 9.7(3.6) & 9.6 & 8.1(4.3) & 7.4\\
\hline SO2 & 11 & 9.1(4.5) & 9 & 10.3(4.2) & 11 & 8.9(4.5) & 8\\
\hline SO3 & 12 & 7.6(4.8) & 7 & 9.5(4.3) & 9 & 7.3(4.9) & 7\\
\hline SO4 & 13 & 8.2(4.7) & 7.4 & 10(4.1) & 10 & 7.9(4.7) & 7\\
\hline SO5 & 14 & 9.1(4.5) & 9 & 10.3(4.2) & 11 & 8.9(4.5) & 8\\
\hline
\hline CR1 & 25 & 1.8(1.8) & 1.2 & 1.6(1.5) & 1.2 & 1.8(1.8) & 1.2\\
\hline CR2 & 26 & 1.8(2.2) & 1.2 & 1.5(1.3) & 1.1 & 1.8(2.3) & 1.3\\
\hline CR3 & 27 & 1.8(2) & 1.2 & 1.6(1.4) & 1.1 & 1.8(2.1) & 1.2\\
\hline CR4 & 28 & 1.7(1.5) & 1.2 & 1.6(1.4) & 1.2 & 1.8(1.6) & 1.2\\
\hline CR5 & 29 & 1.8(2.2) & 1.2 & 1.5(1.3) & 1.1 & 1.8(2.3) & 1.3\\
\hline
\hline BP1 & 47 & 113(17.3) & 114 & 113(14.7) & 113 & 113(17.7) & 114\\
\hline BP2 & 48 &110(18.7) & \colorbox{SkyBlue}{114} & 108(19.5) & \colorbox{SkyBlue}{106} & 110(18.5) & \colorbox{SkyBlue}{114}\\
\hline BP3 & 49 & 114(22.1) & 114 & 114(19.2) & 112 & 114(22.5) & 114\\
\hline BP4 & 50 & 113(19.6) & 114 & 111(17.7) & 110 & 113(19.8) & 114\\
\hline BP5 & 51 & 110(18.7) & \colorbox{SkyBlue}{114} & 108(19.5) & \colorbox{SkyBlue}{106} & 110(18.5) & \colorbox{SkyBlue}{114}\\
\hline
\hline BM1 & 52 & 78.6(12.5) & 79.2 & 77.2(8.8) & 77 & 78.8(13) & 79.2\\
\hline BM2 & 53 & 77.6(14.6) & 79.2 & 74.8(12.1) & 74 & 78(14.9) & 79.2\\
\hline BM3 & 54 & 78.9(15.3) & 79.2 & 77.1(12.8) & 75 & 79.2(15.6) & 79.2\\
\hline BM4 & 55 & 78.4(13.9) & 79.2 & 75.5(11.2) & 73 & 78.8(14.2) & 79.2\\
\hline BM5 & 56 & 77.6(14.6) & 79.2 & 74.8(12.1) & 74 & 78(14.9) & 79.2\\
\hline
\end{tabular}
\caption[Descriptive Statistics for Unbalanced Variables on Timepoints, part 1]{Descriptive statistics for unbalanced variables on timepoints (part 1). The abbreviation of the variable, its $x_i$ subscript, the mean, standard deviation and median of the variable within the \gs, patients who got diuretics and within the patients who did not get diuretics are reported.}
\label{tab:features-summary2}
\end{table}
\begin{table}
\centering
\begin{tabular}{|c|c||c|c|c|c|c|c|}
\hline \multicolumn{2}{|c|}{{\bf Variable}} & \multicolumn{2}{|c|}{$D^+ \cup D^-$ (1522)} & \multicolumn{2}{c|}{$D^+$ (189)} & \multicolumn{2}{c|}{$D^-$ (1333)}\\
\hline Name & $i$ & Mean(SD) & Me & Mean(SD) & Me & Mean(SD) & Me\\
\hline FI1 & 30 & 1.7(1.3) & 1.4 & 1.4(1.2) & 1.1 & 1.7(1.3) & 1.4\\
\hline FI2 & 31 & 2.6(2.3) & 2 & 3(3.2) & 2.2 & 2.5(2.1) & 2\\
\hline FI3 & 32 & 1.1(1.2) & 0.7 & 1(1.1) & 0.6 & 1.1(1.2) & 0.7\\
\hline FI4 & 33 & 1.4(1.4) & 1 & 1.6(1.7) & 1 & 1.4(1.4) & 1\\
\hline FI5 & 34 & 2.6(2.3) & 2 & 3(3.2) & 2.2 & 2.5(2.1) & 2\\
\hline
\hline FO1 & 35 & 1.5(1.1) & 1.3 & 1.8(1.5) & 1.6 & 1.4(1.1) & 1.2\\
\hline FO2 & 36 & 1.6(1) & 1.2 & 1.6(3.4) & 1.1 & 1.5(1.4) & 1.2\\
\hline FO3 & 37 & 1.4(1.3) & 1.1 & 1.9(1.7) & 1.7 & 1.3(1.2) & 1\\
\hline FO4 & 38 & 1.4(1.5) & 1.1 & 1.6(3.1)  & 1.1& 1.3(1.1) & 1.2\\
\hline FO5 & 39 & 1.6(1.8)  & 1.2& 1.6(3.4) & 1.1 & 1.5(1.4) & 1.2\\
\hline
\hline FB1 & 40 & 0.2(1.7) & 0.05 & -0.35(2) & -0.4 & 0.3(1.7) & 0.1\\
\hline FB2 & 41 & 1(2.9) & 0.7 & 1.4(4.8) & 1 & 1(2.5) & 0.7\\
\hline FB3 & 42 & -0.3(1.8) & -0.3 & -1(2.2) & -0.9 & -0.3(1.7) & -0.1\\
\hline FB4 & 43 & 0.001(2.1) & -0.01 & 0.001(3.7) & 0.6 & 0.1(1.8) & -0.05\\
\hline FB5 & 44 & 1(2.9) & 0.7 & 1.4(4.8) & 1 & 1(2.5) & 0.7\\
\hline
\end{tabular}
\caption[Descriptive Statistics of Variables with Timepoints, part 2]{Summary of the values for the timepoint variables related to the fluids measurements. The abbreviation of the variable, its $x_i$ subscript, the mean, standard deviation and median of the variable within the \gs, patients who got diuretics and within the patients who did not get diuretics are reported.}
\label{tab:features-summary3}
\end{table}

Detailed descriptions of those variables and of sepsis are available respectively in Chapter~\ref{cpt:variables} and Appendix~\ref{apx:medicalBackground}. Instead for a more detailed description of the procedures produced to realize the \se see Appendix~\ref{apx:software}.

\clearpage \mbox{} \clearpage 
\chapter{\PA}
\label{cpt:propensityAnalysis}

\section{Introduction}
\label{cpt:propensityAnalysisIntro}
Randomized controlled trials (RCTs) typically compare balance in baseline covariates between treated and untreated subjects. They are a type of scientific experiment, a form of clinical trial, most commonly used in testing the safety (or more specifically, information about adverse drug reactions and adverse effects of other treatments) and efficacy or effectiveness of healthcare services (such as medicine or nursing) or health technologies (such as pharmaceuticals, medical devices or surgery).

An {\it observational study} is an empirical investigation of treatments and of the effects they cause, but it differs from a randomized controlled trial in the fact that the investigators can't control the assignment of the treatments to the subjects. Observational studies are, by nature, non-randomized and retrospective. Therefore, there is no reason to assume that baseline covariates will be balanced in expectation between treated and untreated subjects. Indeed, treated subjects tend to differ systematically from untreated subjects. Consider the comparison between two heart surgeons, both of them have completed 100 surgeries. The first one had 10 deaths, while the second 5. Apparently the second one would seem to be the best, but how can the two surgeons be compared if the patients of the first one were older and had a higher risk compared to those of the second surgeon?

Under the example scenario presented above, it is important and necessary to seek group of patients under both the doctors that are alike in the statistical sense. This could be achieved by forming sub-groups of patients and then assessing balance in the covariates among these sub groups. Several authors have proposed methods for assessing balance in observational studies. Recent efforts to address issues of nonrandom assignment, including a class of methods known as {\it Propensity Scoring}, can reduce bias in the estimation of treatment effects when assignment is not random.

Propensity score techniques are useful in this context, that is when there may be important differences in patient characteristics between treated and not treated groups. In fact, this kind of medical analysis aims to show whether the differences in the outcomes are attributable to the differences in the treatments provided to the patients, when sometimes it is infeasible or unethical to assign patients to different treatments.

\begin{figure}[htbp]
\centering
\includegraphics[width=\textwidth]{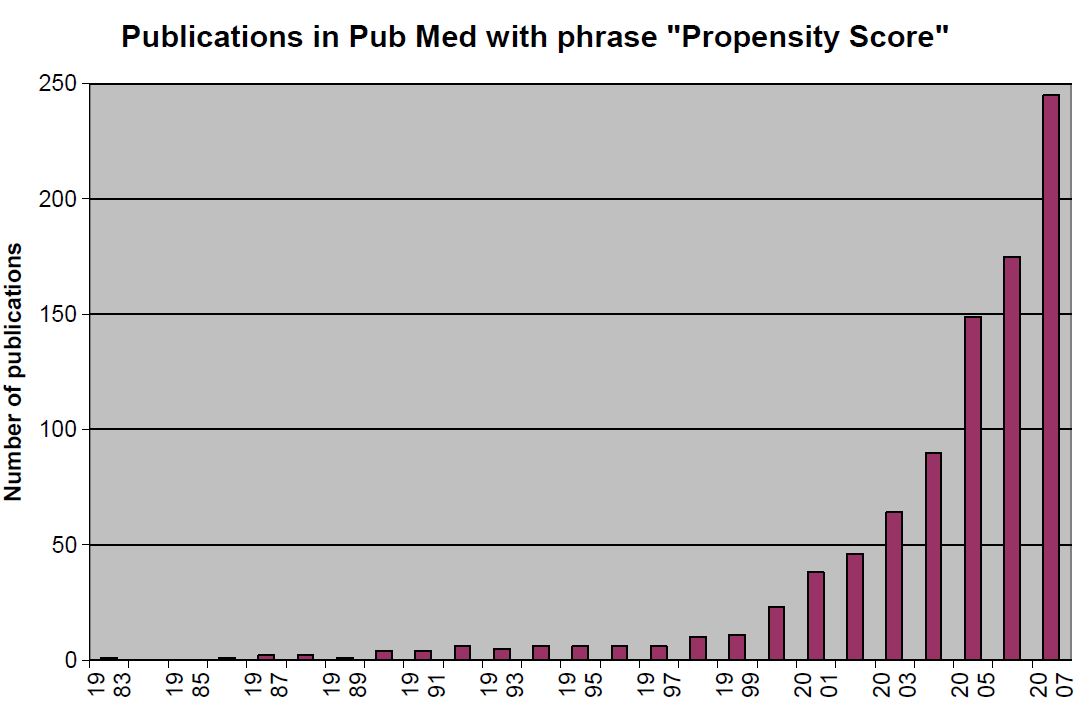}
\caption[Propensity Score Publications]{The publication regarding the propensity score are increasing in the last three decades.}
\label{fig:ps-pub}
\end{figure}

As shown in Figure~\vref{fig:ps-pub}, the propensity score method has gained an increasing interest during the last decade. By counting the publications is shown that the number of papers rose sharply from $<10$ in 1997 to $>200$ in 2007. 
Through the propensity score, it is possible to produce comparable groups under some nonrandomized conditions. It provides a way to summarize covariate\footnote{A covariate is a variable that is possibly predictive of the outcome under study. A covariate may be of direct interest or it may be a confounding or interacting variable. A confounding variable is an extraneous variable in a statistical model that correlates (positively or negatively) with both the dependent variable and the independent variable. An interaction variable may arise when considering the relationship among three or more variables, and describes a situation in which the simultaneous influence of two variables on a third is not additive.} information about treatment selection into a scalar value.

In the next Section the original definition of this technique provided for the first in 1983 time by Rosenbaum and Rubin in \cite{PS-FIRST} will be described.

\section{Summary of the Dataset}
\label{cpt:summary}
A brief summary of the dataset assembled in Chapter~\ref{cpt:datasetExtraction} is now provided. There is a total of 1,522 patients in the \gs. Out of these 189 recieved diuretics and 1333 did not. In the subsequent sections these patients will be referred to as $D^+$ and $D^-$ respectively. For each of these patients there are:
\begin{description} 
\item[(a)] 55 variables which will be referred to as covariates.
\vspace{-3mm}
\item[(b)] 1 diuretics decision variable.
\vspace{-3mm}
\item[(c)] 2 outcomes, i.e., 30-day mortality and length of stay in ICU.
\end{description} 

\section{Propensity score model building and balancing}
\label{cpt:build-bal}
\begin{figure}[htbp]
\centering
\includegraphics[width=\textwidth]{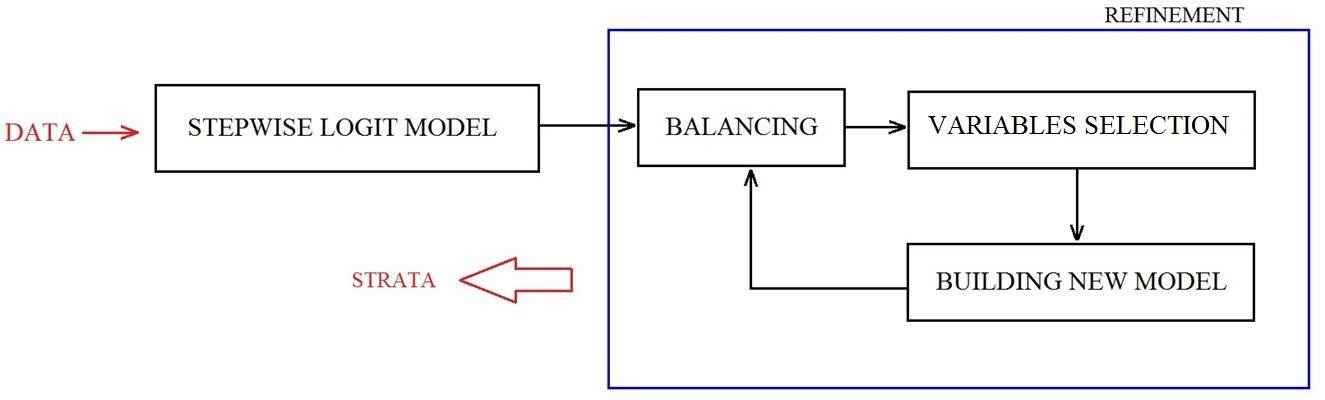}
\caption[Propensity Score Process]{According to Rosenbaum and Rubin, the propensity score method is composed by a first phase where a stepwise discriminant analysis is performed on the whole dataset and a second iterative phase in which the achieved balance is evaluated and improved.}
\label{fig:ps-overall}
\end{figure}

A propensity score is the conditional probability of assignment to a particular treatment given a vector of observed covariates. Consider the \gs in which are compared two treatments, labeled 1 and 0, denoted by the variable $z$ representing the treatment assignment. Each of the patient is represented by a set of covariates $\textbf{x} = \left \{ x_1, x_2, ..., x_{55} \right \}$. The propensity score is the conditional probability that a patient with vector $\mathbf{x}$ of observed covariates will be assigned to treatment 1 given by:
\begin{equation}
e(\textbf{x}) = Pr(z = 1 |\textbf{x}).
\end{equation}
A systematic approach to build the model and refine it is presented in Rubin and Rosenbaum\cite{PS-FIRST}. The method follows the 4 steps:
\begin{description}
\item [Step 1]: Building a propensity model via stepwise logit model.
\vspace{-3mm}
\item [Step 2]: Stratification and balance assessment.
\vspace{-3mm}
\item [Step 3]: Refinement of the model.
\vspace{-3mm}
\item [Step 4]: Decision if the desired balance is achieved or goto Step 2.
\end{description}
An overview of the process is shown in Figure~\vref{fig:ps-overall}. In the following Subsections details about each step will be provided and the results achieved on the dataset created in Chapter~\ref{cpt:datasetExtraction} will be presented.
The goal of propensity score model building and refining is to find the subgroups (a.k.a subclassifications) of the patients along the propensity score axis such that within each subclass,  patients who are $D^+$ and $D^-$,  are statistically similar in the covariate space. Statistical similarity is measured by analyzing the difference in each covariate values between the $D^+$ and $D^-$. Refinement of the propensity score model is achieved by adding variates or their interaction terms resulting in:
\begin{description}
\item[(a)]changing of the propensity score values for the patients.
\vspace{-3mm}
\item[(b)]changing the subclass membership of a few patients (if not all).
\vspace{-3mm}
\item[(c)]improving the statistical similarity of covariates within each group.
\end{description}

\subsection{Step 1: Building a Propensity Score Model via Stepwise Logit Model}
\label{cpt:logit-model}
The propensity score is estimated using a logit model (Cox 1970) for
\begin{equation}
e(x)=  \frac{e(y)}{1 - e(y)} = \alpha + \beta ^Tf(x),
\end{equation}
with $y = log[\frac{e(x)}{1 - e(x)}]$ and where $\alpha$ and $\beta$ being parameters and $f(\cdot)$ a specified function determined with the regression model. To build the first propensity score model first the $55$ covariates have been provided, to choose from to the stepwise discriminant analysis. The stepwise discriminant analysis method selects a subset of variables $\mathbf{x_s} \in \mathbf{x}$. Then the pariwise interaction terms $x_i \cdot x_j$ where $x_i, x_j \in \mathbf{x_s}$ is provide to the stepwise discriminant analysis method. Each time the parameters for the logit model are estimated via maximum likelihood method\cite{PS-FIRST}.
In the \gs, the variables (and their pairwise interaction terms) chosen at the end of this step are shown in Table~\vref{tab:autovars}.
\begin{table}[htdp]
\centering
\begin{tabular}{|c|c|c|}
\hline
Variable & Detail\\
\hline $x_{11}$ & Sofa mean of values during the first day\\
\hline $x_{12}$ & Sofa mean of values during day T1\\
\hline $x_{16}$ & Elixhauser congestive heart failur\\
\hline $x_{17}$ & Elixhauser cardiac arrhythmias \\
\hline $x_{40}$ & Balance average of sums from day 1 to day T1\\
\hline $x_{41}$ & Balance sum of values during the first day (41)\\
\hline $x_{43}$ & Balance sum of values during day T2\\
\hline $x_{45}$ & Use of vasopressors\\
\hline $x_{46}$ & Mechanical ventilation\\
\hline $x_{47}$ & Arterial bp average from day 1 to day T1\\
\hline $x_{55}$ & Arterial bp mean mean of values during day T2\\
\hline $x_{11} \cdot x_{55}$ & -\\
\hline $x_{40} \cdot x_{43}$ & -\\
\hline $x_{40} \cdot x_{46}$ & -\\
\hline $x_{41} \cdot x_{43}$ & -\\
\hline $x_{43} \cdot x_{43}$ & -\\
\hline $x_{46} \cdot x_{55}$ & -\\
\hline
\end{tabular}
\caption[Covariates and Interactions]{Covariates and their interactions selected after first step.}
\label{tab:autovars}
\end{table}

\subsection{Step 2: Stratification and Balance Assessment}
\label{cpt:strata}
The propensity score model built in the previous section provides a score for each patient. Consequently, the patients have been subclassified into 5 groups each group corresponding to a quintile of the distribution of the estimated propensity score. Figure~\vref{fig:ps-\group} illustrates the process of subclassification. 
\begin{figure}[htbp]
\centering
\includegraphics[width=\textwidth]{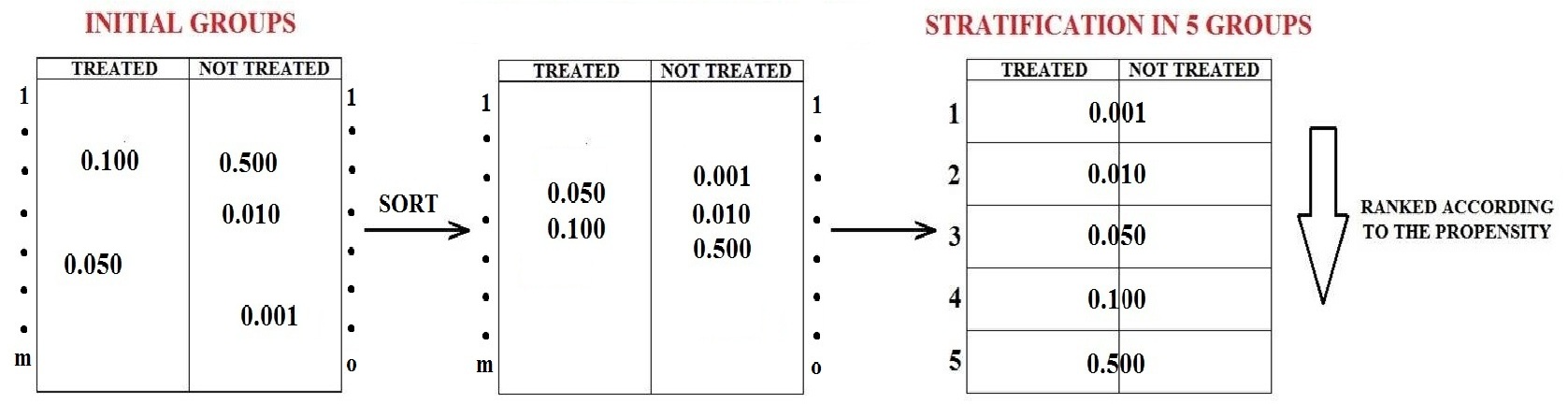}
\caption[Propensity Score Stratification]{The stratification process consists in the ranking of the patients according to their propensity score and then in the division of the whole ranked datasets in 5 \group. Rosenbaum and Rubin suggest that 5 \groups can reduce the 90\% of the bias in the original dataset.}
\label{fig:ps-\group}
\end{figure}
It is suggested in Rubin and Rosenbaum that by performing subclassification into \textit{quintiles} based on the propensity score, one can largely balance all observed covariates.The balance is achieved $\mathbf{x}$, in the sense that within subclasses that are homogeneous in $e(x)$, the distribution of $\mathbf{x}$ is the same for treated and control (not treated) patients. Note that while all the covariates are not included in the propensity score model, the balance is still sought across all the covariates. In fact this is a key point in the propensity score methodology. At this stage the effectiveness of the subclassification due to this specific propensity score model can be measures by following the method in\cite{PS-FIRST}. The effectiveness is quantified by calculating F-Ratios. The statistical technique, which uses F-ratios, used to assess the balance is briefly explained in Section~\ref{cpt:assessing}, but more details will be provided in Appendix~\ref{apx:software}.

\subsection{Assessing the Balance with Subclasses}
\label{cpt:assessing}
To assess balance each of the 55 covariates are subjected to a two-way (2 (treatments) x 5 (subclasses) analysis of variance (ANOVA). In its simplest form, ANOVA provides a statistical test of whether or not the means of several groups are all equal, and therefore generalizes T-test to more than two groups. By doing this two F-values for each covariate are calculated. The first one is for treatment vs no treatment interaction. The second one is for treatment vs subclass interaction. The first value will be called \textit{primary effect} and the second one, \textit{secondary effect}\footnote{In \cite{PS-FIRST} the first F-ratio hs been called as main effect and the second one as interaction effect. In order to avoid confusion between these definitions and the previous definitions of main effects, being the variables themselves, and interactions effects, representing pairwise products $x_i \cdot x_j$, alternative names have been chosen.}.

The achieved balance has been analyzed by comparing a five-number summary (that is minimum, lower quartile, median, upper quartile, maximum) of the 55 F-ratios\footnote{An F-test is any statistical test in which the test statistic has an F-distribution under the null hypothesis. It is most often used when comparing statistical models that have been fit to a data set, in order to identify the model that best fits the population from which the data were sampled.} prior to subclassification with the F-ratios for the primary effect of the treatment and the treatment x subclass interaction in the two-way analysis of variance. The five point summary prior to propensity score modeling and after the first step of propensity score modeling and startification is presented in Figure~\vref{fig:F-test1}.  Note that the summary statistics are the same for primary and secondary effects initially since the groups have not been stratified yet. It can be observed that after this first step, the F-ratios referring to both primary effects and secondary are decreased significantly, indicating the improvement in the balances of the respective groups subsequent to the propensity score method.
\begin{figure}
\centering
\rotatebox{90}{
\stepcounter{figure}
\begin{minipage}{\textheight}
\begin{minipage}{0.50\textheight}
\centering
\includegraphics[width=.9\linewidth]{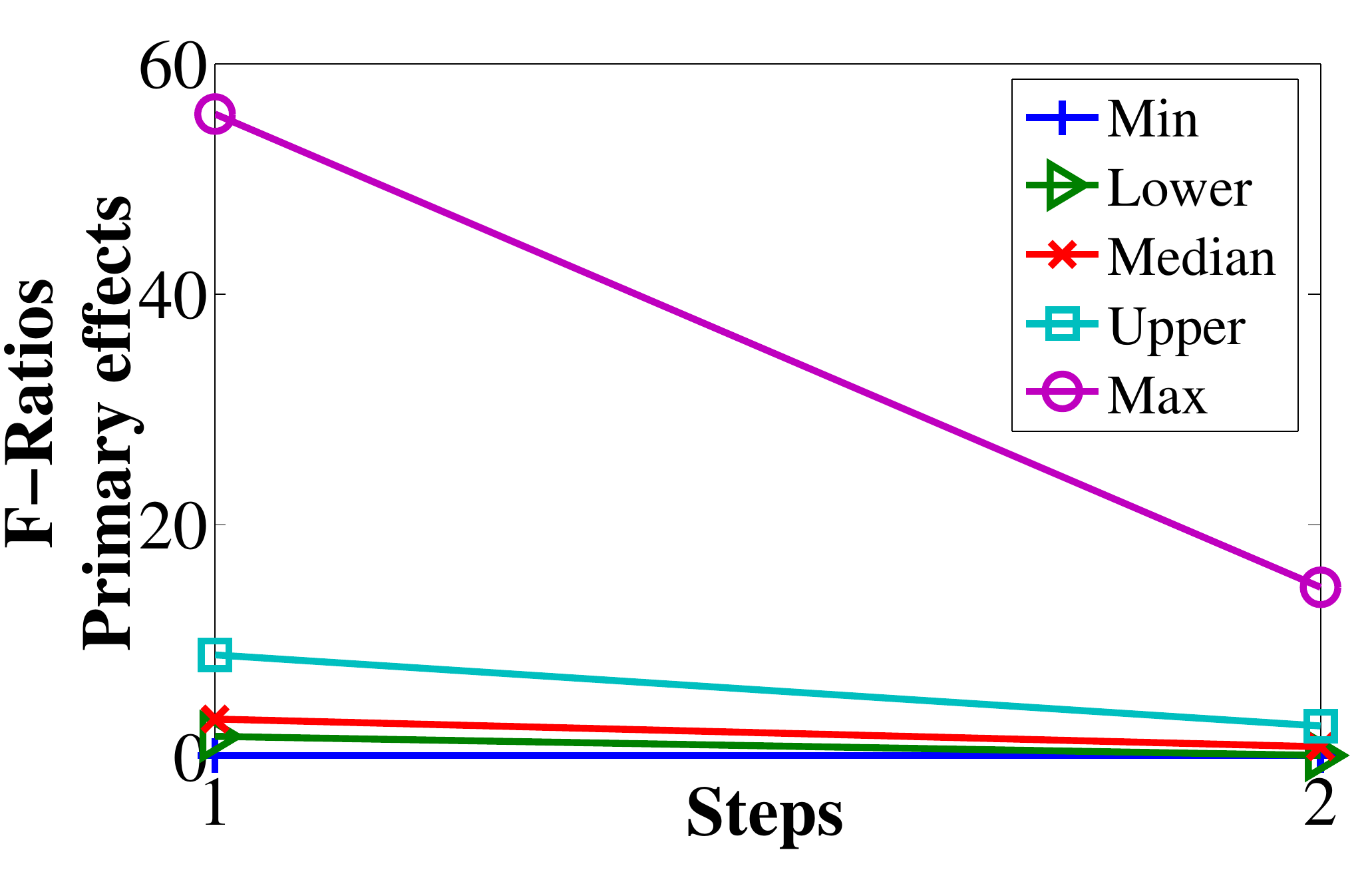}
\captionof{subfigure}[]{The F-Ratio statistics after the first step for primary effects.}
\label{F-test11}
\end{minipage}
\begin{minipage}{0.50\textheight}
\centering
\includegraphics[width=.9\linewidth]{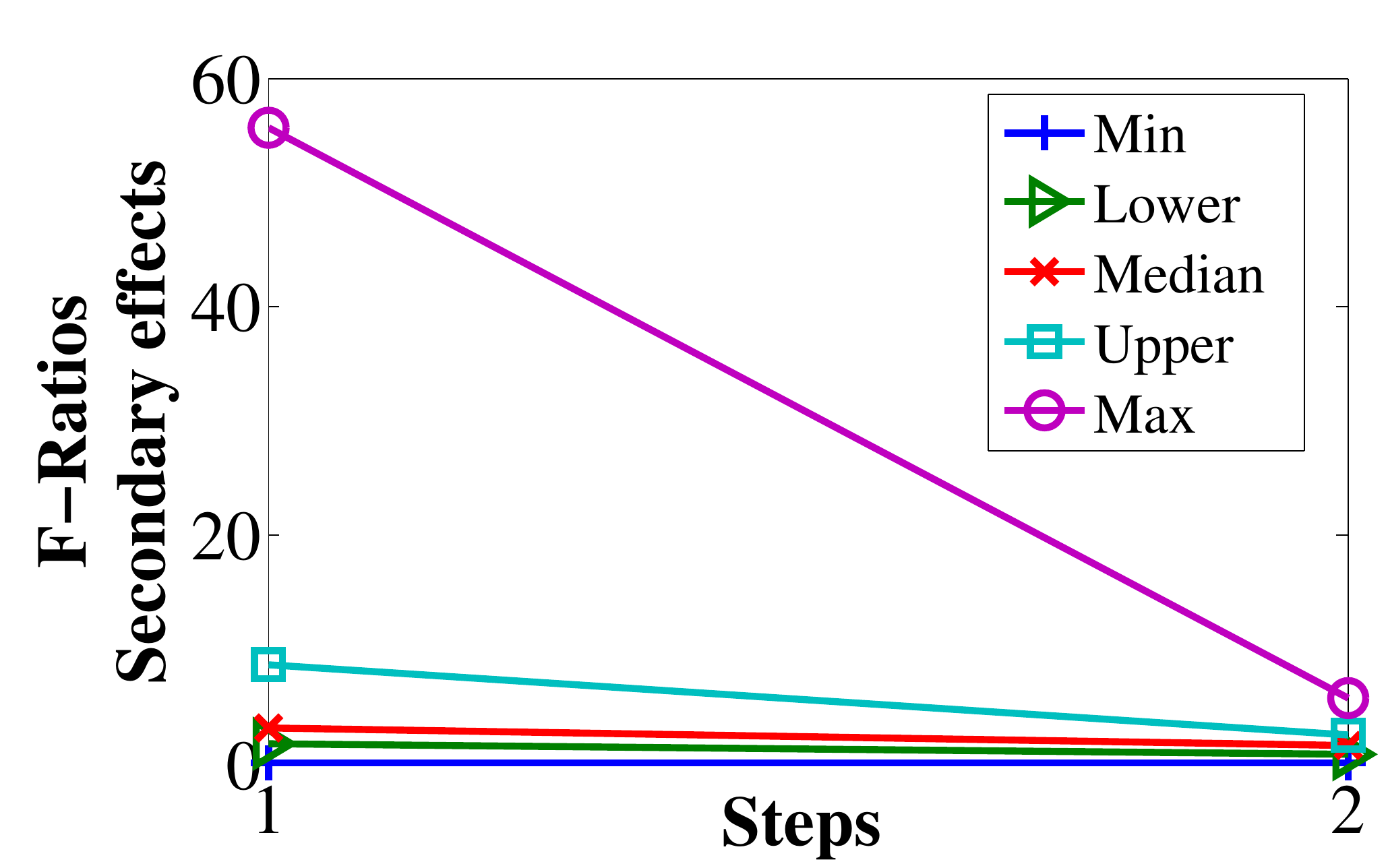}
\captionof{subfigure}[]{The F-Ratio statistics after the first step for secondary.}
\label{fig:F-test12}
\end{minipage}
\addtocounter{figure}{-1}
\captionof{figure}[F-Ratios on Automatic Dataset - Primary Effects (a) and Secondary Effects (b)]{The F-Ratio statistics after the first step. The values pertaining to \textit{NONE} refer to the balance on the original dataset, while the values pertaining to ONE refer to balance after the propensity score method.}
\label{fig:F-test1}
\end{minipage}
}
\end{figure}

\subsection{Step 3: Refinement of the Model}
\label{cpt:refinement}
In this step, the refinements of the existing logistic model for propensity have been performed to improve the covariates balance further. Covariates with large F-ratios that had previously been excluded from the model were considered for addition to the model. Adding these variables changes the propensity scores for the patients, resulting in reassignment of patients to different quintiles (groups or subclasses).  After adding each variable,  a logistic models was fitted by maximum likelihood.  If the variable produced a lower F-ratio, it was kept.  If the variable produced a large F-ratio after inclusion in the model, the square of the variable and cross-products with other variables were instead tried, per the advice of \cite{PS-FIRST}.  

This refinement process added 11 of 44 variables, that is  25\% of them. 
Figure~\vref{fig:f-improvements} shows the improvement made ​​by the inclusion of each variable in the new refined model. It appears that the most important improvements are due to the first few variables which have the biggest F-ratios.
\begin{figure}
\centering
\rotatebox{90}{
\stepcounter{figure}
\begin{minipage}{0.8\textheight}
\centering
\includegraphics[width=.9\linewidth]{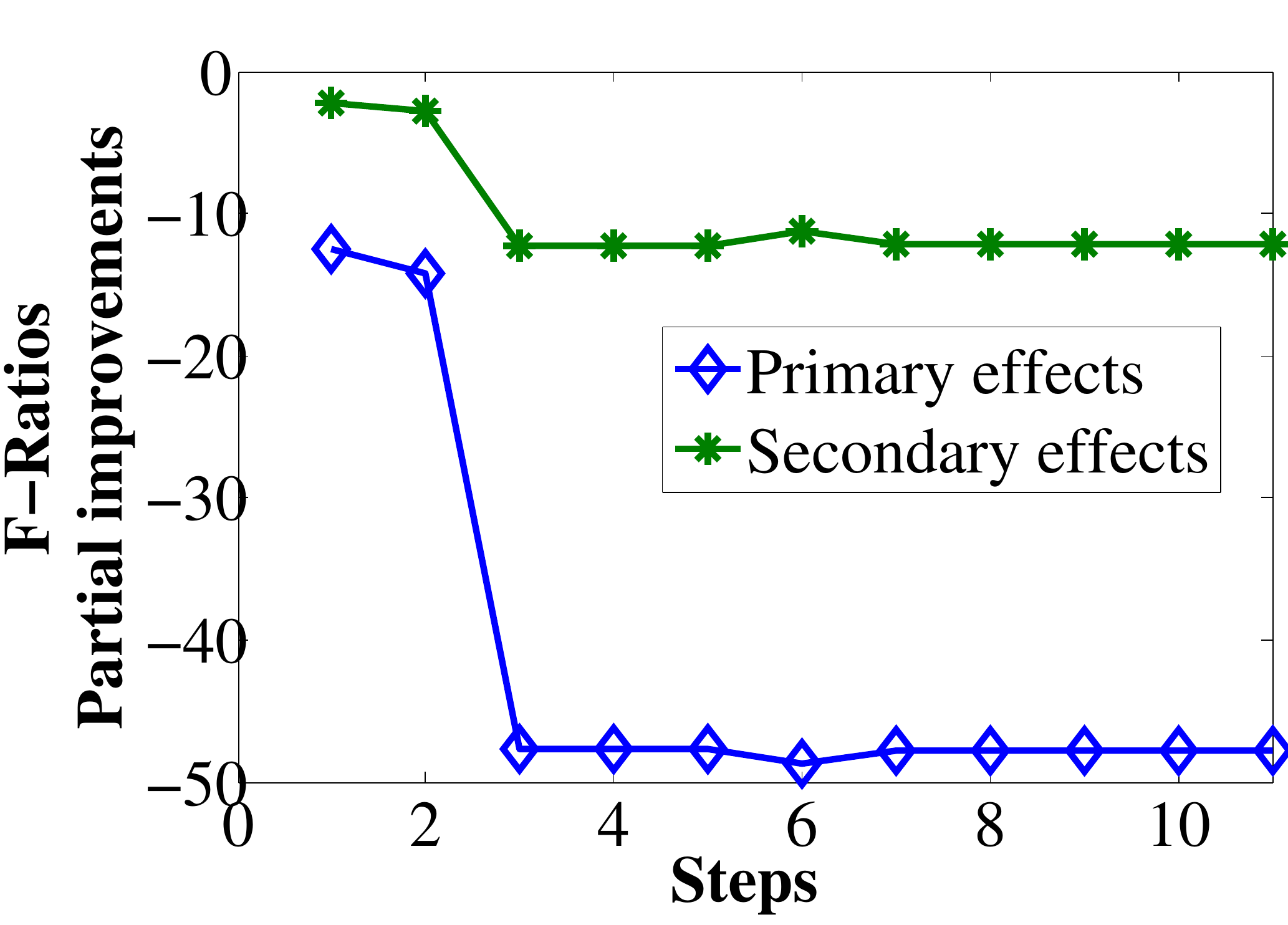}
\addtocounter{figure}{-1}
\captionof{figure}[Refinement Process and F-Ratios Improvements]{The improvement made ​​by the inclusion of each variable in the new refined model.}
\label{fig:f-improvements}
\end{minipage}
}
\end{figure}
Figure~\vref{fig:final1} shows the balance achieved in the final refined model, which is considered satisfactory. Note that,  if the improvements in the maximum quintile are not considered, there is no substantial improvement in the balance comparing the groups formed after the refinement process.
\begin{figure}
\centering
\rotatebox{90}{
\stepcounter{figure}
\begin{minipage}{\textheight}
\begin{minipage}{0.50\textheight}
\centering
\includegraphics[width=.9\linewidth]{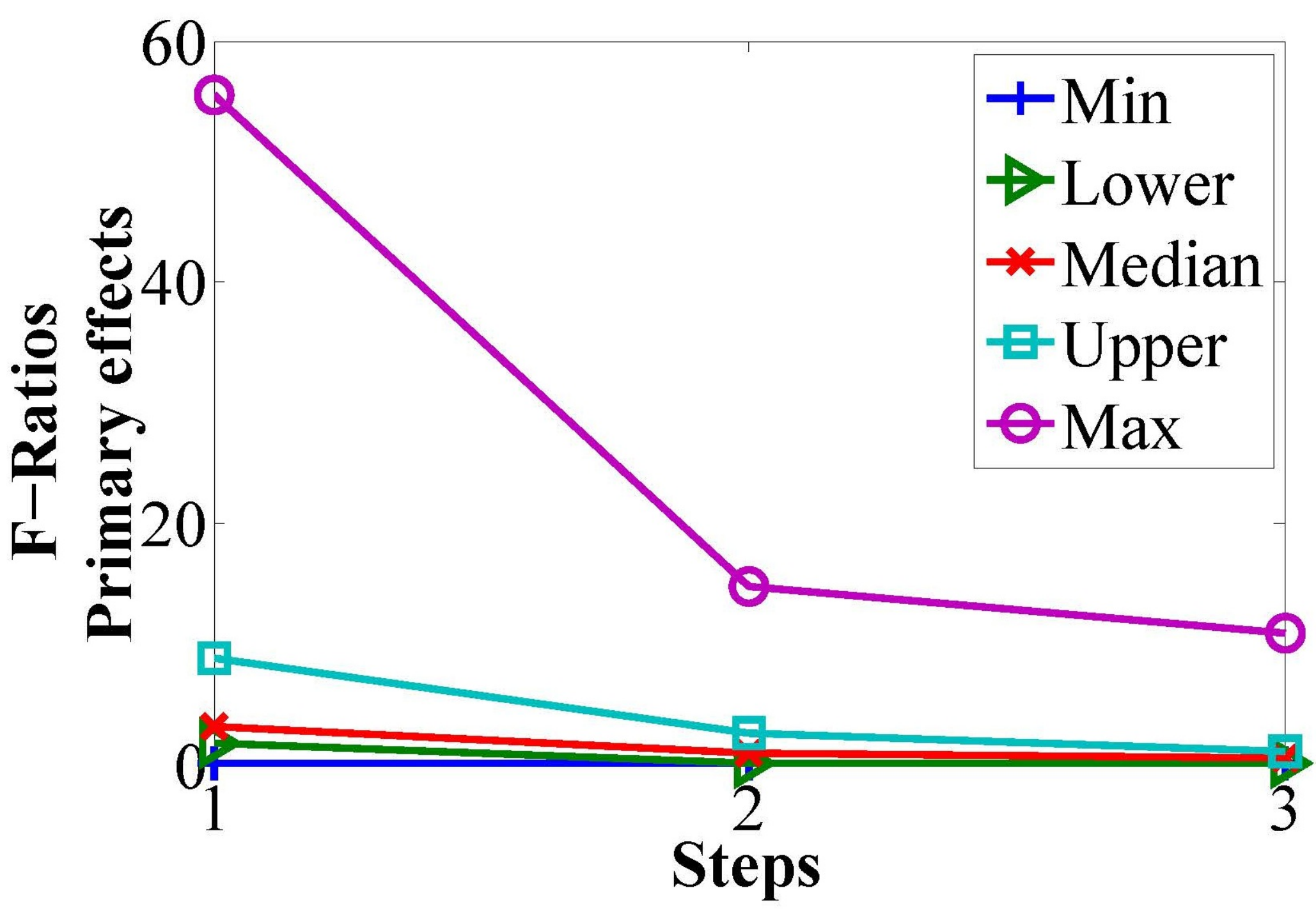}
\captionof{subfigure}[]{The F-Ratio statistics after the first step for primary effects.}
\label{F-test21}
\end{minipage}
\begin{minipage}{0.50\textheight}
\centering
\includegraphics[width=.9\linewidth]{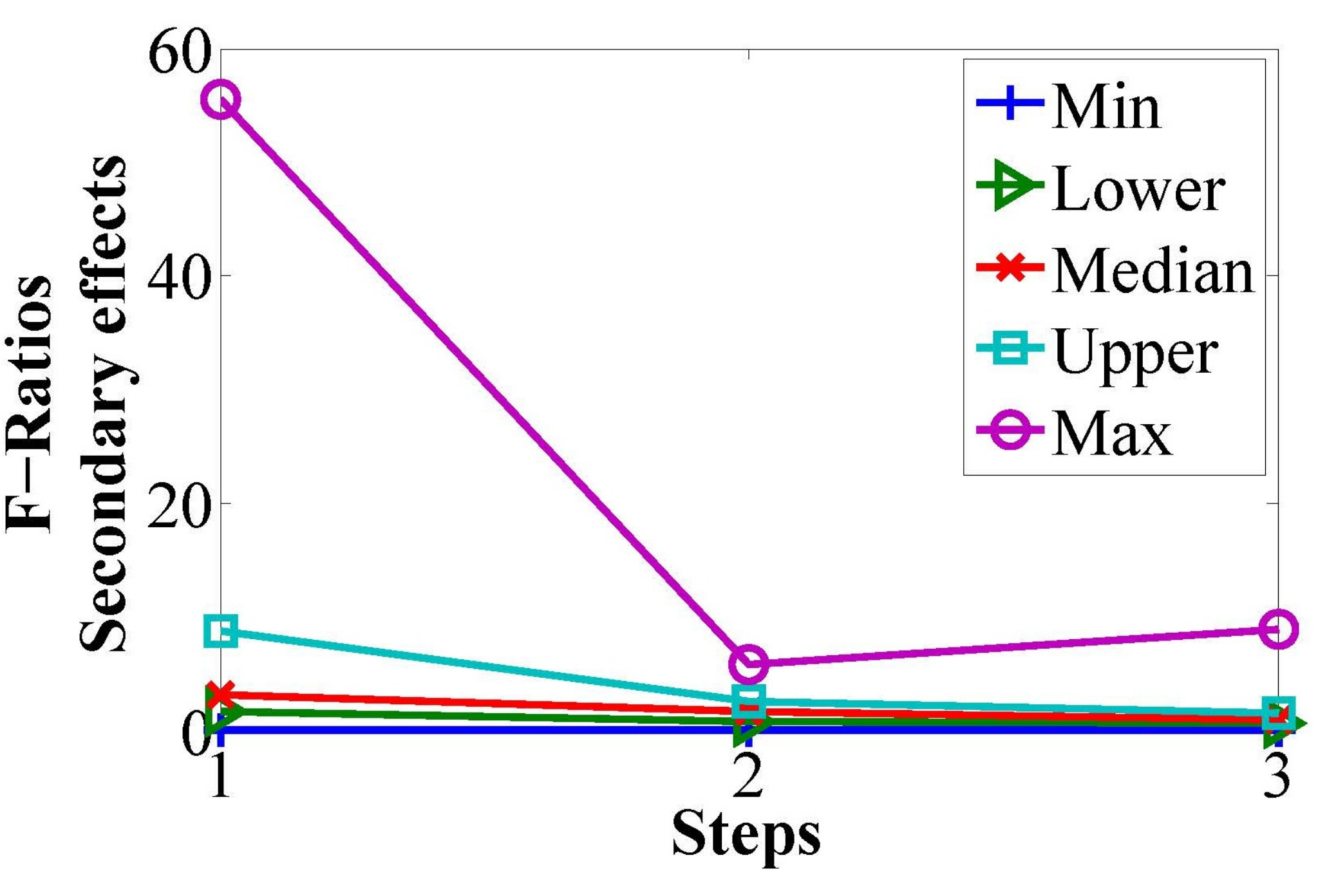}
\captionof{subfigure}[]{The F-Ratio statistics after the first step for secondary.}
\label{fig:F-test22}
\end{minipage}
\addtocounter{figure}{-1}
\captionof{figure}[F-Ratios after Refinement - Primary effects (a), Secondary Effects(b)]{The F-statistics on the refined dataset on the primary and secondary effects after the refinement. The values compare the original balance, the balance after the first logit model and the one in the final model.}
\label{fig:final1}
\end{minipage}
}
\end{figure}

The complete list of variables included in this final model, of which the first 17 variables were already in the logit model, are in Table~\vref{tab:final_list}.
\begin{table}[htdp]
\centering
\begin{tabular}{|c|c|c|}
\hline
Variable & Detail\\
\hline $x_{11}$ & Sofa mean of values during the first day\\
\hline $x_{12}$ & Sofa mean of values during day T1\\
\hline $x_{16}$ & Elixhauser congestive heart failur\\
\hline $x_{17}$ & Elixhauser cardiac arrhythmias \\
\hline $x_{40}$ & Balance average of sums from day 1 to day T1\\
\hline $x_{41}$ & Balance sum of values during the first day (41)\\
\hline $x_{43}$ & Balance sum of values during day T2\\
\hline $x_{45}$ & Use of vasopressors\\
\hline $x_{46}$ & Mechanical ventilation\\
\hline $x_{47}$ & Arterial bp average from day 1 to day T1\\
\hline $x_{55}$ & Arterial bp mean mean of values during day T2\\
\hline $x_{11} \cdot x_{55}$ & -\\
\hline $x_{40} \cdot x_{43}$ & -\\
\hline $x_{40} \cdot x_{46}$ & -\\
\hline $x_{41} \cdot x_{43}$ & -\\
\hline $x_{43} \cdot x_{43}$ & -\\
\hline $x_{46} \cdot x_{55}$ & -\\
\hline $x_{20}$ & Elixhauser diabetes uncomplicated\\
\hline $x_{32}$ & Inputs sum of values during day T1\\
\hline $x_{42}$ & Balance sum of values during day T1\\
\hline $x_{8}$ & Saps mean of values during day T2\\
\hline $x_{12} \cdot x_{5}$ & $x_{12} \cdot$ Saps average of sums from day 1 to day T1\\
\hline
\end{tabular}
\caption[Covariates and Interactions]{Covariates and their interactions selected after first step.}
\label{tab:final_list}
\end{table}

\subsection{Experts' Covariate Sets}
\label{cpt:experts}
Finally, the analysis was repeated for 2 new variable sets based on advice by clinical experts. The details of this analysis are presented in Appendix~\ref{apx:dataset}.  Note that,  if the improvements in the maximum quintile are not considered, there is no substantial improvement in the balance comparing the experts' variable sets to the final refinement-based balance.

\begin{figure}[htbp]
\centering
\includegraphics[height=0.8\textheight]{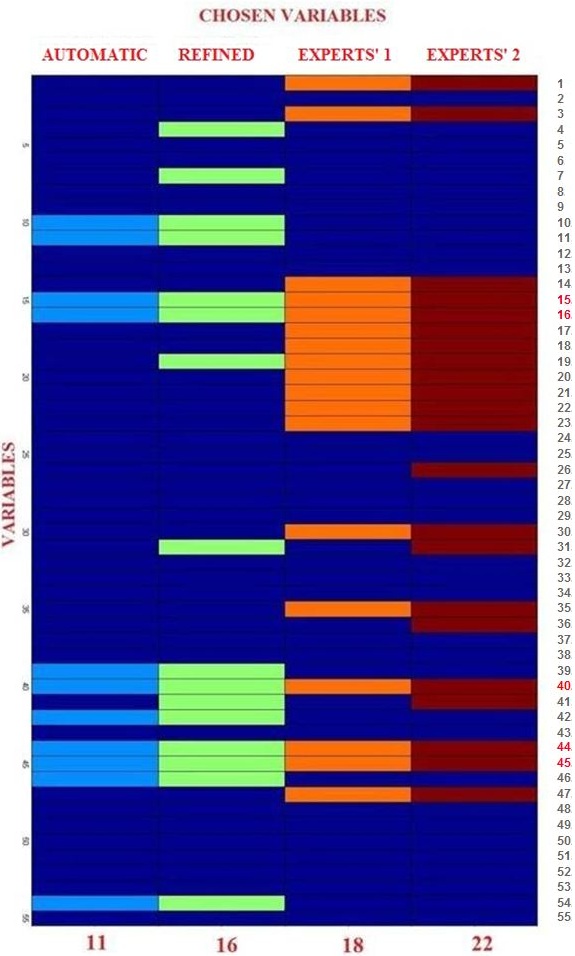}
\caption[Chosen Variables across the Variable Sets]{Five variables were chosen from all the four variable sets: Elixhauser congestive heart failure ($x_{16}$), Elixhauser cardiac arrhythmias ($x_{17}$), Fluids balance sum of values during the first day ($x_{41}$), Use of vasopressors ($x_{45}$) and Mechanical ventilation ($x_{46}$).}
\label{fig:chosen-vars}
\end{figure}

Figure~\vref{fig:chosen-vars} shows the variables which were selected in each of  4 variable sets: the set selected in the \textit{automatic} stepwise discrimination process before iterative refinement, the set selected in the \textit{refined} model and the two sets resulting from  starting with the experts' variable sets. Five variables were select in all four variable sets:  Elixhauser congestive heart failure ($x_{16}$), Elixhauser cardiac arrhythmias ($x_{17}$), Fluids balance sum of values during the first day ($x_{41}$), Use of vasopressors ($x_{45}$) and Mechanical ventilation ($x_{46}$).

\subsection{Estimating the Average of Treatment Effects}
\label{cpt:extimatingOutcome}
In the Rosenbaum and Rubin study,\cite{PS-FIRST}, the groups defined by the propensity analysis, which are now homogeneous, are directly compared with respect to basic statistics on mortality or on other predetermined outcomes.

Austin warns that when using propensity score methods involving pair matching between patients, the matched nature of the pairs should be considered in the analysis of the outcomes \cite{PS-STATISTICS}. But in the creation of the \group, stratification is performed prior to treatment assignment (as in randomized controlled trials) and this implies that there is no reason subjects within a stratum are more similar then randomly selected ones.

Considering this, it is possibile using the stratification approach to estimate the treatment effects of the treatment just by direclty comparing the treated and not treated groups.

\clearpage

\section{Stratification Results}
\label{cpt:resultsDatasets}
Table~\vref{tab:automatic-dataset} shows the quintiles, their propensity score ranges and their split between $D^+$ and $D^-$ in terms of number of patients, mortality rate and mean length of ICU stay for the propensity score model generated automatically by stepwise discrimination, before refinement.  The stepwise logit model in this case maximizes the accuracy of the model, however, in fact, the first two \groups{} are very unbalanced in the number of treated and untreated patients and cannot considered useful for analysis. Quintiles 3 and 4 are less unbalanced even if the number of untreated patients is still a lot bigger then the treated ones. In \group 5 there are balanced numbers.

From the results, it appears that typically the treated patients spend more time in the ICU. The mortality however seems to be more or less the same, except for group 3 where diuretics seem to slightly improve the chances of survival.

Table~\vref{tab:automatic-refined-dataset} shows the quintiles, their propensity score ranges and their split between $D^+$ and $D^-$ in terms of number of patients, mortality rate and mean length of ICU stay for the propensity score model which was subsequently iteratively refined. Even though the variable set went through the refinement, \groups{} 1 and 2 are still very unbalanced. Quintiles 3 and 4 are less unbalanced, while Quintile 5 is balanced. Except for \group{} 5, also in this dataset it seems that patients who got diuretics have a slightly better chance of survival.

\begin{table}
\centering
\begin{tabular}{|c|c|c|}
\hline {\bf \Group~1 $PS \in \left[ 0.00;0.00 \right] $} & Diuretics given & Diuretics not given\\
\hline Number of patients & 2 & 302\\
\hline Deaths & 0\% & 43\%\\
\hline Mean length of stay & 40 days & 3.7 days\\
\hline {\bf \Group~2 $PS \in \left[ 0.00;0.02 \right] $} & Diuretics given & Diuretics not given\\
\hline Number of patients & 5 & 299\\
\hline Deaths & 40\% & 32\%\\
\hline Mean length of stay & 28 days & 4.1 days\\
\hline {\bf \Group~3 $PS \in \left[ 0.02;0.06 \right] $} & Diuretics given & Diuretics not given\\
\hline Number of patients & 20 & 284\\
\hline Deaths & 15\% & 30\%\\
\hline Mean length of stay & 13.3 days & 6 days\\
\hline {\bf \Group~4 $PS \in \left[ 0.07;0.17 \right] $} & Diuretics given & Diuretics not given\\
\hline Number of patients & 27 & 277\\
\hline Deaths & 37\% & 40\%\\
\hline Mean length of stay & 13.5 days & 9.5 days\\
\hline {\bf \Group~5 $PS \in \left[ 0.17;1.00 \right] $} & Diuretics given & Diuretics not given\\
\hline Number of patients & 134 & 170\\
\hline Deaths & 41\% & 44\%\\
\hline Mean length of stay & 15 days & 10 days\\
\hline
\end{tabular}
\caption[Automatic Generation Dataset]{Results on the Automatic generation dataset.}
\label{tab:automatic-dataset}
\end{table}

\begin{table}
\centering
\begin{tabular}{|c|c|c|}
\hline {\bf \Group~1 $PS \in \left[ 0.00;0.01 \right] $} & Diuretics given & Diuretics not given\\
\hline Number of patients & 4 & 300\\
\hline Deaths & 25\% & 30\%\\
\hline Mean length of stay & 46.7 days & 1.3 days\\
\hline {\bf \Group~2 $PS \in \left[ 0.01;0.04 \right] $} & Diuretics given & Diuretics not given\\
\hline Number of patients & 4 & 300\\
\hline Deaths & 25\% & 33\%\\
\hline Mean length of stay & 7 days & 5.2 days\\
\hline {\bf \Group~3 $PS \in \left[ 0.04;0.08 \right] $} & Diuretics given & Diuretics not given\\
\hline Number of patients & 25 & 279\\
\hline Deaths & 24\% & 38\%\\
\hline Mean length of stay & 13.8 days & 7.4 days\\
\hline {\bf \Group~4 $PS \in \left[ 0.08;0.18 \right] $} & Diuretics given & Diuretics not given\\
\hline Number of patients & 27 & 277\\
\hline Deaths & 29\% & 46\%\\
\hline Mean length of stay & 10.7 days & 8.8 days\\
\hline {\bf \Group~5 $PS \in \left[ 0.18;0.99 \right] $} & Diuretics given & Diuretics not given\\
\hline Number of patients & 127 & 177\\
\hline Deaths & 45\% & 42\%\\
\hline Mean length of stay & 15.7 days & 10.5 days\\
\hline
\end{tabular}
\caption[Automatic Generation Refined Dataset]{Results on the Automatic generation dataset after the refinement process.}
\label{tab:automatic-refined-dataset}
\end{table}

\clearpage

\subsection{Comparison between the \Groups}
\label{cpt:comparisonDatasets}
\begin{figure}[htbp]
\centering
\includegraphics[width=\textwidth]{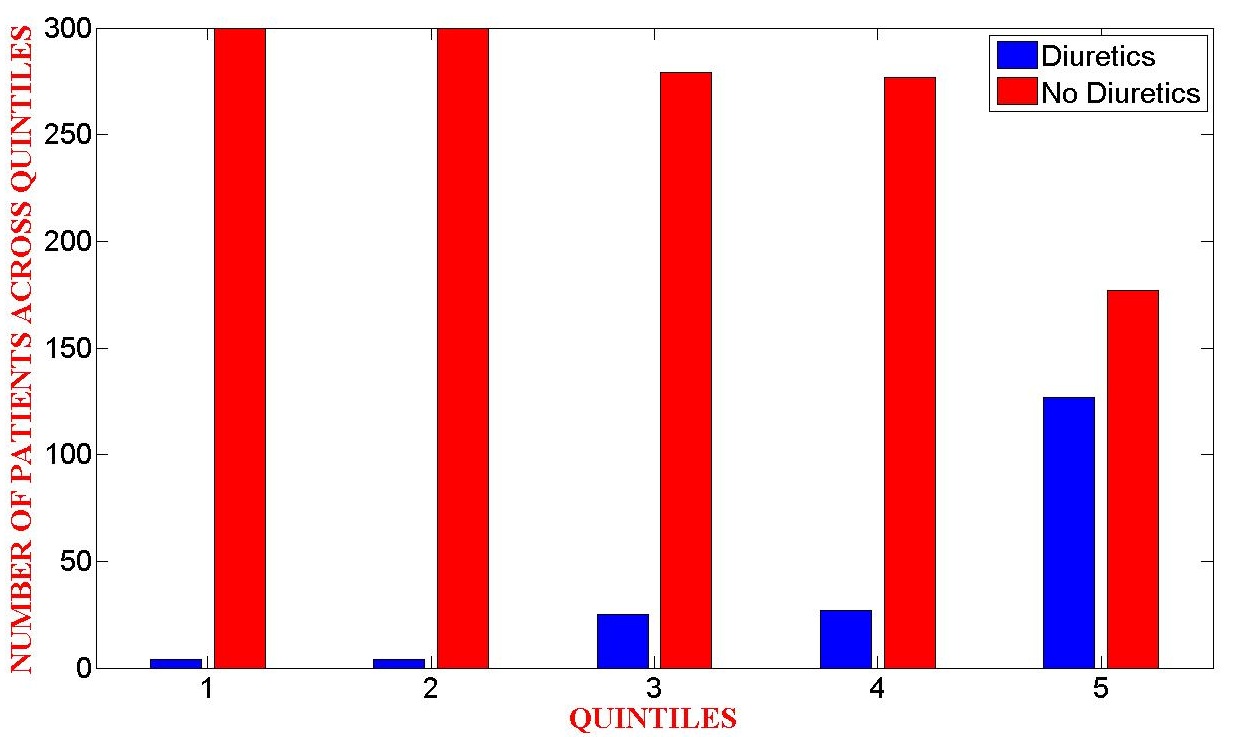}
\caption[Number of Patients Across the Quintiles]{Number of patients in the refined dataset, across the quintiles.}
\label{fig:bal3}
\end{figure}

\begin{figure}[htbp]
\centering
\includegraphics[width=\textwidth]{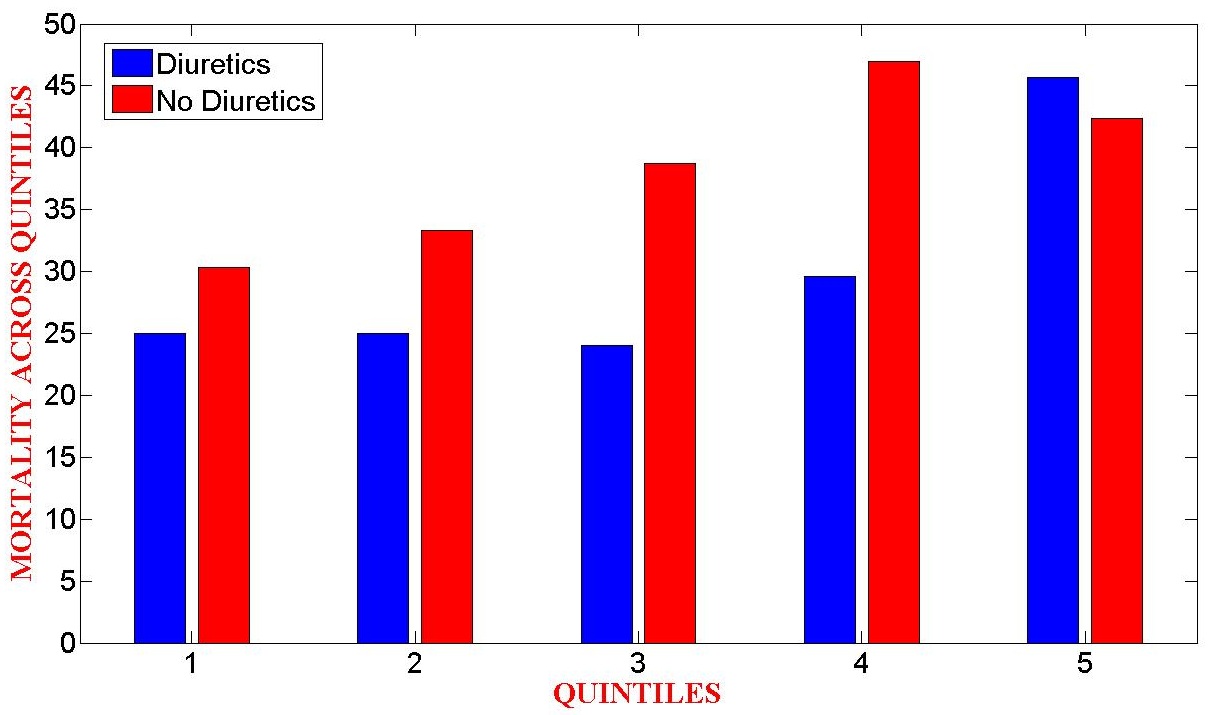}
\caption[Deaths Balance Across \Groups]{Death balance in the 5 \groups .}
\label{fig:bal1}
\end{figure}

\begin{figure}[htbp]
\centering
\includegraphics[width=\textwidth]{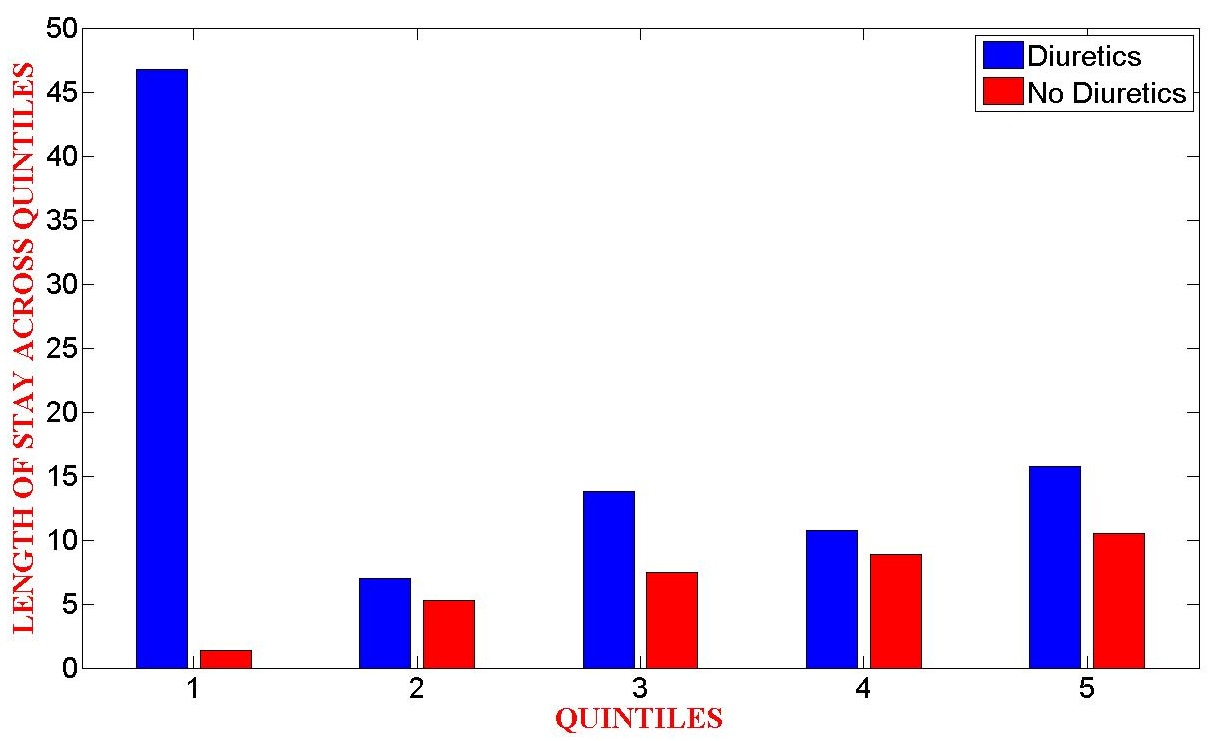}
\caption[\LOS Balance Across the \Groups]{\Los in ICU balance in the 5 \groups.}
\label{fig:bal2}
\end{figure}

Now a comparison between the \groups for the Refined dataset will be discussed. In Figure~\vref{fig:bal3} shows a parallel between the number of patients for the 5 \group. The numbers seem to be consistent as going from \group~1 to 5, there is an increasing number of patients who received diuretics as the propensity of getting them is increasing.
Figure~\vref{fig:bal1} show a comparison between the mortality rate in the 5 \group. In the first 4 \group there is a slightly better chance of survival by \gettingdiuretics , while for \group 5 the chances are similar.
Figure~\vref{fig:bal2} show a comparison between the \los in ICU in the 5 \group. The results seem consistent except for the first \group, where probabily there are outliers or noisy values for the patients who got diuretics. It seems that the patients treated with diuretics have a longer stay in ICU.

\subsection{Comments on the Results}
\label{cpt:commentsDatasets}
In both the 2 datasets, mortality rates are usually similar between treated and untreated patients and looking at those values, it seems that the treated patients have a slightly better chance of survival. But from this analysis is not easy to determine in which cases the diuretics should be given, it can just be concluded that, in case of indecision, they may be of benefit for the patient. Instead, in all the four datasets, seems that the treated patients stay more time in the ICU.

In Table~\vref{tab:analysis-results} are shown the averages of the outcomes in the 2 models. The averages are made considering only \group~3, 4 and 5 as \group~1 and 2 have too few patients to whom diuretics were administrated. The patients to whom diuretics were given seem to have a better chance of survival, while they have a longer stay in the ICU.
\begin{table}
\centering
\begin{tabular}{|c|c|c|}
\hline {\bf Automatic Dataset} & Deaths & Length of Stay\\
\hline Diuretics given & 31\% & 13.9 days\\
\hline Diuretics not given & 38\% & 8.5 days\\
\hline {\bf Automatic Refined Dataset} & Deaths & Length of Stay\\
\hline Diuretics given & 32.6\% & 13.4 days\\
\hline Diuretics not given & 42\% & 8.9 days\\
\hline
\end{tabular}
\caption[Results on \Groups~3, 4 and 5]{Results on \groups~3, 4 and 5. The patients to whom diuretics were given seems to have a better chance of survival, while they have a longer stay in the ICU.}
\label{tab:analysis-results}
\end{table}
In Table~\vref{tab:analysis-results2} is shown the same analysis performed using all 5 \group. Comparing the two tables, it is possibile to see that the values for the length of stay for the patients who got diuretics of the first two datasets seem to be outliers: in these groups the patients who actually got the drugs were too few to have a reliable result.

It is possibile to see another reason to use only \group 3, 4 and 5 by noticing that in the results of the analysis performed on all the 5 \group both the mortality rates and the lengths of stay in the ICU in all the four datasets are better, in fact the chance of survival is higher and the lenght of stay in the ICU shorter. This result is consistent with what was said by doctors involved in the work: clinicians are usually against giving diuretics because in general they are harmful drugs and if they give them, the length of stay in ICU of the patient lengthens. In the tables, going from the \group 1 to 5, the propensity for diuretics of being prescribed increases, and this goes along with the worsening of the conditions of the patients: by using \groups 3, 4 and 5 the analysis is capturing the chances of survival of the patients on the border line, that is the patients whose conditions leave to the doctors the judgment whether giving or not diuretics.
\begin{table}
\centering
\begin{tabular}{|c|c|c|}
\hline {\bf Automatic Dataset} & Deaths & Length of Stay\\
\hline Diuretics given & 26.6\% & 21.9 days\\
\hline Diuretics not given & 37.8\% & 6.6 days\\
\hline {\bf Automatic Refined Dataset} & Deaths & Length of Stay\\
\hline Diuretics given & 29.6\% & 18.7 days\\
\hline Diuretics not given & 37.8\% & 6.6 days\\
\hline
\end{tabular}
\caption[Results on all the 5 \Groups]{Results on all the 5 \groups. The mortality rates are decreasing as a bigger propensity of getting diuretics is related the a worse condition of the patient.}
\label{tab:analysis-results2}
\end{table}
Furthermore, have to be said that in group 5 there are the sicker patients: in this groups seems not to make any difference if diuretics were given or not. This could mean that \group 3 and 4 are the most relevant for the analysis, as in these groups the fact that diuretics were given or not seems to make the difference in the survival of the patients.
\begin{table}
\centering
\begin{tabular}{|c|c|c|}
\hline {\bf Automatic Dataset} & Deaths & Length of Stay\\
\hline Diuretics given & 26\% & 13.4 days\\
\hline Diuretics not given & 35\% & 7.7 days\\
\hline {\bf Automatic Refined Dataset} & Deaths & Length of Stay\\
\hline Diuretics given & 26.5\% & 12.2 days\\
\hline Diuretics not given & 42\% & 8.1 days\\
\hline
\end{tabular}
\caption[Results on \Groups~3 and 4]{Results on \groups~3 and 4. The differences of the chances of survivals between treated and untreated patients are wider in this case.}
\label{tab:analysis-results3}
\end{table}
In Table~\vref{tab:analysis-results3} are shown the results for these two groups.

The percentage of death and length of stay in the ICU have also been calculated on the whole (without the propensity method) dataset. The results on the original dataset are, comparing the patients to whom diuretics were given to the ones who didn't got diuretics, the  32.8\% vs 37.8\% percentage of death and 15.1 vs 6.3 days in the ICU. These results are shown in Table~\vref{tab:analysis-results4}.
\begin{table}
\centering
\begin{tabular}{|c|c|c|}
\hline {\bf Original Dataset} & Deaths & Length of Stay\\
\hline Diuretics given & 32.8\% & 15.1 days\\
\hline Diuretics not given & 37.8\% & 6.3 days\\
\hline
\end{tabular}
\caption[Results on the original dataset]{Results on the original dataset, without propensity analysis and stratification.}
\label{tab:analysis-results4}
\end{table}
As predicted, here the differences of the mortality rates are slightly narrowing because is not considered the fact that usually the patients to whom diuretics are administrated are sicker, and have a worst chance of survival.

\subsection{Conclusions of the \PA on the Diuretics Problem}
\label{cpt:conclusionsDatasets}
In conclusion from this analysis, only comparing the outcomes within the \group, seems that patients who received diuretics have a slightly better chance of survival, while they stay longer in the ICU even if it is not clear how statistically relevant these results are and when exactly diuretics should be given.

The refined dataset seems to be the best one, and on it have been performed a series of statistic tests to decide if the results should be considered statistically significant.

The Chi-squared test\footnote{A Chi-squared test is any statistical hypothesis test in which the sampling distribution of the test statistic is a chi-squared distribution when the null hypothesis is true, or any in which this is asymptotically true, meaning that the sampling distribution (if the null hypothesis is true) can be made to approximate a chi-squared distribution as closely as desired by making the sample size large enough. It can be used for dichotomous variables, as mortality.} has been performed to compare mortality between the 5 \group and the T-test\footnote{A T-test is any statistical hypothesis test in which the test statistic follows a Student's t distribution if the null hypothesis is supported. It is most commonly applied when the test statistic would follow a normal distribution if the value of a scaling term in the test statistic were known. When the scaling term is unknown and is replaced by an estimate based on the data, the test statistic (under certain conditions) follows a Student's t distribution. It can be used for continuous variables, as length of stay.} has been used for length of stay.

From the tests appears that the results for mortality are not statistically significant in all the 5 \group, while the ones for length of stay are significant for \group 1, 3 and 5.

This confirms that is not clear if diuretics are harmful or not: in fact, even if in the 2 datasets, when comparing the percentages of death between the \group between the treated and untreated patients the chances of survival are increasing of 7\% or 8\%, these differences are not statistically significant and they may be randomly happened. Instead it is easier to deduce that the length of stay in the ICU increases for the patients who got diuretics, according to the average results between 5 and 6 days.

From this analysis can be concluded that, in general, diuretics should not be given as they do not seem to make difference in the chances of survival of the patients, while by giving them the length of stay in ICU is lengthened.

This analysis can not provide any information on the conditions when diuretics should be given and this will be the object of Chapter~\ref{cpt:outcomeAnalysis}.

\clearpage \mbox{} \clearpage 
\chapter{\OA}
\label{cpt:outcomeAnalysis}

\section{Introduction}
\label{cpt:outcomeAnalysisIntroduction}
In this Chapter, Step~3 of the analysis, will be described. While the propensity score analysis of Chapter~\ref{cpt:propensityAnalysis} provides balanced quintiles of patients with respect to propensity of \gettingdiuretics , it does not account for confounding factors which might also affect \mort and \los . An obvious confounding factor is a patient's \illness , i.e. underlying illness. In this Chapter the question that have been answered is:
\nibf{Does \gettingdiuretics have a statistically significant effect on \mort or \los? If so, to what extent?}

Therefore, is first described a modeling methodology for:
\begin{description}
\item[A.] Determining if, when \illness is taken into account, \gettingdiuretics has a significant effect on outcome (\mort or \los).
\item[B.] Determining, if \gettingdiuretics has no significant effect in (A), whether the cross-variable interaction of \gettingdiuretics and \illness, has a significant effect on outcome (\mort or \los).
\item[C.] Given (B), i.e. that \gettingdiuretics has a significant effect on outcome, determining if \gettingdiuretics crossed with \illness has a significant effect on outcome when the study group is adjusted according to \illness .
\end{description}
Each determination involves the regression of a model and statistical determination of effect.  There is 1 step in the methodology corresponding to each of the determinations:
\begin{description}
\item[Step A.] Use the study group to regress propensity score, \illness and \diureticsDecision as independent variables for the outcome \mort using a logistic regression.  For the \los  outcome, use generalized linear regression. These models will be labelled by \ModelA{} and append to the model label either '\Mort ' or 'LOS' for \los , e.g. \ModelA.Mortality and \ModelA.LOS .

Set up a null hypothesis that \diureticsDecision has no significant effect on outcome.  Examine the \pvalue of \gettingDiureticsvar.  If the \pvalue $<0.05$, accept the null hypothesis. If the null hypothesis is rejected, revisit the balanced propensity quintiles and consider the result of a chi-squared test for significance of difference in outcome between patients \gettingdiuretics and not with \gettingdiuretics to be conclusive.

For more information on statistical hypothesis tests and use of \pvalue see Appendix~\ref{apx:stats}.
\item[Step B.] If the effect of \gettingdiuretics is \textit{NOT} significant, use the study group to regress the same variables as Step~A while adding a new variable \SAPSD. Such a model will be labelled as \ModelB. Set up a null hypothesis that the new cross-variable \SAPSD has no significant effect on outcome. Examine the \pvalue of this variable. If the \pvalue$<0.05$, reject the null hypothesis and proceed to Step~C.
\item[Step~C.] Divide the study group into 2 subsets according to \illness by using \SAPSTZ 's median value as a threshold. Repeat Step~A with each subset and evaluate the null hypothesis that \diureticsDecision has no significant effect on outcome, in the cases of a subset of sick and another of less sick patients.
\end{description}
In the following sections the modeling methodology will be demonstrated.

\clearpage

\section{Confounding Factors}
\label{cpt:confounding}
A confounding factor in a study is a variable which is related to one or more of the variables defined in the study. A confounding factor may mask an actual association or falsely demonstrate an apparent association between the study variables where no real association between them exists. If confounding factors are not measured and considered, bias may result in the conclusion of the study\cite{CONFOUNDING}.

\begin{figure}[htbp]
\centering
\includegraphics[width=\textwidth]{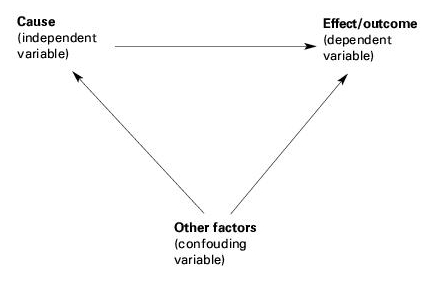}
\caption[Confounding Factors]{A confounding factor in a study is a variable which is related to one or more of the variables defined in a study. A confounding factor may mask an actual association or falsely demonstrate an apparent association between the study variables where no real association between them exists. If confounding factors are not measured and considered, bias may result in the conclusion of the study.}
\label{fig:conf-factors}
\end{figure}

\clearpage

\section{Step~A, \ModelA: \illness and propensity adjustment.}
\label{cpt:modelA}
The purpose of Step~A is to determine, when \illness is taken into account, whether \gettingdiuretics has a significant effect on outcome (\mort or \los). For the regression of both \ModelA.Mortality and \ModelA.LOS , the patient's \diureticsDecision as \textbf{$x_1$} and his/her propensity score (as calculated in Chapter~\ref{cpt:datasetExtraction}) as \textbf{$V_1$} have been included. To express \illness the following independent variables \textbf{$x_2$}, \textbf{$x_3$}, \textbf{$x_5$}, \textbf{$x_{10}$} and \textbf{$x_{15}$} have been chosen: Age, Gender, \SAPSTZ , \SOFATZ , \Elix.
\begin{table}
\centering
\begin{tabular}{|c|c|c|c|c|c|}
\hline \multicolumn{2}{|c|}{\textbf{ \ModelA}} & \Mort & \Mort & LOS & LOS\\
\multicolumn{2}{|c|}{\textbf{}} & \pvalue & $\beta_{i,1}$ & \pvalue & $\beta_{i,2}$\\
\hline
\hline $x_1$ & \textbf{\DiureticsDecision} & \textbf{0.075} & \textbf{-0.189} & \colorbox{yellow}{$<0.001$} & \textbf{2.626}\\
\hline $x_2$ & Age & \colorbox{yellow}{$<0.001$} & 0.023 & \colorbox{yellow}{$<0.001$} & -0.078\\
\hline $x_3$ & Gender & 0.410 & 0.048  & 0.753 & -0.092\\
\hline $x_5$ & \SAPSTZ & \colorbox{yellow}{$<0.001$} & 0.053 & \colorbox{yellow}{0.004} & 0.202\\
\hline $x_{10}$ & \SOFATZ & \colorbox{yellow}{$<0.001$} & 0.125 & 0.649 & -0.039\\
\hline $x_{15}$ & \Elix & 0.095  & 0.058 & \colorbox{yellow}{0.019} & -0.409\\
\hline $V_1$ & Propensity Score & 0.478  & -0.289 & \colorbox{yellow}{$<0.001$} & 11.795\\
\hline
\end{tabular}
\caption[Effects in \ModelA.Mortality and \ModelA.LOS]{Effects of variables in \ModelA.Mortality and \ModelA.LOS. For both \mort and \los outcomes, \illness variables $x_2$, $x_3$, and $x_4$ (Age, \SAPSTZ, and \SOFATZ) have statistically significant effects and thus are highlighted with yellow cells. \GettingDiuretics instead appears to have a significant effect only for \los. The positive $\beta$ coefficient sign implies \los increases when \diureticsDecision is true.}
\label{tab:ModelA}
\end{table}
Table~\vref{tab:ModelA} shows the \pvalue and \betacoeff analyses for both outcomes. With respect to \mort, \illness variables $x_2$, $x_3$, and $x_10$ (Age, \SAPSTZ, and \SOFATZ) have statistically significant effects. The \diureticsDecision does not have a statistically significant effect on \mort. With respect to \los , illness variables $x_2$, $x_4$, $x_{15}$ (Age, \SAPSTZ , and \Elix) have statistically significant effects, as does propensity score. Importantly, and in contrast to \mort, the null hypothesis that the effect of \diureticsDecision is not significant on \los , is rejected (\pvalue $<0.001$). For \los outcomes, these findings imply \illness is not a confounding factor and \diureticsDecision is independently significant in its effect on \LOS. One can go back to each propensity quintile and, where there is sufficient data, examine the T-test of the difference between the \los outcome distribution for patients with \gettingdiuretics and those without. In this case, the test indicates for \group~ 1, \group~ 3 and \group~ 5 a significant difference, leading to the conclusion, qualified for this study group, that \gettingdiuretics increases a patient's \los in the ICU.

\section{Step~B}
\label{cpt:modelB}
For \mort outcome, the \diureticsDecision is \textit{NOT} independently significant in its effect. However, it may be there is an interactive effect of \diureticsDecision with \illness that is more than random. Therefore, what happens in modeling when the cross-interaction variable \SAPSD, \textbf{$x_1 \cdot x_5$}, is included will now be examined. Table~\vref{tab:ModelB} columns 2 and 3 show the \pvalue and \betaCoeff analyses for this \mort outcome logistic regression. The null hypothesis that \SAPSD has no significant cross-dependent effect on \mort is rejected given the \pvalue$=0.013$. Therefore, Step~C have been performed and two models generated: \ModelCLessSick and \ModelCSicker using two subsets divided by relative \SAPSTZ median within the study group.

\section{Step~C, \ModelC: \Illness Split and New Adjustment Models}
\label{cpt:modelC}

\subsection{Splitting the Study Group by \Illness}
\label{cpt:splittingGroup}
The 2 \illness subsets are divided across the median \SAPSTZ score of 17. The less sick subset is composed of 816 patients and the sicker subset has 706 patients. Descriptive statistics of the two subsets in terms of Age, \SAPSONE and \SOFAONE are provided in Figures~\vref{fig:histo_1_less_sick} for the less sick subset and Figures~\vref{fig:histo_2_sicker} for the sicker one. Other descriptive statistics of the subsets are provided in Appendix~\ref{apx:dataset}.

\begin{figure}
\centering
\rotatebox{90}{
\stepcounter{figure}
\begin{minipage}{\textheight}
\begin{minipage}{0.33\textheight}
\centering
\includegraphics[width=.9\linewidth]{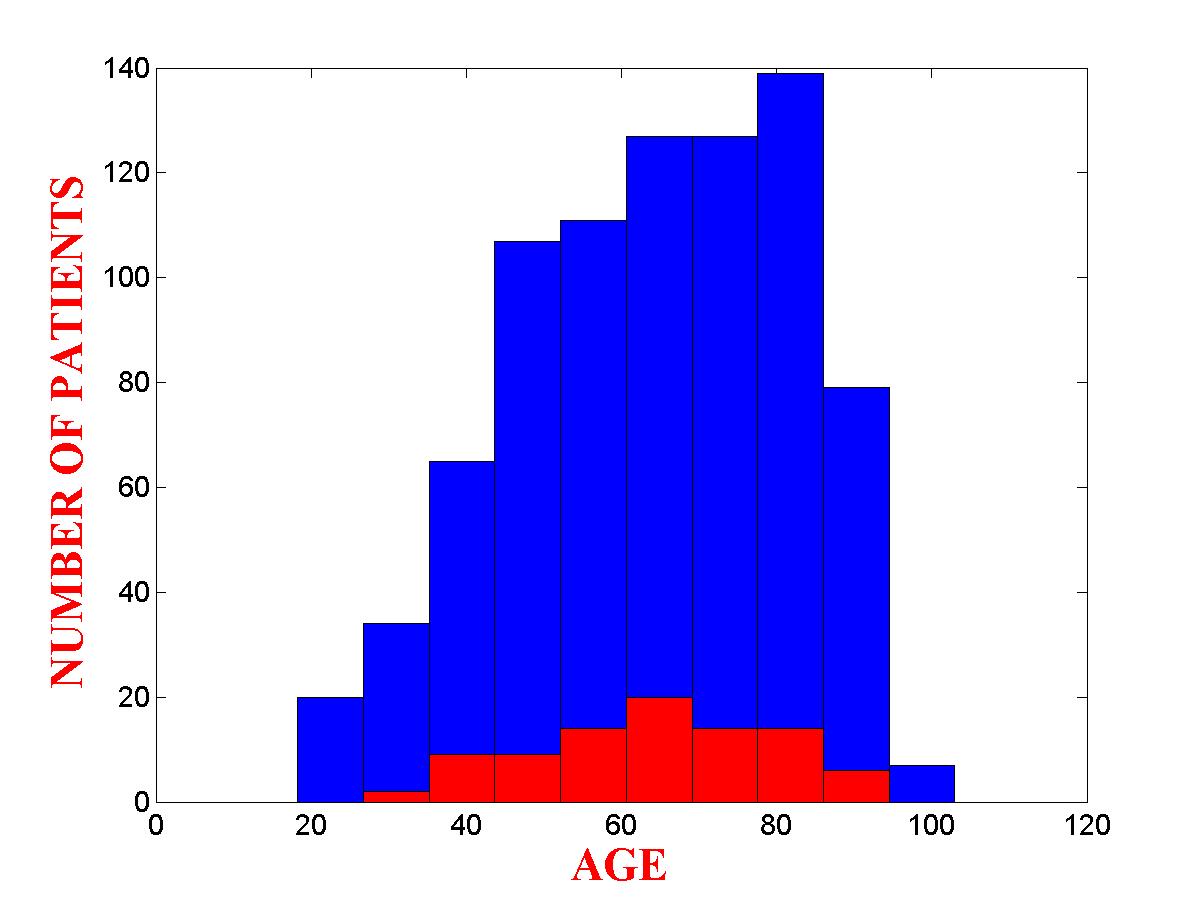}
\captionof{subfigure}[]{Age is centered on 65 years.}
\label{fig:age_1}
\end{minipage}
\begin{minipage}{0.33\textheight}
\centering
\includegraphics[width=.9\linewidth]{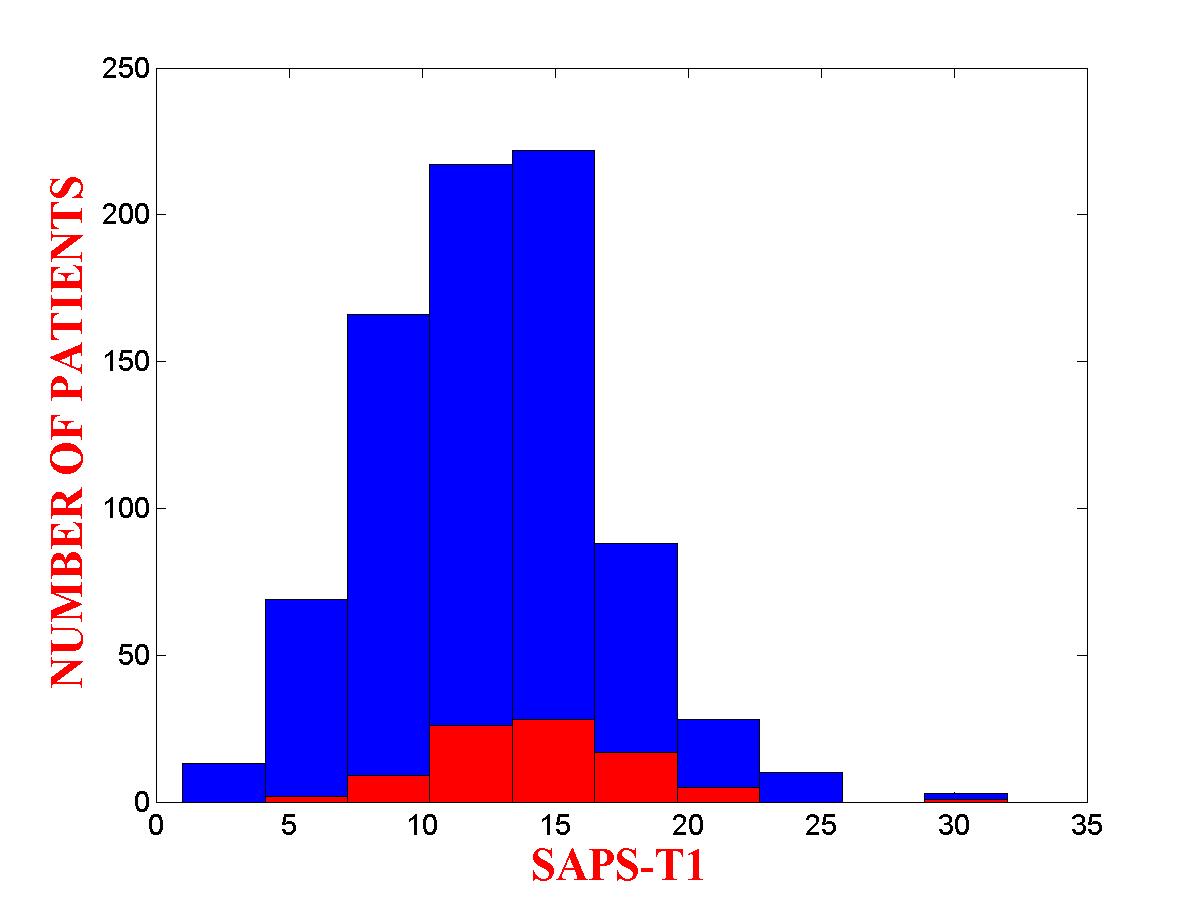}
\captionof{subfigure}[]{\SAPSONE is centered on 13.}
\label{fig:saps_1}
\end{minipage}
\begin{minipage}{0.33\textheight}
\centering
\includegraphics[width=.9\linewidth]{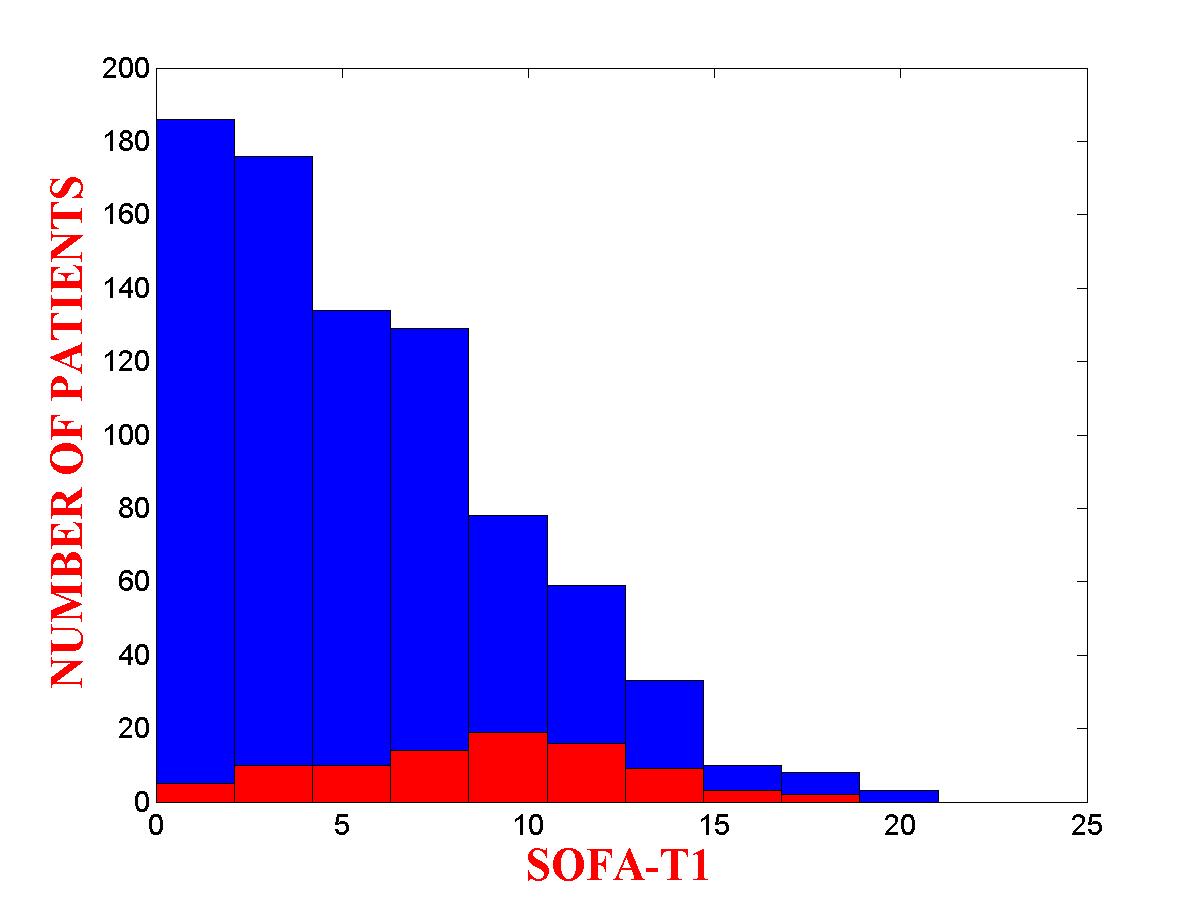}
\captionof{subfigure}[]{\SOFAONE is centered on 5.}
\label{fig:sofa_1}
\end{minipage}
\addtocounter{figure}{-1}
\captionof{figure}[Age, \SAPSONE and \SOFAONE in the subsets formed by \SAPSTZ median for \ModelCLessSick .]{Histogram of Age, \SAPSONE and \SOFAONE in the subsets divided by \illness for \ModelCLessSick . In red the values for the $D^+$ patients only.}
\label{fig:histo_1_less_sick}
\end{minipage}
}
\end{figure}
\begin{figure}
\centering
\rotatebox{90}{
\stepcounter{figure}
\begin{minipage}{\textheight}
\begin{minipage}{0.33\textheight}
\centering
\includegraphics[width=.9\linewidth]{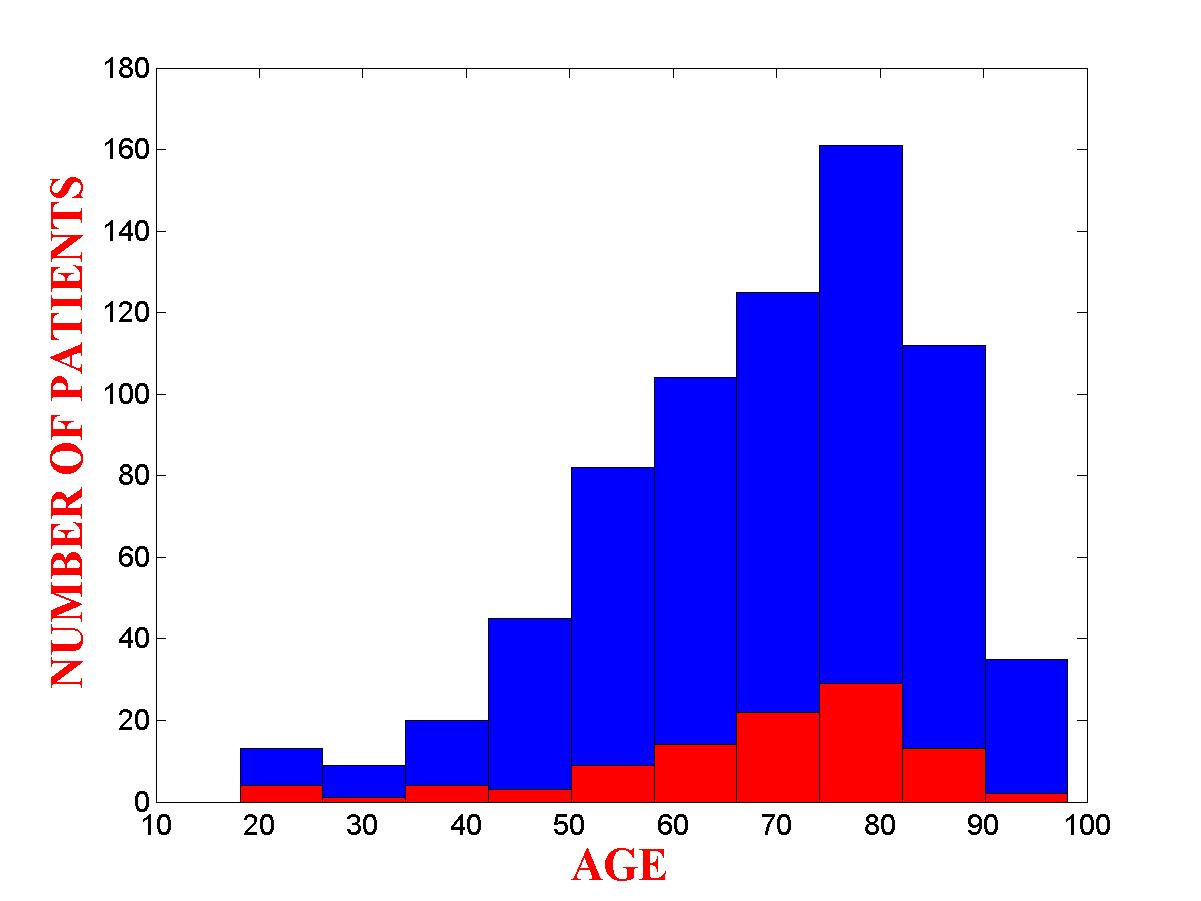}
\captionof{subfigure}[]{Age is centered on 71 years.}
\label{fig:age_2}
\end{minipage}
\begin{minipage}{0.33\textheight}
\centering
\includegraphics[width=.9\linewidth]{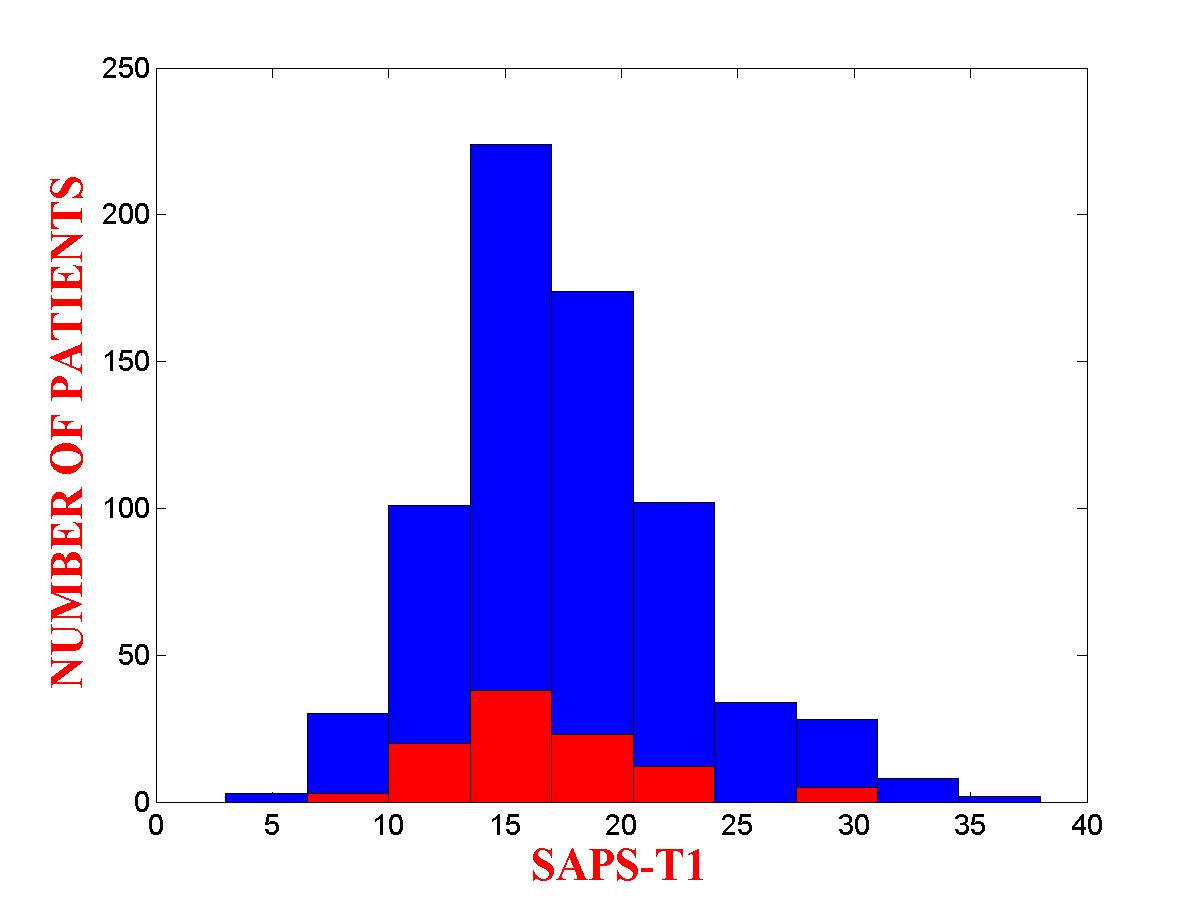}
\captionof{subfigure}[]{\SAPSONE is centered on 17.}
\label{fig:saps_2}
\end{minipage}
\begin{minipage}{0.33\textheight}
\centering
\includegraphics[width=.9\linewidth]{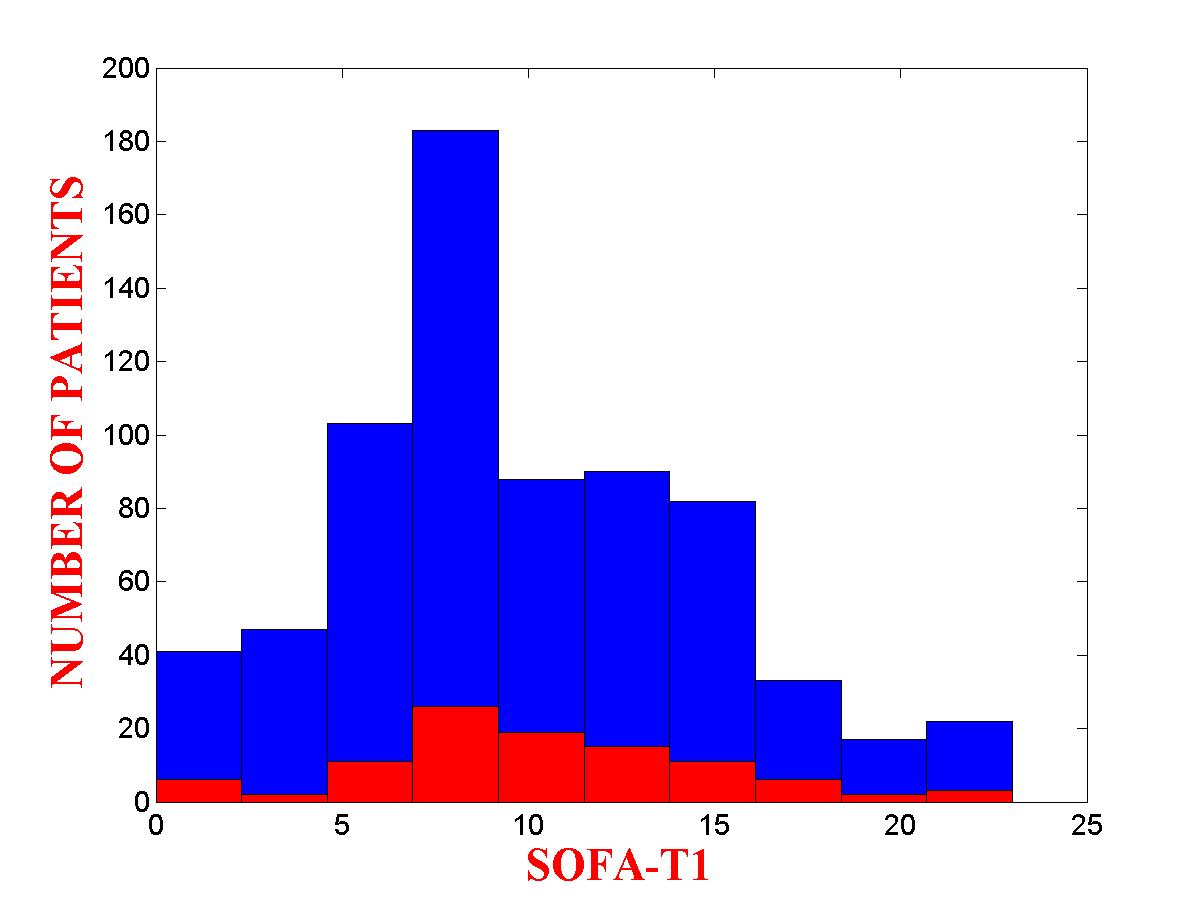}
\captionof{subfigure}[]{\SOFAONE is centered on 9.}
\label{fig:sofa_2}
\end{minipage}
\addtocounter{figure}{-1}
\captionof{figure}[Age, \SAPSONE and \SOFAONE in the subsets formed by \SAPSTZ median for \ModelCSicker .]{Histogram of Age, \SAPSONE and \SOFAONE in the subsets divided by \illness for \ModelCSicker . In red the values for the $D^+$ patients only.}
\label{fig:histo_2_sicker}
\end{minipage}
}
\end{figure}
As predicted, all the clinical values for the sicker group are generally worst.

\subsection{\ModelCLessSick and \ModelCSicker: New Adjustment Models}
\label{cpt:modelCNew}
Table~\vref{tab:ModelB} columns 4 and 5 show the \pvalue and \betaCoeff analyses for the \mort outcome regression on the less sick subset. The  null hypothesis that \diureticsDecision has a no significant cross-dependent effect on \mort in the less sick subset is rejected. Columns 6 and 7 show the \pvalue and \betaCoeff analyses for the \mort outcome regression on the sick subset. The null hypothesis that \diureticsDecision has a no significant cross-dependent effect on \mort in the sicker subset is \textit{NOT} rejected.
\begin{table}
\centering
\begin{tabular}{|c|c|c||c|c||c|c|}
\hline {\bf Var} & {\bf Model~B} &  & {\bf Model~C} &  & {\bf Model~C} &\\
{\bf } & \pvalue & $\beta_{i,0}$ & Less Sick & $\beta_{i,1}$ & Sicker & $\beta_{i,2}$\\
\hline $x_1$ & 0.069 & 0.602 & \colorbox{yellow}{0.004} & 1.842 & \textcolor{red}{0.531} & -0.50\\
\hline $x_2$ & $<0.001$ & 0.023 & $<0.001$ & 0.0234 & $<0.001$ & 0.025\\
\hline $x_3$ & 0.347 & 0.054 & 0.445 & 0.064 & 0.760 & 0.025\\
\hline $x_5$ & 0.278 & 0.021 & 0.270 & -0.054  & 0.244 & 0.043\\
\hline $x_{10}$ & $<0.001$ & 0.123 & $<0.001$ & 0.094 & $<0.001$ & 0.141\\
\hline $x_{15}$ & 0.100 & 0.057 & 0.645 & 0.023 & 0.067 & 0.092\\
\hline $V_1$ & 0.576 & -0.224 & 0.093 & 1.036 & 0.038 & -1.148\\
\hline $x_1 \cdot x_5$ & \colorbox{yellow}{0.013} & -0.043 & 0.001 & -0.145 & 0.709 & 0.013\\
\hline
\end{tabular}
\caption[\ModelB and \ModelC Analysis]{\ModelB Analysis (columns 2 and 3) indicates that the cross product variable \SAPSD \textbf{$(x_1 \cdot x_5)$} has a significant effect on \mort.  \ModelC analysis (columns 4 and 5 for the less sick subset, and columns 6 and 7 for the sicker subset. In the less sick subset, \diureticsDecision is a significant independent variable effect, whereas in the sicker subset, it is not (red font).}
\label{tab:ModelB}
\end{table}

\subsection{Stratification Analysis with Adjustment for Confounding Factor of \Illness}
\label{cpt:strataAnalysis}

Returning to quintile analysis: the \gs has been divided by \SAPSTZ median threshold into 2 groups.  

Descriptive statistics of \groups~4 and 5 in terms of Age, \SAPSONE and \SOFAONE is provided in Figures~\vref{fig:histo_3_less_sick} for the less sick subset and Figures~\vref{fig:histo_4_sicker} for the sicker one.

All patients of a group are ranked by propensity score and divide the ranked group into 5 quintiles of equal size.   In a \group, in each \illness subset, the \mort rate for those patients with \gettingdiuretics and those which did not were compared. In this case the null hypothesis is that the outcomes come from the same distribution. To test the null hypothesis, the Chi-Squared test has been used.

The \illness adjusted stratification analysis is summarized in Table~\vref{tab:lessSick}. In the less sick subset of \group~4 ($PS \in \left[ 0.06;0.15 \right] $) \mort rate is significantly less for the patients with \gettingdiuretics compared to those without. It is not significantly different for \group~5 ($PS \in \left[ 0.15;0.99 \right] $). The \mort rate is not significant in \group{s} 1, 2 and 3 either.

For the sicker group, see Table~\vref{tab:Sick}, \mort rate is not significantly different for \groups~1 to 5.

\clearpage

\begin{figure}
\centering
\rotatebox{90}{
\stepcounter{figure}
\begin{minipage}{\textheight}
\begin{minipage}{0.33\textheight}
\centering
\includegraphics[width=.9\linewidth]{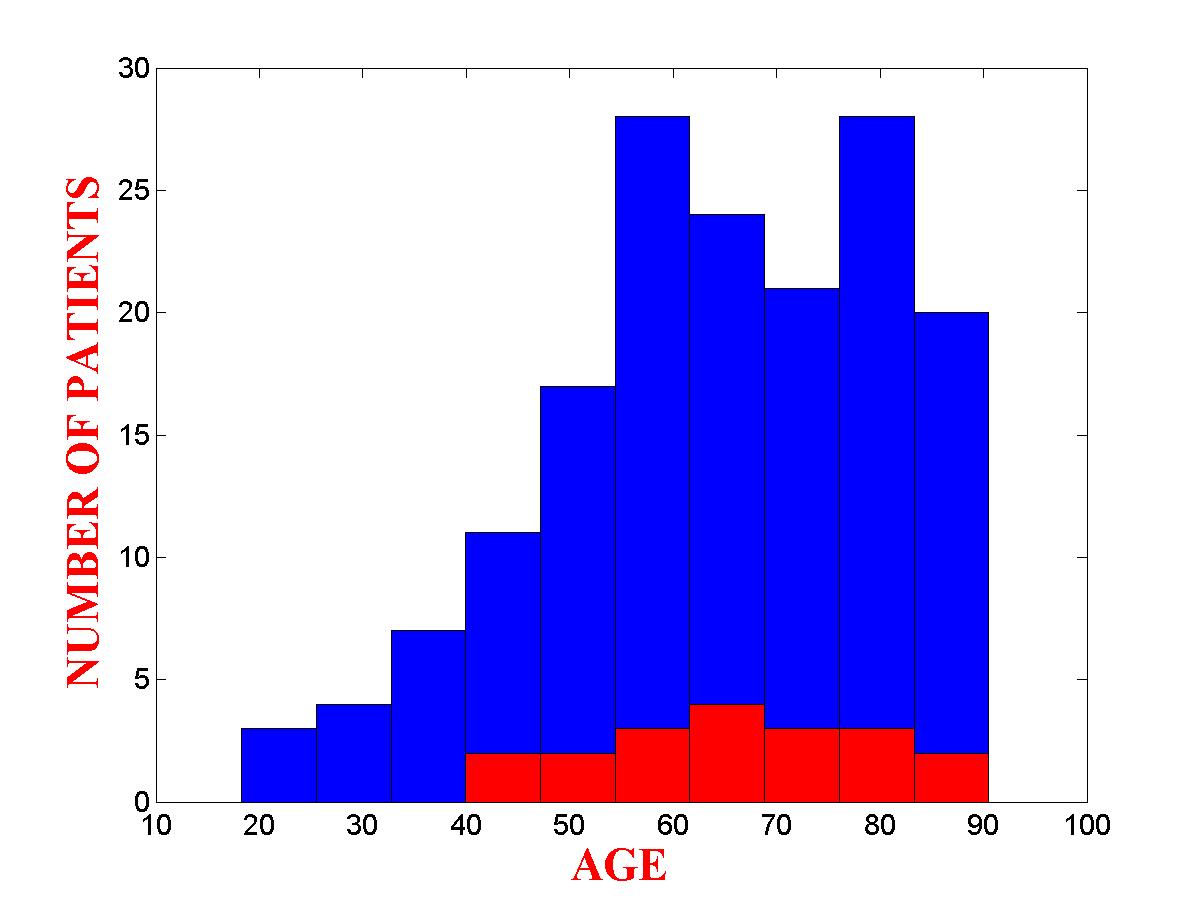}
\captionof{subfigure}[]{Age is centered on 65 years.}
\label{fig:age_1_2}
\end{minipage}
\begin{minipage}{0.33\textheight}
\centering
\includegraphics[width=.9\linewidth]{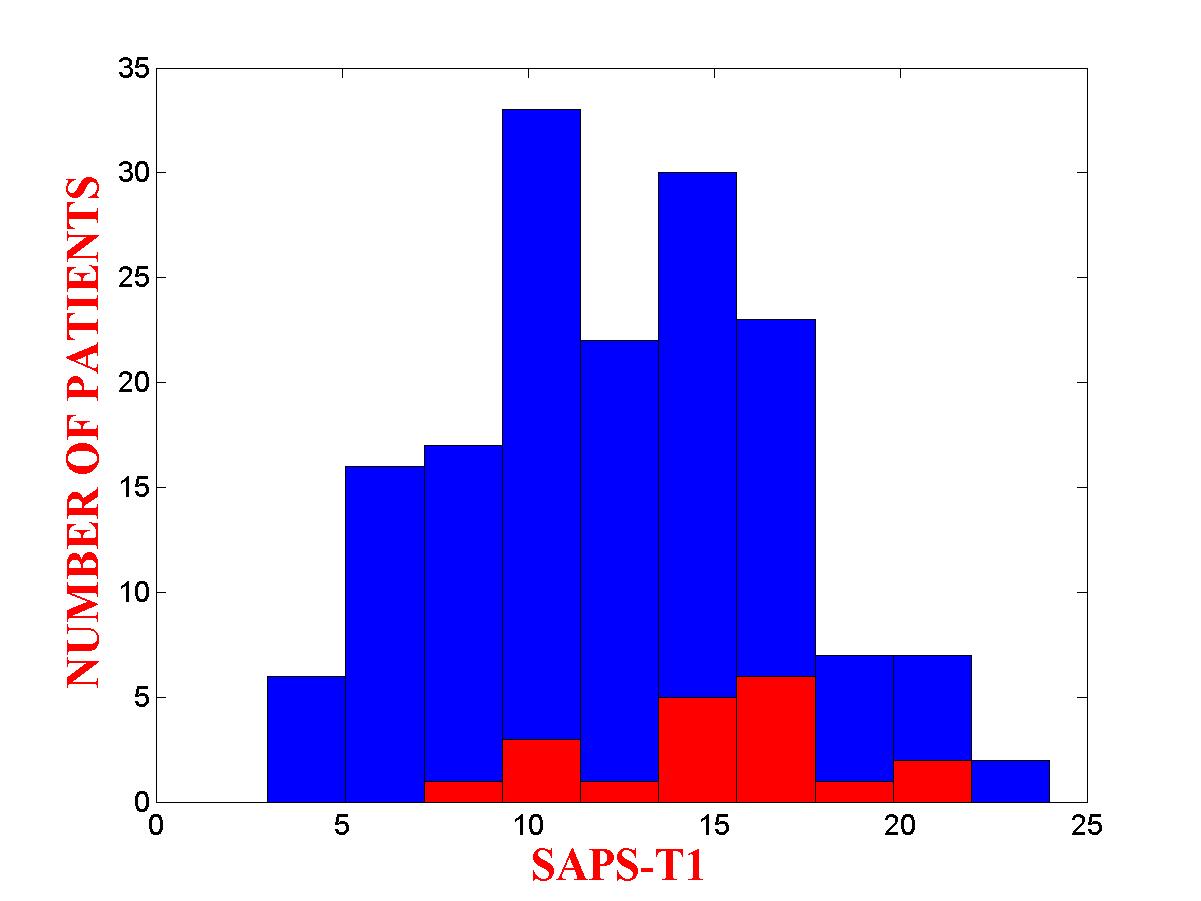}
\captionof{subfigure}[]{\SAPSONE is centered on 12.}
\label{fig:saps_1_2}
\end{minipage}
\begin{minipage}{0.33\textheight}
\centering
\includegraphics[width=.9\linewidth]{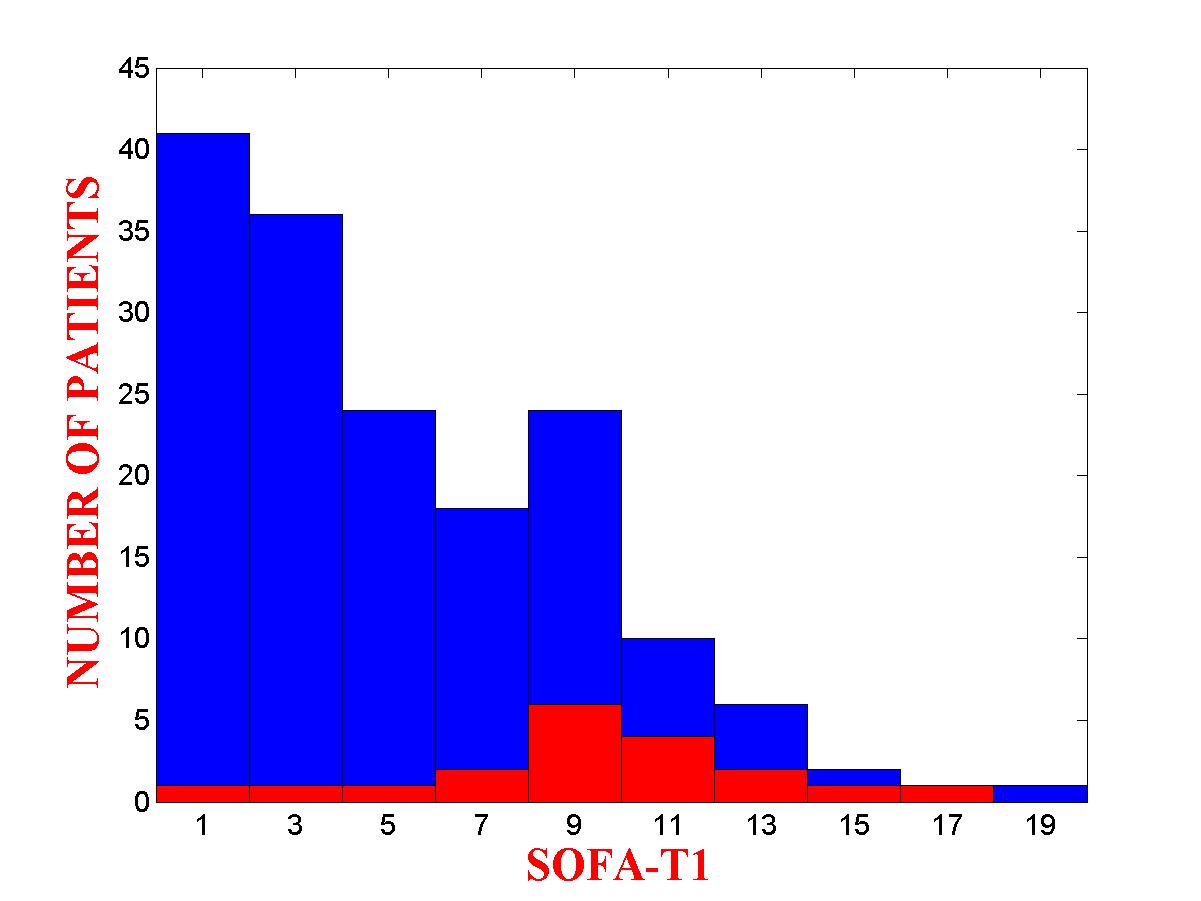}
\captionof{subfigure}[]{\SOFAONE is centered on 5.}
\label{fig:sofa_1_2}
\end{minipage}
\begin{minipage}{0.33\textheight}
\centering
\includegraphics[width=.9\linewidth]{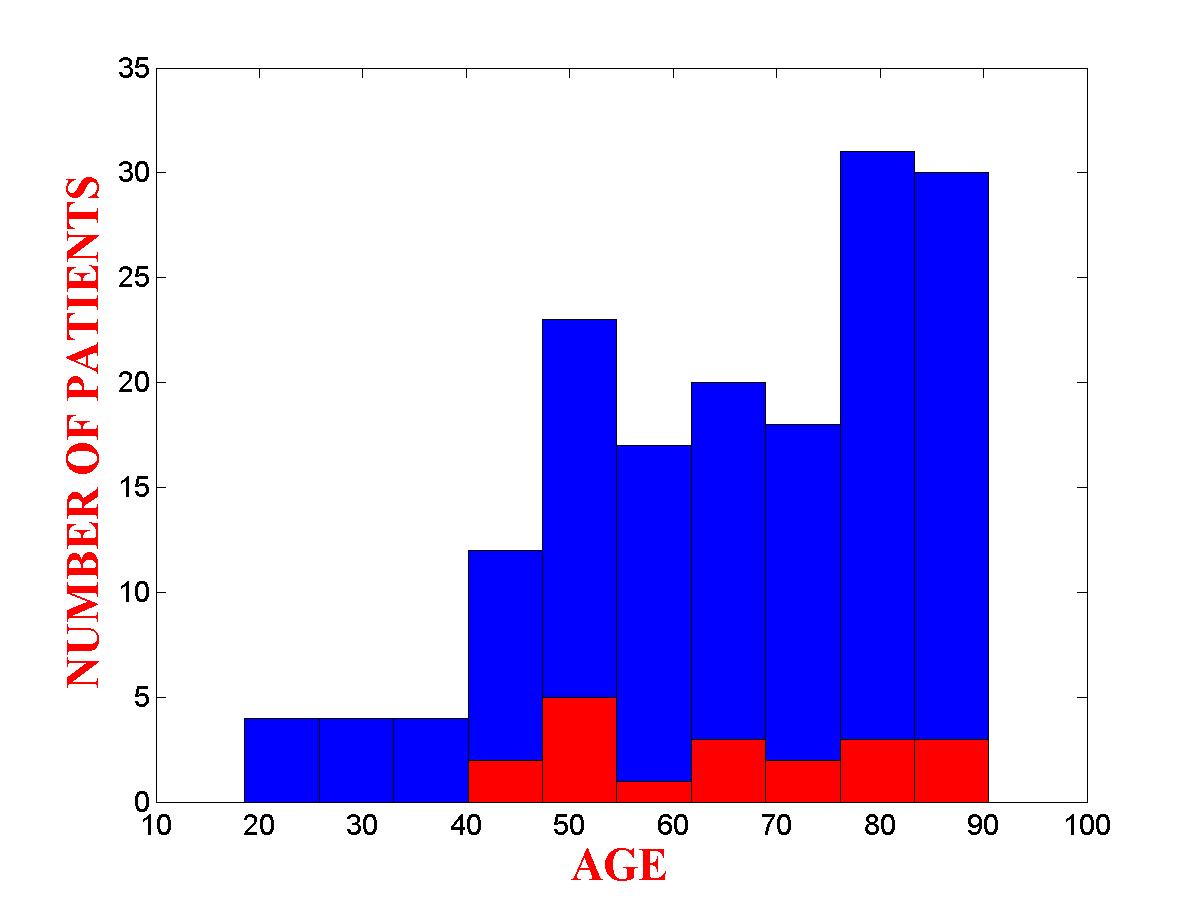}
\captionof{subfigure}[]{Age is centered on 72 years.}
\label{fig:age_1_3}
\end{minipage}
\begin{minipage}{0.33\textheight}
\centering
\includegraphics[width=.9\linewidth]{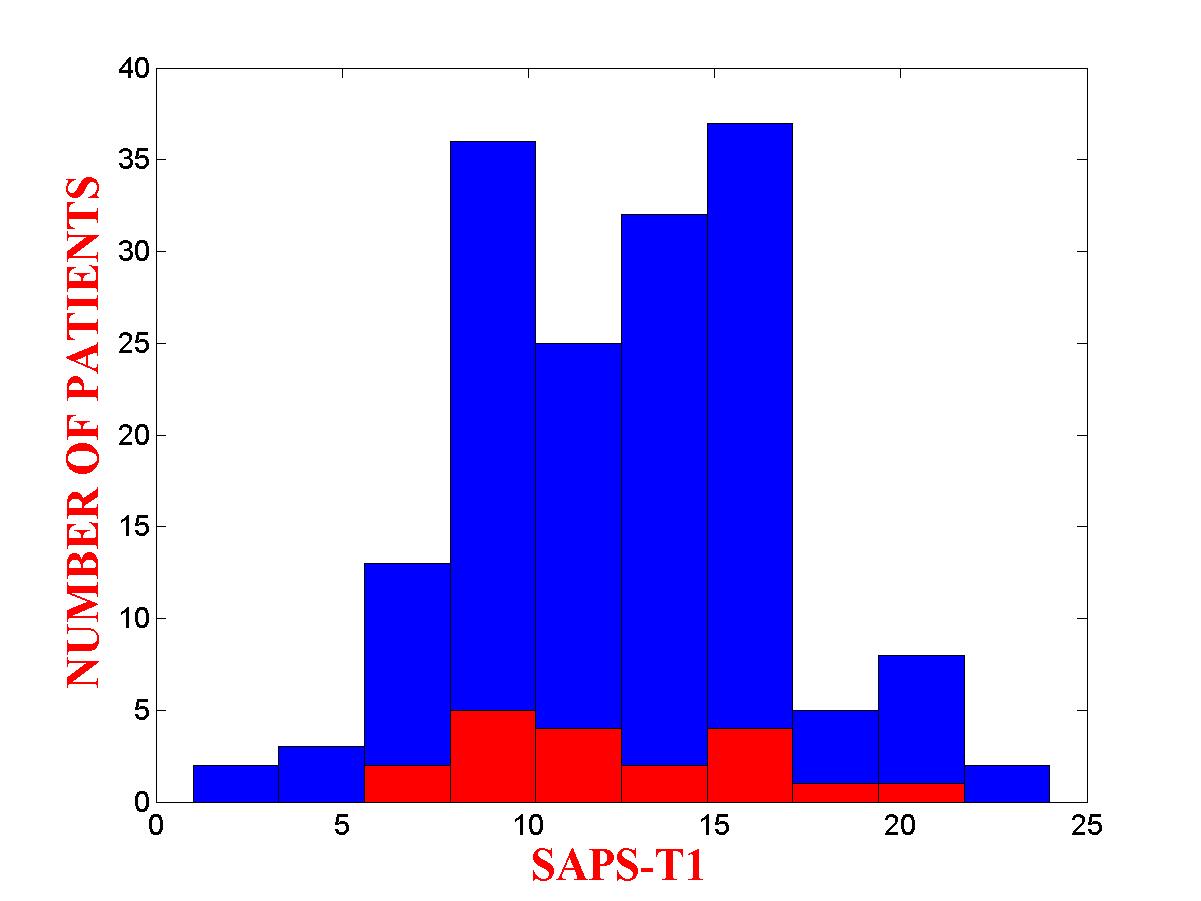}
\captionof{subfigure}[]{\SAPSONE is centered on 17.}
\label{fig:saps_1_3}
\end{minipage}
\begin{minipage}{0.33\textheight}
\centering
\includegraphics[width=.9\linewidth]{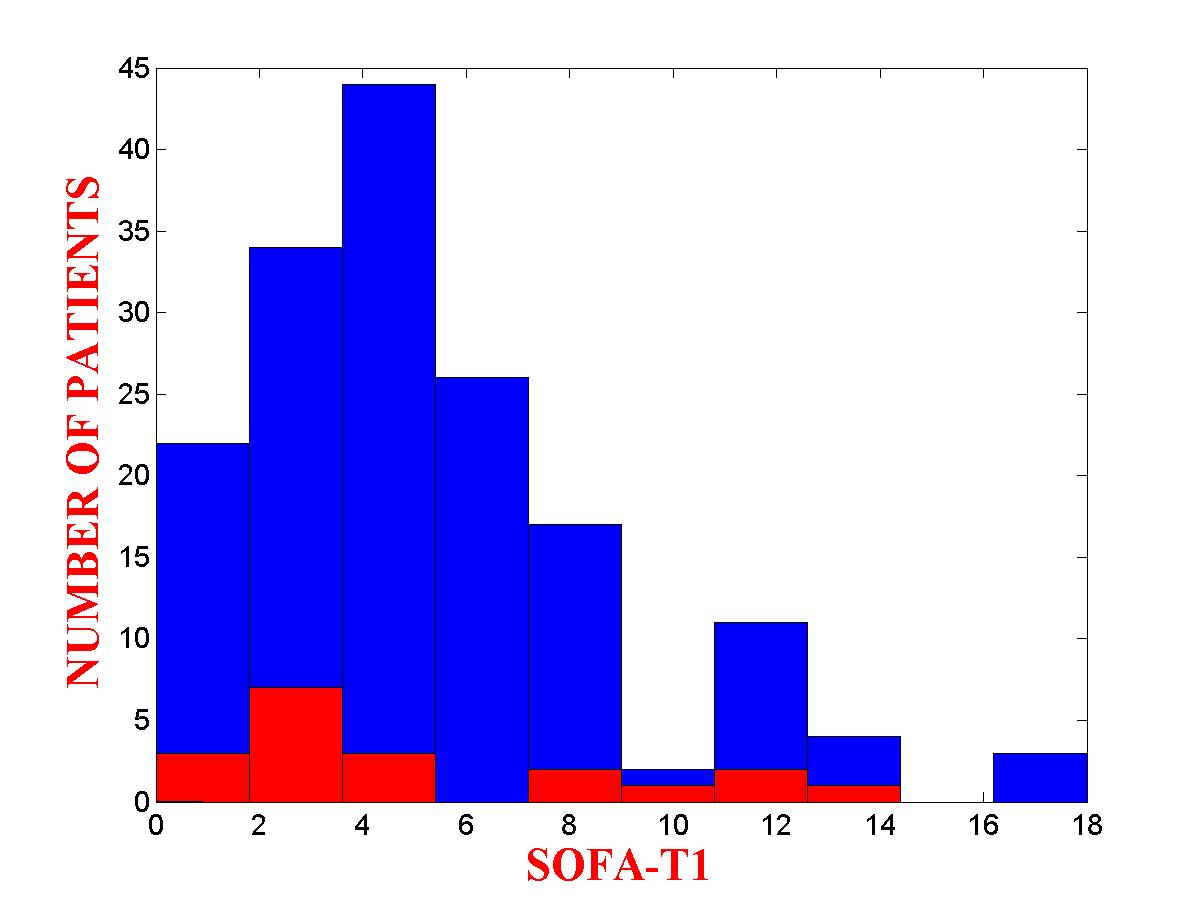}
\captionof{subfigure}[]{\SOFAONE is centered on 10.}
\label{fig:sofa_1_3}
\end{minipage}
\addtocounter{figure}{-1}
\captionof{figure}[Age and \SAPSTZ in \Groups~4 and 5 formed by \SAPSTZ median for \ModelCLessSick .]{Histogram of Age and \SAPSTZ in \groups~4 and 5 formed by \SAPSTZ median for \ModelCLessSick . In red the values for the $D^+$ patients only.}
\label{fig:histo_3_less_sick}
\end{minipage}
}
\end{figure}
\begin{figure}
\centering
\rotatebox{90}{
\stepcounter{figure}
\begin{minipage}{\textheight}
\begin{minipage}{0.33\textheight}
\centering
\includegraphics[width=.9\linewidth]{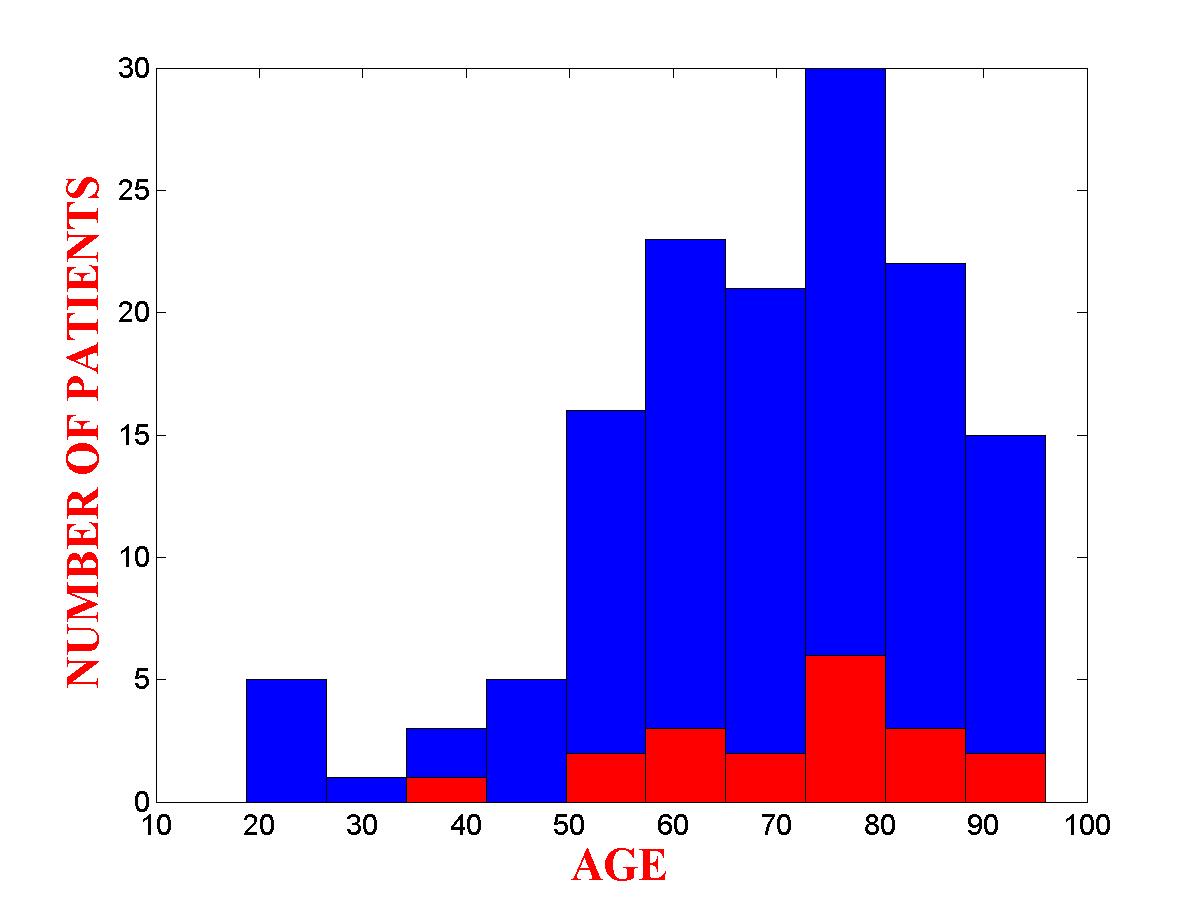}
\captionof{subfigure}[]{Age is centered on 68 years.}
\label{fig:age_2_2}
\end{minipage}
\begin{minipage}{0.33\textheight}
\centering
\includegraphics[width=.9\linewidth]{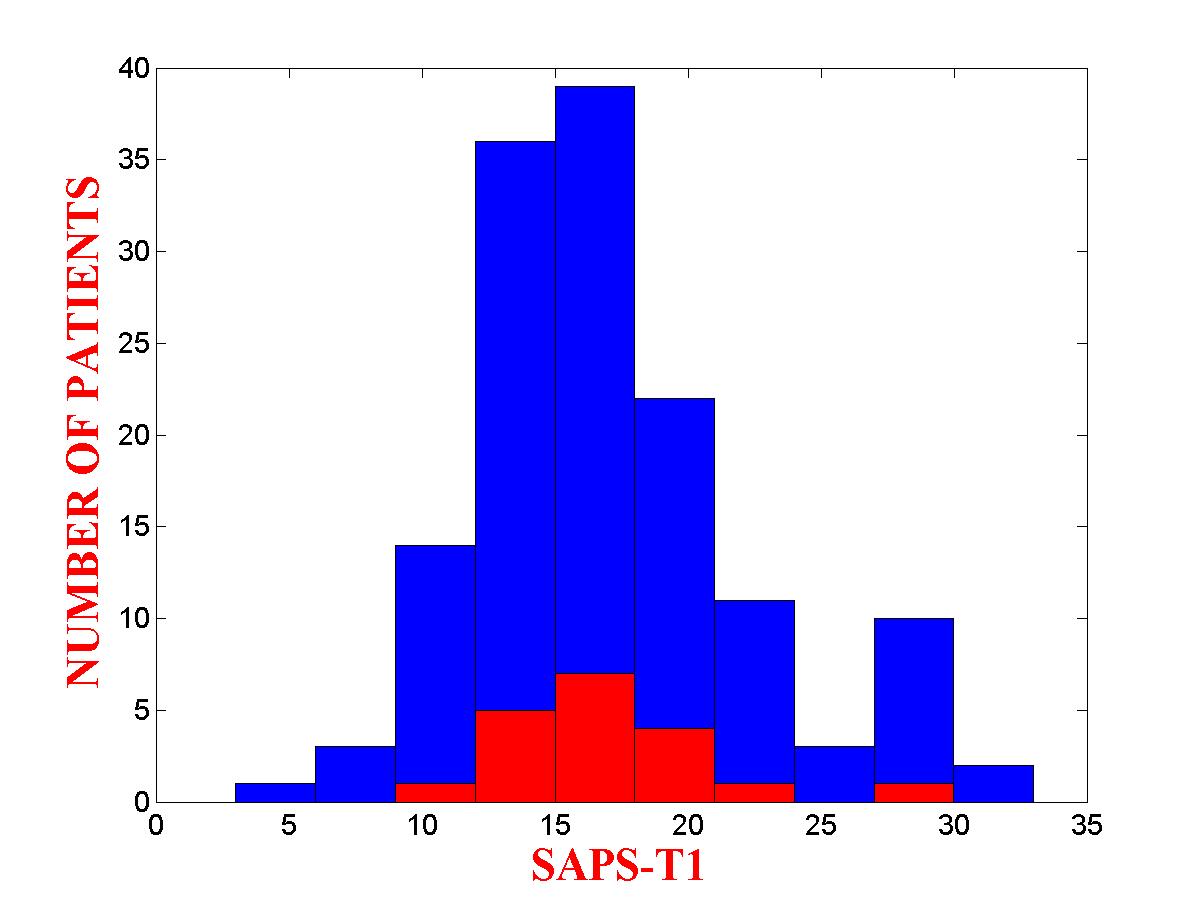}
\captionof{subfigure}[]{\SAPSONE is centered on 13.}
\label{fig:saps_2_2}
\end{minipage}
\begin{minipage}{0.33\textheight}
\centering
\includegraphics[width=.9\linewidth]{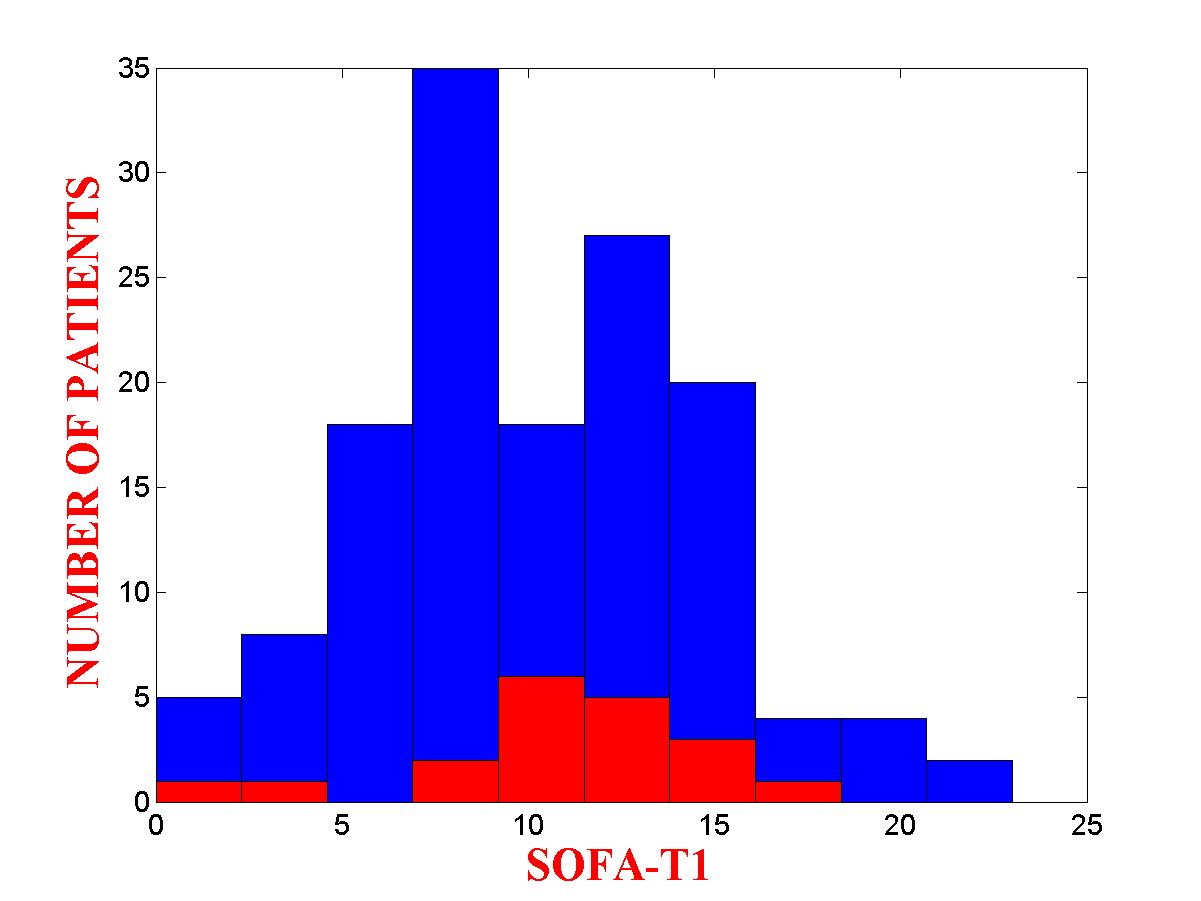}
\captionof{subfigure}[]{\SOFAONE is centered on 4.}
\label{fig:sofa_2_2}
\end{minipage}
\begin{minipage}{0.33\textheight}
\centering
\includegraphics[width=.9\linewidth]{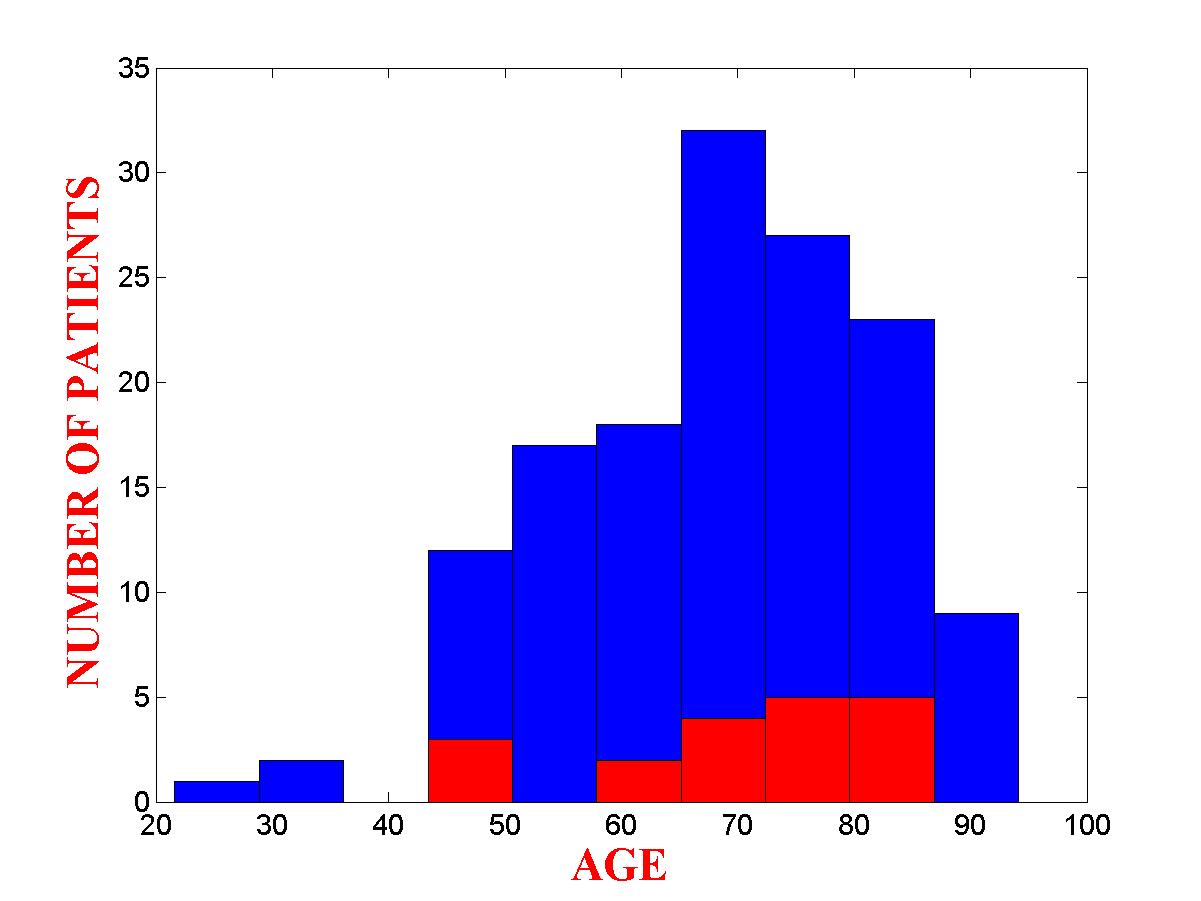}
\captionof{subfigure}[]{Age is centered on 69 years.}
\label{fig:age_2_3}
\end{minipage}
\begin{minipage}{0.33\textheight}
\centering
\includegraphics[width=.9\linewidth]{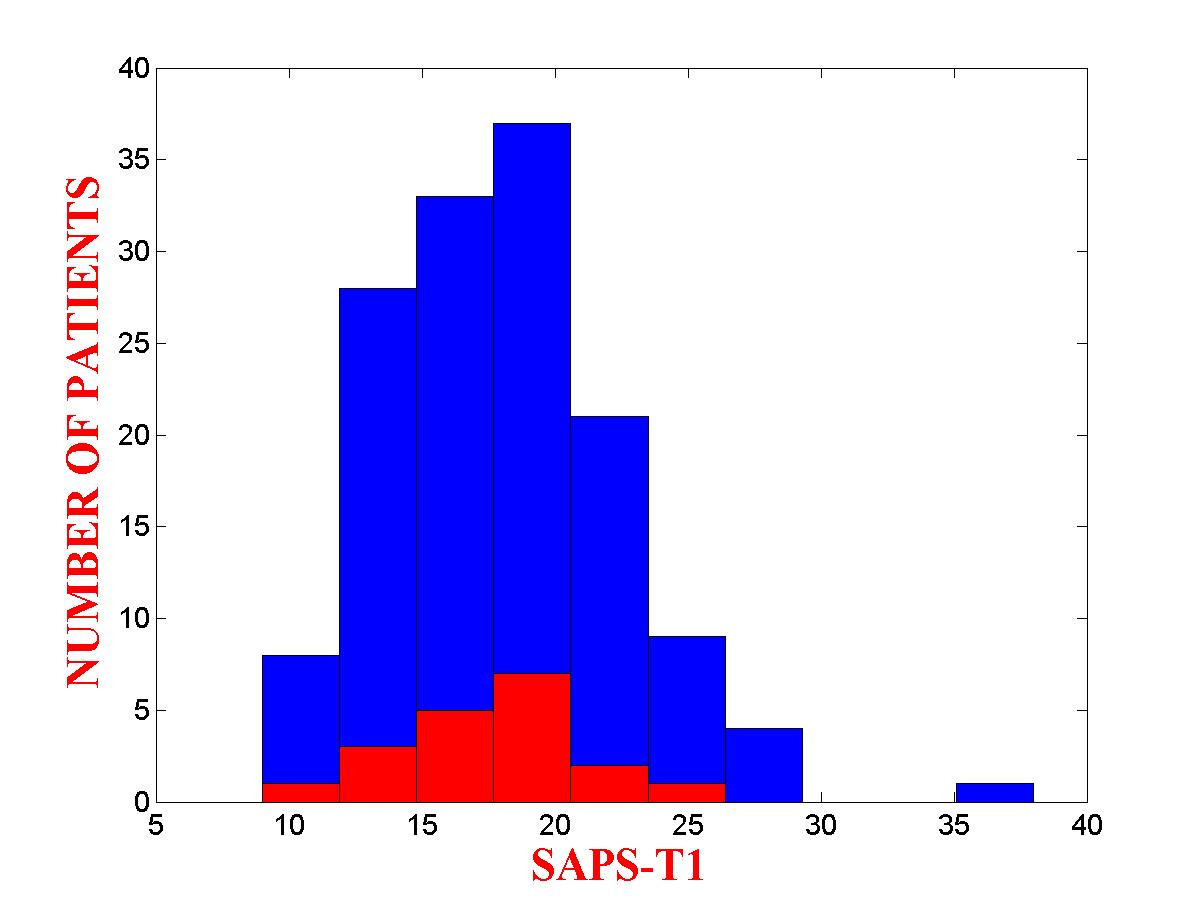}
\captionof{subfigure}[]{\SAPSONE is centered on 18.}
\label{fig:saps_2_3}
\end{minipage}
\begin{minipage}{0.33\textheight}
\centering
\includegraphics[width=.9\linewidth]{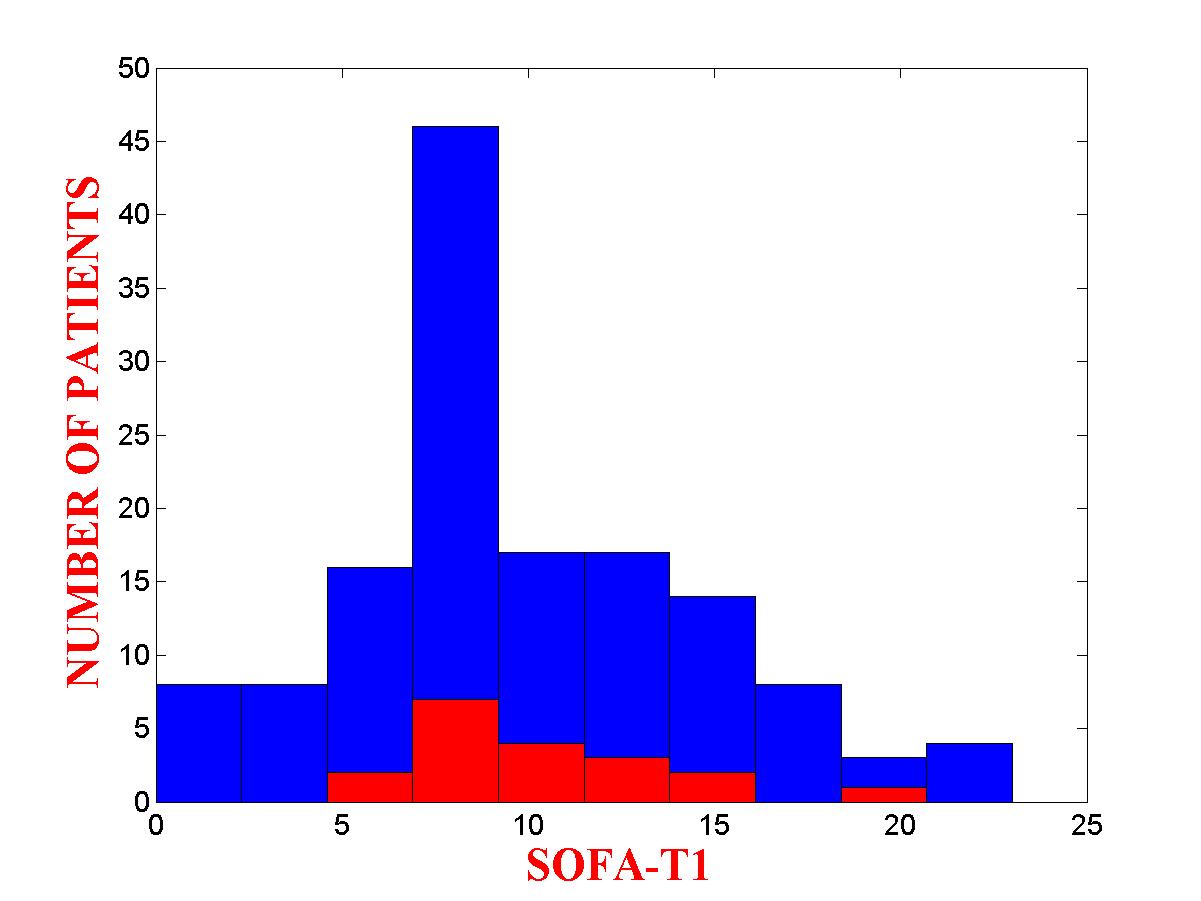}
\captionof{subfigure}[]{\SOFAONE is centered on 9.}
\label{fig:sofa_2_3}
\end{minipage}
\addtocounter{figure}{-1}
\captionof{figure}[Age and \SAPSTZ in \Groups~4 and 5 formed by \SAPSTZ median for \ModelCSicker .]{Histogram of Age and \SAPSTZ in \groups~4 and 5 formed by \SAPSTZ median for \ModelCSicker . In red the values for the $D^+$ patients only.}
\label{fig:histo_4_sicker}
\end{minipage}
}
\end{figure}

\clearpage

\begin{table}
\centering
\begin{tabular}{|c|c|c|}
\hline
\hline {\bf \group~ 1 $PS \in \left[ 0.00;0.01 \right] $} & $D^+$ & $D^-$\\
\hline Number of patients & 0 & 163\\
\hline Deaths & 0\% & 14\% \\
\hline
\hline {\bf \group~ 2 $PS \in \left[ 0.01;0.02 \right] $} & $D^+$ & $D^-$\\
\hline Number of patients & 4 & 159 \\
\hline Deaths & 25\% & 24\% \\
\hline
\hline {\bf \group~ 3 $PS \in \left[ 0.02;0.06 \right] $} & $D^+$ & $D^-$\\
\hline Number of patients & 6 & 157 \\
\hline Deaths & 33\% & 24\% \\
\hline
\hline {\bf \group~ 4 $PS \in \left[ 0.06;0.15 \right] $} & $D^+$ & $D^-$\\
\hline Number of patients & 21 & 142 \\
\hline \colorbox{yellow}{Deaths} & \colorbox{yellow}{14\%} & \colorbox{yellow}{28\%}\\
\hline
\hline {\bf \group~ 5 $PS \in \left[ 0.15;0.99 \right] $} & $D^+$ & $D^-$\\
\hline Number of patients & 57 & 106 \\
\hline Deaths & 28\% & 40\% \\
\hline
\hline
\end{tabular}
\caption[Mortality Outcomes after Propensity and Less Sick Stratification]{Mortality outcomes after propensity and less sick stratification.  The difference in \mort is statistically significant (null hypothesis of Chi-Squared test is rejected) in \group~4 only. The difference in \mort is not statistically significant (null hypothesis of Chi-Squared test is not rejected) in \groups~1, 2, 3 and 5.}
\label{tab:lessSick}
\end{table}
\begin{table}
\centering
\begin{tabular}{|c|c|c|}
\hline
\hline {\bf \group~ 1 $PS \in \left[ 0.00;0.02 \right] $} & $D^+$ & $D^-$\\
\hline Number of patients & 4 & 137\\
\hline Deaths & 25\% & 61\% \\
\hline
\hline {\bf \group~ 2 $PS \in \left[ 0.02;0.05 \right] $} & $D^+$ & $D^-$\\
\hline Number of patients & 4 & 137\\
\hline Deaths & 25\% & 46\% \\
\hline
\hline {\bf \group~ 3 $PS \in \left[ 0.05;0.1 \right] $} & $D^+$ & $D^-$\\
\hline Number of patients & 9 & 132\\
\hline Deaths & 22\% & 53\% \\
\hline
\hline {\bf \group~ 4 $PS \in \left[ 0.1;0.21 \right] $} & $D^+$ & $D^-$\\
\hline Number of patients & 15 & 126\\
\hline Deaths & 53\% & 54\%\\
\hline
\hline {\bf \group~ 5 $PS \in \left[ 0.21;0.99 \right] $} & $D^+$ & $D^-$\\
\hline Number of patients & 68 & 73\\
\hline Deaths & 55\% & 45\% \\
\hline
\hline
\end{tabular}
\caption[Mortality Outcomes after Propensity and Sicker Stratification]{Mortality outcomes after propensity and sicker stratification. The difference in \mort is not statistically significant (null hypothesis of Chi-Squared test is not rejected) in \groups~1, 2, 3, 4 and 5.}
\label{tab:Sick}
\end{table}

\clearpage \mbox{} \clearpage 
\chapter{Machine Learning with GP Analysis}
\label{cpt:GPML}

\section{Introduction}
\label{cpt:GPMLIntroduction}
In this Chapter the analysis performed by using GP techniques will be discussed. A description of Machine Learning and Genetic Programming techniques is available in Appendix~\ref{apx:machineLearning}. This analysis used GPLAB, A Genetic Programming Toolbox for MATLAB produced by Sara Silva\footnote{Sara Silva is currently, Summer 2012, senior researcher of the KDBIO group at INESC-ID Lisboa, IST / UTL.}. GPLAB is a genetic programming toolbox for MATLAB and its architecture follows a highly modular and parameterized structure. For a description of the toolbox see\cite{GPLAB-GUIDE}.

GP was used to classify the \gs on \mort and to evolve symbolic regression to predict \los . For this analysis the 8 variables in Table~\vref{tab:ModelGP} were used.
\begin{table}
\centering
\begin{tabular}{|c|c|}
\hline {\bf Var} & Name\\
\hline $x_1$ & \DiureticsDecision\\
\hline $x_2$ & Age\\
\hline $x_3$ & Gender\\
\hline $x_5$ & \SAPSTZ\\
\hline $x_{10}$ & \SOFATZ\\
\hline $x_{15}$ & Elixhauser Score\\
\hline $V_1$ & Propensity Score\\
\hline $x_1 \cdot x_5$ & -\\
\hline
\end{tabular}
\caption[Variables for GP Analysis]{The variables used in the GP analysis.}
\label{tab:ModelGP}
\end{table}

The presented results are preliminary and need further study to tune the GP method properly. This should be considered an initial exploration.

The goal of this analysis was the use of GP-based machine learning (ML) for predictive outcome modeling with the diuretics study as startup demostration context. The envisioned approach to helping a new patient, is to:
\begin{description}
\item[A:] identify the cluster were the new patient is placed.
\item[B:] push each new patient's variables into the identified cluster model.
\end{description}

The used approach is divided into 2 steps:
\begin{description}
\item[Step 1:] use an unsupervised ML technique (optional) to cluster the patients in the \gs . For each cluster follow step 2.
\item[Step 2:] is divided in 2 parts:
\begin{itemize}
\item[(a)] evolve a GP classifier to predict \mort as a classification problem.
\item[(b)] evolve a GP model for predicting \los .
\end{itemize}
\end{description}

\subsection{Step 1: Unsupervised Learning of Clusters}
\label{cpt:GPMLStep1}
To reduce variance, clusters was (optionally) performed by using K-means\cite{k-means}.

Four clusters have been generated using the k-means\footnote{K-means clustering is a method of cluster analysis which aims to partition $n$ observations into $k$ clusters in which each observation belongs to the cluster with the nearest mean.} clustering method. The method was applied on a subset of the variables which describe the clinical conditions of the patients. The chosen variables were: Age, Sex, \SAPSTZ , \SOFATZ and \ELIX .

The 4 generated clusters grouped the patients according to their conditions as follows:
\begin{itemize}
\item {\bf Cluster 1:} this cluster is composed by 221 patients. Age and \SAPSTZ are shown in Figures~\vref{fig:cluster_1}.
\begin{figure}
\centering
\rotatebox{90}{
\stepcounter{figure}
\begin{minipage}{\textheight}
\begin{minipage}{0.5\textheight}
\centering
\includegraphics[width=.9\linewidth]{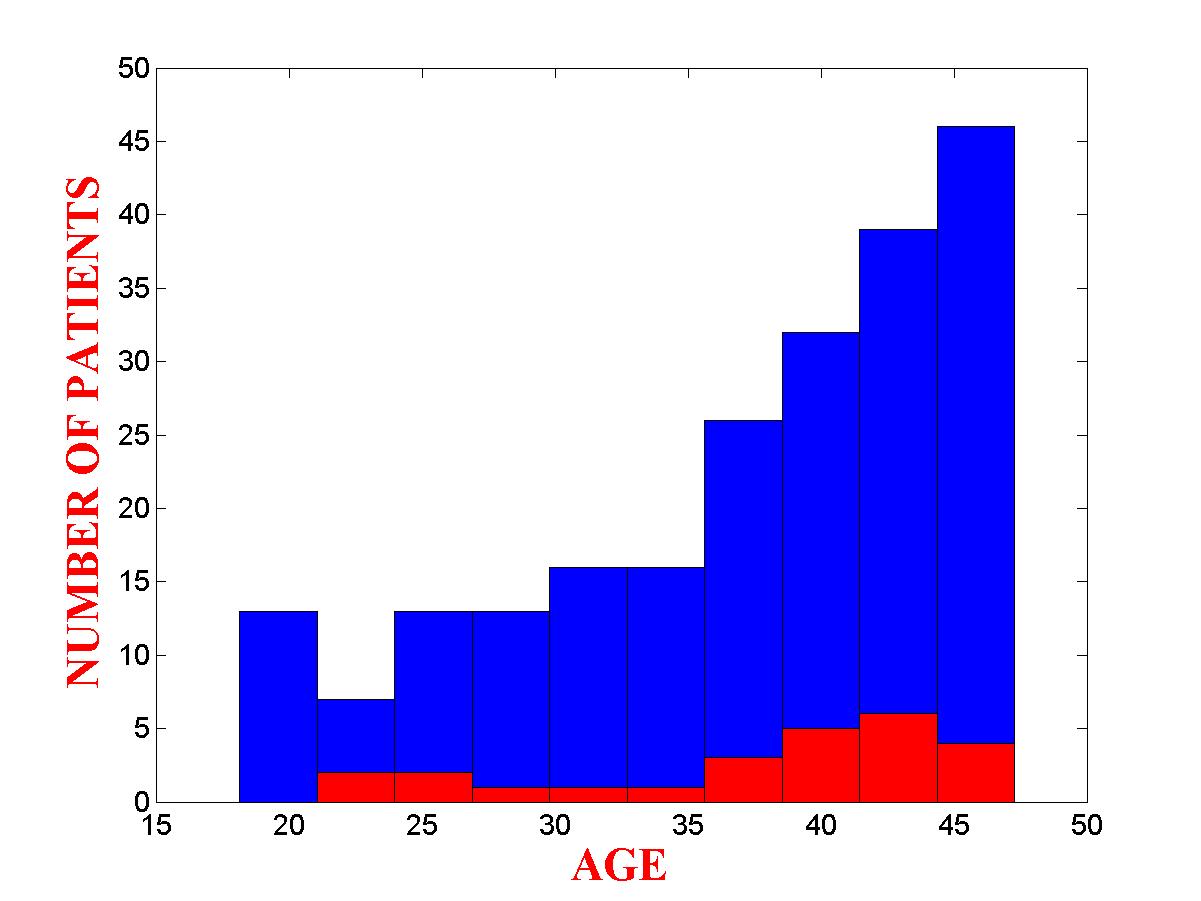}
\captionof{subfigure}[]{Histogram of Age for Cluster 1. The values are centered on 39 years old.}
\label{fig:age_11}
\end{minipage}
\begin{minipage}{0.5\textheight}
\centering
\includegraphics[width=.9\linewidth]{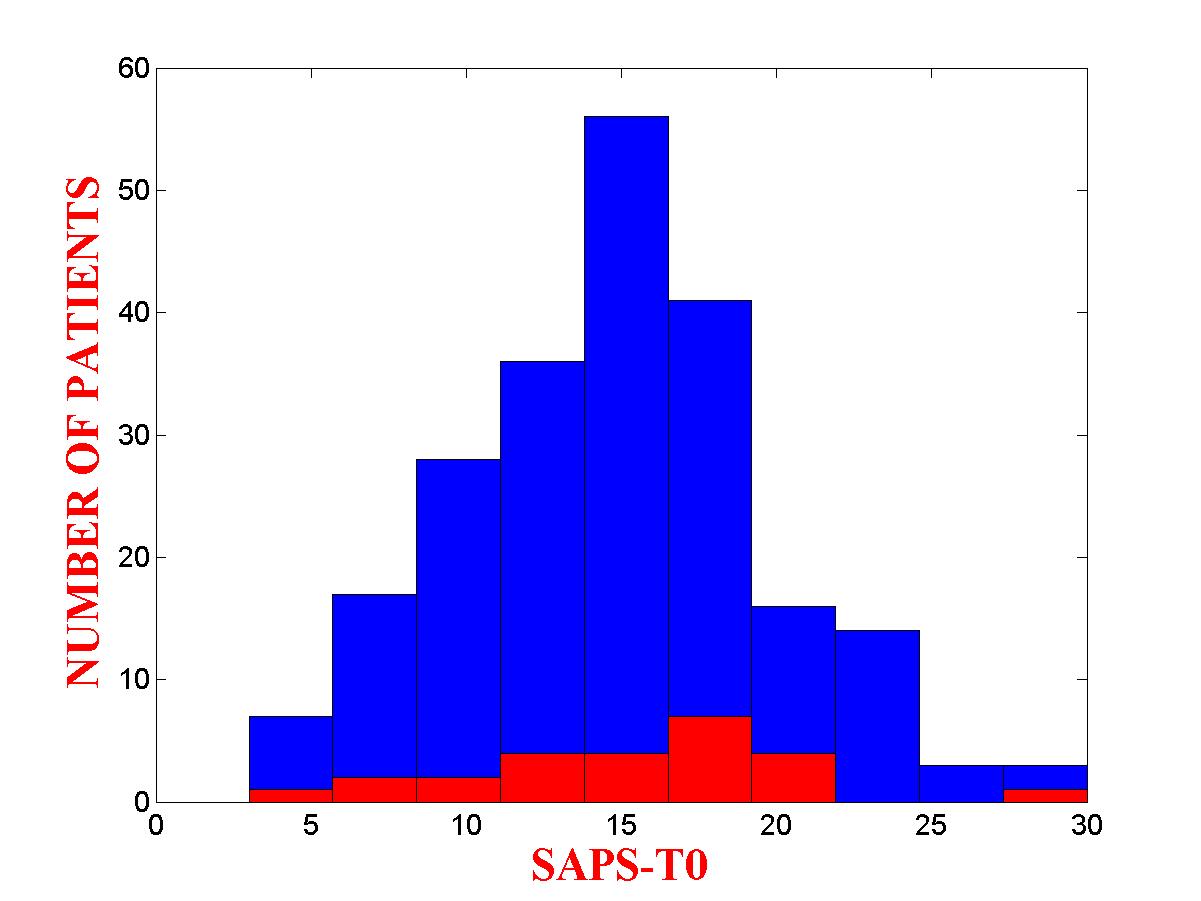}
\captionof{subfigure}[]{Histogram of \SAPSTZ for Cluster 1. The values are centered on 15.}
\label{fig:saps_11}
\end{minipage}
\addtocounter{figure}{-1}
\captionof{figure}[Age and \SAPSTZ for Cluster 1.]{Histograms of Age and \SAPSTZ for Cluster 1. In red the values for the $D^+$ patients only.}
\label{fig:cluster_1}
\end{minipage}
}
\end{figure}
\item {\bf Cluster 2:} this cluster is composed by 426 patients. Age and \SAPSTZ are shown in Figures~\vref{fig:cluster_2}.
\begin{figure}
\centering
\rotatebox{90}{
\stepcounter{figure}
\begin{minipage}{\textheight}
\begin{minipage}{0.5\textheight}
\centering
\includegraphics[width=.9\linewidth]{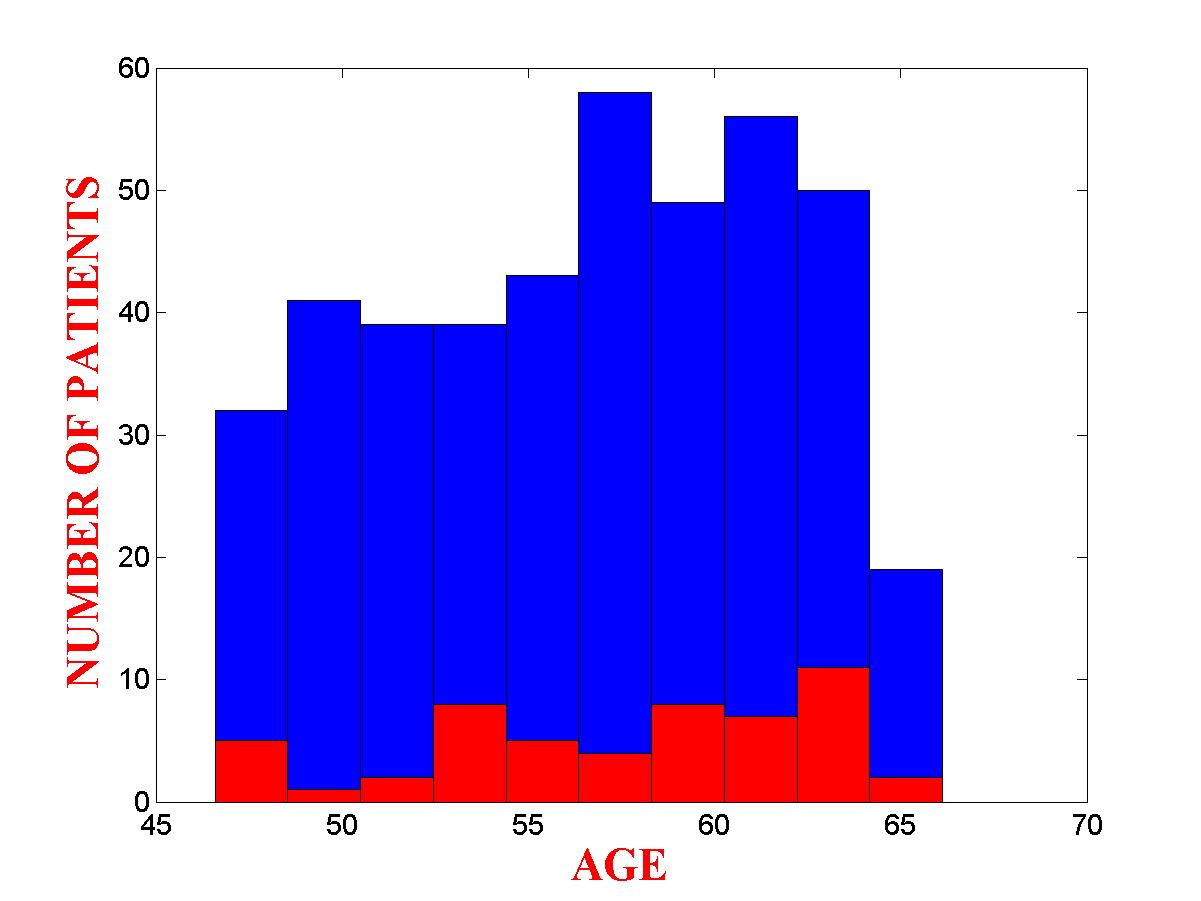}
\captionof{subfigure}[]{Histogram of Age for Cluster 2. The values are centered on 57 years old.}
\label{fig:age_22}
\end{minipage}
\begin{minipage}{0.5\textheight}
\centering
\includegraphics[width=.9\linewidth]{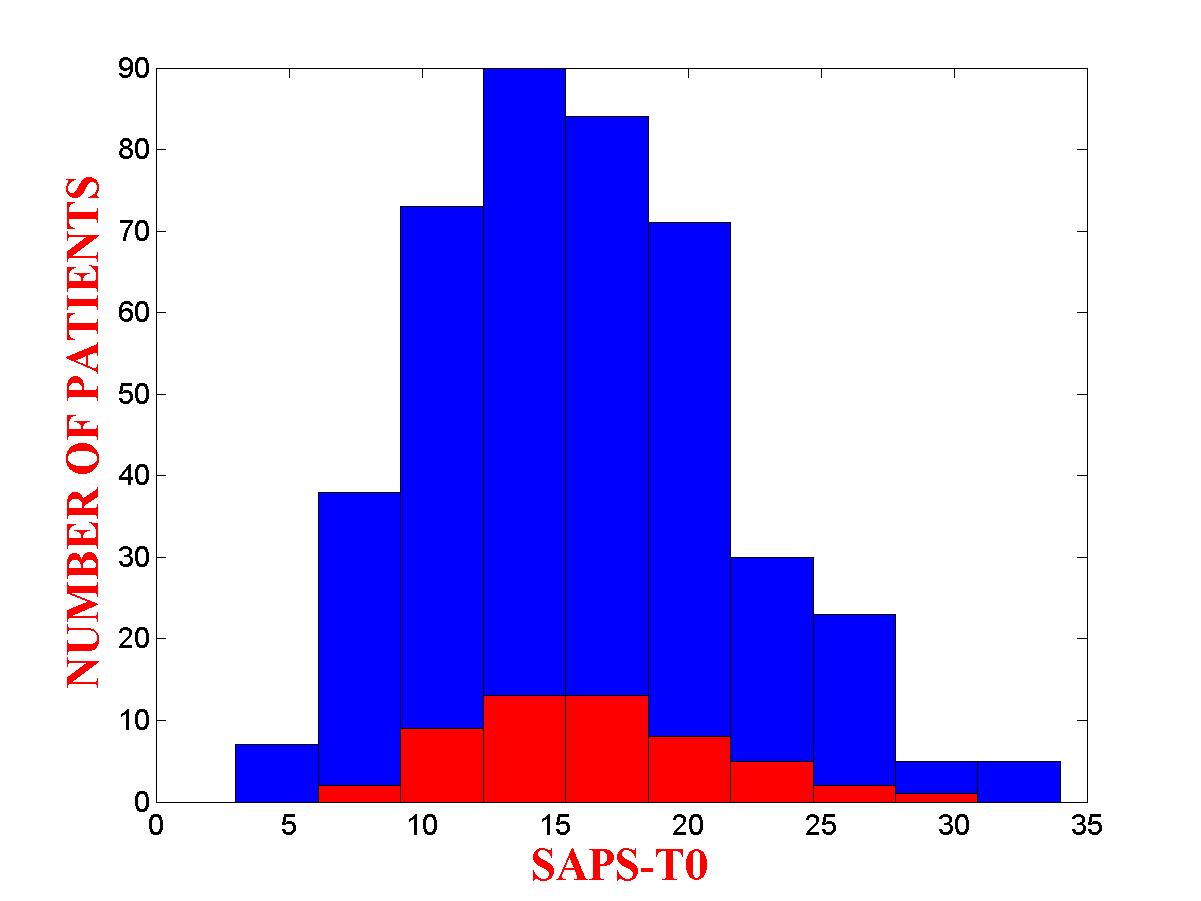}
\captionof{subfigure}[]{Histogram of \SAPSTZ for Cluster 2. The values are centered on 16.}
\label{fig:saps_22}
\end{minipage}
\addtocounter{figure}{-1}
\captionof{figure}[Age and \SAPSTZ for Cluster 2.]{Histograms of Age and \SAPSTZ for Cluster 2. In red the values for the $D^+$ patients only.}
\label{fig:cluster_2}
\end{minipage}
}
\end{figure}
\item {\bf Cluster 3:} this cluster is composed by 435 patients. Age and \SAPSTZ are shown in Figures~\vref{fig:cluster_3}.
\begin{figure}
\centering
\rotatebox{90}{
\stepcounter{figure}
\begin{minipage}{\textheight}
\begin{minipage}{0.5\textheight}
\centering
\includegraphics[width=.9\linewidth]{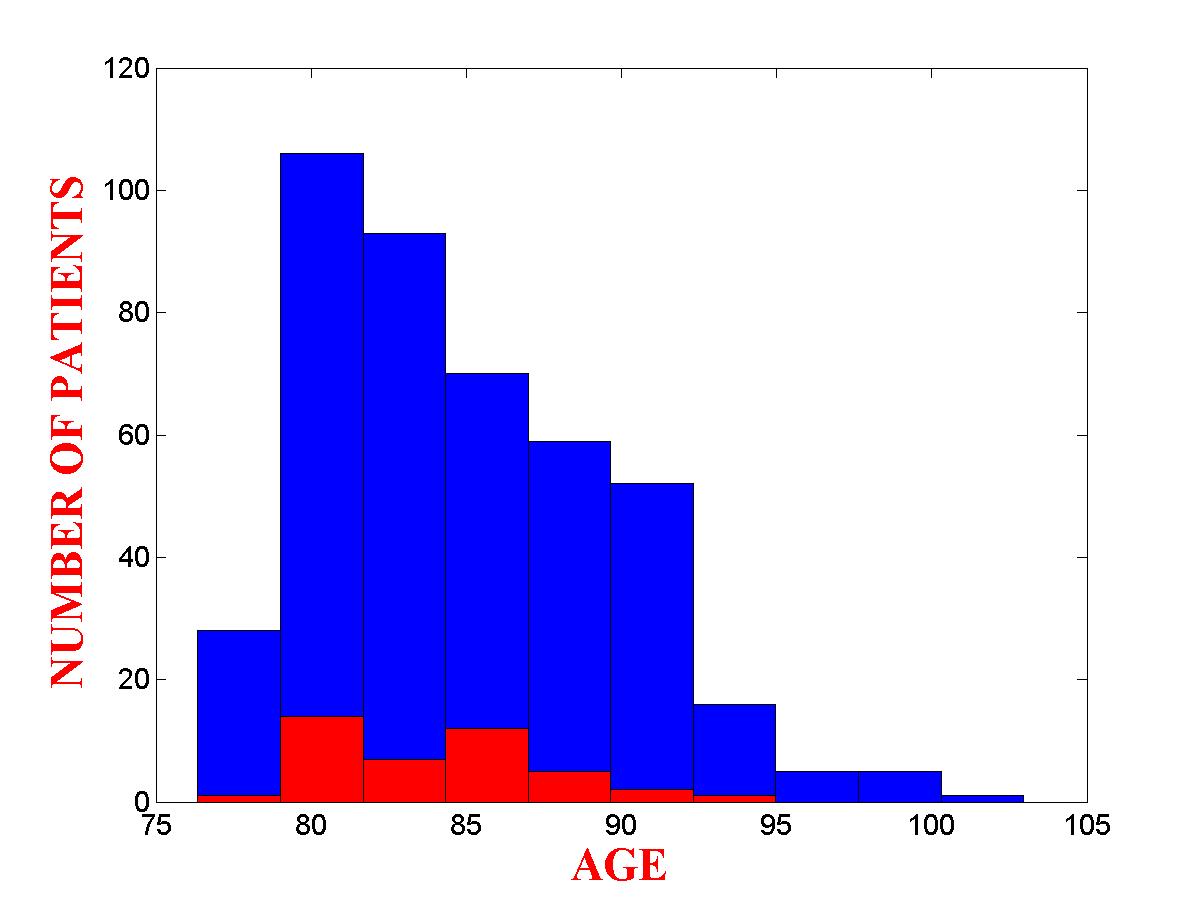}
\captionof{subfigure}[]{Histogram of Age for Cluster 3. The values are centered on 84 years old.}
\label{fig:age_33}
\end{minipage}
\begin{minipage}{0.5\textheight}
\centering
\includegraphics[width=.9\linewidth]{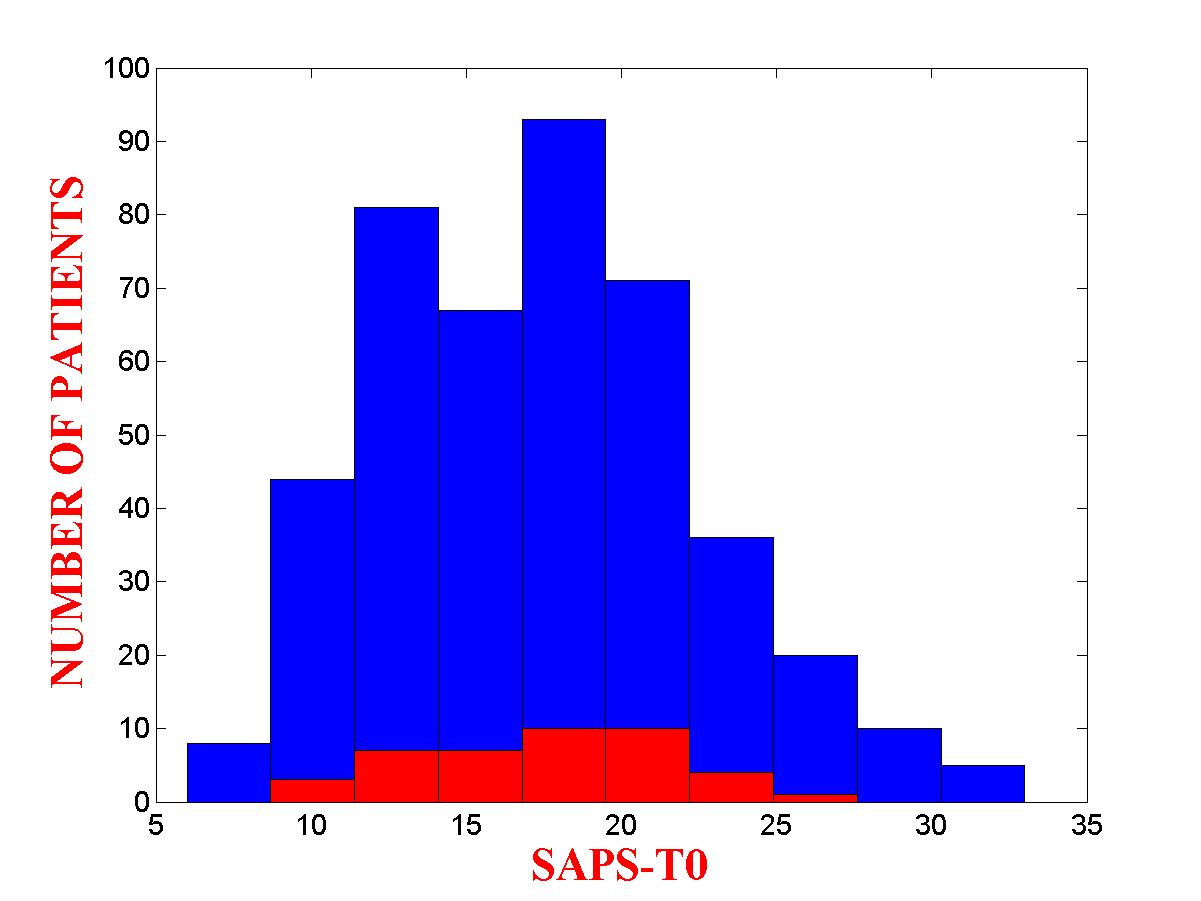}
\captionof{subfigure}[]{Histogram of \SAPSTZ for Cluster 3. The values are centered on 17.}
\label{fig:saps_33}
\end{minipage}
\addtocounter{figure}{-1}
\captionof{figure}[Age and \SAPSTZ for Cluster 3.]{Histograms of Age and \SAPSTZ for Cluster 3. In red the values for the $D^+$ patients only.}
\label{fig:cluster_3}
\end{minipage}
}
\end{figure}
\item {\bf Cluster 4:} this cluster is composed by 440 patients. Age and \SAPSTZ are shown in Figures~\vref{fig:cluster_4}.
\begin{figure}
\centering
\rotatebox{90}{
\stepcounter{figure}
\begin{minipage}{\textheight}
\begin{minipage}{0.5\textheight}
\centering
\includegraphics[width=.9\linewidth]{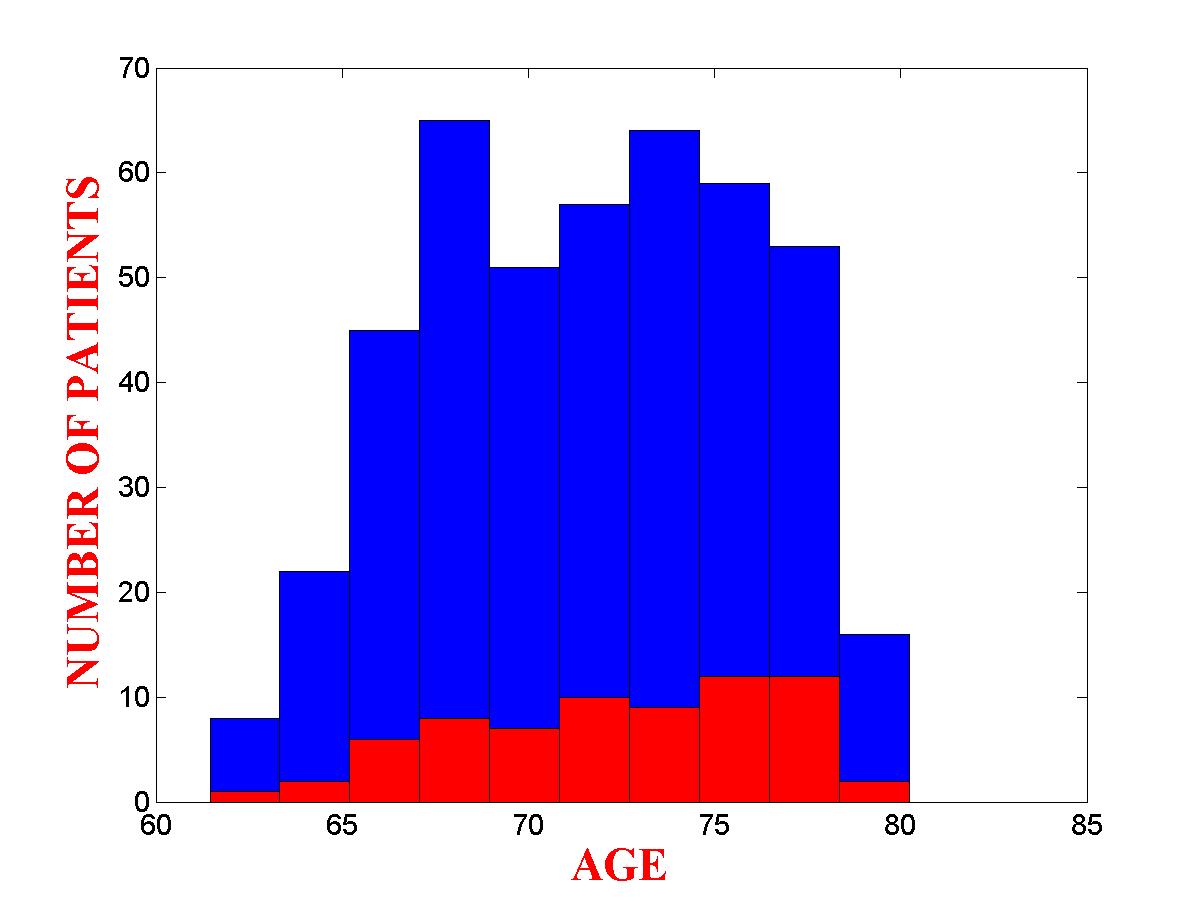}
\captionof{subfigure}[]{Histogram of Age for Cluster 4. The values are centered on 72 years old.}
\label{fig:age_44}
\end{minipage}
\begin{minipage}{0.5\textheight}
\centering
\includegraphics[width=.9\linewidth]{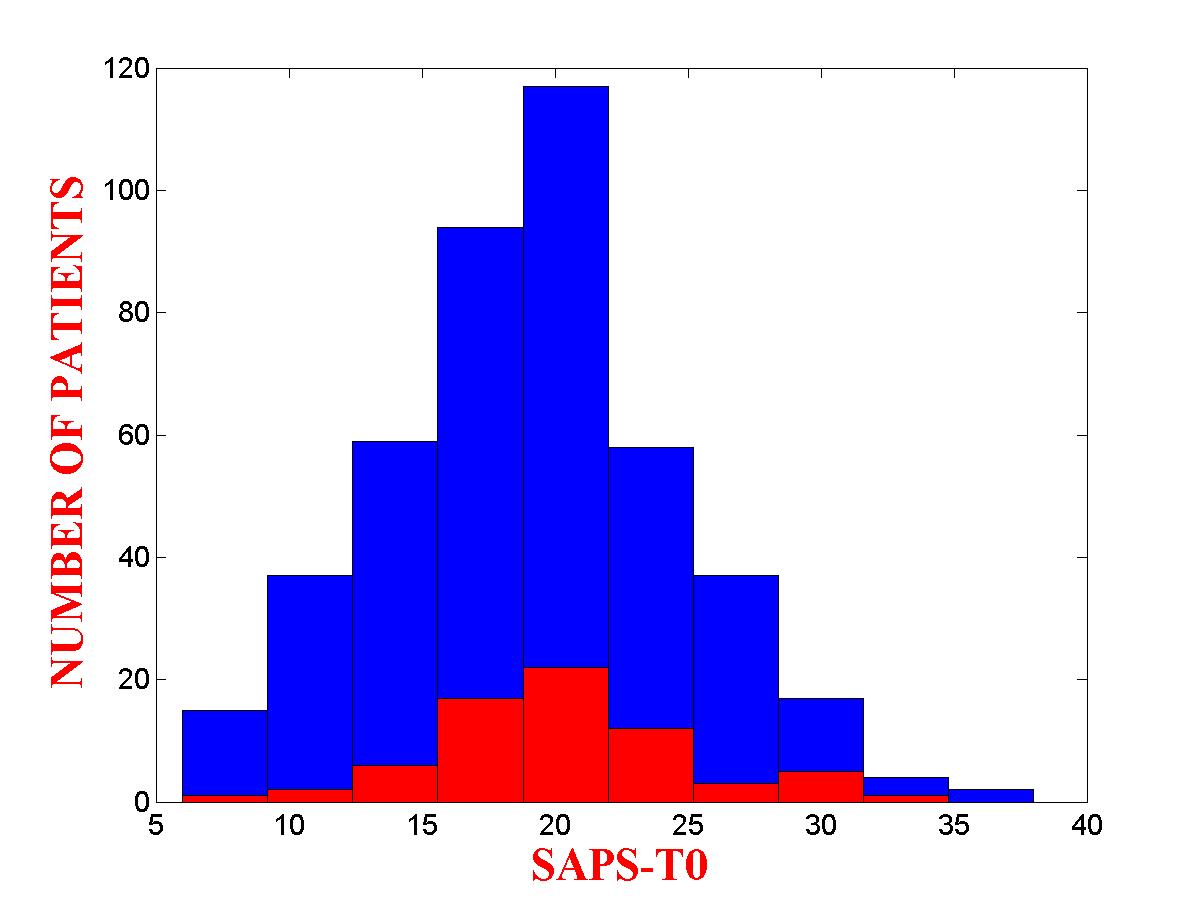}
\captionof{subfigure}[]{Histogram of \SAPSTZ for Cluster 4. The values are centered on 19.}
\label{fig:saps_44}
\end{minipage}
\addtocounter{figure}{-1}
\captionof{figure}[Age and \SAPSTZ for Cluster 4.]{Histograms of Age and \SAPSTZ for Cluster 4. In red the values for the $D^+$ patients only.}
\label{fig:cluster_4}
\end{minipage}
}
\end{figure}
\end{itemize}

Clustering does not and should not create logically separated groups. However very roughly can be observed that these clusters have approximate charaterization by Age and \illness . A description of the clusters is provided in Table~\vref{tab:k-clusters}.
\begin{table}
\centering
\begin{tabular}{|c|c||c|c|c|c|c|c|}
\hline {\bf Cluster} & \# & $D^+$ & $D^-$ & Mean(SD) & Median & Mean(SD) & Median\\
\hline {\bf} & & & & Age & Age & \SAPSTZ & \SAPSTZ\\
\hline 1 & 221 & 25 & 196 & 37(8) & 39 & 14.8(4.9) & 15\\
\hline 2 & 426 & 53 & 373 & 56.5(5) & 57 & 16(5.3) & 16\\
\hline 3 & 435 & 42 & 393 & 84.8(4.6) & 84 & 17.4(5) & 17\\
\hline 4 & 440 & 79 & 371 & 71.5(4.2) & 72 & 19.2(5.5) & 19\\
\hline
\end{tabular}
\caption[Description of the 4 Clusters]{Clusters create by 1 execution of K-means with k=4 on variables: Age, Sex, \SAPSTZ , \SOFATZ and \ELIX .}
\label{tab:k-clusters}
\end{table}

\section{Step 2: GP modeling}
\label{cpt:GPMLStep2}
For both the two outcomes, the step 2 of the analysis was performed on the whole dataset (1,522 patients), on a series of subsets composed by 4 clusters and on the less sick and sicker groups used in the analysis discussed in Chapter~\ref{cpt:outcomeAnalysis}. In Table~\vref{tab:gp-params} are shown the parameters for the GP executions.
\begin{table}
\centering
\begin{tabular}{|c|c|}
\hline {\bf Operator} & Value\\
\hline Population Size & 100\\
\hline \# of Generations & 10\\
\hline Operators & $\left \{ +, -, *, /, log_2, \sqrt{ } \right \}$\\
\hline Probability of Reproduction $p_m$ & 0.1\\
\hline Initial Probability of Crossover $p_c$ & 0.5\\
\hline Initial Probability of Mutation $p_m$ & 0.5\\
\hline Initialization Type & Tournament\\
\hline Maximum Depth of the Trees & 17\\
\hline
\end{tabular}
\caption[Parameters of the GP Executions]{Parameters of the GP executions.}
\label{tab:gp-params}
\end{table}

In Table~\vref{tab:ill-groups} is shown a description of the less sick and sicker groups. Median \SAPSTZ on which the groups were splitted is 17.
\begin{table}
\centering
\begin{tabular}{|c|c||c|c|c|c|c|c|}
\hline {\bf Cluster} & \# & $D^+$ & $D^-$ & Mean(SD) & Median & Mean(SD) & Median\\
\hline {\bf} & & & & Age & Age & \SAPSTZ & \SAPSTZ\\
\hline Less Sick & 816 & 88 & 728 & 63.8(17.7) & 60 & 13(2.9) & 14\\
\hline Sicker & 706 & 101 & 605 & 68.8(15.5) & 71 & 21.9(3.5) & 21\\
\hline
\end{tabular}
\caption[Description of the Less Sick and Sicker Groups]{Clusters chosen according to \illness \textit{WITHOUT} any machine learning technique.}
\label{tab:ill-groups}
\end{table}

\subsection{Results on \Mort}
\label{cpt:resMor}
For all the groups described above 10 indipendent GP runs have been performed using each time the 70\% of the patients current group for training and the remaining 30\% for testing. Each time the patients for both training and testing were randomly chosen. For all the analyzed groups the median values of the 10 runs for success rate, true positive, true negative, false positive, false negative, sensitivity ($\frac{TP}{TP + FP}$) and specificity ($\frac{TN}{TN + FN}$) will be given and discussed. All the results refer to the whole set of each group as the obtained results on the respective training and the testing sets are always similar between each others.

\subsubsection{Results on the Original Dataset}
\label{cpt:resAllSet}
Table~\vref{tab:gp_res1} shows the overall results on the whole dataset. The models have the 46\% of success on average more or less equaly divided between true positive and true negative even if the models could have problems to evaluate false positive, as the false negative are a lot. 
\begin{table}
\centering
\begin{tabular}{|c|c|c|c|c|c|c|c|}
\hline {\bf Dataset} & Success Rate & TP & TN & FP & FN & SEN & SPE\\
\hline 1522 Patients & 47\% & 411 & 293 & 663 & 155 & 0.38 & 0.65\\
\hline
\end{tabular}
\caption[GP Overall Results on the Dataset for Mortality]{GP Overall Results on the Dataset for Mortality.}
\label{tab:gp_res1}
\end{table}
Table~\vref{tab:gp_res2} shows the two best results on the whole dataset. The best model has the 63\% of success.
\begin{table}
\centering
\begin{tabular}{|c|c|c|c|c|c|c|c|}
\hline {\bf Dataset} & Success Rate & TP & TN & FP & FN & SEN & SPE\\
\hline 1522 Patients & 63\% & 0 & 956 & 0 & 566 & 0 & 0.63\\
\hline 1522 Patients & 59\% & 62 & 829 & 127 & 504 & 0.32 & 0.62\\
\hline
\end{tabular}
\caption[GP Best Results on the Dataset for Mortality]{GP Best Results on the Dataset for Mortality.}
\label{tab:gp_res2}
\end{table}

\subsubsection{Results on the Less Sick and Sicker groups}
\label{cpt:resSick}
Table~\vref{tab:gp_res3} shows the overall results on the less sick and sicker groups. The models have the bad rate of success on average on the less sick group. In this groups the models have big problems to evaluate false positive.
\begin{table}
\centering
\begin{tabular}{|c|c|c|c|c|c|c|c|c|}
\hline {\bf Group} & Size & SR & TP & TN & FP & FN & SEN & SPE\\
\hline Less Sick & 816 Patients & 26\% & 209 & 3 & 604 & 0 & 0.26 & 1\\
\hline Sicker & 706 Patients & 50\% & 2.5 & 342 & 7 & 354.5 & 0.26 & 0.49\\
\hline
\end{tabular}
\caption[GP Overall Results on the Less Sick and Sicker Groups for Mortality]{GP Overall Results on the Less Sick and Sicker Groups for Mortality.}
\label{tab:gp_res3}
\end{table}
Table~\vref{tab:gp_res4} shows the two best results on the less sick and sicker groups. The best model is for the less sick group with a success rate of 73\%.
\begin{table}
\centering
\begin{tabular}{|c|c|c|c|c|c|c|c|c|}
\hline {\bf Group} & Size & SR & TP & TN & FP & FN & SEN & SPE\\
\hline Less Sick & 816 Patients & 73\% & 1 & 597 & 10 & 208 & 0.09 & 0.74\\
\hline Less Sick & 816 Patients & 48\% & 116 & 272 & 335 & 93 & 0.26 & 0.74\\
\hline Sicker & 706 Patients & 54\% & 320 & 64 & 285 & 37 & 0.53 & 0.63\\
\hline Sicker & 706 Patients & 54\% & 320 & 64 & 285 & 37 & 0.53 & 0.63\\
\hline
\end{tabular}
\caption[GP Best Results on the Less Sick and Sicker Groups for Mortality]{GP Best Results on the Less Sick and Sicker Groups for Mortality.}
\label{tab:gp_res4}
\end{table}

\subsubsection{Results on the 4 Clusters}
\label{cpt:resuClusters}
Table~\vref{tab:gp_res5} shows the overall results on the 4 clusters. The models have more or less the 40\% of success on average. The detection of false negative is a problem in this case too.
\begin{table}
\centering
\begin{tabular}{|c|c|c|c|c|c|c|c|c|}
\hline {\bf Cluster} & Size & SR & TP & TN & FP & FN & SEN & SPE\\
\hline Cluster 1 & 221 Patients & 44\% & 36.5 & 59.5 & 114.5 & 10.5 & 0.24 & 0.85\\
\hline Cluster 2 & 426 Patients & 37\% & 114.5 & 42.5 & 253.5 & 15.5 & 0.31 & 0.73\\
\hline Cluster 3 & 435 Patients & 47\% & 203 & 0.5 & 230.5 & 1 & 0.47 & 0.33\\
\hline Cluster 4 & 440 Patients & 44\% & 182.5 & 11 & 244 & 2.5 & 0.43 & 0.81\\
\hline
\end{tabular}
\caption[GP Overall Results on the 4 Clusters for Mortality]{GP Overall Results on the 4 Clusters for Mortality.}
\label{tab:gp_res5}
\end{table}
Table~\vref{tab:gp_res6} shows the two best results on the 4 clusters. The best models are the ones of clusters 1 and 2 where the success rate is better then 64\%.
\begin{table}
\centering
\begin{tabular}{|c|c|c|c|c|c|c|c|c|}
\hline {\bf Cluster} & Size & SR & TP & TN & FP & FN & SEN & SPE\\
\hline Cluster 1 & 426 Patients & 71\% & 7 & 149 & 25 & 40 & 0.22 & 0.79\\
\hline Cluster 1 & 426 Patients & 66\% & 0 & 145 & 29 & 47 & 0 & 0.75\\
\hline Cluster 2 & 440 Patients & 68\% & 0 & 291 & 5 & 130 & 0 & 0.69\\
\hline Cluster 2 & 440 Patients & 65\% & 0 & 275 & 21 & 130 & 0 & 0.68\\
\hline Cluster 3 & 221 Patients & 53\% & 15 & 214 & 17 & 189 & 0.47 & 0.53\\
\hline Cluster 3 & 221 Patients & 49\%  & 122 & 92 & 139 & 82 & 0.47 & 0.53\\
\hline Cluster 4 & 435 Patients & 53\% & 9 & 226 & 29 & 176 & 0.24 & 0.56\\
\hline Cluster 4 & 435 Patients & 50\% & 136 & 83 & 172 & 49 & 0.44 & 0.63\\
\hline
\end{tabular}
\caption[GP best Results on the 4 Clusters for Mortality]{GP Best Results on the 4 Clusters for Mortality.}
\label{tab:gp_res6}
\end{table}

\subsection{Results on \LOS in ICU}
\label{resLOS}
The analysis was performed in the same way of the one for mortality, but this time the mean absolute error is the only result presented. Even in this case, all the results refers to the whole analyzed groups as the results on training and testing sets are similar between each others.

\subsubsection{Results on the Original Dataset}
\label{cpt:resLOSAllSet}
Table~\vref{tab:gp_res7} shows the overall results on the whole dataset. The mean absolute error is of 7.5 days on average.
\begin{table}
\centering
\begin{tabular}{|c|c|}
\hline {\bf Dataset} & Median Mean Absolute Error\\
\hline 1522 Patients & 7.5 days\\
\hline
\end{tabular}
\caption[GP Overall Results on the Dataset for LOS]{GP Overall Results on the Dataset for LOS.}
\label{tab:gp_res7}
\end{table}
Table~\vref{tab:gp_res8} shows the two best results on the whole dataset. The mean absolute error is more or less of 7 days on average.
\begin{table}
\centering
\begin{tabular}{|c|c|}
\hline {\bf Dataset} & Median Mean Absolute Error\\
\hline 1522 Patients & 6.7 days\\
\hline 1522 Patients & 7 days\\
\hline
\end{tabular}
\caption[GP Best Results on the Dataset for LOS]{GP Best Results on the Dataset for LOS.}
\label{tab:gp_res8}
\end{table}

\subsubsection{Results on the Less Sick and Sicker groups}
\label{cpt:resLOSIllGroup}
Table~\vref{tab:gp_res9} shows the overall results on the less sick and sicker groups. The mean absolute error goes between 6 to 11 days on average.
\begin{table}
\centering
\begin{tabular}{|c|c|c|}
\hline {\bf Group} & Patiens & Median Mean Absolute Error\\
\hline Less Sick & 816 & 6.4 days\\
\hline Sicker & 706 & 11.6 days\\
\hline
\end{tabular}
\caption[GP Overall Results on the Less Sick and Sicker Groups for LOS]{GP Overall Results on the Less Sick and Sicker Groups for LOS.}
\label{tab:gp_res9}
\end{table}
Table~\vref{tab:gp_res10} shows the two best results on the less sick and sicker groups. The mean absolute error goes between 5 to 8 days on average.
\begin{table}
\centering
\begin{tabular}{|c|c|c|}
\hline {\bf Group} & Patiens & Median Mean Absolute Error\\
\hline Less Sick & 816 & 5.8 days\\
\hline Less Sick & 816 & 5.8 days\\
\hline Sicker & 706 & 8.1 days\\
\hline Sicker & 706 & 8.6 days\\
\hline
\end{tabular}
\caption[GP Best Results on the Less Sick and Sicker Groups for LOS]{GP Best Results on the Less Sick and Sicker Groups for LOS.}
\label{tab:gp_res10}
\end{table}

\subsubsection{Results on the 4 Clusters}
\label{cpt:resLOSClusters}
Table~\vref{tab:gp_res11} shows the overall results on the 4 clusters. The mean absolute error goes between 5 to 10 days on average.
\begin{table}
\centering
\begin{tabular}{|c|c|c|}
\hline {\bf Cluster} & Patiens & Median Mean Absolute Error\\
\hline Cluster 1 & 426 Patients & 10.9 days\\
\hline Cluster 2 & 440 Patients & 7.9 days\\
\hline Cluster 3 & 221 Patients & 5.1 days\\
\hline Cluster 4 & 435 Patients & 8.2 days\\
\hline
\end{tabular}
\caption[GP Overall Results on the 4 Clusters for LOS]{GP Overall Results on the 4 Clusters for LOS.}
\label{tab:gp_res11}
\end{table}
Table~\ref{tab:gp_res12} shows the two best results on the 4 clusters. The mean absolute error goes between 4 to 7 days on average.
\begin{table}
\centering
\begin{tabular}{|c|c|c|}
\hline {\bf Cluster} & Patiens & Median Mean Absolute Error\\
\hline Cluster 1 & 426 Patients & 7.3 days\\
\hline Cluster 1 & 426 Patients & 7.5 days\\
\hline Cluster 2 & 440 Patients & 6 days\\
\hline Cluster 2 & 440 Patients & 6.5 days\\
\hline Cluster 3 & 221 Patients & 4.2 days\\
\hline Cluster 3 & 221 Patients & 5 days\\
\hline Cluster 4 & 435 Patients & 6.5 days\\
\hline Cluster 4 & 435 Patients & 6.7 days\\
\hline
\end{tabular}
\caption[GP Best Results on the 4 Clusters for LOS]{GP Best Results on the 4 Clusters for LOS.}
\label{tab:gp_res12}
\end{table}

\subsubsection{Comment on the GP Results}
\label{cpt:resComment}
Both for \mort and \los in ICU the prediction results are not satisfactory. Especially for length of stay the error is big. This probably is due to the difficult of the problem. The fact that the models generate on average a lot false negative indicates the difficulty of evaluating the chance of survival of the patients, expecially of the sickest ones. Furthermore, these are preliminary results and the GP models could be tuned in a better way.

\subsection{Simulated Outcomes}
\label{cpt:resSimulated}
In this final analysis the chances of survival with or without diuretics have been evaluated by using the two best models produced with GP for each dataset. All the available values for the patients have been used except for the diuretics variable. Then the chances of survival for each patients have been evaluated by inserting the two possibile values, given and not given, for the diurecits variables. In this way two perfectly paired patients have been created for each actual patient.

This analysis try to overcome the real problem of an observational study that is the fact that investigators can not control the assignment of the treatments to patients and, hence, the experiment is non-randomized. With an approach of this type instead, there is the possibility of duplicate the dataset and then confront perfectly paired patients. The results of this analysis are presented in this Section. But it should be borne in mind that these results are influenced a lot by the models accuracy. So, as the models accuracy is not satisfactory, they should be considered preliminary and may be object of further analysis.

\subsubsection{Results on \Mort}
\label{cpt:resSimulatedMort}
In Table~\vref{tab:gp_res13} are shows the results for mortality.
\begin{table}
\centering
\begin{tabular}{|c|c|c|c|c|}
\hline {\bf Group} & MOR 1 D- & MOR 1 D+ & MOR 2 D- & MOR 2 D+\\
\hline Dataset & 0\% & 0\% & 0\% & 1\% \\
\hline Cluster 1 & 14\% & 14\% & 13\% & 13\%\\
\hline Cluster 2 & 11\% & 11\% & 4\% & 4\%\\
\hline Cluster 3 & 60\% & 60\% & 7\% & 7\%\\
\hline Cluster 4 & 8\% & 8\% & 70\% & 70\%\\
\hline Less Sick & 13\% & 13\% & 58\% & 40\%\\
\hline Sicker & 1\% & 0\% & 1\% & 1\%\\
\hline
\end{tabular}
\caption[GP Simulated Results for Mortality]{GP Simulated Results for Mortality.}
\label{tab:gp_res13}
\end{table}
An other problem that an analysis of this type could have it that the diuretics variable is binary, hence it is possible that by only flipping it, the model could not capture any difference: this could be because either the diuretics are not actually making difference or because the model is not accurate enough. In fact by analyzing the results for mortality, two things stand out: a)the models are generating a lot of false negative and the mortality rates are low, b)in the most of the models diuretics do not seem to make difference, for one of the reason defined above. The last thing that stands out is curiously the result for the less sick group: only in this case in fact, for the second best model, the results seems to give a better chance of survival for the patients who are getting diuretics and this goes along with the results obtained by the analysis of the first Section of this Chapter. Obviously what said is to be taken with caution, given the low accuracy of the models used.

\subsubsection{Results on \LOS}
\label{cpt:resSimulatedLOS}
In Table~\vref{tab:gp_res14} are shows the results for length of stay.
\begin{table}
\centering
\begin{tabular}{|c|c|c|c|c|}
\hline {\bf Group} & LOS 1 D+ & LOS 1 D- & LOS 2 D+ & LOS 2 D-\\
\hline Dataset & 2 days & 2 days & 1.3 days & 1.3 days \\
\hline Cluster 1 & 3.6 days & 3.6 days & 5 days & 5.6 days\\
\hline Cluster 2 & 1.9 days & 1.8 days & 5.7 days & 5.7 days\\
\hline Cluster 3 & 0 days & 0 days & 1.5 days & 1 days\\
\hline Cluster 4 & 2.2 days & 2.2 days & 1.9 days & 0.9 days\\
\hline Less Sick & 0.5 days & 0.5 days & 0.3 days & 0.3 days\\
\hline Sicker & 0 days & 0 days & 6.6 days & 6.6 days\\
\hline
\end{tabular}
\caption[GP Simulated Results for LOS]{GP Simulated Results for LOS.}
\label{tab:gp_res14}
\end{table}
In this case, the results are not conclusive, possibly because of the poor accuracy of the models used.

\subsubsection{Comments on the Results on the Simulated Outcomes}
\label{cpt:resSimulatedComments}
As already said, the tecnique used in this Section relies on the accurancy of the used models. In this case, the models have not a satisfactory accurancy, hence the presented results should be taken with caution. But this method could be the object of further analysis.

\clearpage \mbox{} \clearpage 
\chapter{Conclusions}
\label{cpt:conclusions}

\section{Summary of Findings}
\label{cpt:summaryFinal}
A brief summary of the findings woven through Chapter~\ref{cpt:outcomeAnalysis} is now provided:
\begin{itemize}
\item \nibf{Finding 1: Length of stay and \gettingdiuretics:}\\
With respect to \los, \illness variables $x_2$, $x_5$ and $x_{15}$ (Age, \SAPSTZ, and \Elix) have statistically significant effects, as does propensity score. The null hypothesis that the independent effect of \diureticsDecision is not significant on \los, is rejected (\pvalue $<0.001$).\\
For \los outcome, these findings imply \illness is not a confounding factor and \diureticsDecision is independently significant in its effect on \los.\\
This validates the findings of the \group~analysis for \los in Chapter~\ref{cpt:propensityAnalysis}. They indicate a statistically significant difference in \los for \group~ 1 ($PS \in \left[ 0.00;0.01 \right] $),  \group~ 3 ($PS \in \left[ 0.04;0.08 \right] $) and  \group~ 5 ($PS \in \left[ 0.19;0.99 \right] $), leading to the conclusion, qualified for this study group, that \gettingdiuretics increases a patient's \los in the ICU.
\item \nibf{Finding 2: Independent Effect of \gettingdiuretics on \mort:}\\
The null hypothesis that the independent effect of \diureticsDecision is not significant on \mort, is accepted (\pvalue $>0.05$). Hence, the \diureticsDecision does not have a statistically significant independent effect on \mort.
\item \nibf{Finding 3: Cross-Dependent Effect of \gettingdiuretics and \illness on \mort:}\\
The null hypothesis that \SAPSD has a not significant cross-dependent effect on \mort is rejected given the \pvalue$=0.013$. Through adjusted regression analysis with \ModelCSick and \ModelCSicker, (see Table~\vref{tab:ModelB} columns 4-5 and 6-7), the null hypothesis that \diureticsDecision has a not significant cross-dependent effect on \mort in the less sick subset is rejected. The null hypothesis that \diureticsDecision has a not significant cross-dependent effect on \mort in the sicker subset is \textit{NOT} rejected.\\
Furthermore, through \illness adjusted stratification analysis, see Table~\vref{tab:lessSick}, in the less sick subset of \group~ 4 ($PS \in \left[ 0.06;0.15 \right] $) \mort rate is significantly less for the patients with \gettingdiuretics compared to those without. It is not significantly less for \group~5 ($PS \in \left[ 0.15;0.99 \right] $). Per Table~\vref{tab:Sick} \mort rate is not significantly different for either of \groups~~4 or 5.
\end{itemize}

In Chapter~\ref{cpt:GPML} a preliminary analysis using genetic programming is described. The Chapter's contribution is to outline the method, whereas the produced results are not reliable.

\section{Future Work}
\label{cpt:futureWorks}

\subsection{\PA}
\label{cpt:futureWorksPA}
The primary objective of this work was to develop a statistical  methodology based on  \pa and logistic or linear regression. A \gs was selected for the analysis and results on it were produced. A first set of possible future work would involve reformulating the \gs both with the aim of studying a larger better selection of patients who exhibit high fluid levels (than has been done in this work) but a different pathology.

It has to be said that during the definition of the \gs and during the subsequent extraction from the \db, certain choices were made, as described in the respective Chapters. This was necessary given the vastness of the topic. Hence, it could be of interest to further study possible alternatives to the already explored choices. While the study of other diseases with the same method would be of obvious interest: for this purpose all the developed procedures were designed to be easily reused in this context.

\subsection{GP Analysis}
\label{cpt:futureWorksGP}
As regards the analysis carried out by genetic programming, its aim was to be only a first exploration of the method. This entire analysis would lend itself to further study. First of all, it requires a precise study of the configuration used for the executions of the algorithm with the aim to improve the reliability of the produced models.

If this is done, the updated results could be re-evaluated and potentially prompt new experiments or method refinements. For example, generating different models using separately the group of patients to which diuretic was administered and the one who did not get diuretics and even further dividing the groups by illness in the case of \mort .

Furthermore, different machine learning techniques for both the generation of the clusters and of the models should be tried and compared.

\clearpage \mbox{} \clearpage 
\addcontentsline{toc}{chapter}{Bibliography} 
\bibliography{biblio_cpt1,biblio_cpt2,biblio_cpt3,biblio_cpt4,biblio_cpt5,biblio_cpt6,biblio_apxA,biblio_apxB,biblio_apxC,biblio_apxD,biblio_apxE}
\bibliographystyle{unsrt} 
\clearpage \mbox{} \clearpage 

\pagenumbering{roman} 
\setcounter{page}{1}
\appendix 
\chapter{Medical Backgrounds}
\label{apx:medicalBackground}

This Appendix aims at providing an overview on medical concepts usefull to understand the analysis discussions. In particular will be defined what sepsis is. The Appendix does not aim at being a full medical guide on the topic, but intends to provide some useful basic medical knowledge.

\section{Definition of Septis}
\label{apxA-1}
Sepsis is the leading cause of death in noncoronary intensive care units (ICU) in the United States\cite{SEPSIS-CHEST}.\\
It is a potentially deadly medical condition that is characterized by a whole-body inflammatory state, called a systemic inflammatory response syndrome or SIRS, and the presence of a known or suspected infection. The body may develop this inflammatory response by the immune system to microbes in the blood, urine, lungs, skin, or other tissues. Severe sepsis is the systemic inflammatory response, plus infection, plus the presence of organ dysfunction\footnote{Organ dysfunction is a condition where an organ does not perform its expected function. When the organ dysfunction gets bad to such a degree that normal homeostasis cannot be maintained without external clinical intervention, occurs organ failure.}.\\
The core of the current definition of sepsis arose from the 1991 American College of Chest Physicians / Society of Critical Care Medicine (ACCP / SCCM) Consensus Conference. This definition was revisited and slightly modified by the 2001 Internal Sepsis Definition Conference.

\subsection{1991 ACCP / SCCM Consensus Conference}
\label{apxA-1-1}
\begin{figure}[htbp]
\centering
\includegraphics[width=\textwidth]{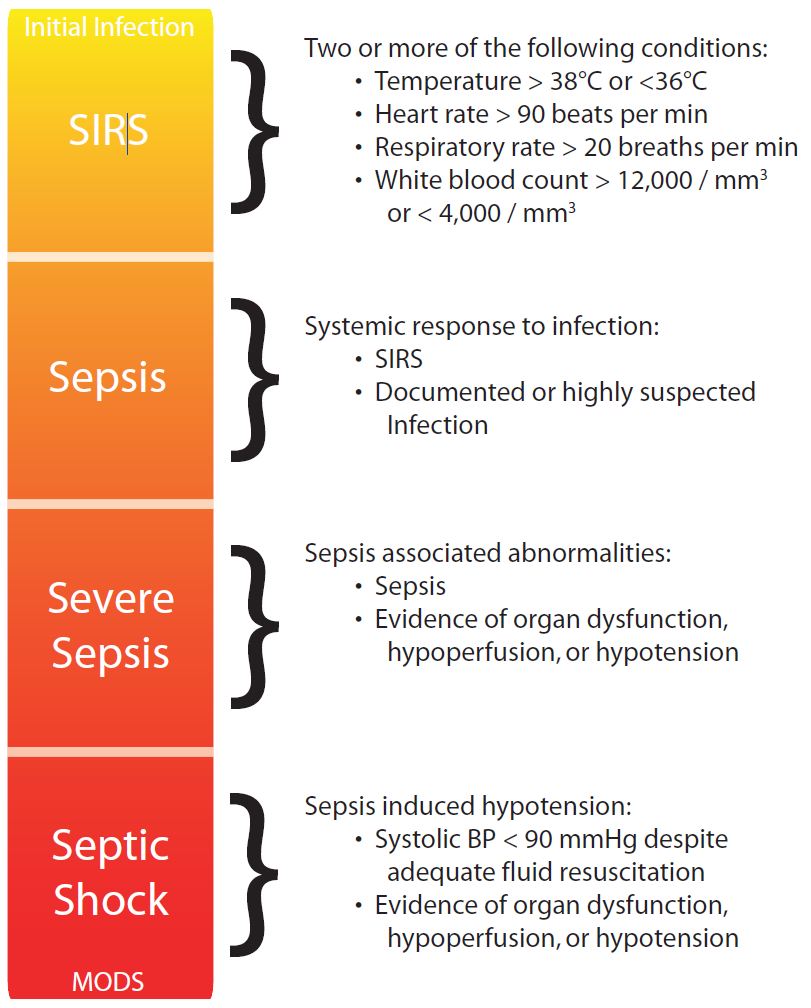}
\caption[Progression of Sepsis Symptoms]{The clinical spectrum of sepsis begins with the nonspecific systemic inflammatory response syndrome and progresses through increasing inflammatory response stages. The spectrum ultimately ends in septic shock and/or multiple organ dysfunction syndrome (MODS).}
\label{fig:Sepsis progression}
\end{figure}

An American College of Chest Physicians / Society of Critical Care Medicine Consensus Conference was held in Chicago in August 1991 with the goal of agreeing on a set of definitions that could be applied to patients with sepsis and its sequelae\cite{SEPSIS-CHEST}. The conference provided a set of definitions used to characterize the progression of the disorder.\\
Sepsis refers to a clinical spectrum of complications starting with the initial infection and ultimately progressing to septic shock. It initially manifests as the nonspecific systemic inflammatory response syndrome (SIRS). SIRS is diagnosed when a patient has two or more of the clinical abnormalities provided in Table~\vref{tab:SIRS-1991}.
\begin{table}
\centering
\begin{tabular}{|c|p{9cm}|}
\hline {\bf Abnormalities} & \centerline{{\bf Values}}\\
\hline Temperature & \centerline{\textless $36\,^{\circ}\mathrm{C}$ ($96.8\,^{\circ}\mathrm{F}$) or \textgreater $38\,^{\circ}\mathrm{C}$ ($100.4\,^{\circ}\mathrm{F}$)}\\
\hline Heart rate & \centerline{\textgreater 90/mins}\\
\hline Respiratory rate & \centerline{\textgreater 20/min or PaCO$_2$\textless 32 mmHg (4.3 kPa)}\\
\hline WBC & \centerline{\textless 4x$10^9$/L (\textless 4000/mm$^3$), \textgreater 12x$10^9$/L} \centerline{(\textgreater 12,000/mm$^3$) or 10\% bands}\\
\hline
\end{tabular}
\caption[1991 ACCP/SCCM Consensus Conference]{According to the 1991 ACCP / SCCM definition, SIRS is diagnosed when a patient has two or more of the clinical abnormalities.}
\label{tab:SIRS-1991}
\end{table}
The patient must present at least two of the following SIRS abnormalities: temperature, heart rate, respiratory rate, WBC\footnote{Total white blood cell count.}.\\
As it has been said, according to the American College of Chest Physicians / Society of Critical Care Medicine, there are different levels of sepsis:
\begin{itemize}
\item Sepsis: defined when SIRS occurs and there is a documented or highly suspected infection.
\item Severe sepsis: defined as sepsis with organ dysfunction, hypoperfusion\footnote{Decreased blood flow through an organ.}, or hypotension\footnote{Abnormally low blood pressure, especially in the arteries of the systemic circulation.}.
\item Septic shock: defined as sepsis with refractory arterial hypotension or hypoperfusion abnormalities in spite of adequate fluid resuscitation.
\end{itemize}
The progression of sepsis symptoms is shown in Figure~\vref{fig:Sepsis progression}.

\subsection{2001 Internal Sepsis Definition Conference}
\label{apxA-1-2}
Ten years after the 1991 ACCP / SCCM Consensus Conference was held to establish to uniform definitions for sepsis and the associated spectrum of progressive injurious processes, the 2001 Internal Sepsis Definition Conference revisited these definitions to evaluate their efficacy and suggest improvements. In the conference was stated that there had been an impetus from experts in the field to modify these definitions to reflect the current understanding of the pathophysiology of these syndromes\cite{SEPSIS-CARE_MED}.\\
Participants of the 2001 Internal Sepsis Definition Conference agreed that in the 1991 ACCP / SCCM Consensus Conference,  SIRS definition was overly sensitive and provided little clinical utility in the initial diagnosis of sepsis. Clinicians did not make the diagnosis of sepsis based on the 1991 SIRS criteria, but rather by analyzing the host of symptoms and deciding the patient “looks septic”—regardless of a documented source of infection\cite{SEPSIS-CARE_MED}.\\
Thus in hopes to increase utility in making the sepsis diagnosis, a more comprehensive list of SIRS criteria was established as provided in Table~\vref{tab:SIRS-2001}.
\begin{table}
\centering
\begin{tabular}{|p{12cm}|}
\hline {\bf Diagnostic criteria for sepsis}\\
{\bf Infection, documented or suspected, and some of the following}\\
\hline {\bf General variables:}\\
Fever (core temperature \textgreater $38.3\,^{\circ}\mathrm{C}$)\\
Hypothermia (core temperature \textless $36\,^{\circ}\mathrm{C}$)\\
Heart rate \textgreater 90min$^{-1}$ or \textgreater 2 {\it SD} above the normal value for age\\
Tachypnea\footnote{Tachypnea, meaning rapid breathing, is a ventilatory rate greater than 20 breaths per minute.}\\
Altered mental status\\
Significant edema or positive fluid balance (\textgreater 20 mL/kg over 24 hrs)\\
Hyperglycemia (plasma glucose \textgreater 120 mg/dL or 7.7 mmol/L) in the absence of diabetes\\
\hline {\bf Inflammatory variables:}\\
Leukocytosis (WBC count \textgreater 12,000 $\mu$L$^{-1}$)\\
Leukopenia (WBC count \textless 4000 $\mu$L$^{-1}$)\\
Normal WBC count with \textgreater 10\% immature forms\\
Plasma C-reactive protein \textgreater 2 {\it SD} above the normal value\\
Plasma procalcitonin \textgreater 2 {\it SD} above the normal value\\
\hline {\bf Hemodynamic variables:}\\
Arterial hypotension (SBP \textless 90 mm Hg, MAP \textless 70, or an SBP decrease \textgreater 40 mm Hg in adults or \textless 2 {\it SD} below normal for age)\\
S$\bar{\nu}O2$ \textgreater 70\%b\\
Cardiac index \textgreater 3.5 L$\cdot$ min$^{-1}\cdot$ M$^{-23}$\\
\hline {\bf Organ dysfunction variables:}\\
Arterial hypoxemia (PaO$_2$/FIO$_2$\textless 300)\\
Acute oliguria (urine output \textless 0.5 mL$\cdot$ kg$^{-1}\cdot$ hr$^{-1}$ or 45 mmol/L for at least 2 hrs)\\
Creatinine increase \textgreater 0.5 mg/dL\\
Coagulation abnormalities (INR \textgreater 1.5 or aPTT \textgreater 60 secs)\\
Ileus (absent bowel sounds)\\
Thrombocytopenia (platelet count \textless 100,000 $\mu$L$^{-1}$)\\
Hyperbilirubinemia (plasma total bilirubin \textgreater 4 mg/dL or 70 mmol/L)\\
\hline {\bf Tissue perfusion variables:}\\
Hyperlactatemia (\textgreater 1 mmol/L)\\
Decreased capillary refill or mottling\\
\hline
\end{tabular}
\caption[2001 Internal Sepsis Definition Conference]{In the 2001 Internal Sepsis Definition Conference the definition of SIRS was updated.}
\label{tab:SIRS-2001}
\end{table}
Except for expanding the SIRS list, the conference found no evidence to support any need for changes in the 1991 ACCP / SCCM Consensus Conference definition.

\section{Epidemiology}
\label{apxA-2}
In the United States, sepsis is the second-leading cause of death in non-coronary Intensive Care Unit (ICU) patients and the tenth most common cause of death overall according to data from the Centers for Disease Control and Prevention (the first being heart disease)\cite{SEPSIS-EPIDEMIOLOGY}. Sepsis is common and also more dangerous in elderly, immunocompromised, and critically ill patients. It occurs in 1–2\% of all hospitalizations and accounts for as much as 25\% of ICU bed utilization. It is a major cause of death in intensive-care units worldwide, with mortality rates that range from 20\% for sepsis, through 40\% for severe sepsis, to over 60\% for septic shock.\\
It is importante to note that the results in the studies of sepsis are highly sensitive to the case definition for sepsis used in the study. Additionally, retrospective studies (for examples using discharge summaries) are at the mercy of clinicians to make diagnoses and most of those are made on the basis of a gut feeling that the patient is “looking septic”.

\clearpage \mbox{} \clearpage 
\chapter{Software}
\label{apx:software}

All the procedures used in this work will be described in this Appendix. The first Section refers to the extraction from the \db of the variables for the patients in the \gs and expands the discussion made in Chapter~\ref{cpt:datasetExtraction}. The second Section describes the variables preparation modules and it is also an extension of the discription provided in Chapter~\ref{cpt:datasetExtraction}.

The third Section describes all the procedures used to perform the \pa described in Chaper~\ref{cpt:propensityAnalysis}.

The last Section briefly summarize the procedures regarding the \oa and the machine learning with gp analysis of Chapters~\ref{cpt:outcomeAnalysis} and~\ref{cpt:GPML}.

\section{Dataset Extraction}
\label{apx:softwareExtraction}
As already anticipated in Chapter~\ref{cpt:datasetExtraction}, the extraction of the records for the analysis has been performed by three Matlab Scripts: a)SQL Script, b)Diuretics Naive Condition and c)Data Filtering. In this Section a deeper description of the produced code will be made.

\subsection{SQL Script}
\label{apx:softwareSQL}
The SQL Script module consists in 21 queries on a PostgreSQL database containing an updated image of the \db. The schema of the database is well defined in\cite{MIMIC-GUIDE}, so refers to it for a deeper description. The next contents are also drawn from\cite{MIMIC-GUIDE}.

\begin{figure}[htbp]
\centering
\includegraphics[scale=0.3]{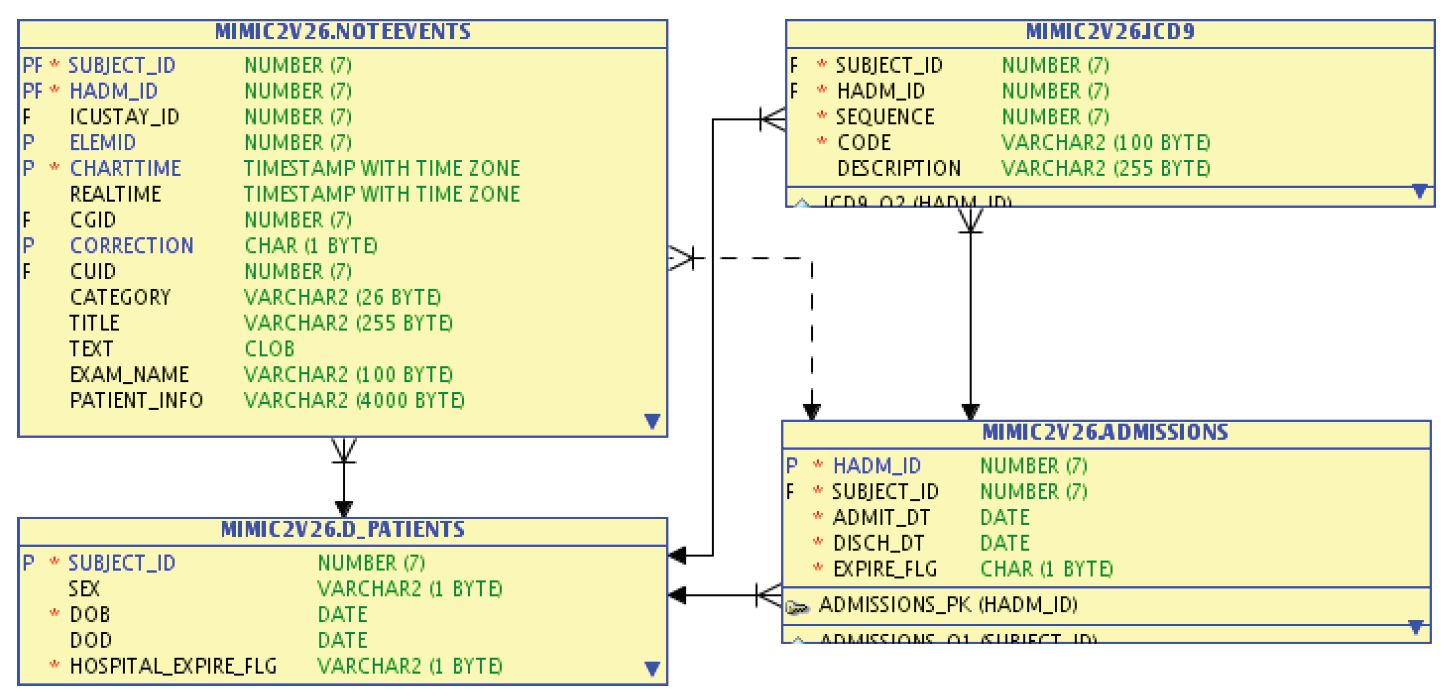}
\caption[Relationship to Patients Table]{Relationship between the table containing the patients' data and hospital admissions, ICD9 codes and note events tables.}
\label{fig:mimic-adm}
\end{figure}

\begin{figure}[htbp]
\centering
\includegraphics[scale=0.5]{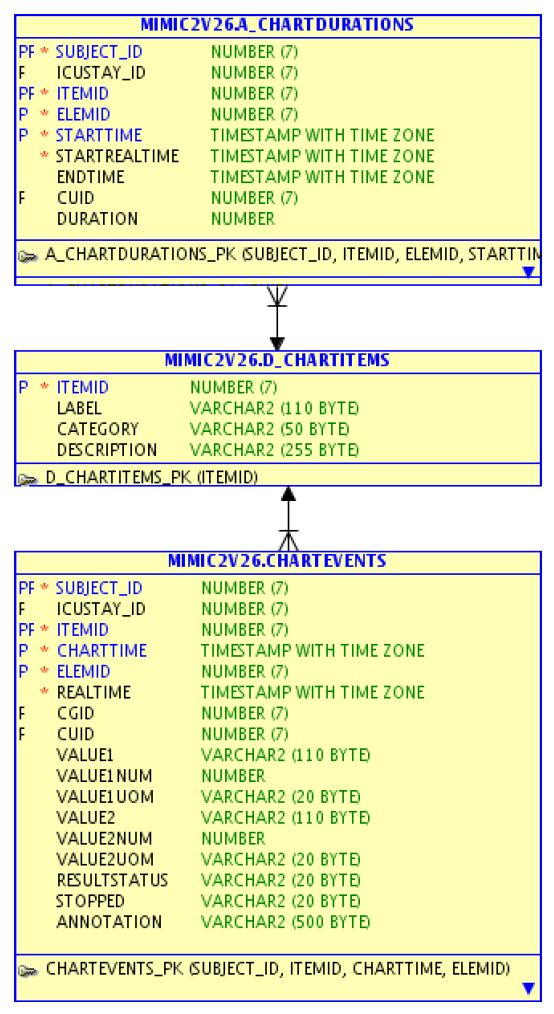}
\caption[Chart Events]{Patients' chart values are stored in 3 tables: chartevents, d\_chartitems and a\_chartdurations. The events' timeline is in the chartevents table while the related elements and durations are in the other tables}
\label{fig:mimic-chart}
\end{figure}

\begin{figure}[htbp]
\centering
\includegraphics[scale=0.45]{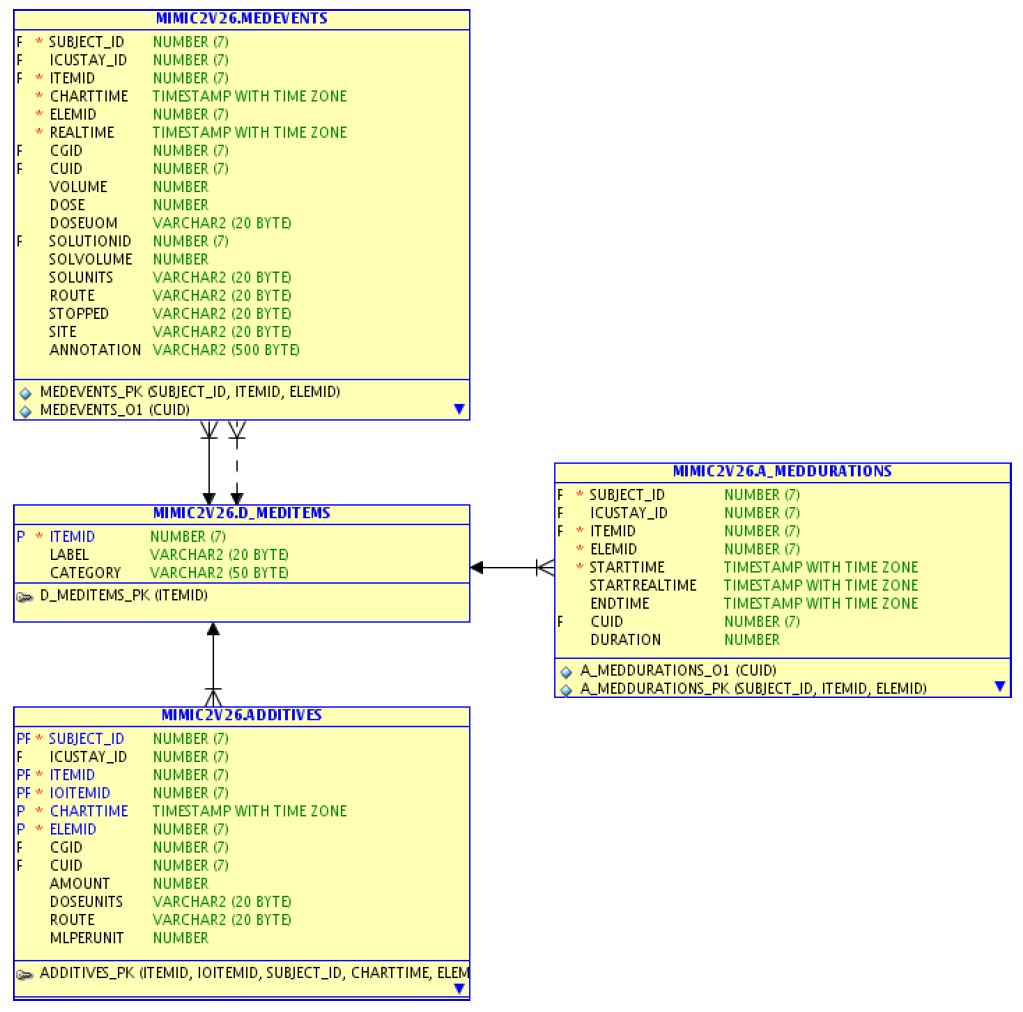}
\caption[Medication Events]{Patients' medications are stored in 4 tables: medevents, d\_meditems, a\_meddurations and additives. The events' timeline is in the medevents table while the related elements and durations are in the other tables.}
\label{fig:mimic-meds}
\end{figure}

\begin{figure}[htbp]
\centering
\includegraphics[scale=0.45]{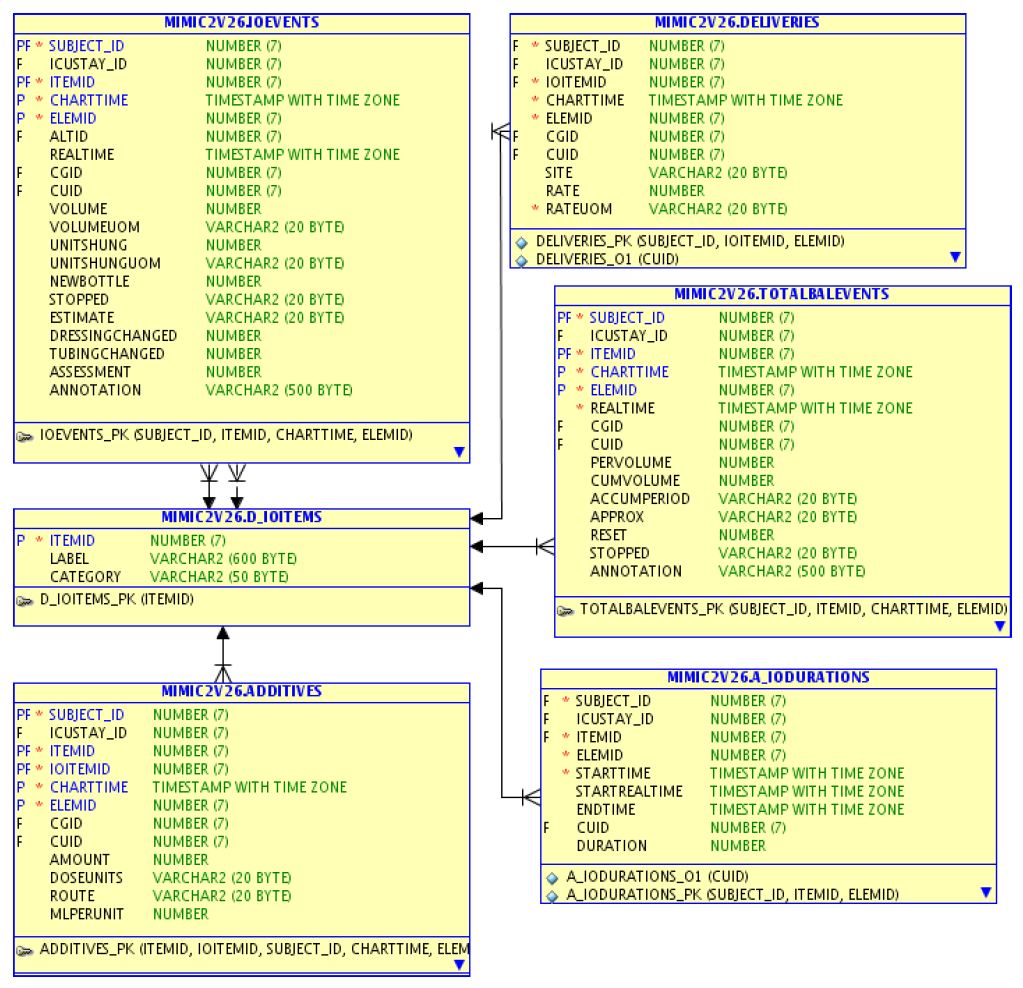}
\caption[Input/Output Events]{Patients' IO values are stored in 6 tables: ioevents, d\_ioitems, a\_iodurations, deliveries, totalbalevents and additives. The events' timeline is in the ioevents table while the related elements and durations are in the other tables.}
\label{fig:mimic-io}
\end{figure}

In Figure~\vref{fig:mimic-adm} are shown the relationships between the tables of the database which identify a patient.
The clinical conditions and related exams of each patient are stored for four significant contexts each of these in a separated series of tables: chart events, see~\vref{fig:mimic-chart}, medication events, see~\vref{fig:mimic-meds}, input/output events, see~\vref{fig:mimic-io} and lab events, see~\vref{fig:mimic-lab}.

\begin{figure}[htbp]
\centering
\includegraphics[scale=0.3]{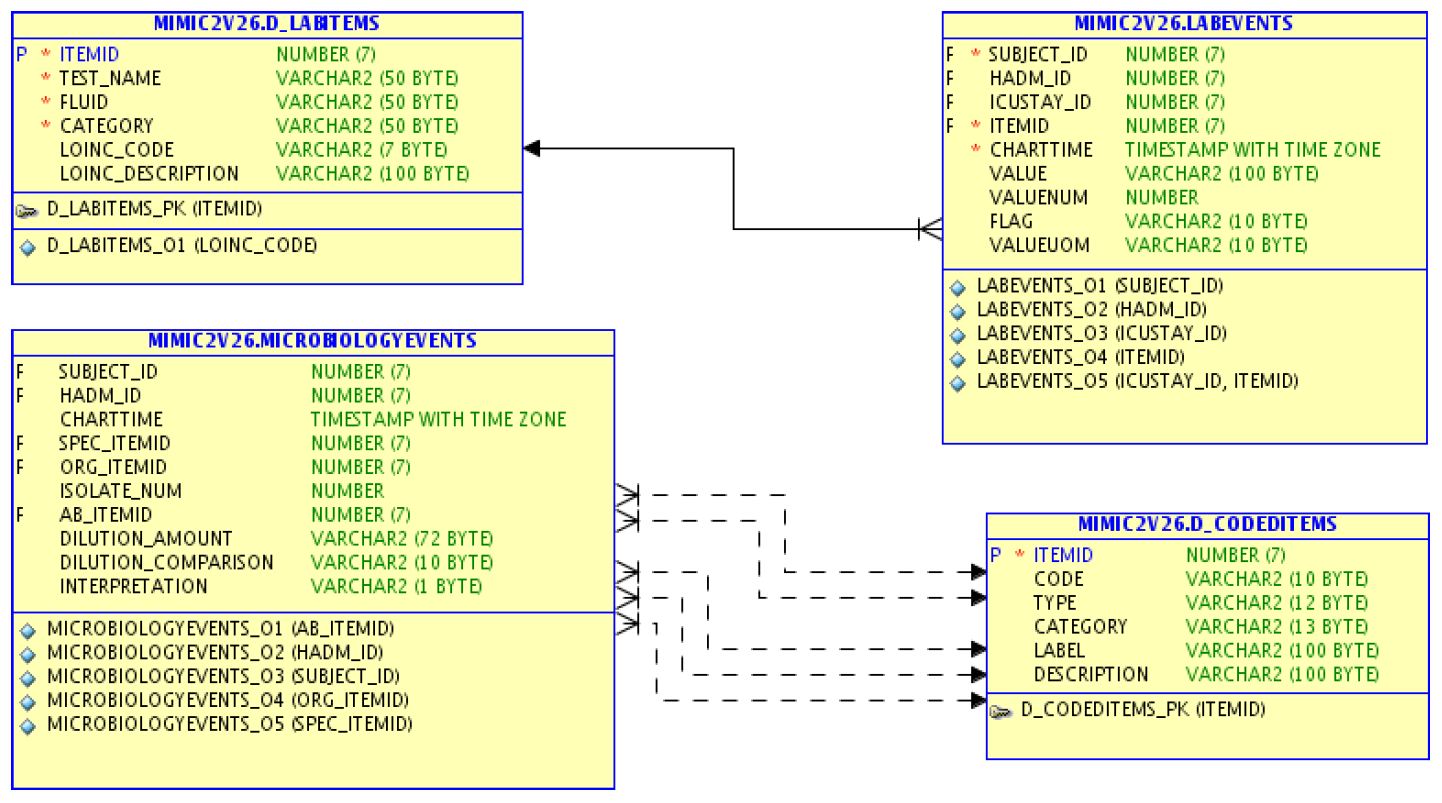}
\caption[Lab Events]{Laboratory and microbiology tests are stored in 4 tables: labevents, microbiologyevents, d\_labitems and d\_coded items. The events' timeline is in the labevents and microbiologyevents tables while the related elements, containing full descriptions of the lab tests (with LOINC codes, a database and universal standard for identifying medical laboratory observations) and microbiology tests (specimin, organism and antibiotic), are in the other tables.}
\label{fig:mimic-lab}
\end{figure}
Should be noted that even if different tables are made for each context, all of them have a central table where the timeline of the events for each patients are saved and a series of other tables where are saved the descriptions of these events, for instance the kind of performed medication or the duration of an exam.

The queries are performed for the following tasks and realize the extraction of the values for the patients on the cohort of study:
\begin{enumerate}
\item {\bf triples ordered by subject\_id:} performs the steps of the filtering and order the results by subject\_id;
\item {\bf triples ordered by hadm\_id:} performs the steps of the filtering and order the results by hadm\_id;
\item {\bf triples ordered by icustay\_id:} performs the steps of the filtering and order the results by icustay\_id;
\item {\bf discharges summaries ordered by hadm\_id:} extracts the discharge summaries and order the results by hadm\_id;
\item {\bf diuretics ordered by icustay\_id:} extracts the IDs of the patients who got diuretics at least one time during the stay in ICU and order the results by icustay\_id;
\item {\bf diuretics first time ordered by icustay\_id:} extracts the time when the patients got diuretics for the first time  and order the results by icustay\_id;
\item {\bf demographic data ordered by icustay\_id:} extracts age and gender for each patient and order the results by icustay\_id;
\item {\bf race ordered by hadm\_id:} extracts the race for each patient and order the results by hadm\_id;
\item {\bf saps ordered by icustay\_id:} extracts the saps score for each patient and order the results by icustay\_id;
\item {\bf sofa ordered by icustay\_id:} extracts the sofa score for each patient and order the results by icustay\_id;
\item {\bf elixhauser ordered by hadm\_id:} extracts the elixhauser score for each patient and order the results by hadm\_id;
\item {\bf elixhauser binary ordered by hadm\_id:} extracts the elixhauser parameters for each patient and order the results by hadm\_id;
\item {\bf creatinine ordered by icustay\_id:} extracts the creatinine values for each patient and order the results by icustay\_id;
\item {\bf fluids inputs ordered by icustay\_id:} extracts the fluids inputs values for each patient and order the results by icustay\_id;
\item {\bf fluids outputs ordered by icustay\_id:} extracts the fluids outputs values for each patient and order the results by icustay\_id;
\item {\bf use of vasopressors ordered by icustay\_id:} extract a binary value representing if vasopressors were given to the patient during the ICU. For the positive records are save the IDs of the patients order by icustay\_id; 
\item {\bf mechanical ventilation ordered by icustay\_id:} extract a binary value representing if patient was on mechanical ventilation during the ICU. For the positive records are save the IDs of the patients order by icustay\_id;
\item {\bf arterial bp ordered by icustay\_id:} extract the bloop pressure values and order the results by icustay\_id;
\item {\bf arterial bp mean ordered by icustay\_id:} extract the bloop pressure mean values and order the results by icustay\_id;
\item {\bf mortality within 30 days ordered by icustay\_id:} extract the mortality value and order the results by icustay\_id;
\item {\bf length of stay ordered by icustay\_id:} extract the lengths of stay in ICU and order the results by icustay\_id;
\end{enumerate}
The first three queries perform the filtering steps and order the results by the three IDs. This have been done to reduce the complexity of the next procedure, that will have to perform some kind of searching in a list of sorted files. The three files generated by these queries save the three IDs of the records of the dataset and in this sense represent the individual of a possibile GP population applied to the problem. The query four extract the discharge summaries used in the next procedure.

The following queries extracts the data for each variable used in the further the analysis. For the variables which have a timeline (see queries 6, 9, 10, 13, 14, 15, 18, 19), the time when a single value is refferring to is saved has an offset with respect to when the respective patient entered the ICU. For those variables with timeline the sampling rate of when the values are saved is irregular, to normalize the rate to a daily one have been computed the average of the values considered by day. Furthermore, it happens that sometimes there are more values at the same time. To decrease the weight of outlayes, have been computated the median value of those values.

The queries five and six extract the needed values for the input, that is diuretics given or not in the ICU, variable. The time of the first dose of diuretics is extracted for statistical analysis on when this first dose occurs. Queries seven and eigh extract pieces of information about age, gender and race.

Queries nine, ten and eleven extract the values for saps, sofa and elixhauser, the  twelfth query extract a binary value for 9 of the 30 parameters of the elixhauser score. Query thirteen extract the values for the creatinine variable.

Queries fourteen and fifteen extracts the amounts of fluids inputs and outputs administrated to each patients. As these values are amount, the single values available are summed daily instead of computing the average.

Queries sixteen and seventeen extract two binary values. The first one is the use of vasopressors during the stay in ICU and the second one capture a binary value regarding the usage of mechanical ventilation for a patient during the ICU stay. This second value is not directly available in the \db , therefore to obtain this value has been used an euristic procedure: if there are two changes in the ventilator's parameter for a patient at a distance longer than 6 hours, have been assumed that the current patient went from extubated to intubated, hence the mechanical ventilation was considered occurred. Queries eighteen and nineteen extract the values for the bood pressure.

The last two queries, the twenty and twenty first, extract the values for the two identified outcomes: \mort within 30 days and the \los in the ICU.

\subsection{Diuretics Naive Condition}
\label{apx:softwareNaive}
The diuretics naive condition refers to the fact that a certain patients received diuretics before the admission in the ICU. A patients is considered naive if didn't receive any kind of diuretics before entering the ICU, all the patients that were not naive, were discarded from the dataset.

The condition was verified by parsing a text like field extracted from the \db , the discharge summary. Everytime a patient leave the hospital, a summary is stored which a series of information regarding the stay of the patient in the hospital, the drugs the patients declared to have received before entering the hospital and the medications given to the patient while leaving.

However, the discharge summaries don't have a standard form: the Section of the summary are usualy, even if not always, demarcated by Section titles. These titles, though, are not always the same. For instance the diuretics informstion regarding the drugs given to the patient before entering the ICU could have been demarcated with {\it DRUGS ON ADMISSION} or with {\it ON ADMISSION} or in other different ways.

The parsing process was made with a Perl script that gets as an imput a file with the HADM\_ID and the summary of each patient and provides as an output a file with the list of HADM\_ID of the naive patients.

\subsection{Data Filtering}
\label{apx:softwareFilter}
Aim of this procedure is to create a list of records combining the SQL filtering to the diurecits naive condition. Plus the procedure set a series of variables to be mandatory for a patient, for instance the fluids inputs or outputs, and discard the records without a value for them.

The objective is achived in two steps. The first one go through the files considered mandatory and save three files containing the triples ordered by one of the three IDs each. The diuretics naive condition is combined with the mandatory variables. The second step go throught the files provided by step one and save in series of files the data for all the variales.

The input of this procedure are the files provided by the previous ones except for the discharge summary, which are analized in the diretics navice condition procedure. The output is a series of files all of them ordered by the respective ID, three with the triples (that is defining the dataset's records) and the others with all the variables.

An easy method to go through the files for the seach would be to use two nested loops for each file, one going through the triples file ordered by the ID of the current variable, and one going through the file of the variable: this approach has a complexity $O[N\cdot (n \cdot m)]$, being N the number of variables to be analized, n dimention of input file that is the current variable file, and m the dimension of the dataset. In this way the extraction process was too computationally expensive.

The complexity was then reduce to $O[N\cdot (n + m)]$ by exploiting the fact that the files were in numerical acending order. In this way every line of the files were read to most one time each. The pseudo-code of the algorithm is shown in Algorithm~\vref{sorted-search}.
\begin{algorithm}
\caption[Sorted Search]{Pseudocode of the algorithm used to perform the merging process.}
\begin{algorithmic}
\STATE external loop go through the IDs file ($F1$)
\STATE internal loop go through the current variable file ($F2$)
\WHILE{(Both files have lines)}
\IF{$F1 == F2$}
\STATE go on reading one line from $F2$ file. Then write one line to output
\ENDIF
\IF{$F1 != F2 AND F1 < F2$}
\STATE save $F2$ data and go on one step with $F1$
\ENDIF
\IF{$F1 != F2 AND F1 > F2$}
\STATE go on reading from file $F2$
\ENDIF
\ENDWHILE
\end{algorithmic}
\label{sorted-search}
\end{algorithm}

\section{Variables Preparation}
\label{apx:softwareVarPreparation}
The variables preparation consists in a transformation of the dataset's values in a format straight forward usable in the analysis procedures.

Before the variables preparation itself, are executed a few script as follow:
\begin{itemize}
\item run a script to get age and gender values: in the \db all patients over 90 years old have been saved as 200 years old, the script get the true age. Then the value for gender, that is M for male and F for female, is modified in a binary one, 0 for male and 1 for female;
\item run a script to get the elixhauser values: extracts in single files the binary values for the exlixhauser parameters;
\item run a script to compute the balance values defined as $inputs - outputs$;
\item run a script to define the times where to save certain values, that is $T1$, $T2$ and $T3$, the times discussed previously in Chapter~\ref{cpt:datasetExtraction}.
\end{itemize}
Completed these steps, the variables preparation starts and each variable is saved in a file, in particulare the following procedures have been realized:
\begin{itemize}
\item average processing: to save the values for the variables that require to calculate the average (or the sums) till a certain day;
\item binary processing: to save the values for the binary variables;
\item numeric processing: to save the values for the numeric variables;
\item time processing: to save the values for the variables for the variables with timelines.
\end{itemize}
The scripts define above, also perform an analysis on the results and save the in a file.

After the variables preparation the procedure save the results in two files directly usable for the analysis. The first file contains the labels for the computated variables, the second the data itself.

\section{Propensity Analysis}
\label{apx:softwarePA}
The propensity analysis, following\cite{PS-FIRST}, have been performed by the following procedures:
\begin{enumerate}
\item {\bf Fitting the Propensity Score:} the propensity score, that is the the conditional probability of assignment to a particular treatment given a vector of observed covariates, is calculated with a logit model.
\item {\bf Generating the five \Group:} the patients are ranked according their propensity score and then divided in five \groups .
\item {\bf Assessing the Balance:} the balance in each of the five \groups is evaluated with the ANOVA test for primary effects and secondary effects. 
\item {\bf Refining the \Group:} the balance in the \group is improved by inserting the variables with large F-ratio and the computing a new logit model with them.
\end{enumerate}
In Figure~\vref{fig:ps-phases} the basic elements of the code produced to implement the propensity method.

Now a description of these procedures will be provided. These Sections extend what have already been discussed in Chapter~\ref{cpt:propensityAnalysis}.
\begin{figure}
\centering
\rotatebox{90}{
\stepcounter{figure}
\begin{minipage}{\textheight}
\centering
\includegraphics[width=0.7\textheight]{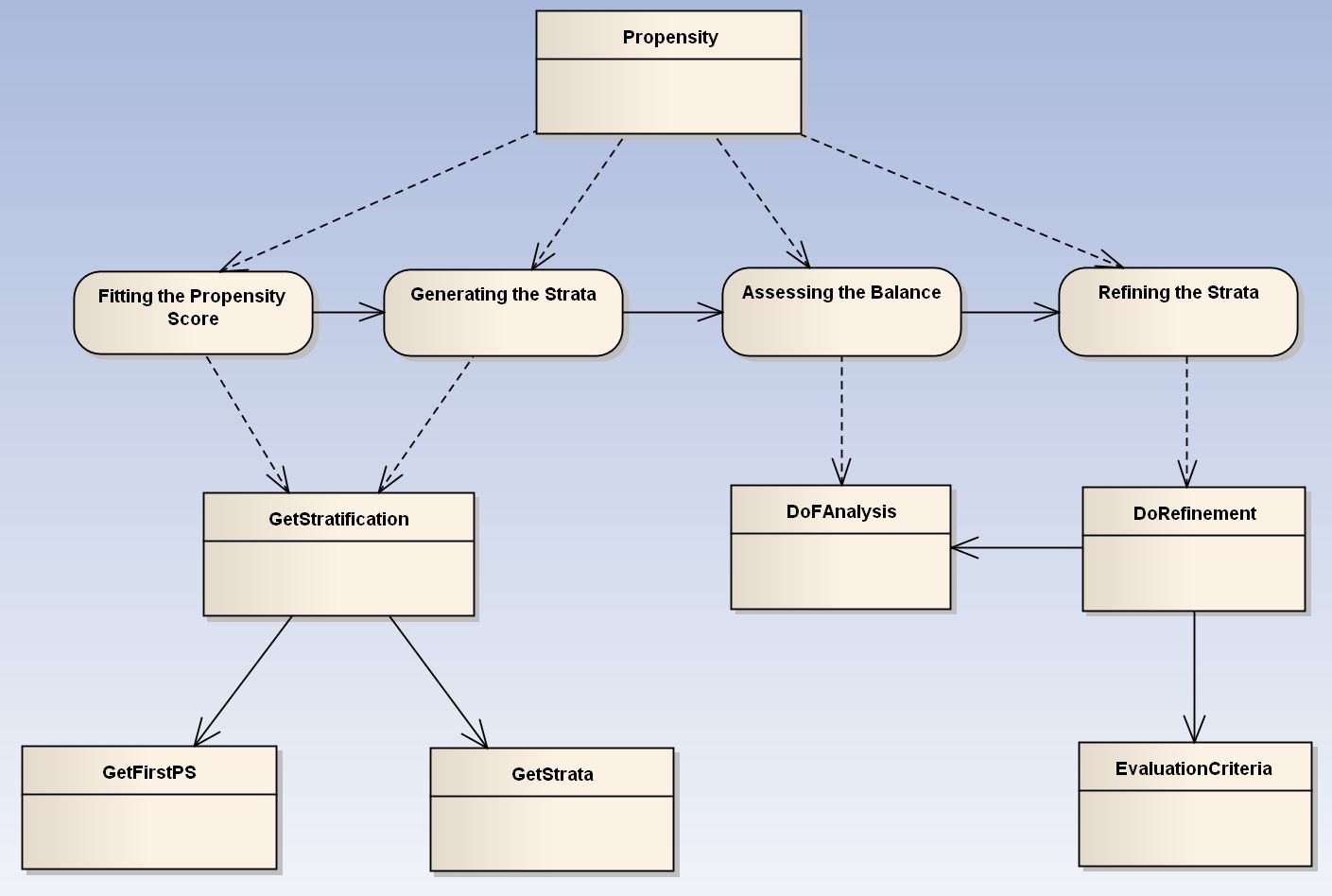}
\addtocounter{figure}{-1}
\captionof{figure}[Propensity Method Process]{Steps of the propensity analysis. It is composed by four importants parts: a)fitting the propensity score, b)generating the five \group, c)assessing the balance and d)refining the \group.}
\label{fig:ps-phases}
\end{minipage}
}
\end{figure}
In Algorithm~\vref{ps-process} is shown an overview of all the propensity score process.
\begin{algorithm}
\caption[Propensity Score]{Pseudocode of the Propensity Score process.}
\begin{algorithmic}
\STATE Generate the stepwise logit model $M1$ on main effects
\STATE Generate the stepwise logit model $M2$ including interactions
\STATE Based on $M2$, compute the propensity score $PS$
\STATE Rank the patients based on $PS$ and perform the subclassification
\STATE Evaluate the balance according to the F-ratios
\STATE Rank the variables based on their F-ratios
\STATE Chose the first variable in the ranked list, $X$
\WHILE{(All the variables not in $M2$ have not considered)}
\IF{(Including $X$), its F-ratio improves}
\STATE include $X$ in the model and proceed to the next variable
\ENDIF
\IF{(Including $X^2$), its F-ratio improves}
\STATE include $X^2$ in the model and proceed to the next variable
\ENDIF
\IF{(Including interaction of $X$), its F-ratio improves}
\STATE include interaction of $X$ in the model and proceed to the next variable
\ENDIF
\STATE Chose the next variable in the ranked list, $X = X_i$
\ENDWHILE
\end{algorithmic}
\label{ps-process}
\end{algorithm}

\subsection{Fitting the Propensity Score}
\label{apx:softwareFit}
The propensity score is calculated by performing a two-phases stepwise discriminant analysis with a logit model: first it is generated a model evaluating the main effects of all the variables in the dataset. The output of this process is a list of variables chosen as main effects. In the study 11 variables were chosen in this first logit model.

After that, a second stepwise discriminant analysis is performed considering only the variables whose main effects were chosen in the first logit model and in this second analysis the interactions between these variables are considered: the result of this second logit model was a model of 17 variables, 11 main effects defined in the first stepwise discriminant analysis and 6 interactions between them added in the second stepwise discriminant analysis. The full list of those variables is presented in Chapter~\ref{cpt:propensityAnalysis}.

At this point, using this model it is possibile to calculate a probability of getting the diuretics. Being $x$ the resulting values available for each patient using the model on their data, $p(x)$ of receiving diuretics is: $p(x) = \frac{e^x}{1 + e^x}$. 

\subsection{Generating the five \Group}
\label{apx:softwareQuintiles}
The records of the patients are then ranked in an increasing order according to the propensity score calculated with the second logit model. Based on this ranked order, are define 5 groups, called \group, and the patients with a lower probability of being administrated with diuretics are in group 1 while the ones with the higher one are in group 5.

\subsection{Assessing the Balance}
\label{apx:softwareAssesing}
The balance in each \group is evaluated using the ANOVA test. ANOVA provides a statistical test of whether or not the means of several groups are all equal.

The test produces a ration $F = \frac{variance-between-groups}{variance-within-groups}$, it is based on the partitioning of the total sum of squares $S$ into components related to the effects used in the model.

Will be defined as $S_1$ the sum of squares of the differences between the means in each group, $m_i$, and the overall mean $m$, that is: $S_1 = \sum_{i} n_i \cdot (m_i - m)^2$, with $n_i$ number of elements of group $i$.

The will be defined as $S_2$ the sum of squares of the differences between the means in each group, $m_i$, and the value of a certain element of that group, $x_{i,j}$. That is $S_2 = \sum_{i} \sum_{j} n_i \cdot (x_{i,j} - m_i)^2$.

At this point, being $k$ the number of groups to be evaluated and $n$ all the elements in the groups, the F-ratio defined by ANOVA is $F = \frac{(\frac{S_1}{k-1})}{(\frac{S_2}{n-k})}$.

\subsubsection{Evaluating the Balance}
\label{apx:softwareEvaluating}
At this point, the one-way ANOVA test have been performed on the original dataset: the test compares the treated vs not treated patients and, of course, the F-values are high.

Then the two-ways ANOVA is calculated on the dataset divided in the 5 \group. The test produces two values: the first one is made considering the main effect of the diuretics (given vs not given) variable. Consider for instance the comparison between two binary variables, if the combinations of their values are listed in a table, the result is a $2 \times 2$ matrix: the main effect values that the two-way ANOVA would calculate on this table are two, the first one considering the rows and the second one considering the columns.
\begin{figure}[htbp]
\centering
\includegraphics[scale=0.64]{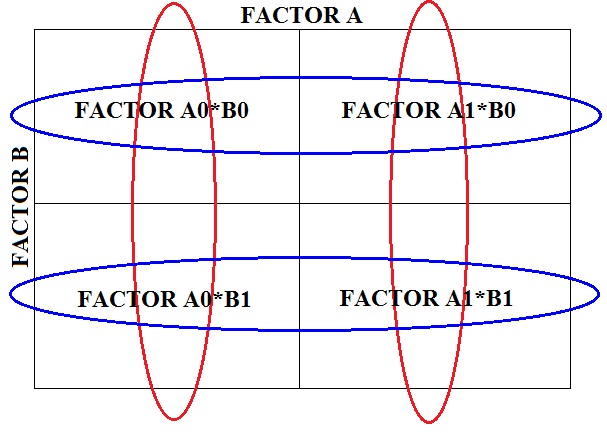}
\caption[Main Effect for a two-ways ANOVA]{The groups for a two-way ANOVA on the main effects. In red are the 2 groups considering the columns and in blue the 2 groups considering the rows.}
\label{fig:table-main-effects}
\end{figure}
The main effect values in the ANOVA test consider as gourps to be compared only the rows or the columns, mixed combinations are not allowed. In Figure~\vref{fig:table-main-effects} the groups for main effects are shown.

The second values considers the interaction effects, that is the effects of one factor on the others. An $A \cdot B$ interaction is a change in the simple main effect of $B$ over levels of $A$ or the change in the simple main effect of $A$ over levels of $B$. Here are involved mixed combinations. In Algorithm~\vref{2-ways-anova} are shown all the steps to perform a two-ways ANOVA.

In the plots in Chapter 4 are shown the improvements of the balance both of main effects and interactions.
\begin{algorithm}
\caption[Two-Ways ANOVA]{Pseudocode of the Two-Ways ANOVA with 2 binary variables.}
\begin{algorithmic}
\STATE {\bf Step 1:} Consider the main effects, that is rows and columns of the two-ways binary factors $A$ and $B$
\STATE {\bf Step 2:} Calculate the overall mean $m$
\STATE {\bf Step 3:} Considering the rows (first factor, $A$), calculate the means of the 2 rows ($i = 1,2$), $m_{a,i}$
\STATE {\bf Step 4:} Considering the rows (first factor, $A$), calculate the elements of the 2 rows ($i = 1,2$), $n_{a,i}$
\STATE {\bf Step 5:} Calculate the between-groups sum of squares for $A$, $S_{1,a}$
\STATE {\bf Step 6:} Considering each element of the groups $x_{i,j}$ calculate the within-groups sum of squares for $A$, $S_{2,a}$
\STATE {\bf Step 7:} Calculate the number of groups $k_a = 2$ and the elements in all the groups $n$
\STATE {\bf Step 8:} Calculate the ratio $F_a$ for factor $A$ using $S_{1,a}$, $S_{2,a}$, $k_a$ and $n$
\STATE {\bf Step 9:} Repeat the process for the columns (second factor, $B$), obteining $S_{1,b}$, $S_{2,b}$ and $F_b$
\STATE {\bf Step 10:} Calculate the between-groups sum of squares, $S_{bw}$, considering all the 4 groups (2 rows and 2 columns)
\STATE {\bf Step 11:} Calculate $S_{1,a \cdot b} = S_{bw} - S_{1,a}, - S_{1,b}$
\STATE {\bf Step 12:} Calculate the within-groups sum of squares, $S_{wi}$, considering all the 4 groups (2 rows and 2 columns)
\STATE {\bf Step 13:} Calculate $S_{2,a \cdot b} = S_{wi} - S_{2,a}, - S_{2,b}$
\STATE {\bf Step 14:} Calculate the number of groups $k_{a \cdot b} = 2 \cdot 2$ and the elements in all the groups $n$
\STATE {\bf Step 15:} Calculate the ratio $F_{a \cdot b}$ for interaction $A \cdot B$ using $S_{1,a \cdot b}$, $S_{2,a \cdot b}$, $k_{a \cdot b}$ and $n$
\end{algorithmic}
\label{2-ways-anova}
\end{algorithm}

\subsection{Refining the \Group}
\label{apx:softwareRefinement}
The refinement is an iterative process. First all the variables excluded by the model are ranked according to their F-ratios and of those the 25\% is inserted one by one in a new model. If after their inclusion, the balance for the current variable is not improving, the square of this variable and then the interactions of it with the variables in the model generated in the fitting phase are tried. If none of this possibilieties improves the F-ratio of the analyzed variable, it is discarded. If it improves it is included in the model.

\section{\OA and Machine Learning with GP Analysis}
\label{apx:softwareOAandML}
After performing the procedures of the previous steps, the ones relating to these two types of analysis are simple.

As regards the \oa , were carried out a series of regressions performed using the methods provided by Matlab and a series of new stratifications performed using procedures similar to the ones previously described.

As regards the machine learning with gp analysis, as already said, the executions of the genetic programming were carried out using GPLAB.

\clearpage \mbox{} \clearpage 
\chapter{Details on the Datasets}
\label{apx:dataset}

In this Appendix details on the datasets will be provided.

\section{List of Diuretics}
\label{apx:diuretics}
The diuretics variable was computated by looking in the \db for the following list of drugs:
\begin{itemize}
\item acetazolamide (Diamox), dichlorphenamide (Daranide);
\item methazolamide(Glauctabs, MZM, Neptazane), torsemide (Demadex), furosemide (Lasix);
\item pironolactone (Aldactone), amiloride (Midamor), triamterene (Dyrenium);
\item hydrochlorothiazide (HCTZ, HydroDIURIL, Aquazide H, Esidrix, Microzide), metolazone (Mykrox, Zaroxolyn);
\item methyclothiazide (Enduron, Aquatensen), chlorothiazide (Diuril), indapamide (Lozol);
\item bendroflumethiazide (Naturetin), polythiazide (Renese), hydroflumethiazide (Saluron), chlorthalidone (Thalitone).
\end{itemize}
This list has been suggested by the medical experts.

\section{List of Fluids}
\label{apx:fluids}
The fluids inputs variable has be computated looking in the \db for the following list of items:
\begin{itemize}
\item 106 - Lactated Ringers, 107 - .9\% Normal Saline, 130 - D5/.45NS, 131 - D5/.45NS 10000.0ml;
\item 134 - .9\% Normal Saline 1000.0ml, 142 - Lactated Ringers 1000.0ml, 151 - 45\% Normal Saline 1000.0ml, 152 - D5/.45NS 1000.0ml;
\item 154 - D5NS, 165 - D5W 1000.0ml, 180 - .45\% Normal Saline, 187 - .9\% Normal Saline 500.0ml;
\item 214 - D5 Normal Saline, 219 - D5RL 1000.0ml, 249 - .9\% Normal Saline 250.0ml, 297 - D5NS 1000.0ml;
\item 299 - D5 Normal Saline 1000.0ml, 309 - .9\% Normal Saline 100.0ml, 615 - D5/.45NS 2000.0ml, 631 - .9\% Normal Saline 2000.0ml.
\end{itemize}
The number next to each fluid is the corresponding identifier in the database.

\section{Variables Descriptive Statistics}
\label{apx:varHistos}
In this Section histograms of the values of all the variables are provided.

Figure~\vref{fig:histo_in-out} show the histograms for diuretics, \mort and \los .
\begin{figure}
\centering
\rotatebox{90}{
\stepcounter{figure}
\setcounter{subfigure}{0}
\begin{minipage}{\textheight}
\begin{minipage}{0.33\textheight}
\centering
\includegraphics[width=.9\linewidth]{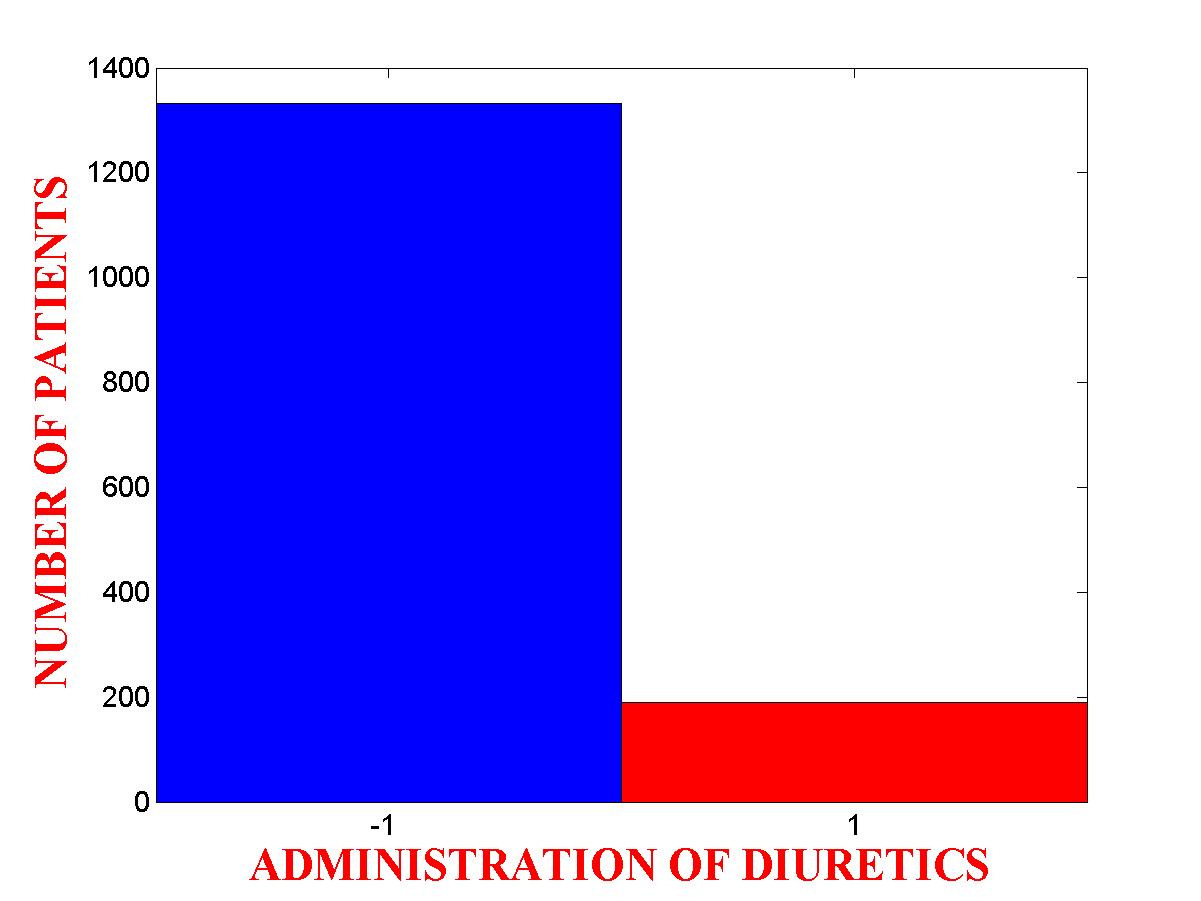}
\captionof{subfigure}[]{189 patients got diuretics and 133 didn't.}
\label{fig:hist_1}
\end{minipage}
\begin{minipage}{0.33\textheight}
\centering
\includegraphics[width=.9\linewidth]{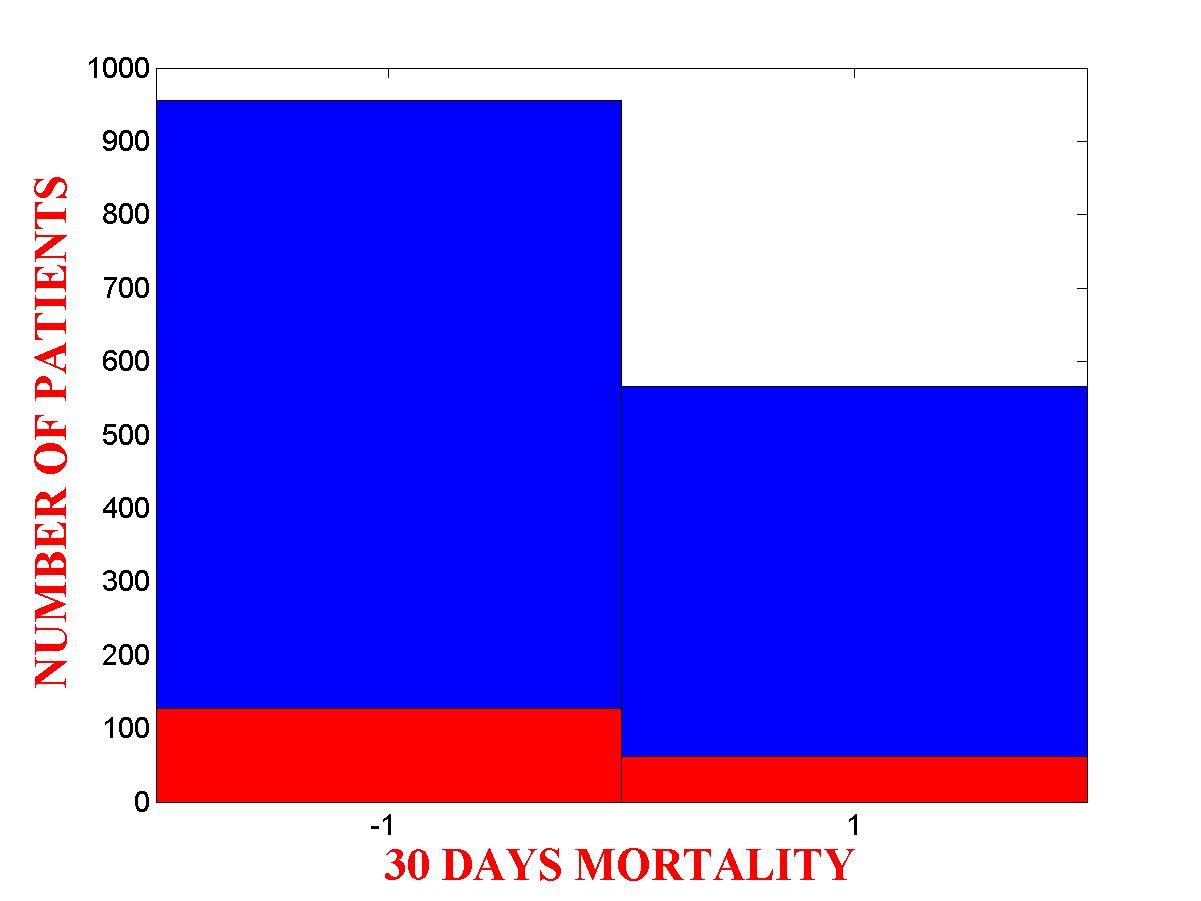}
\captionof{subfigure}[]{566 of 1522 patients died within 30 days.}
\label{fig:hist_2}
\end{minipage}
\begin{minipage}{0.33\textheight}
\centering
\includegraphics[width=.9\linewidth]{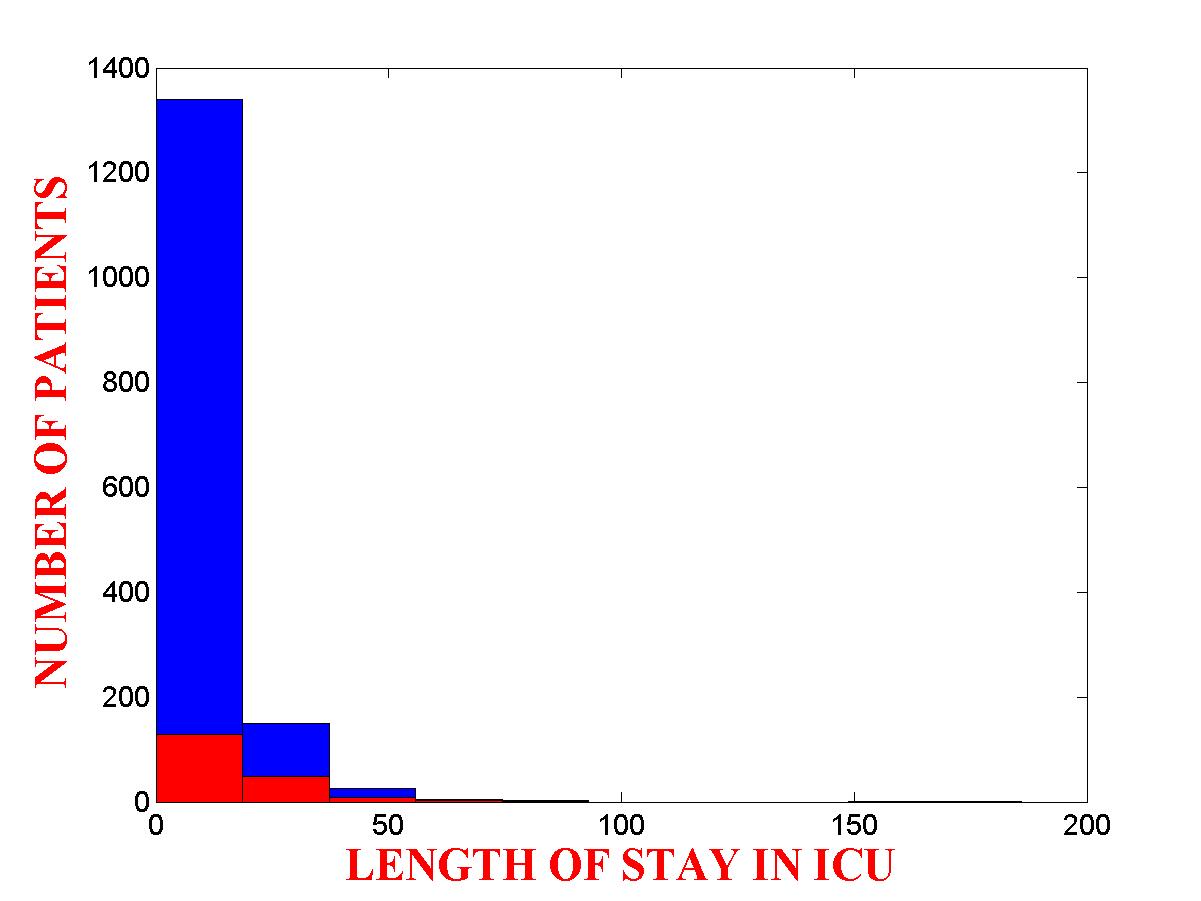}
\captionof{subfigure}[]{\LOS is centered on 7.4 days.}
\label{fig:hist_3}
\end{minipage}
\addtocounter{figure}{-1}
\captionof{figure}[Histograms of Diuretics, \Mort and \LOS]{Histograms of diuretics, \mort and \los .}
\label{fig:histo_in-out}
\end{minipage}
}
\end{figure}

Figures~\vref{fig:histo_binary_1} and~\vref{fig:histo_binary_1}  show the histograms for gender, race, use of vasopressor and mechanical ventilation.
\begin{figure}
\centering
\rotatebox{90}{
\stepcounter{figure}
\setcounter{subfigure}{0}
\begin{minipage}{\textheight}
\begin{minipage}{0.45\textheight}
\centering
\includegraphics[width=.9\linewidth]{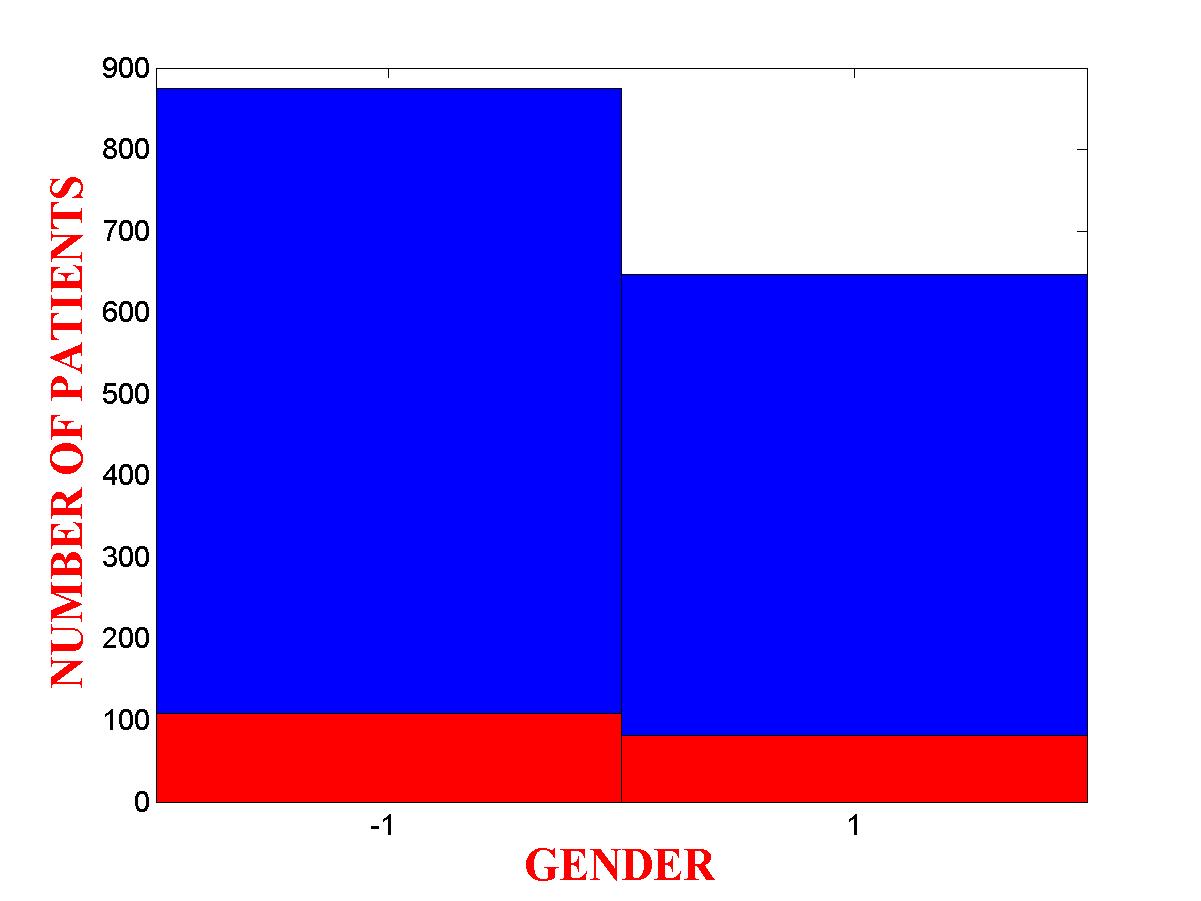}
\captionof{subfigure}[]{647 of 1522 patients are female and 875 male.}
\label{fig:hist_4}
\end{minipage}
\begin{minipage}{0.45\textheight}
\centering
\includegraphics[width=.9\linewidth]{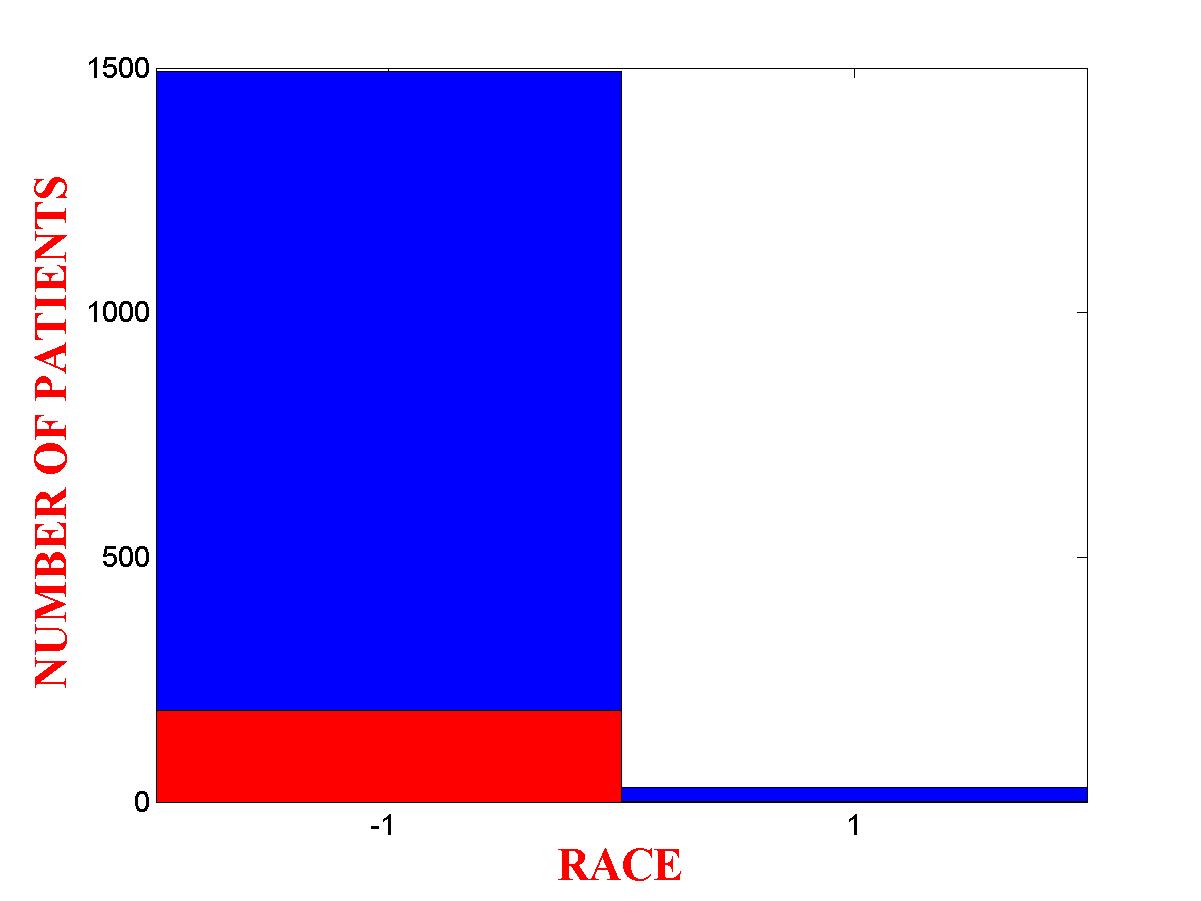}
\captionof{subfigure}[]{1492 of 1522 patients are not white.}
\label{fig:hist_5}
\end{minipage}
\addtocounter{figure}{-1}
\captionof{figure}[Histograms of Gender and Race]{Histograms of gender and race.}
\label{fig:histo_binary_1}
\end{minipage}
}
\end{figure}
\begin{figure}
\centering
\rotatebox{90}{
\stepcounter{figure}
\setcounter{subfigure}{0}
\begin{minipage}{\textheight}
\begin{minipage}{0.45\textheight}
\centering
\includegraphics[width=.9\linewidth]{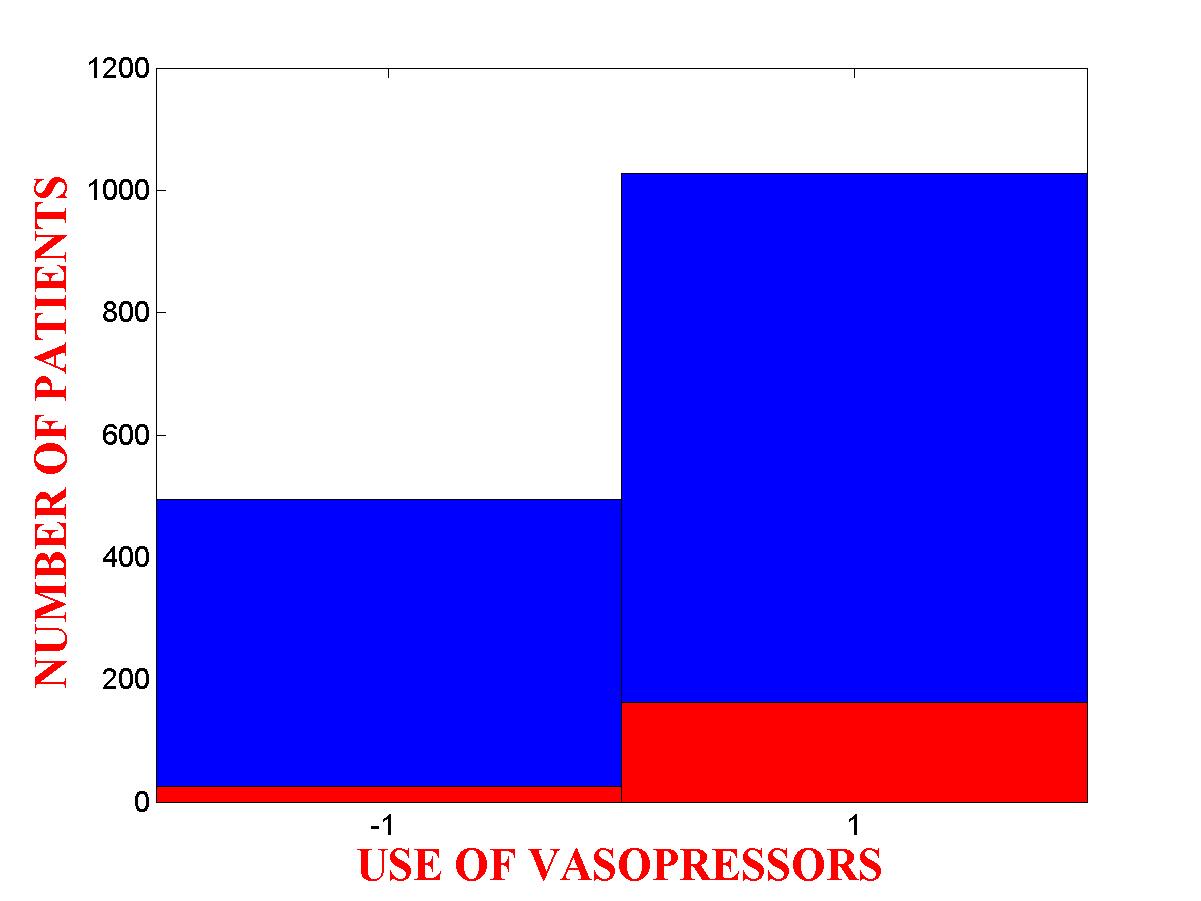}
\captionof{subfigure}[]{1028 of 1522 patients are on vasopressors.}
\label{fig:hist_15}
\end{minipage}
\begin{minipage}{0.45\textheight}
\centering
\includegraphics[width=.9\linewidth]{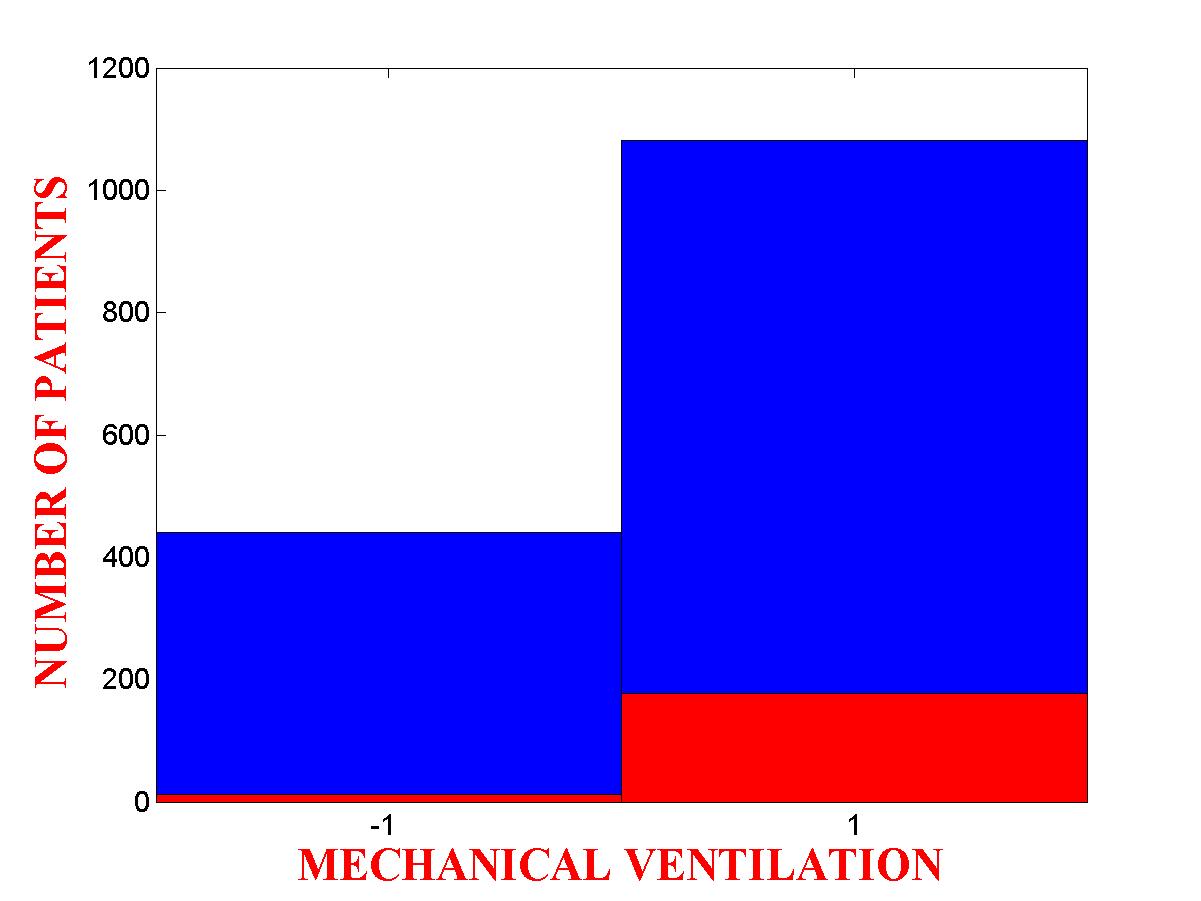}
\captionof{subfigure}[]{1081 of 1522 patients are on mechanical ventilation}
\label{fig:hist_16}
\end{minipage}
\addtocounter{figure}{-1}
\captionof{figure}[Histograms of Use of Vasopressors and Mechanical ventilation]{Histograms of use of vasopressor and mechanical ventilation.}
\label{fig:histo_binary_2}
\end{minipage}
}
\end{figure}

Figures~\vref{fig:histo_eli_1} and ~\vref{fig:histo_eli_2} show the histograms for the 9 Elixhauser parameters.
\begin{figure}
\centering
\rotatebox{90}{
\stepcounter{figure}
\setcounter{subfigure}{0}
\begin{minipage}{\textheight}
\begin{minipage}{0.33\textheight}
\centering
\includegraphics[width=.9\linewidth]{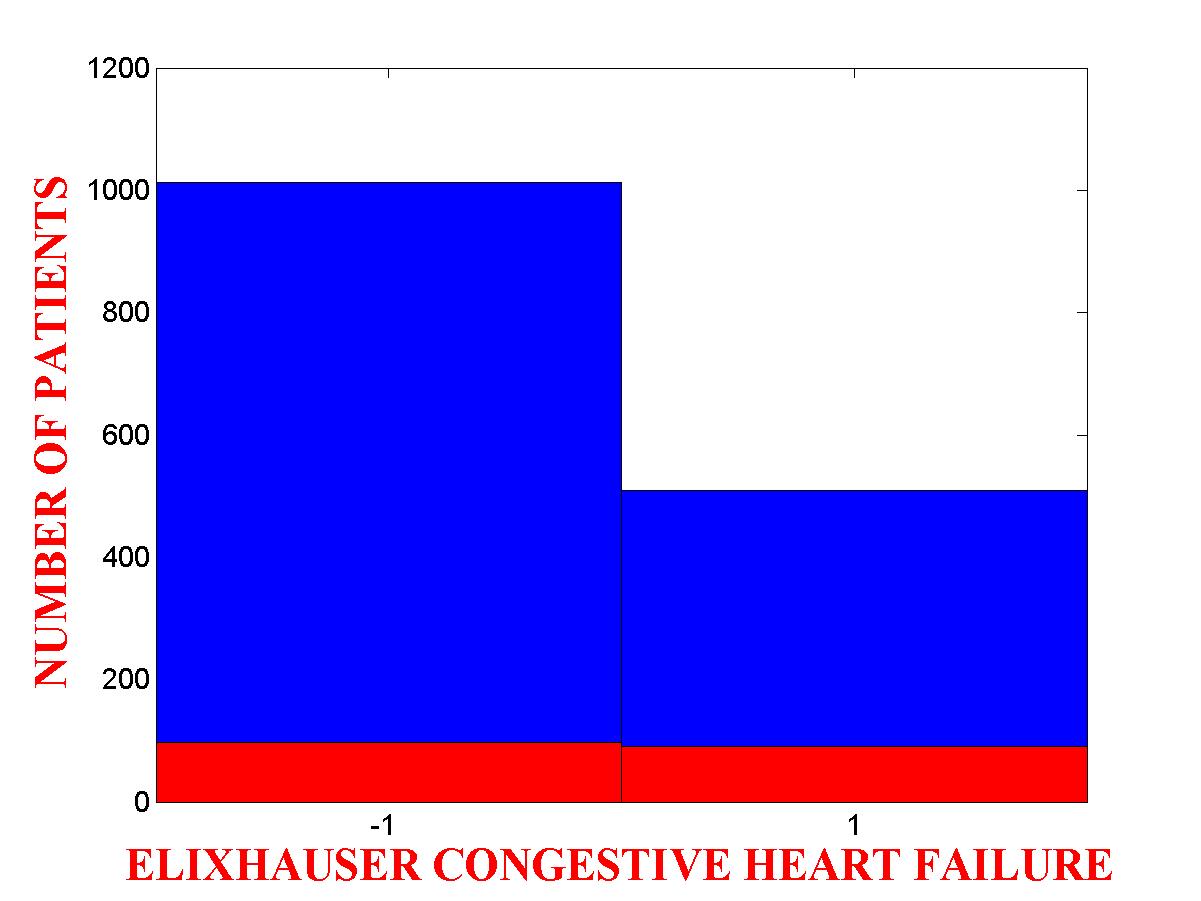}
\captionof{subfigure}[]{509 patients had congestive heart failure.}
\label{fig:hist_6}
\end{minipage}
\begin{minipage}{0.33\textheight}
\centering
\includegraphics[width=.9\linewidth]{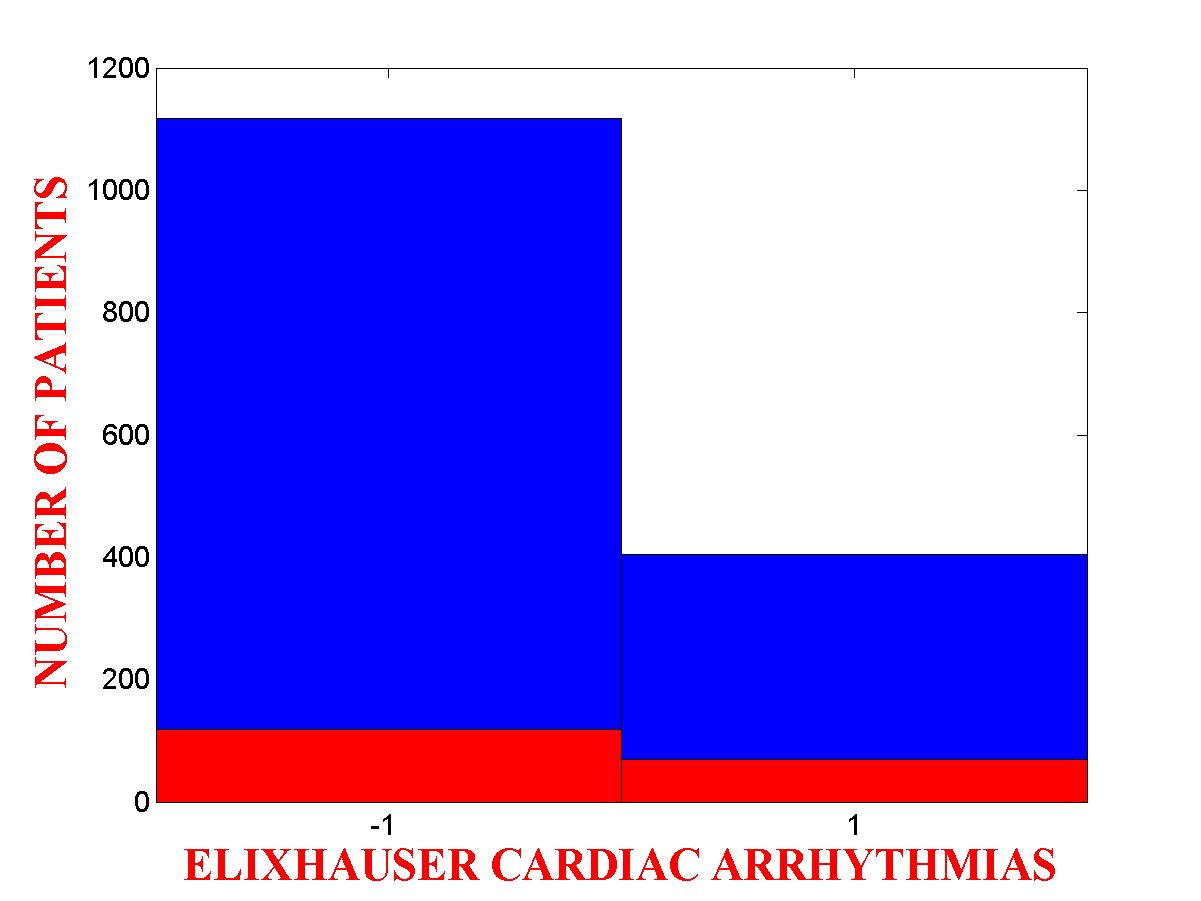}
\captionof{subfigure}[]{405 patients had cardiac arrhythmias.}
\label{fig:hist_7}
\end{minipage}
\begin{minipage}{0.33\textheight}
\centering
\includegraphics[width=.9\linewidth]{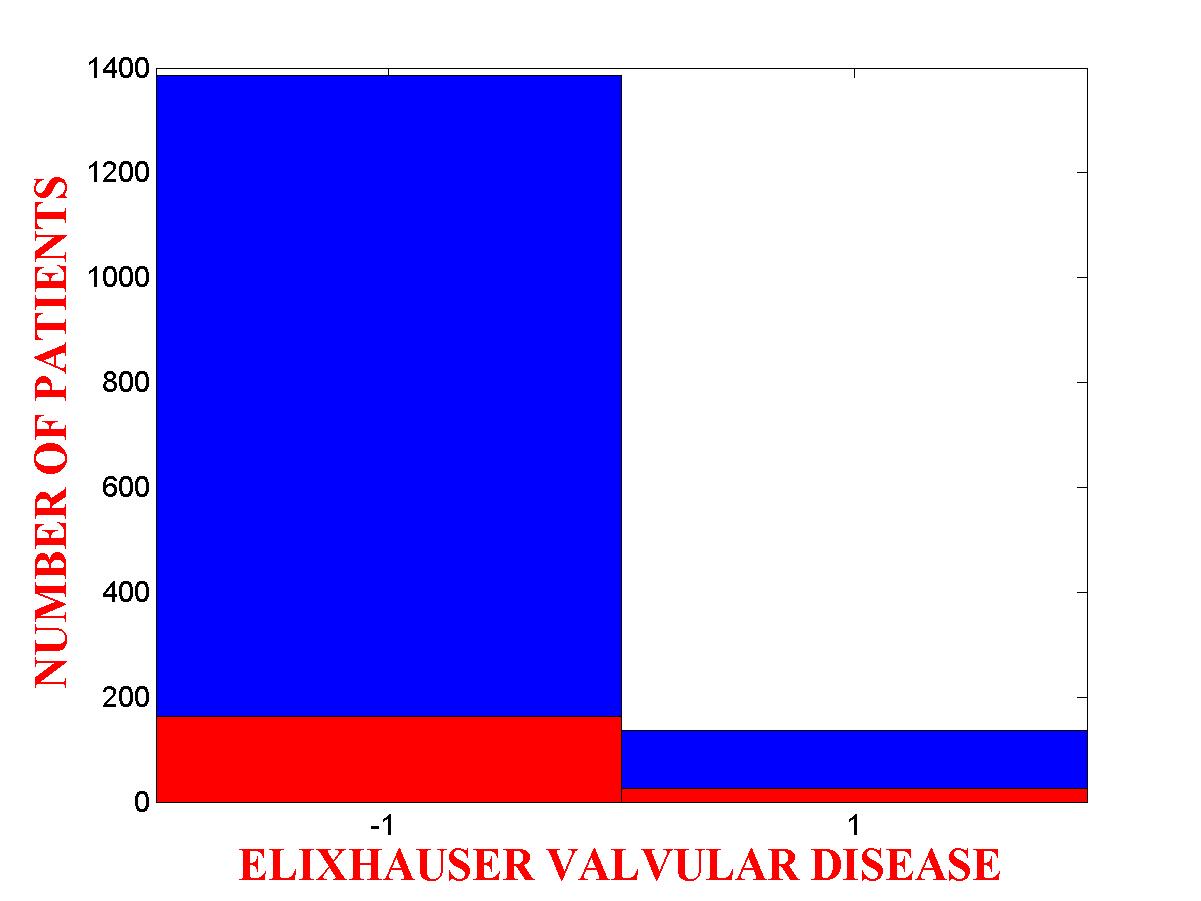}
\captionof{subfigure}[]{136 patients had valvular disease.}
\label{fig:hist_8}
\end{minipage}
\addtocounter{figure}{-1}
\captionof{figure}[Histograms of the Elixahuser Parameters, part 1]{Histograms of the Elixahuser parameters, part 1.}
\label{fig:histo_eli_1}
\end{minipage}
}
\end{figure}
\begin{figure}
\centering
\rotatebox{90}{
\stepcounter{figure}
\setcounter{subfigure}{0}
\begin{minipage}{\textheight}
\begin{minipage}{0.33\textheight}
\centering
\includegraphics[width=.9\linewidth]{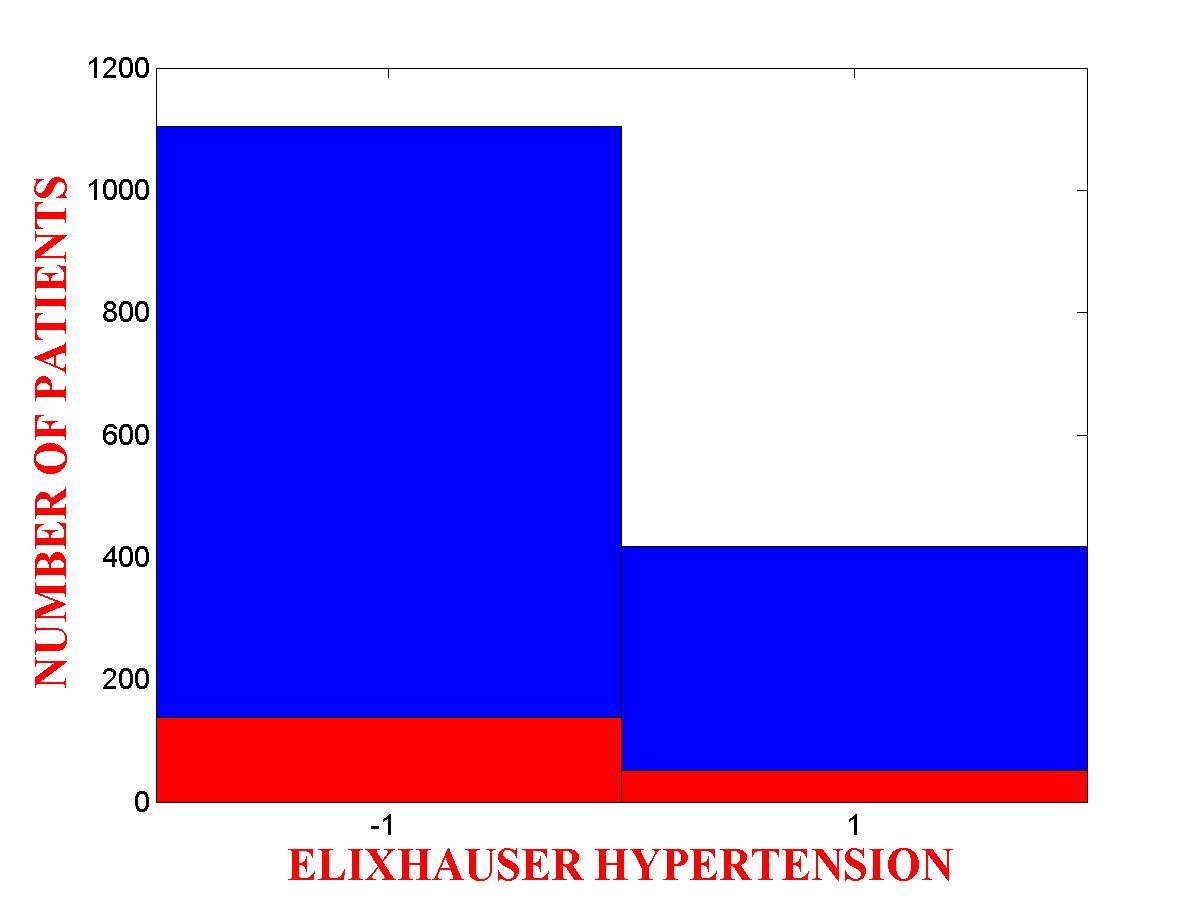}
\captionof{subfigure}[]{417 patients had hypertension.}
\label{fig:hist_9}
\end{minipage}
\begin{minipage}{0.33\textheight}
\centering
\includegraphics[width=.9\linewidth]{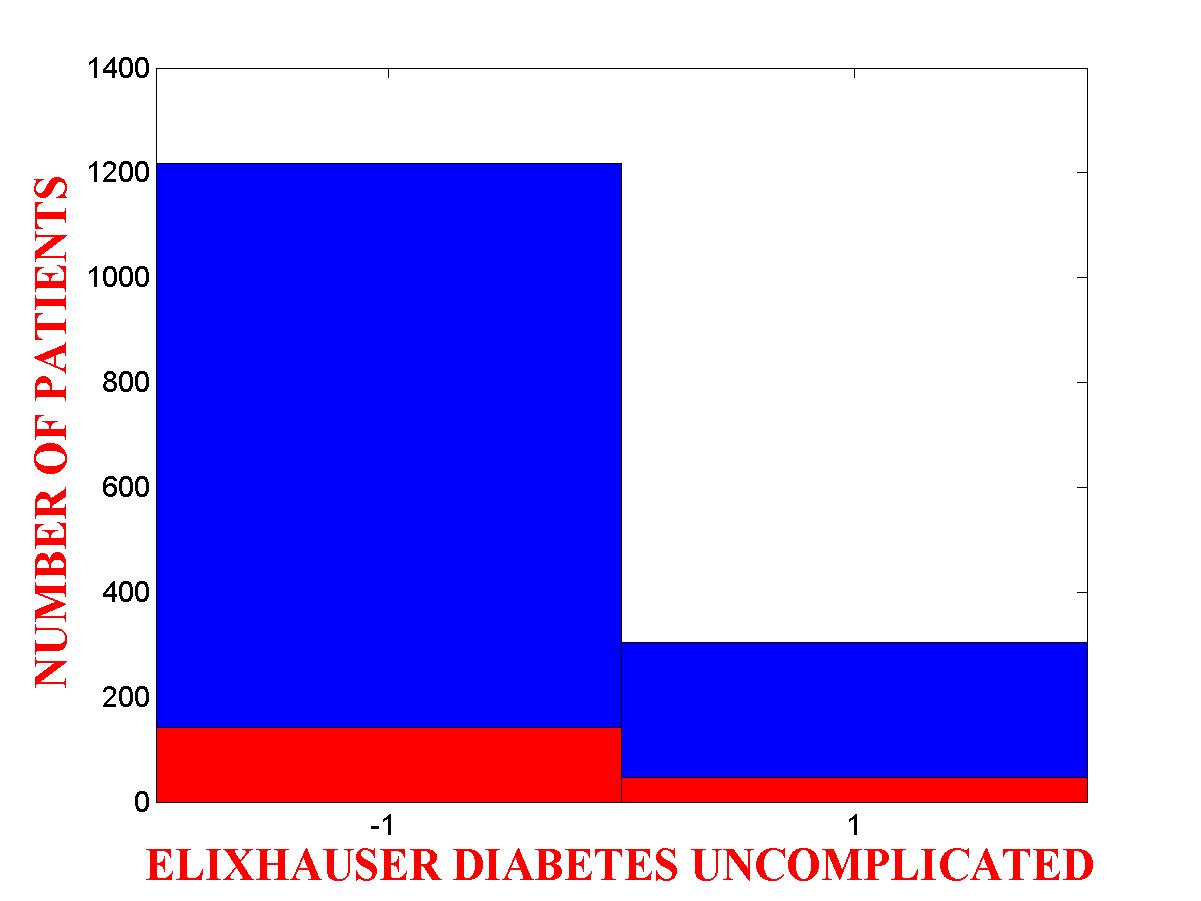}
\captionof{subfigure}[]{305 patients had diabetes uncomplicated.}
\label{fig:hist_10}
\end{minipage}
\begin{minipage}{0.33\textheight}
\centering
\includegraphics[width=.9\linewidth]{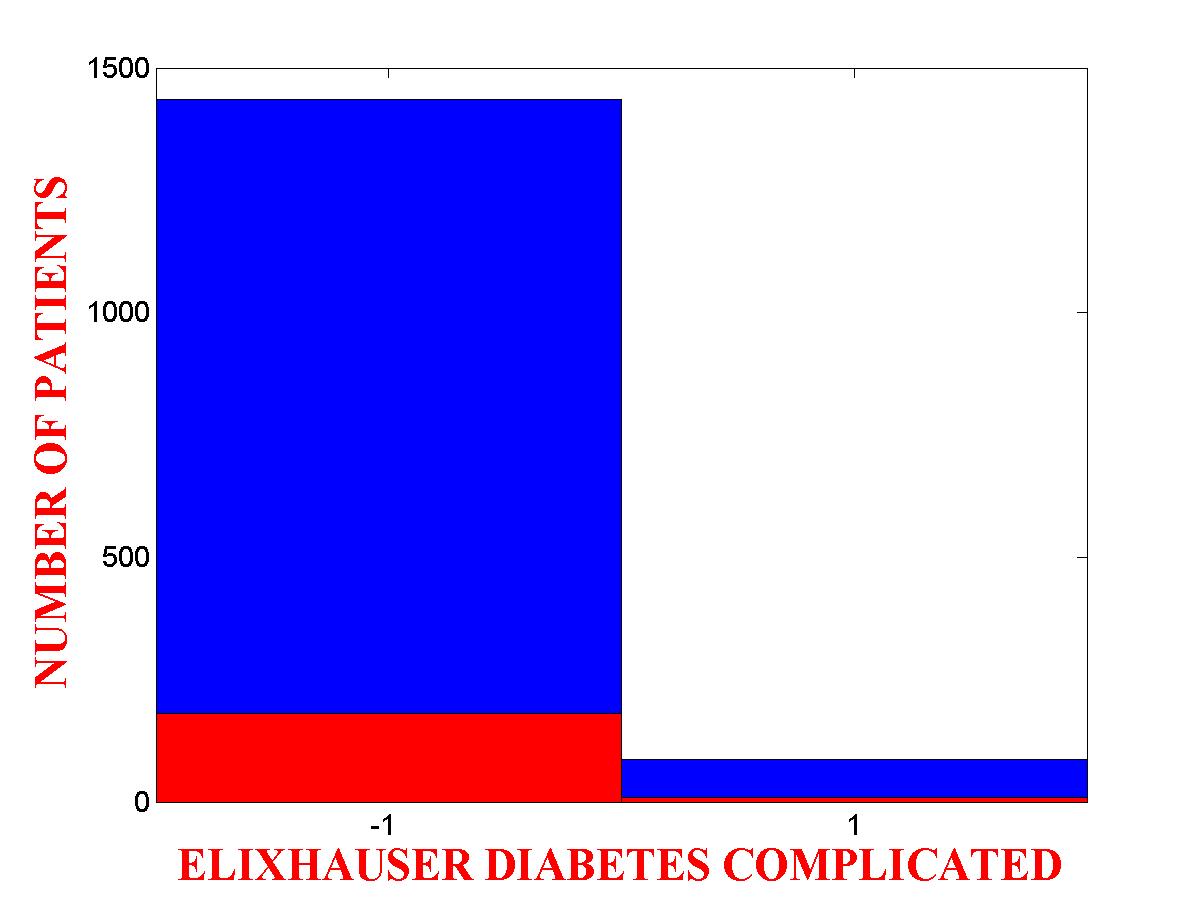}
\captionof{subfigure}[]{86 patients had diabetes complicated.}
\label{fig:hist_11}
\end{minipage}
\begin{minipage}{0.33\textheight}
\centering
\includegraphics[width=.9\linewidth]{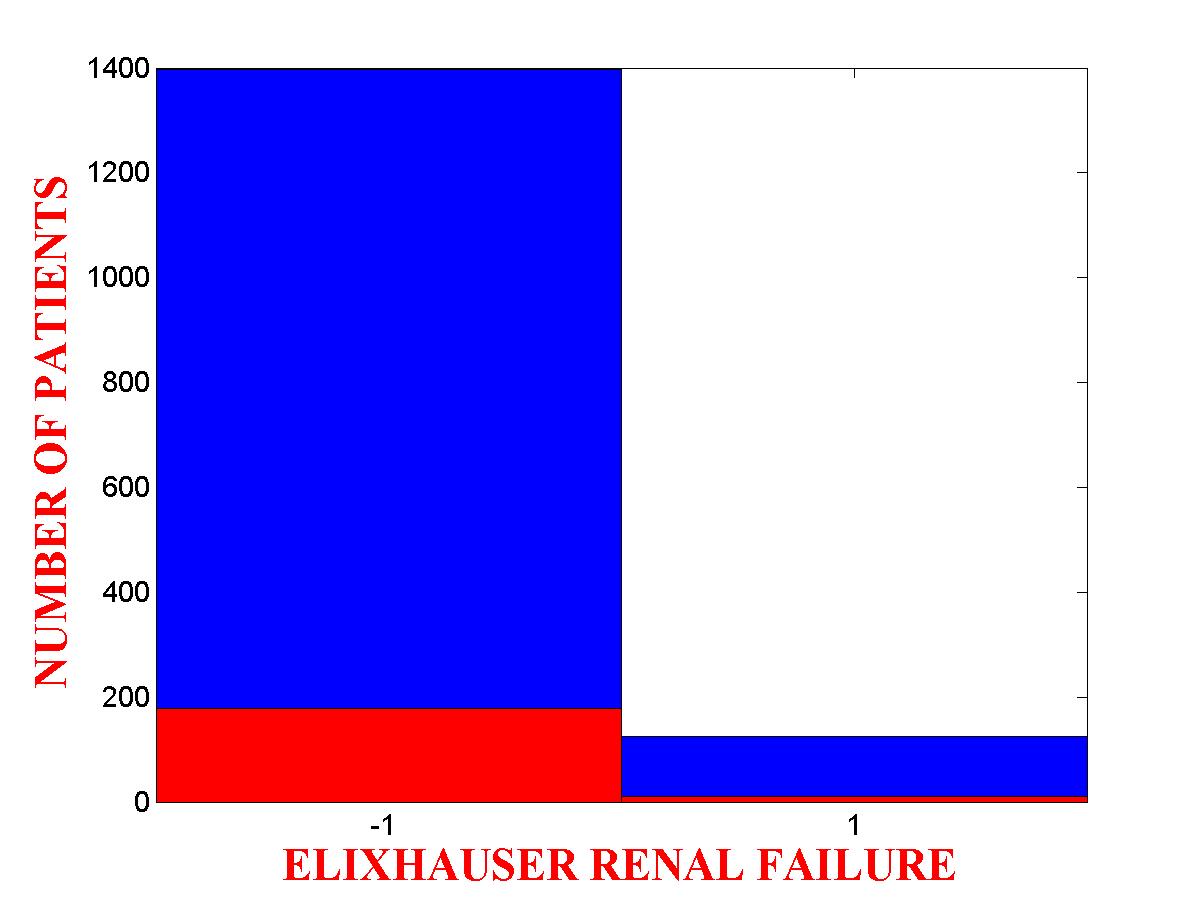}
\captionof{subfigure}[]{125 patients had renal failure.}
\label{fig:hist_12}
\end{minipage}
\begin{minipage}{0.33\textheight}
\centering
\includegraphics[width=.9\linewidth]{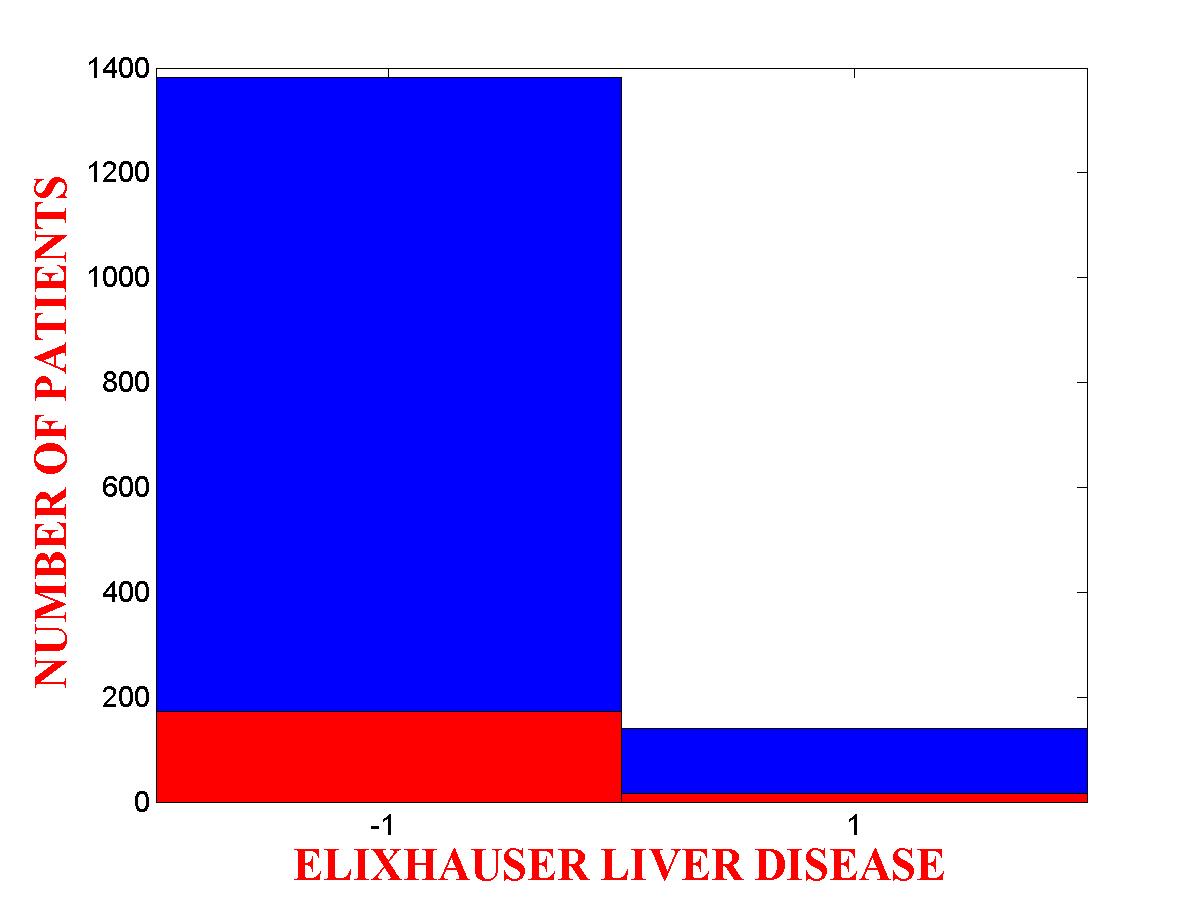}
\captionof{subfigure}[]{141 patients had liver disease..}
\label{fig:hist_13}
\end{minipage}
\begin{minipage}{0.33\textheight}
\centering
\includegraphics[width=.9\linewidth]{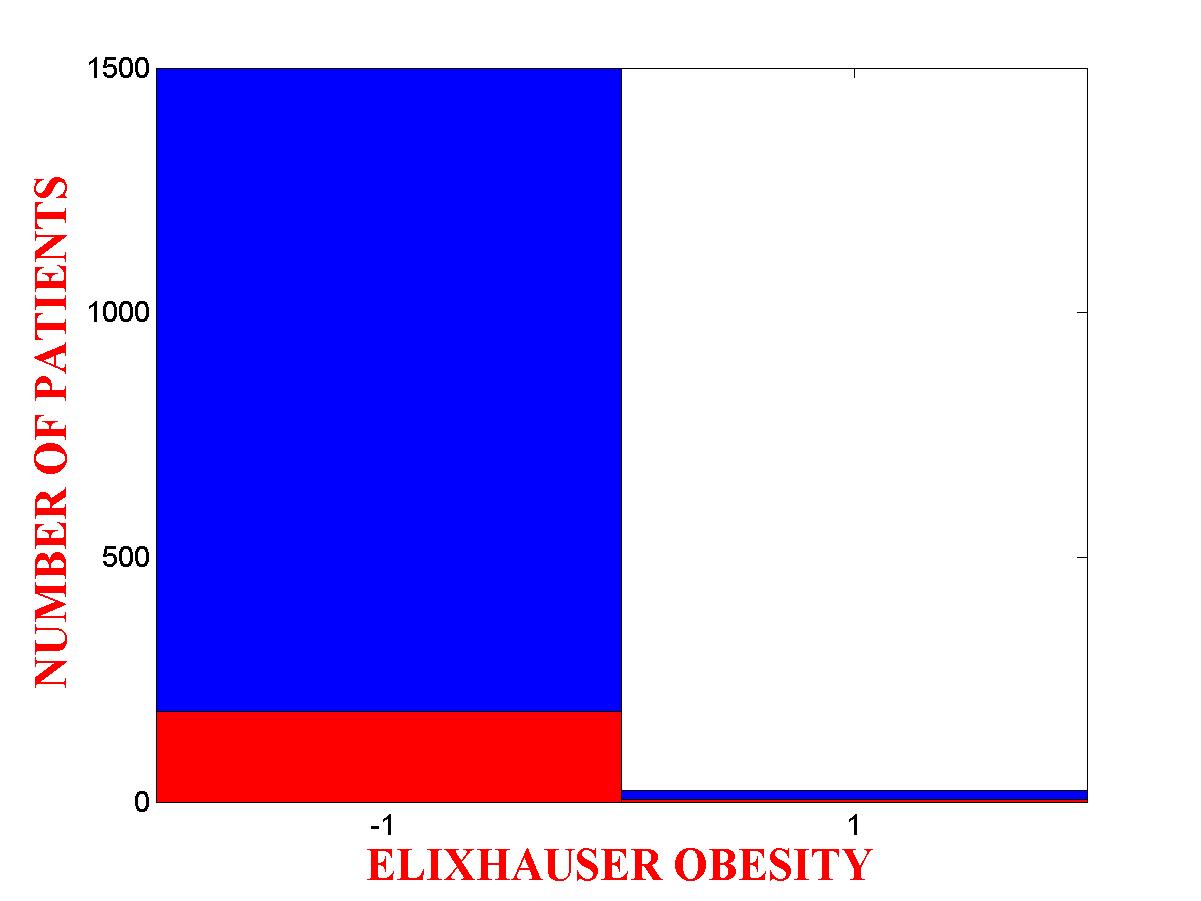}
\captionof{subfigure}[]{23 patients had obesity.}
\label{fig:hist_14}
\end{minipage}
\addtocounter{figure}{-1}
\captionof{figure}[Histograms of the Elixahuser Parameters, part 2]{Histograms of the Elixahuser parameters, part 2.}
\label{fig:histo_eli_2}
\end{minipage}
}
\end{figure}

Figures~\vref{fig:histo_group_1},~\vref{fig:histo_group_2},~\vref{fig:histo_group_3},~\vref{fig:histo_group_4},~\vref{fig:histo_group_5},~\vref{fig:histo_group_6},~\vref{fig:histo_group_7} show the histograms of all the numeric variables.
\begin{figure}
\centering
\rotatebox{90}{
\stepcounter{figure}
\setcounter{subfigure}{0}
\begin{minipage}{\textheight}
\begin{minipage}{0.33\textheight}
\centering
\includegraphics[width=.9\linewidth]{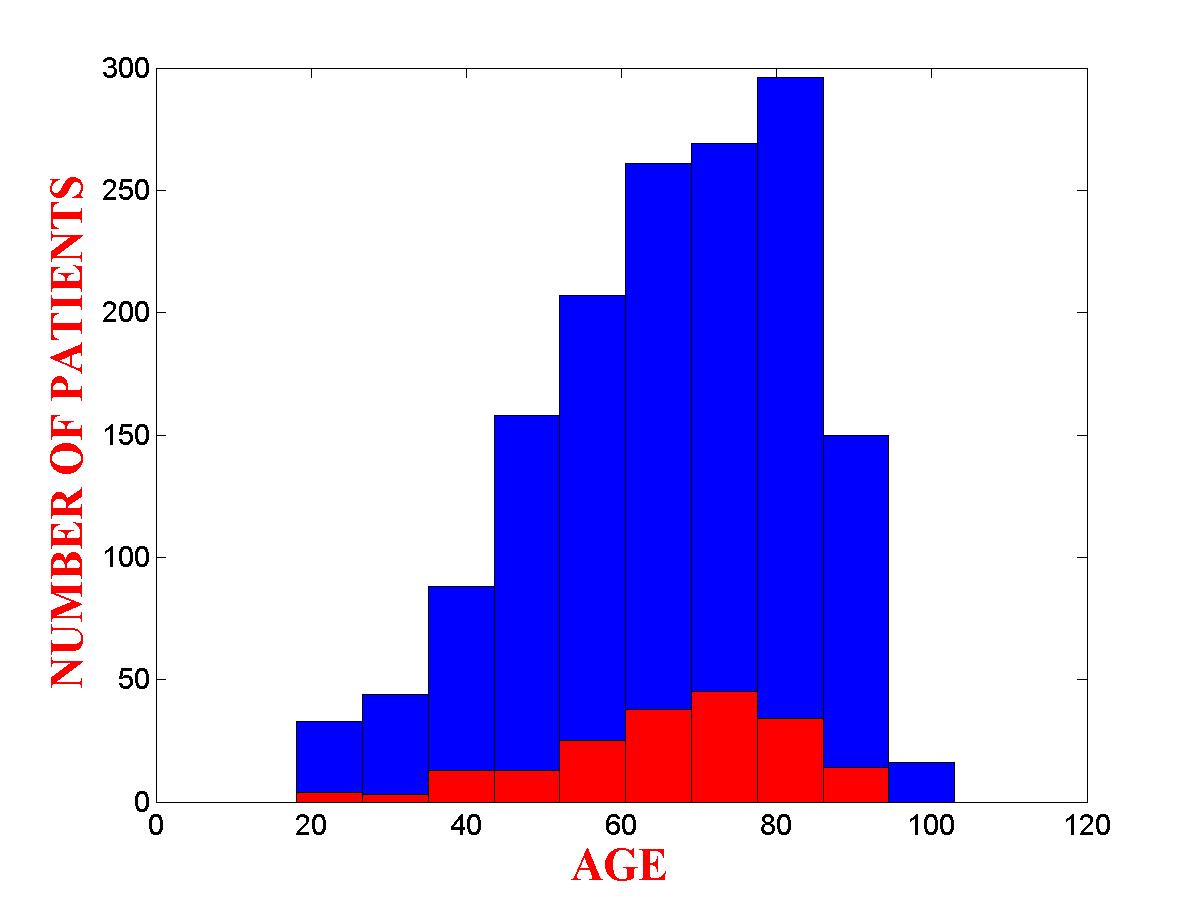}
\captionof{subfigure}[]{Age is centered on 68.3 years.}
\label{fig:hist_17}
\end{minipage}
\begin{minipage}{0.33\textheight}
\centering
\includegraphics[width=.9\linewidth]{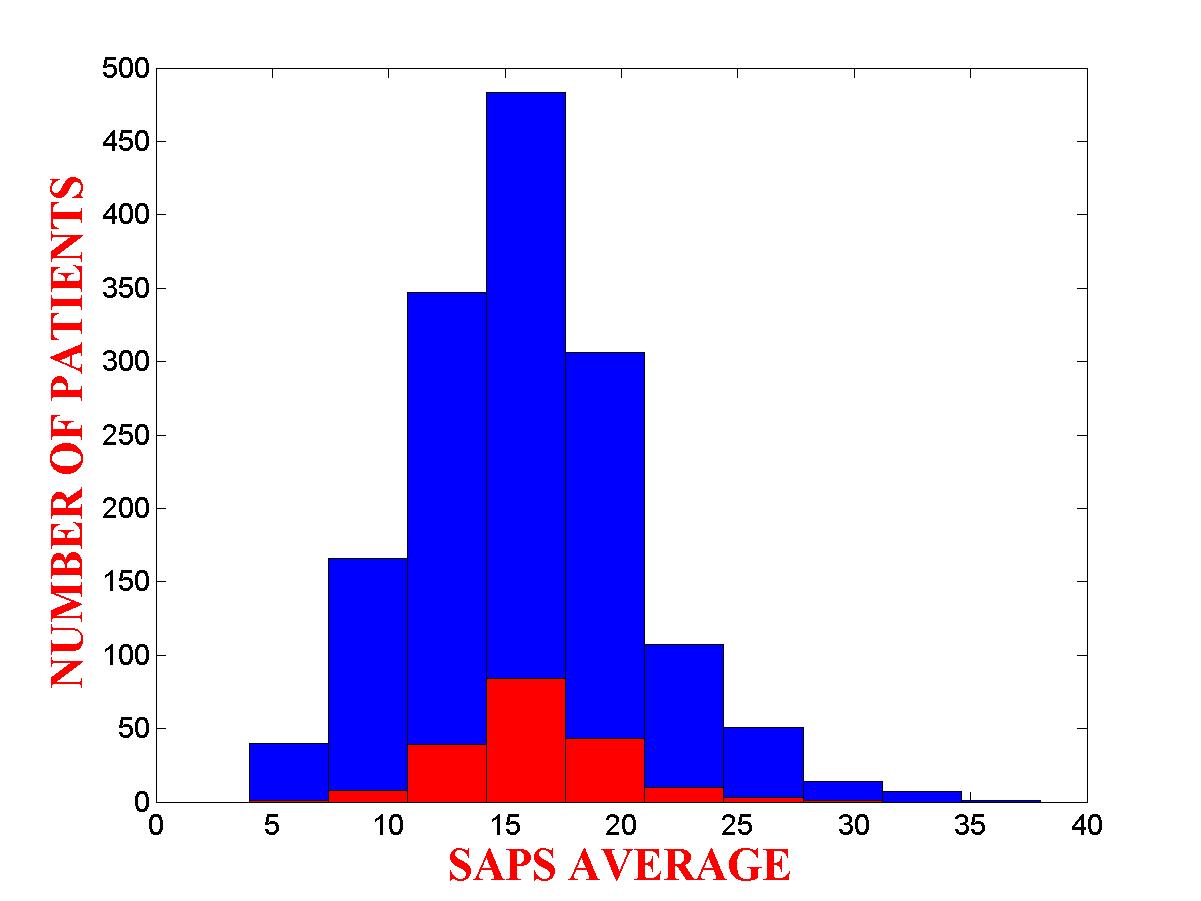}
\captionof{subfigure}[]{SAPS average is centered on 15.4.}
\label{fig:hist_18}
\end{minipage}
\begin{minipage}{0.33\textheight}
\centering
\includegraphics[width=.9\linewidth]{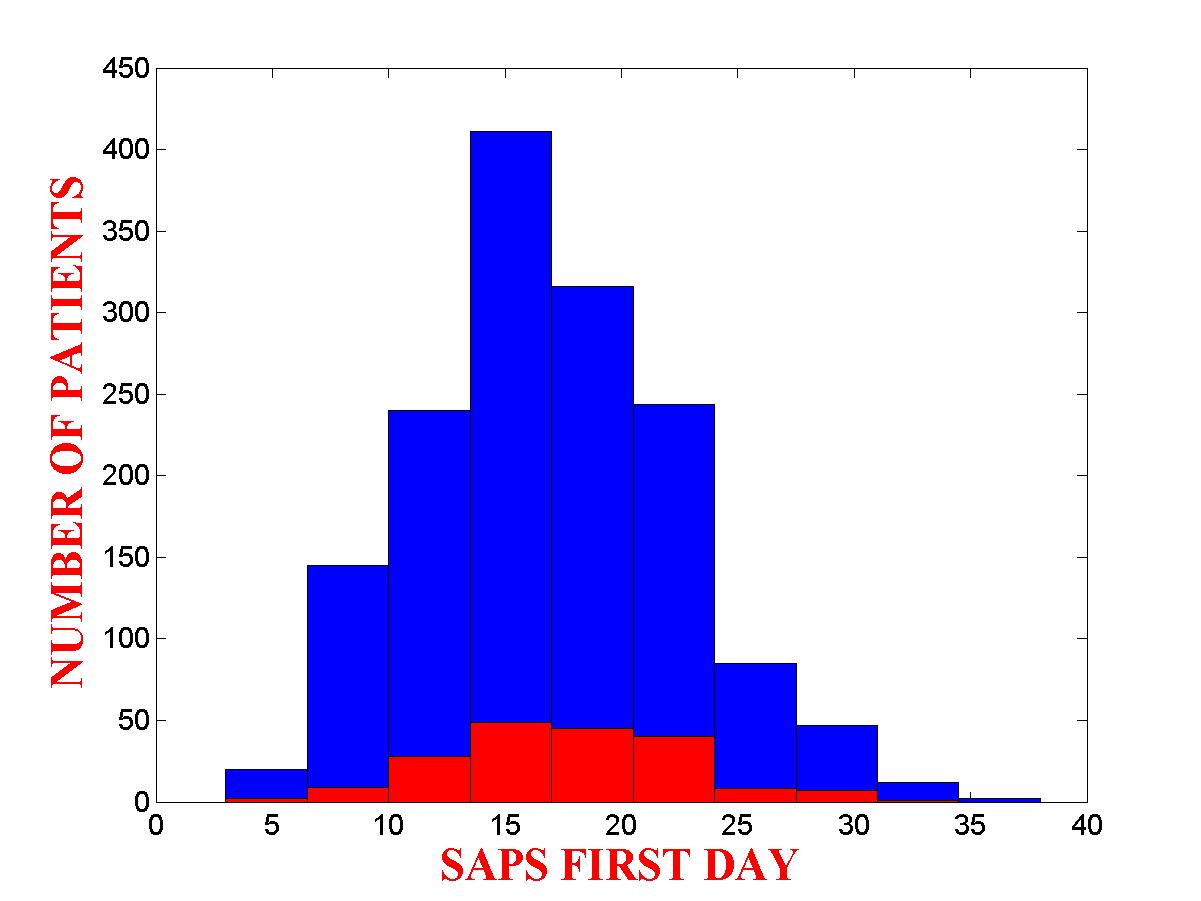}
\captionof{subfigure}[]{SAPS first day is centered on 17.}
\label{fig:hist_19}
\end{minipage}
\begin{minipage}{0.33\textheight}
\centering
\includegraphics[width=.9\linewidth]{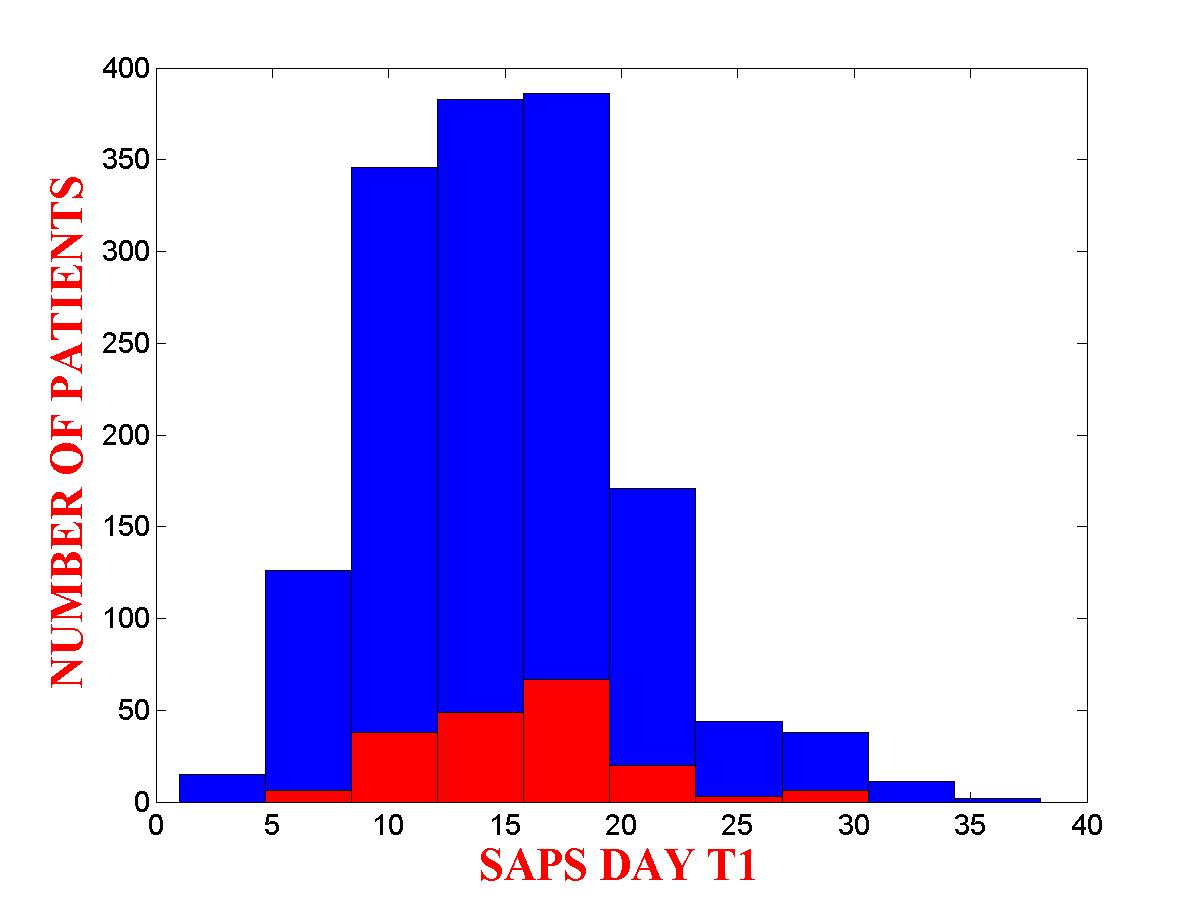}
\captionof{subfigure}[]{SAPS day T1 is centered on 15.}
\label{fig:hist_20}
\end{minipage}
\begin{minipage}{0.33\textheight}
\centering
\includegraphics[width=.9\linewidth]{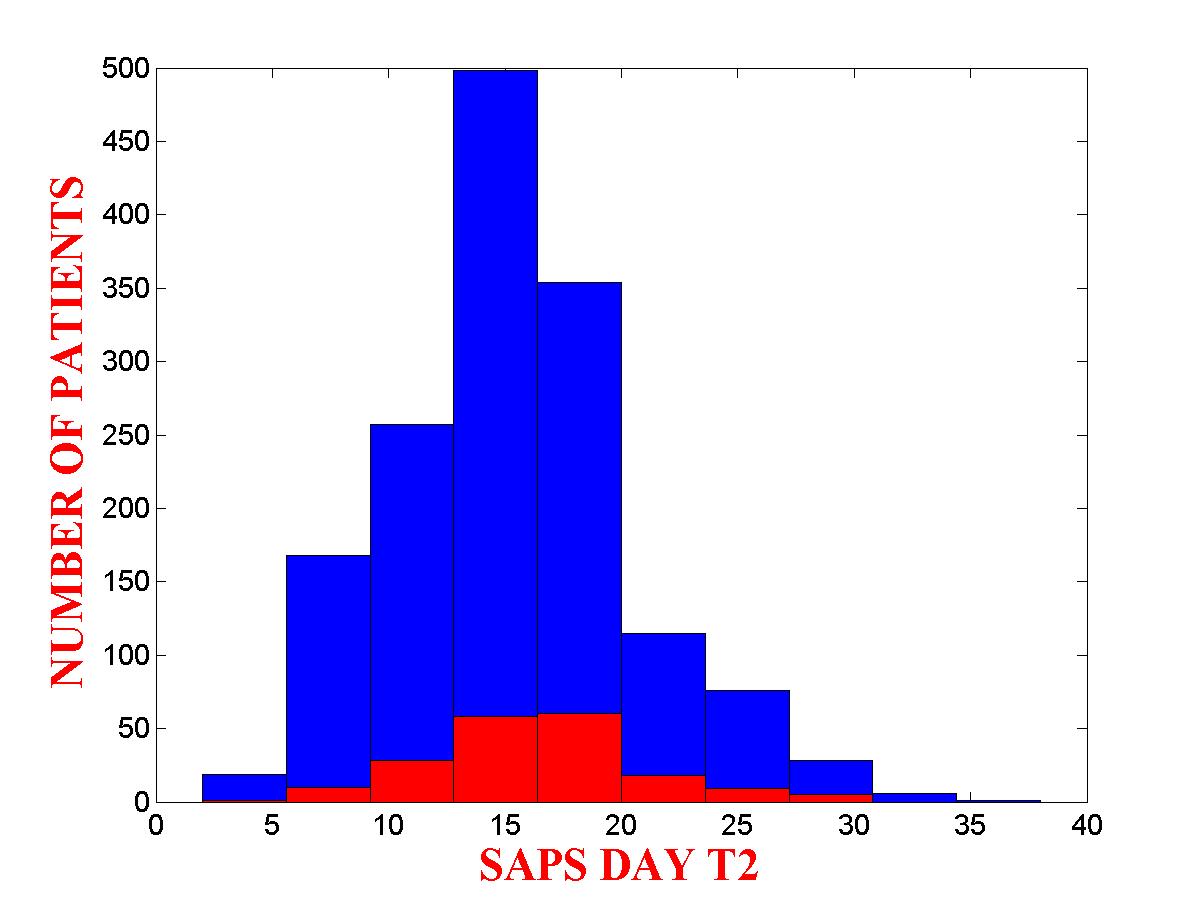}
\captionof{subfigure}[]{SAPS day T2 is centered on 15.}
\label{fig:hist_21}
\end{minipage}
\begin{minipage}{0.33\textheight}
\centering
\includegraphics[width=.9\linewidth]{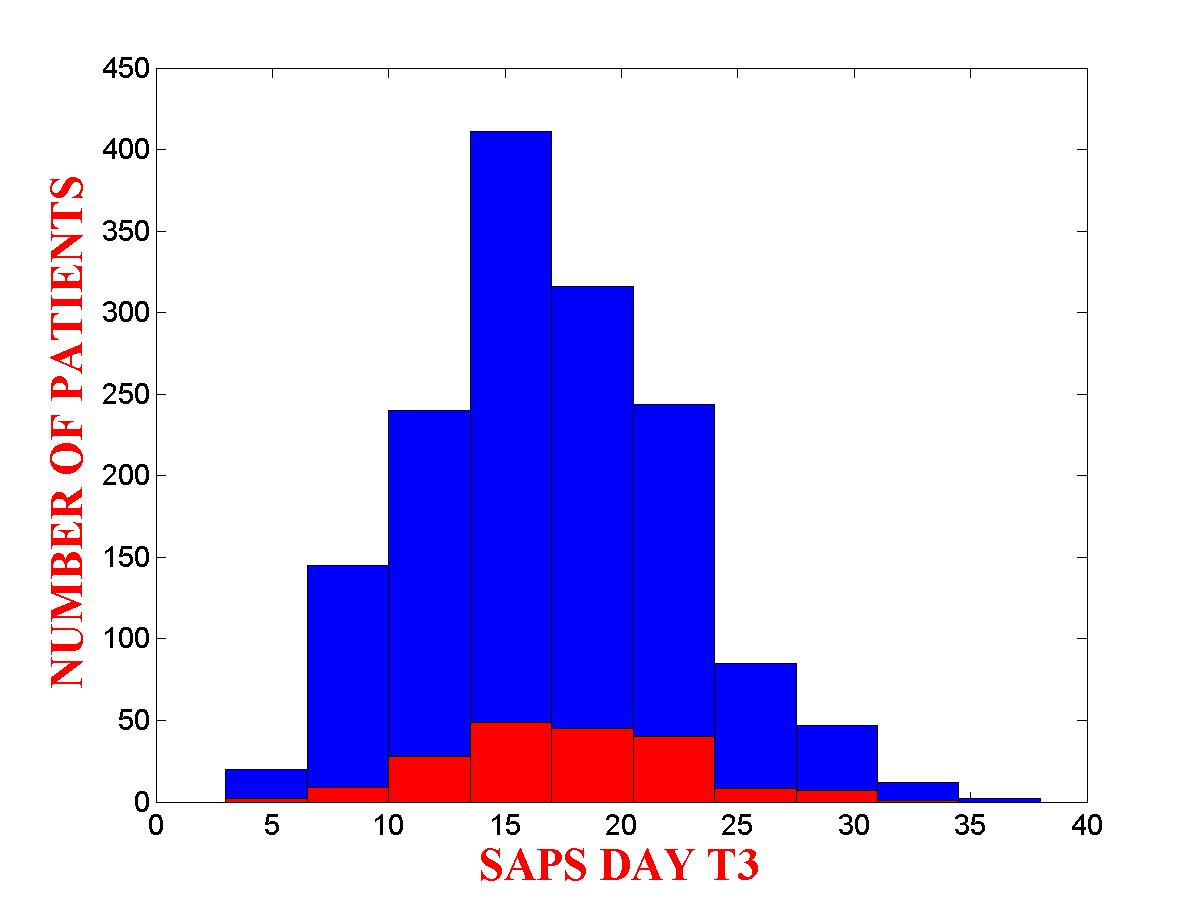}
\captionof{subfigure}[]{SAPS day T3 is centered on 17.}
\label{fig:hist_22}
\end{minipage}
\addtocounter{figure}{-1}
\captionof{figure}[Histograms of the Numeric Variables, part 1]{Histograms of the numeric variables, part 1}
\label{fig:histo_group_1}
\end{minipage}
}
\end{figure}
\begin{figure}
\centering
\rotatebox{90}{
\stepcounter{figure}
\setcounter{subfigure}{0}
\begin{minipage}{\textheight}
\begin{minipage}{0.33\textheight}
\centering
\includegraphics[width=.9\linewidth]{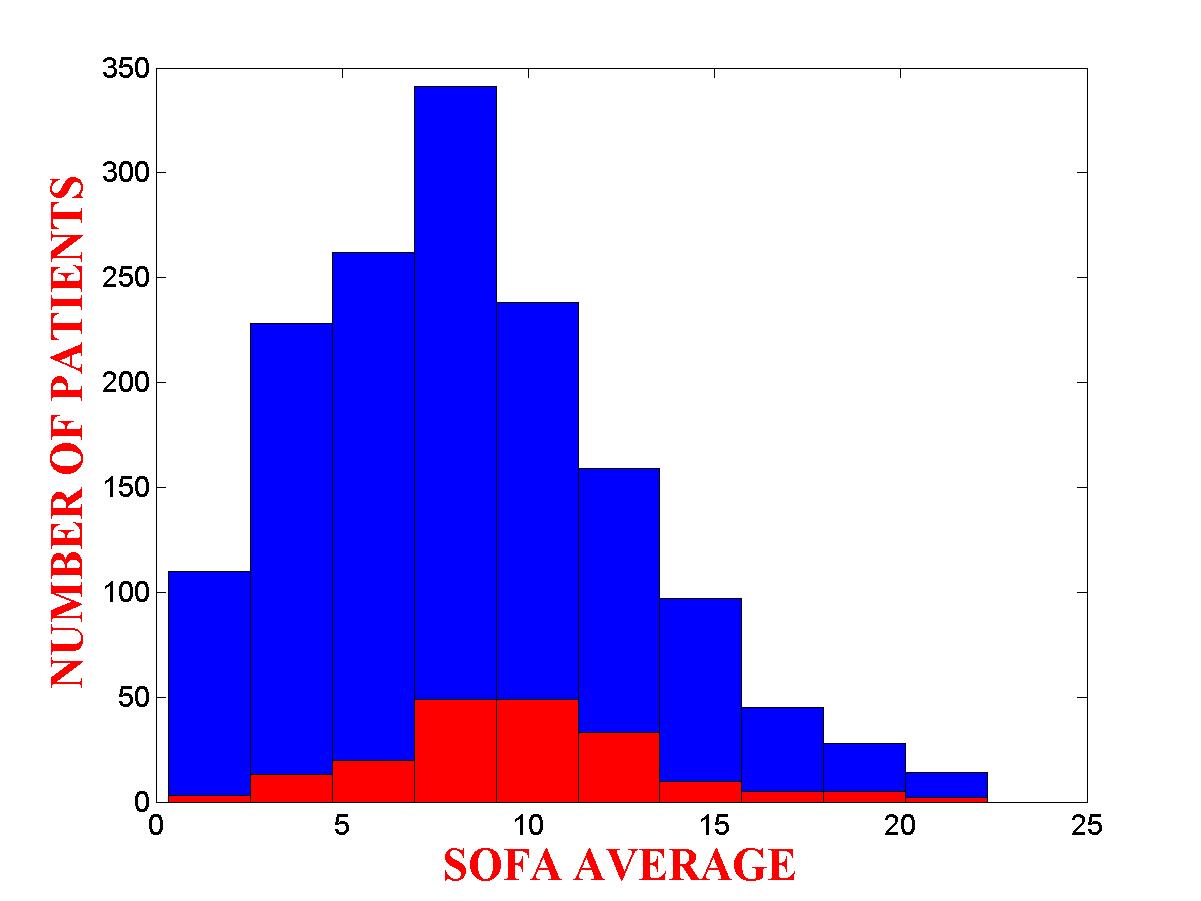}
\captionof{subfigure}[]{SOFA average is centered on 7.6.}
\label{fig:hist_23}
\end{minipage}
\begin{minipage}{0.33\textheight}
\centering
\includegraphics[width=.9\linewidth]{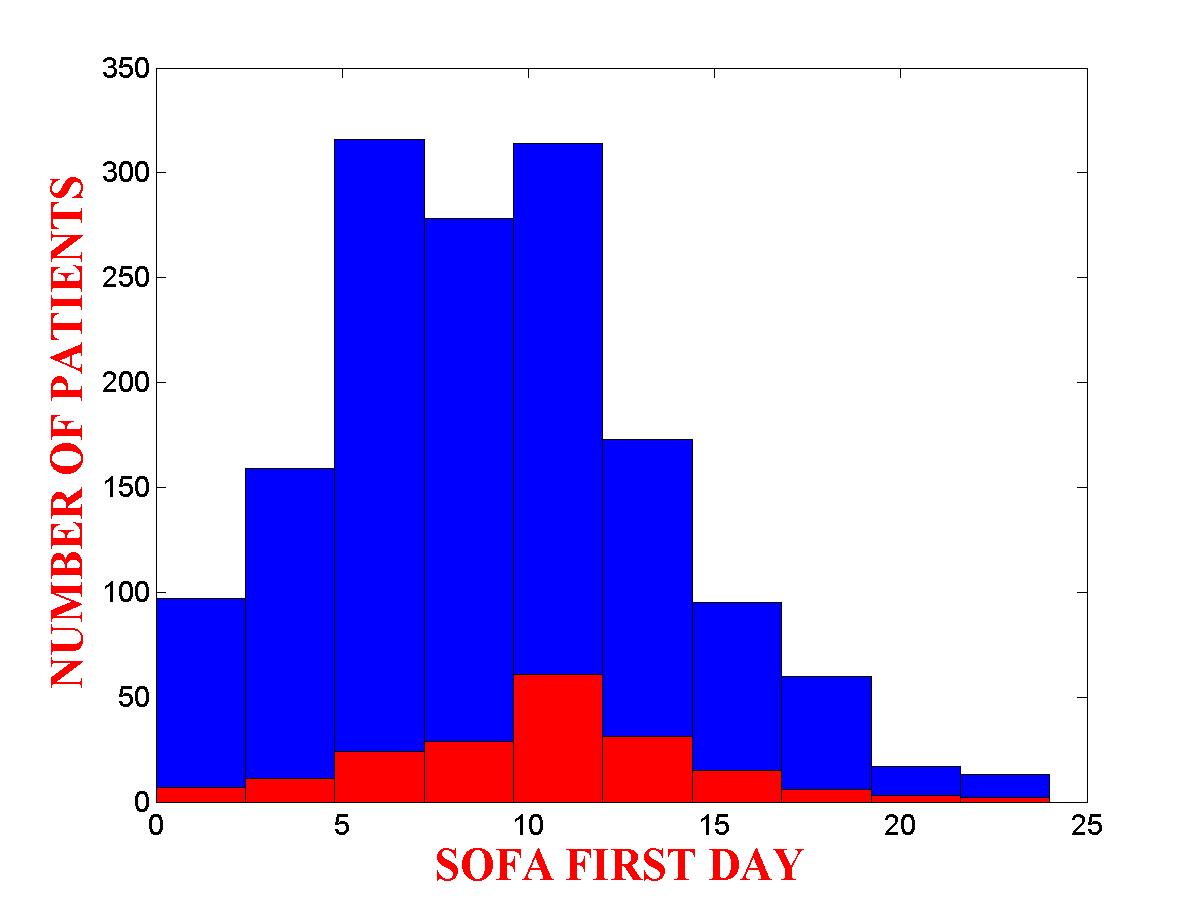}
\captionof{subfigure}[]{SOFA first day is centered on 9.}
\label{fig:hist_24}
\end{minipage}
\begin{minipage}{0.33\textheight}
\centering
\includegraphics[width=.9\linewidth]{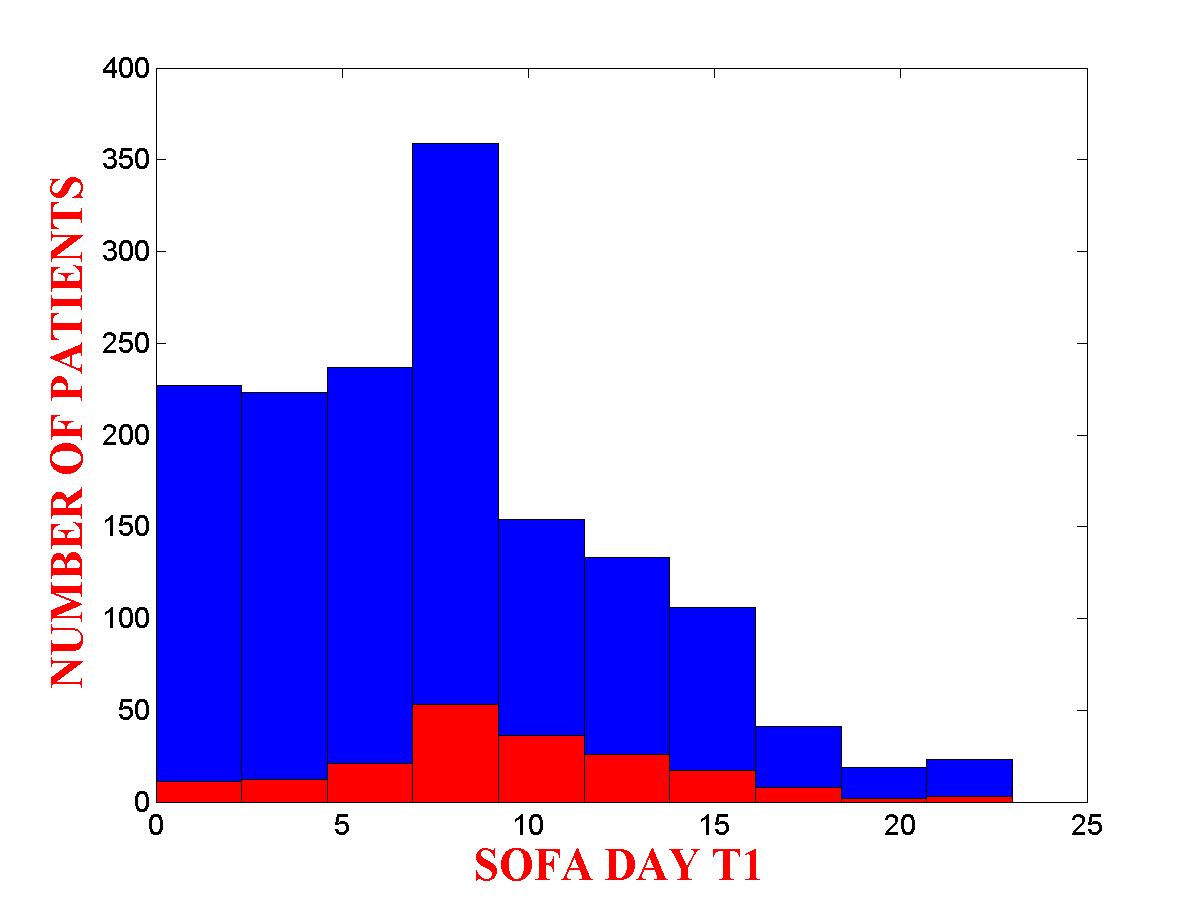}
\captionof{subfigure}[]{SOFA day T1 is centered on 7.}
\label{fig:hist_25}
\end{minipage}
\begin{minipage}{0.33\textheight}
\centering
\includegraphics[width=.9\linewidth]{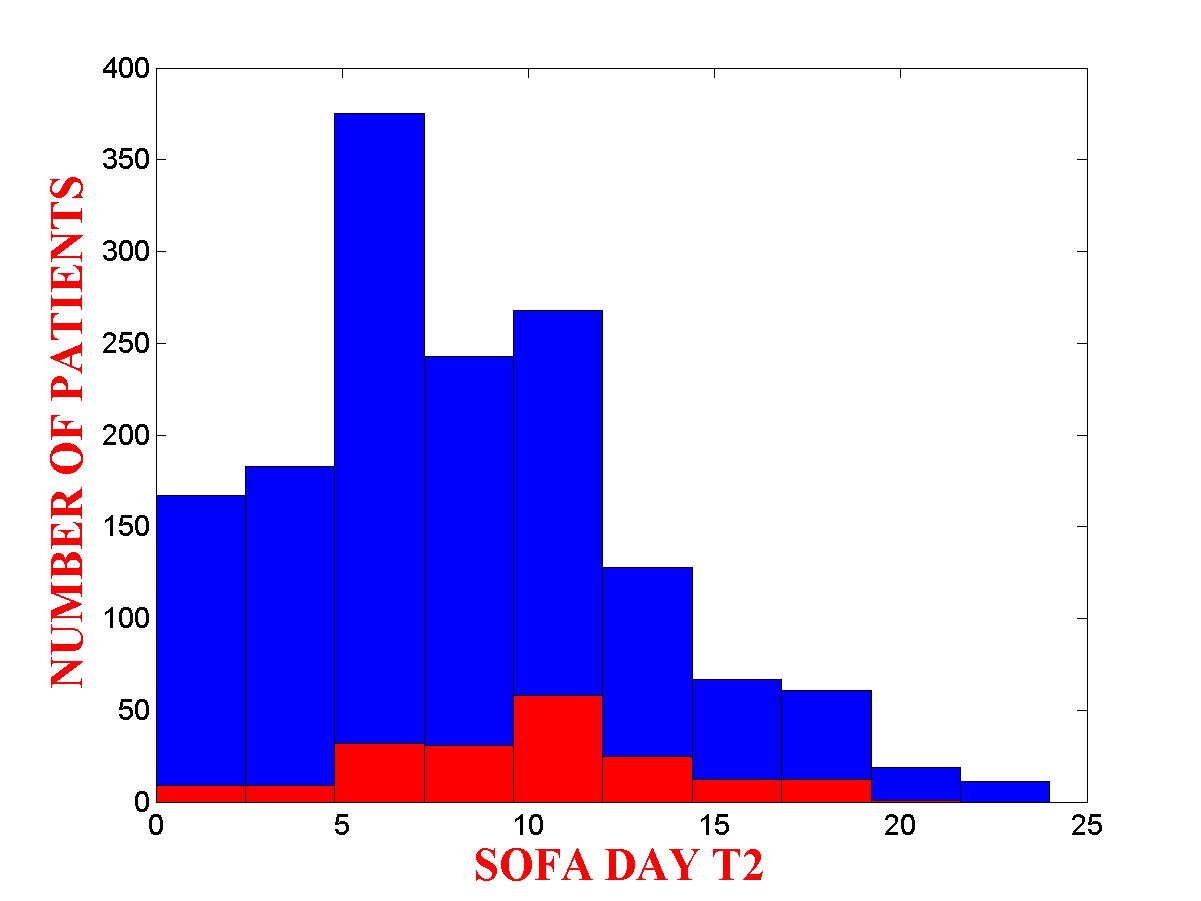}
\captionof{subfigure}[]{SOFA day T2 is centered on 7.4.}
\label{fig:hist_26}
\end{minipage}
\begin{minipage}{0.33\textheight}
\centering
\includegraphics[width=.9\linewidth]{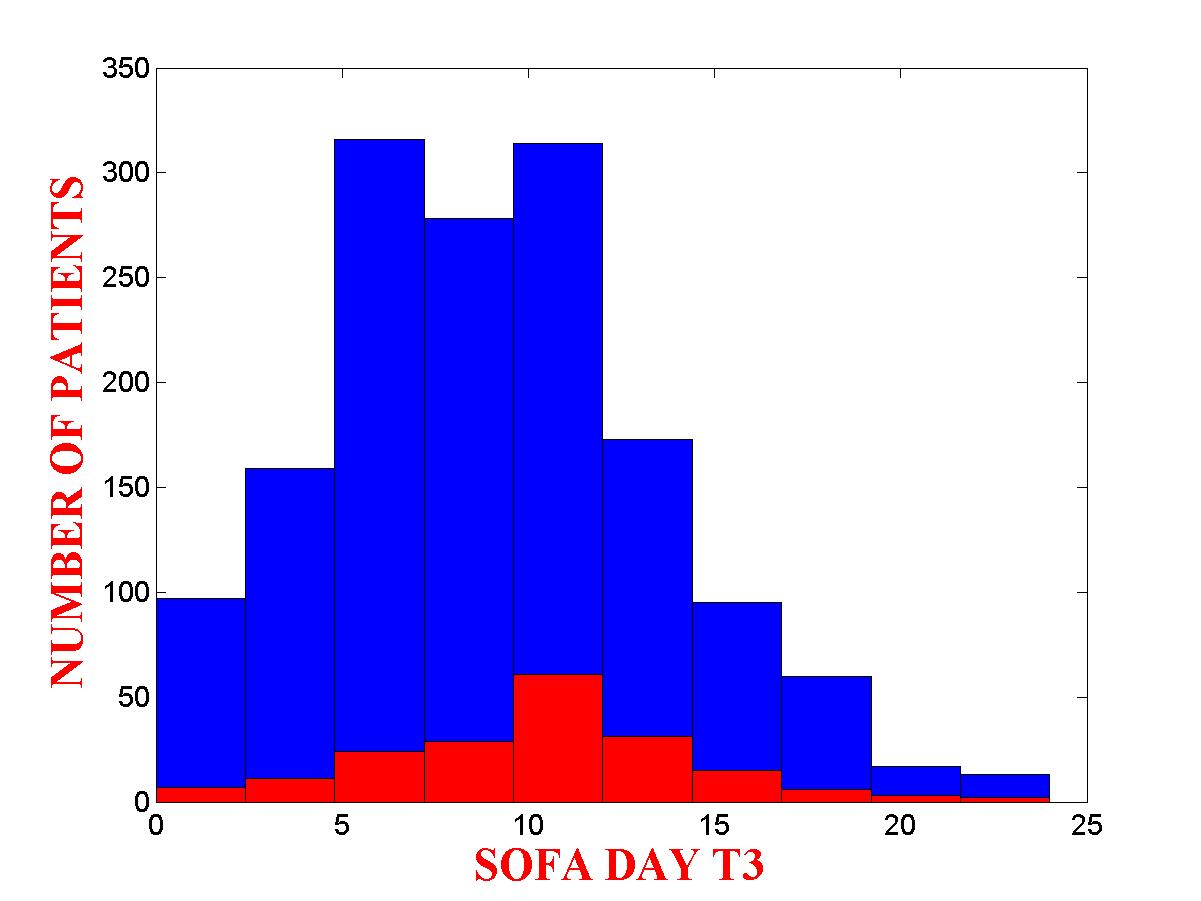}
\captionof{subfigure}[]{SOFA day T3 is centered on 9.}
\label{fig:hist_27}
\end{minipage}
\begin{minipage}{0.33\textheight}
\centering
\includegraphics[width=.9\linewidth]{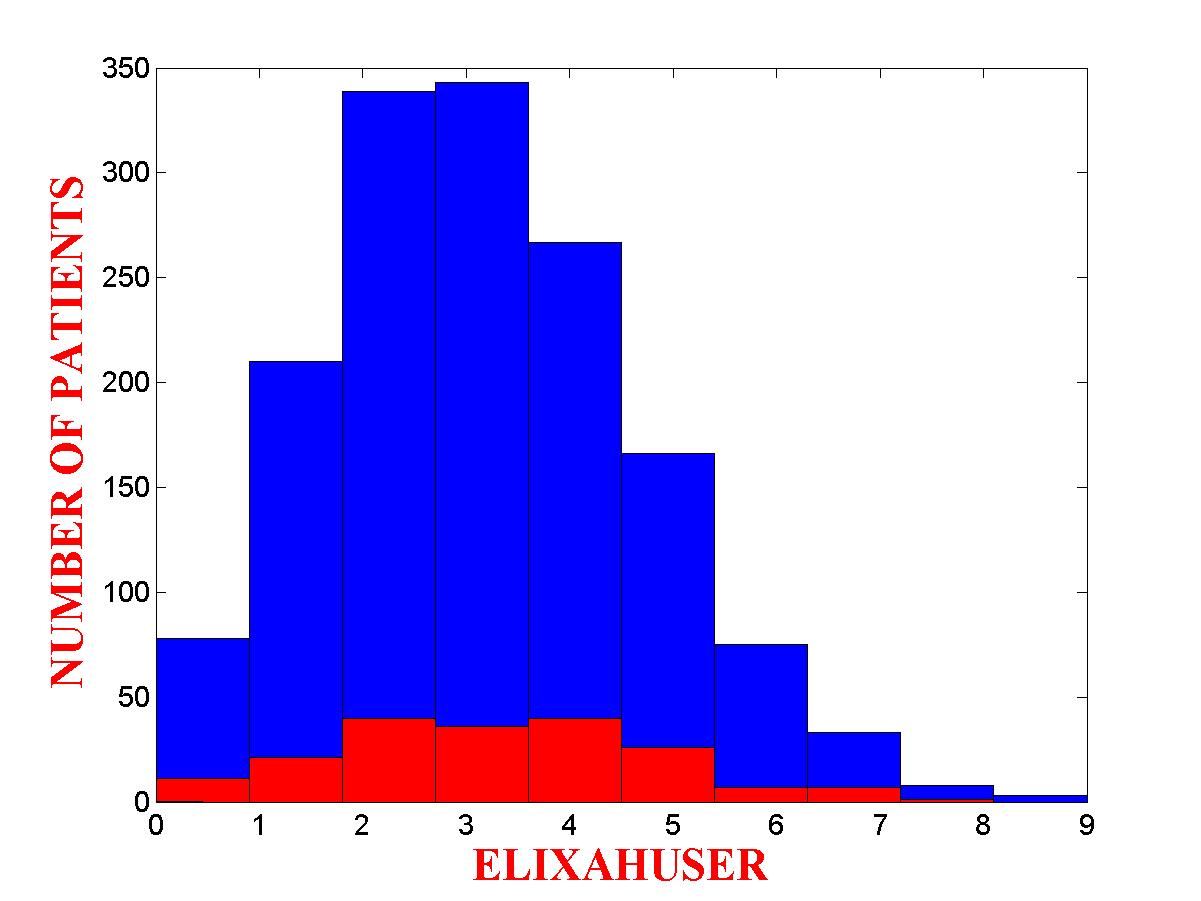}
\captionof{subfigure}[]{Elixahuser is centered on 3.}
\label{fig:hist_28}
\end{minipage}
\addtocounter{figure}{-1}
\captionof{figure}[Histograms of the Numeric Variables, part 2]{Histograms of the numeric variables, part 2}
\label{fig:histo_group_2}
\end{minipage}
}
\end{figure}
\begin{figure}
\centering
\rotatebox{90}{
\stepcounter{figure}
\setcounter{subfigure}{0}
\begin{minipage}{\textheight}
\begin{minipage}{0.33\textheight}
\centering
\includegraphics[width=.9\linewidth]{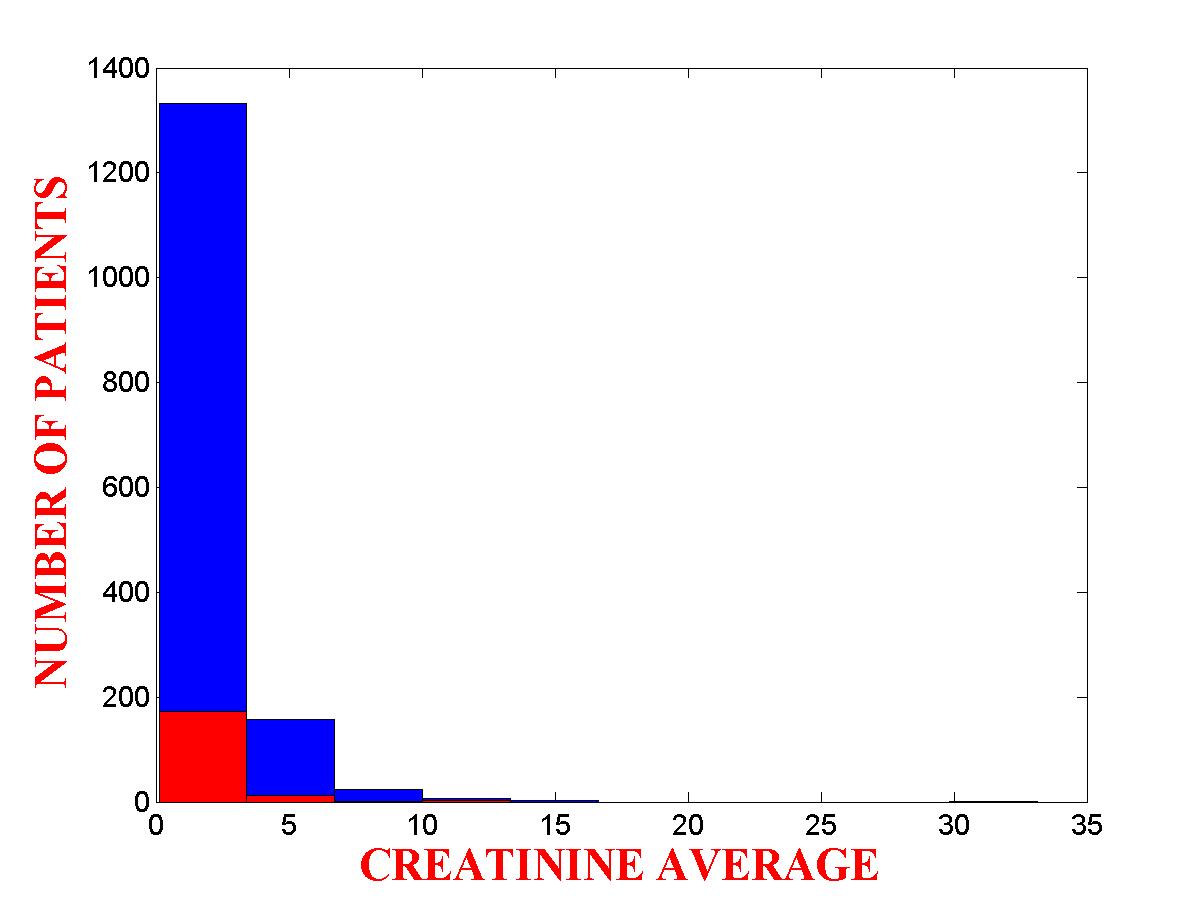}
\captionof{subfigure}[]{Creatinine average is centered on 1.25.}
\label{fig:hist_29}
\end{minipage}
\begin{minipage}{0.33\textheight}
\centering
\includegraphics[width=.9\linewidth]{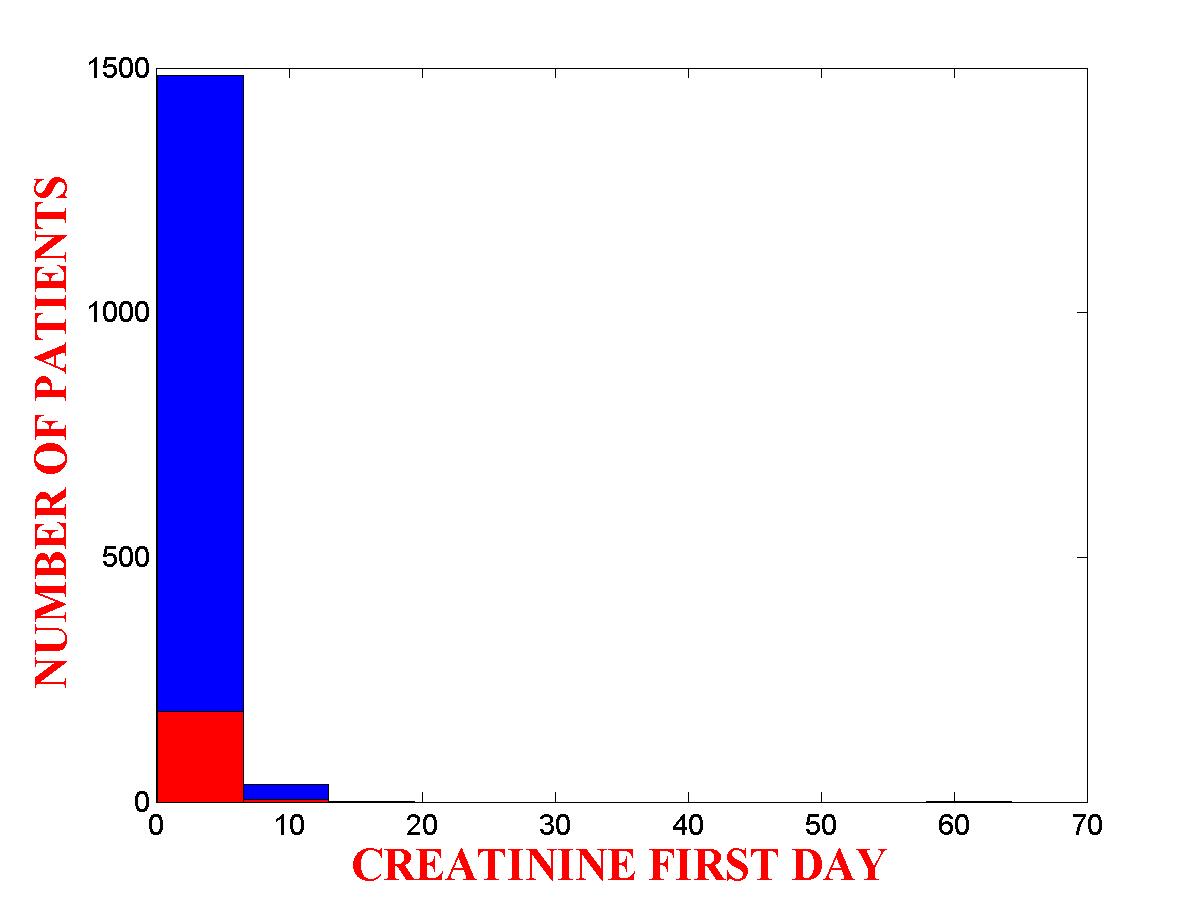}
\captionof{subfigure}[]{Creatinine first day is centered on 1.25.}
\label{fig:hist_30}
\end{minipage}
\begin{minipage}{0.33\textheight}
\centering
\includegraphics[width=.9\linewidth]{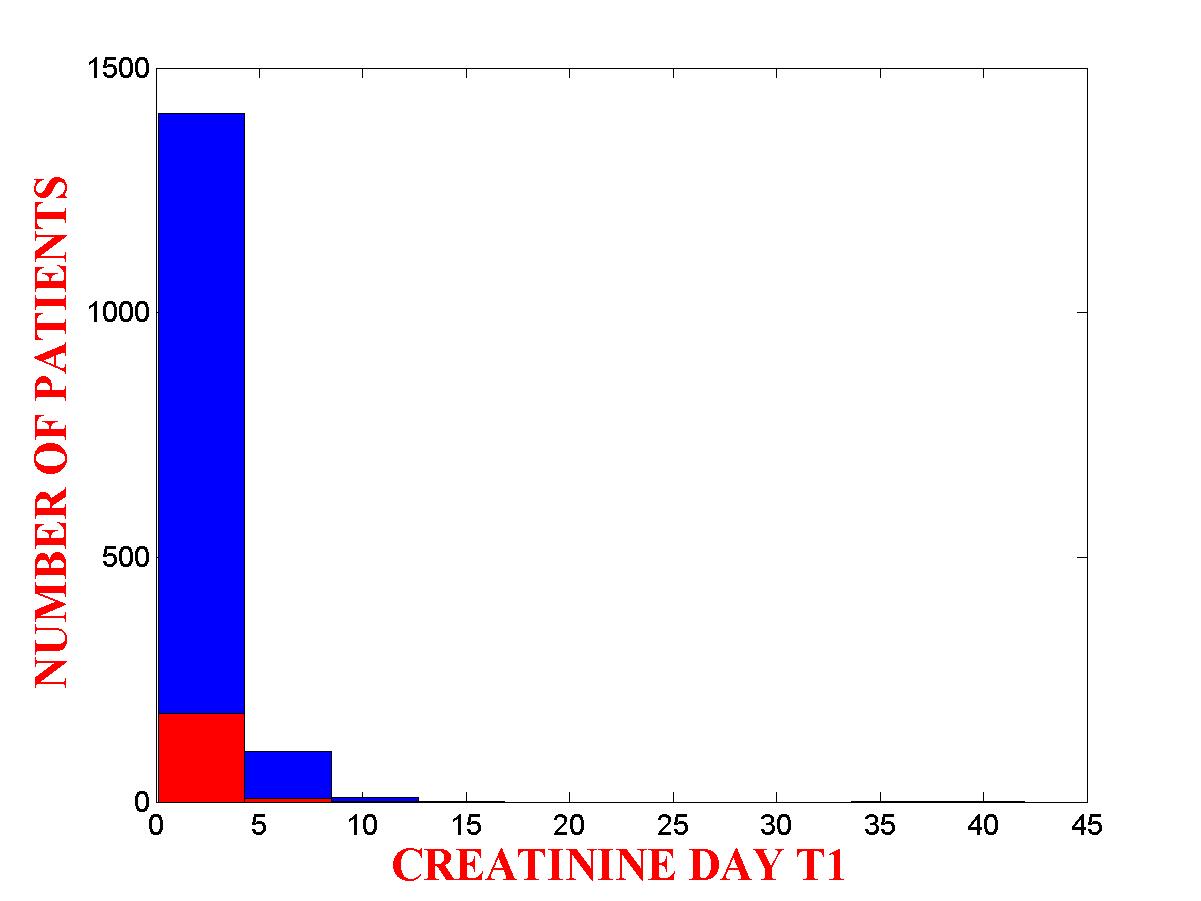}
\captionof{subfigure}[]{Creatinine day T1 is centered on 1.2.}
\label{fig:hist_31}
\end{minipage}
\begin{minipage}{0.33\textheight}
\centering
\includegraphics[width=.9\linewidth]{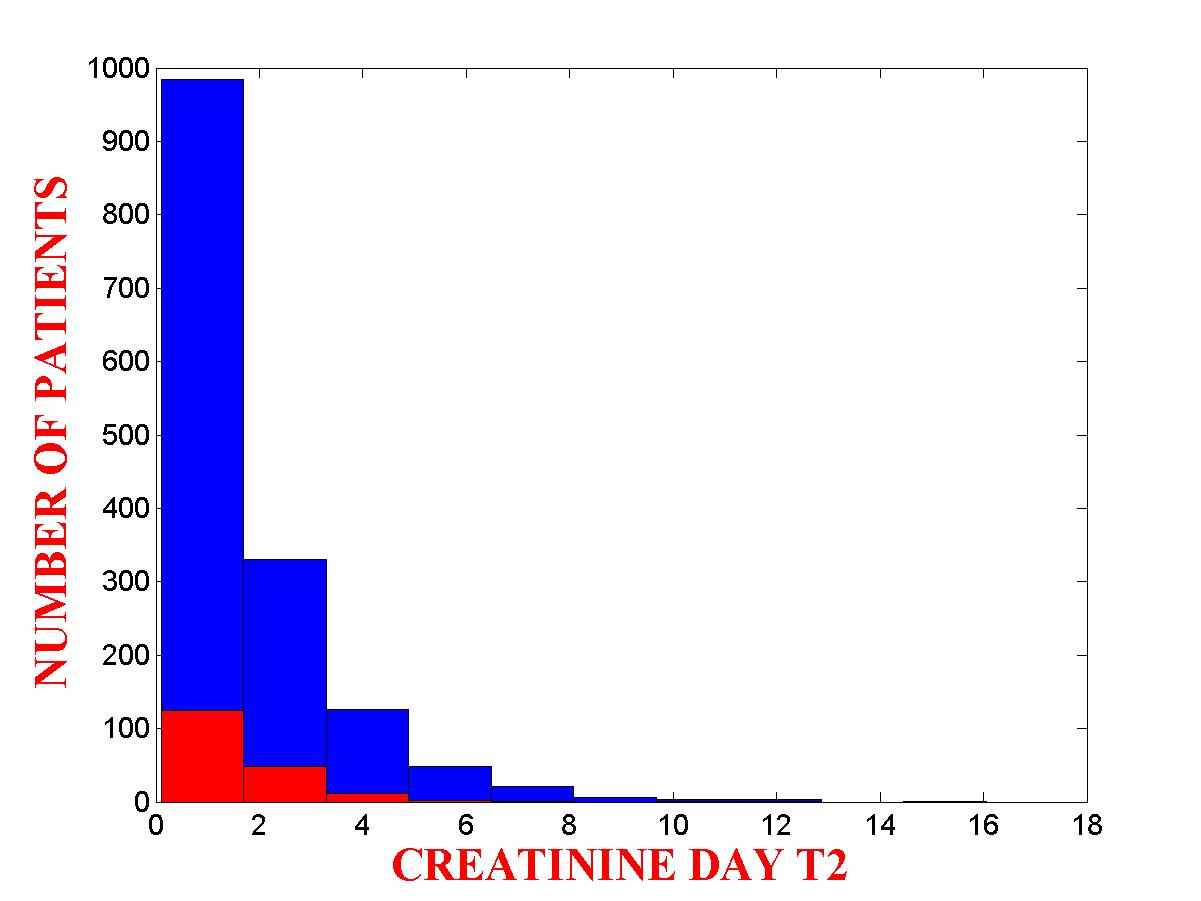}
\captionof{subfigure}[]{Creatinine day T2 is centered on 1.2.}
\label{fig:hist_32}
\end{minipage}
\begin{minipage}{0.33\textheight}
\centering
\includegraphics[width=.9\linewidth]{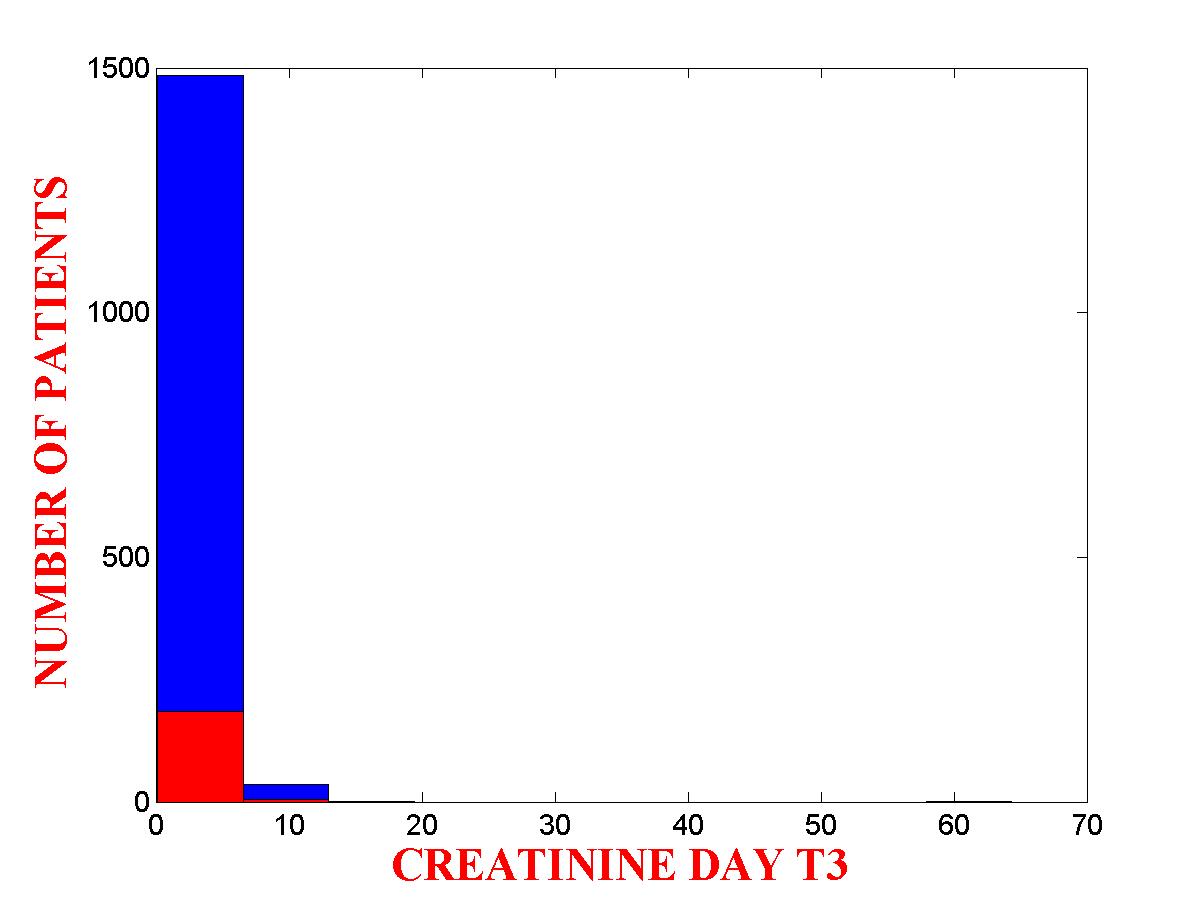}
\captionof{subfigure}[]{Creatinine day T3 is centered on 1.25.}
\label{fig:hist_33}
\end{minipage}
\begin{minipage}{0.33\textheight}
\centering
\includegraphics[width=.9\linewidth]{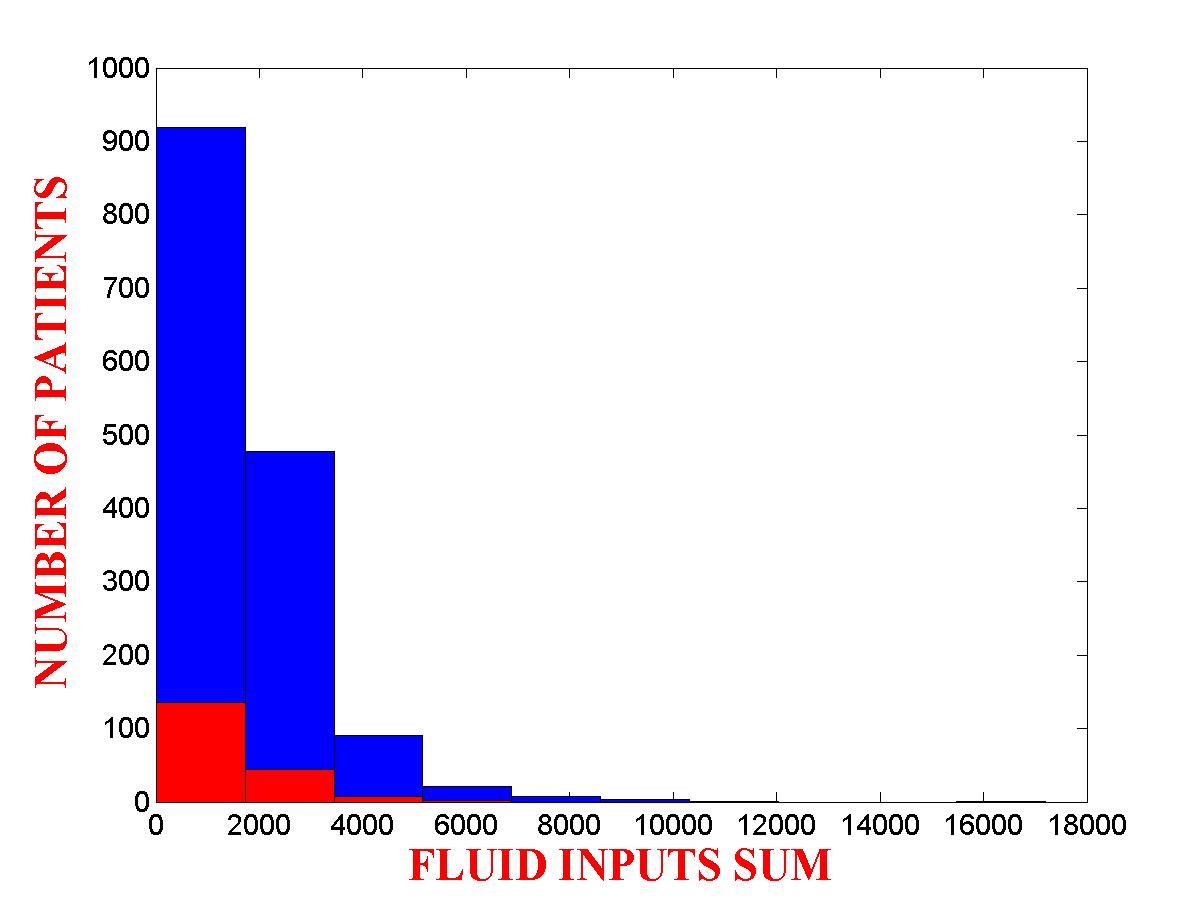}
\captionof{subfigure}[]{Fluids inputs sum is centered on 1.4 liters.}
\label{fig:hist_34}
\end{minipage}
\addtocounter{figure}{-1}
\captionof{figure}[Histograms of the Numeric Variables, part 3]{Histograms of the numeric variables, part 3}
\label{fig:histo_group_3}
\end{minipage}
}
\end{figure}
\begin{figure}
\centering
\rotatebox{90}{
\stepcounter{figure}
\setcounter{subfigure}{0}
\begin{minipage}{\textheight}
\begin{minipage}{0.33\textheight}
\centering
\includegraphics[width=.9\linewidth]{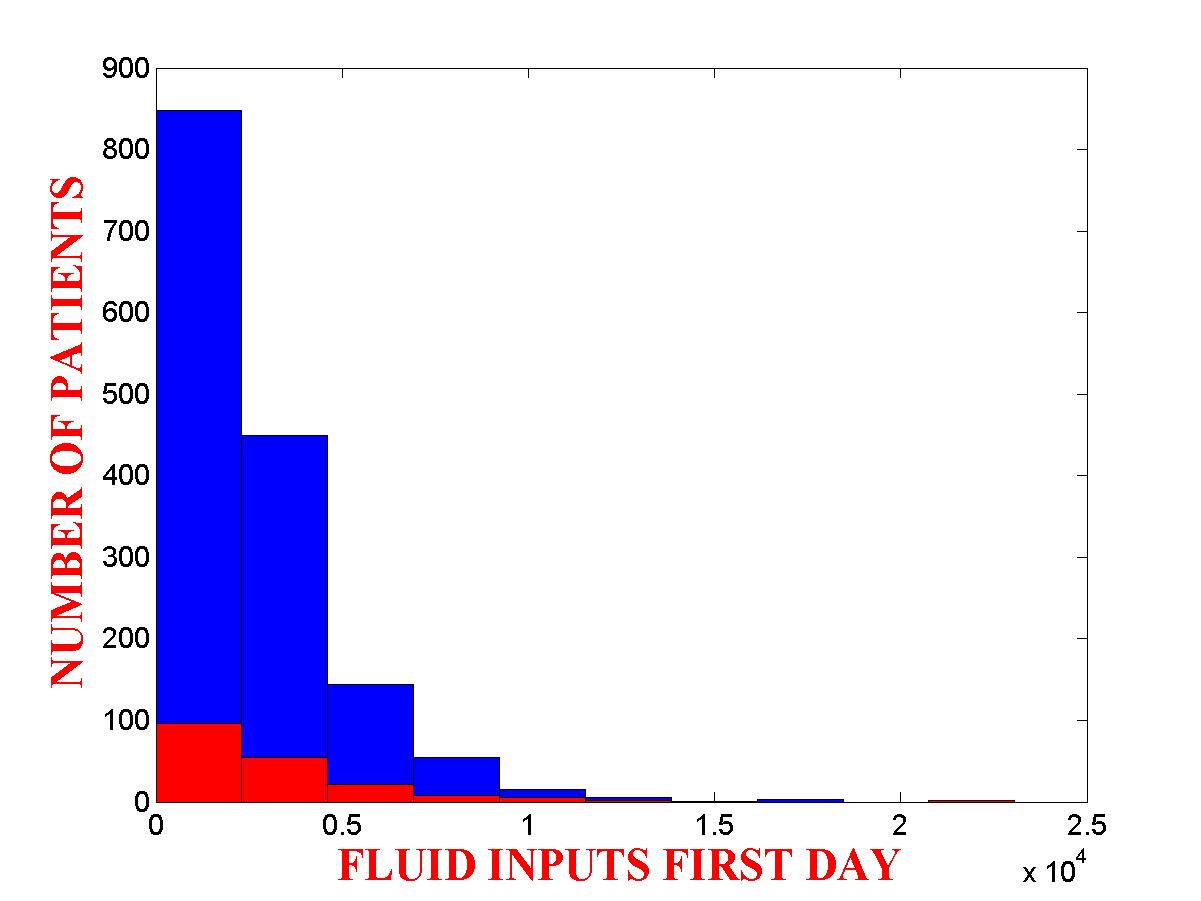}
\captionof{subfigure}[]{Fluids inputs first day is centered on 2 liters.}
\label{fig:hist_35}
\end{minipage}
\begin{minipage}{0.33\textheight}
\centering
\includegraphics[width=.9\linewidth]{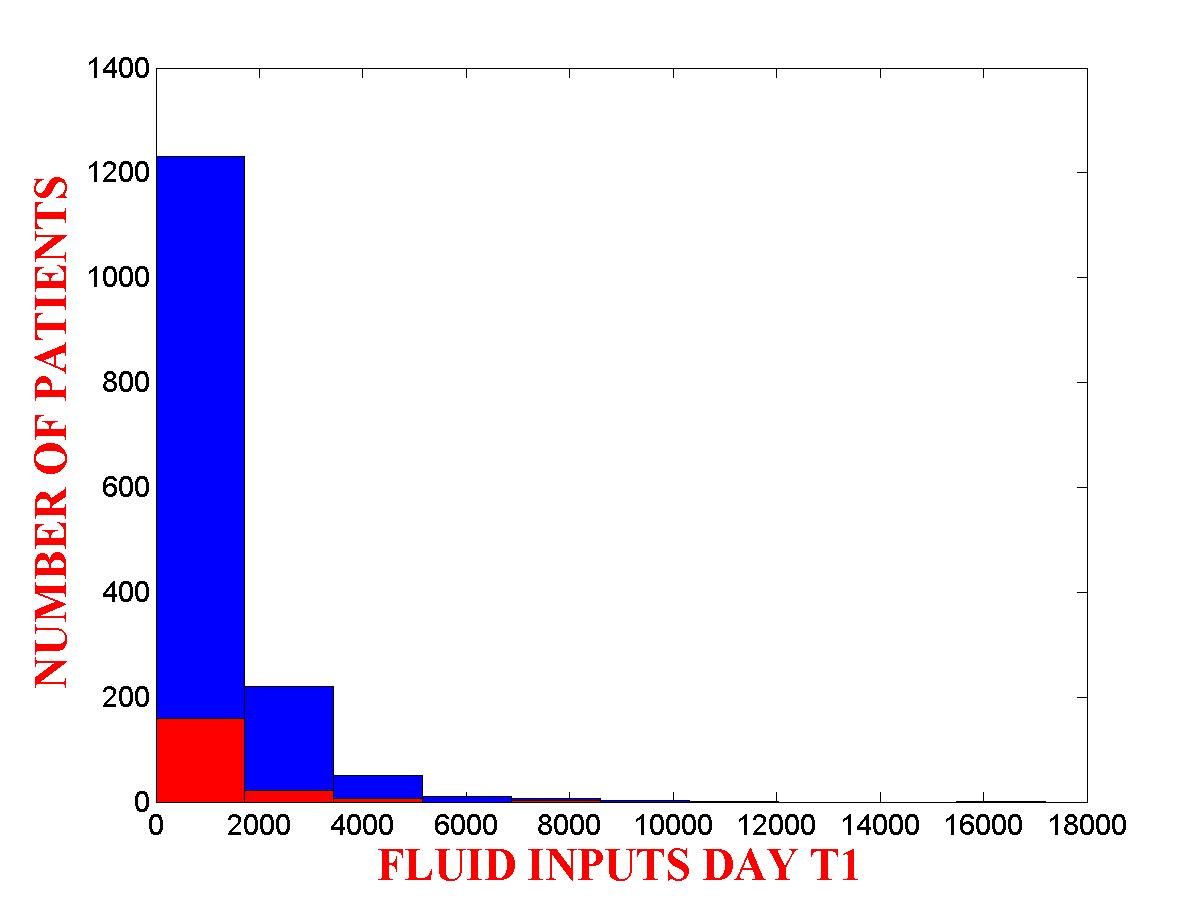}
\captionof{subfigure}[]{Fluids inputs day T1 is centered on 0.7 liters.}
\label{fig:hist_36}
\end{minipage}
\begin{minipage}{0.33\textheight}
\centering
\includegraphics[width=.9\linewidth]{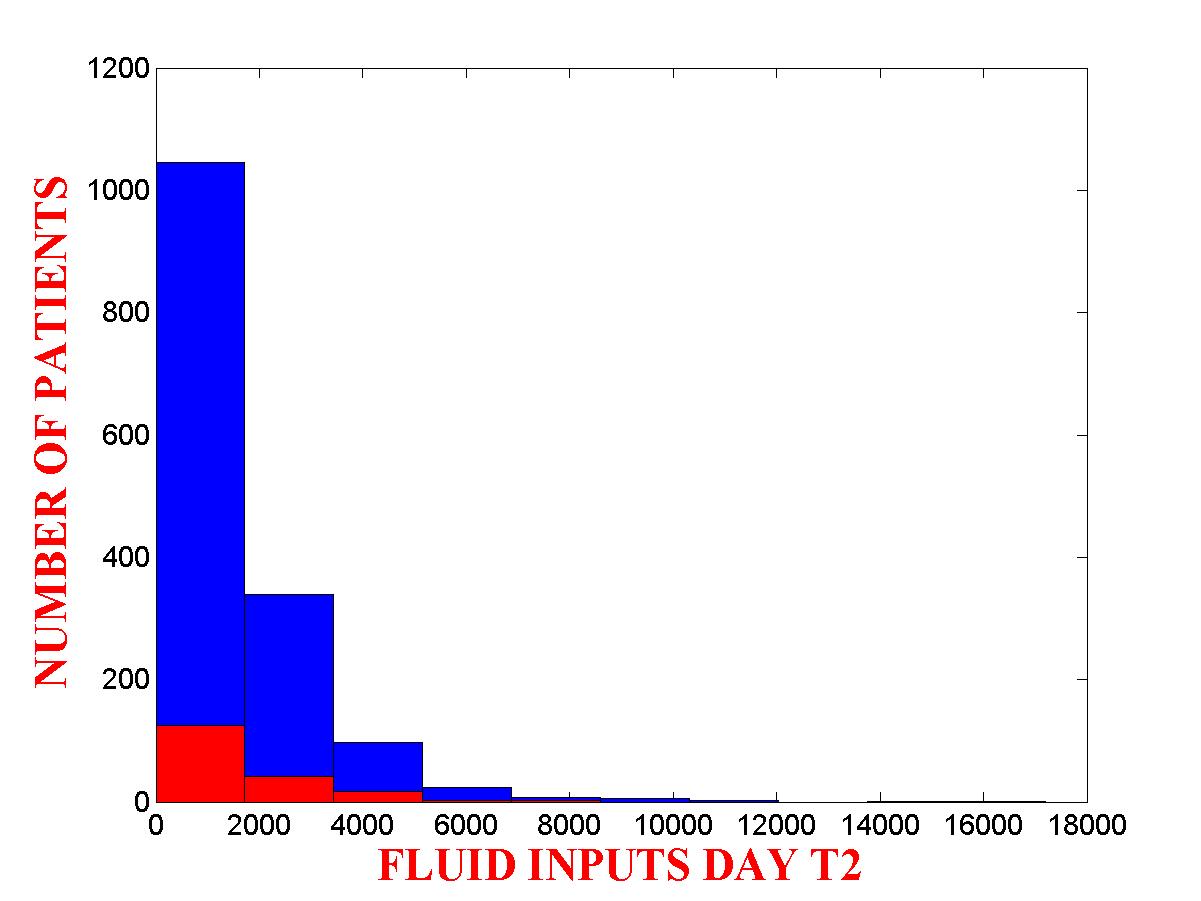}
\captionof{subfigure}[]{Fluids inputs day T2 is centered on 1.0 liters.}
\label{fig:hist_37}
\end{minipage}
\begin{minipage}{0.33\textheight}
\centering
\includegraphics[width=.9\linewidth]{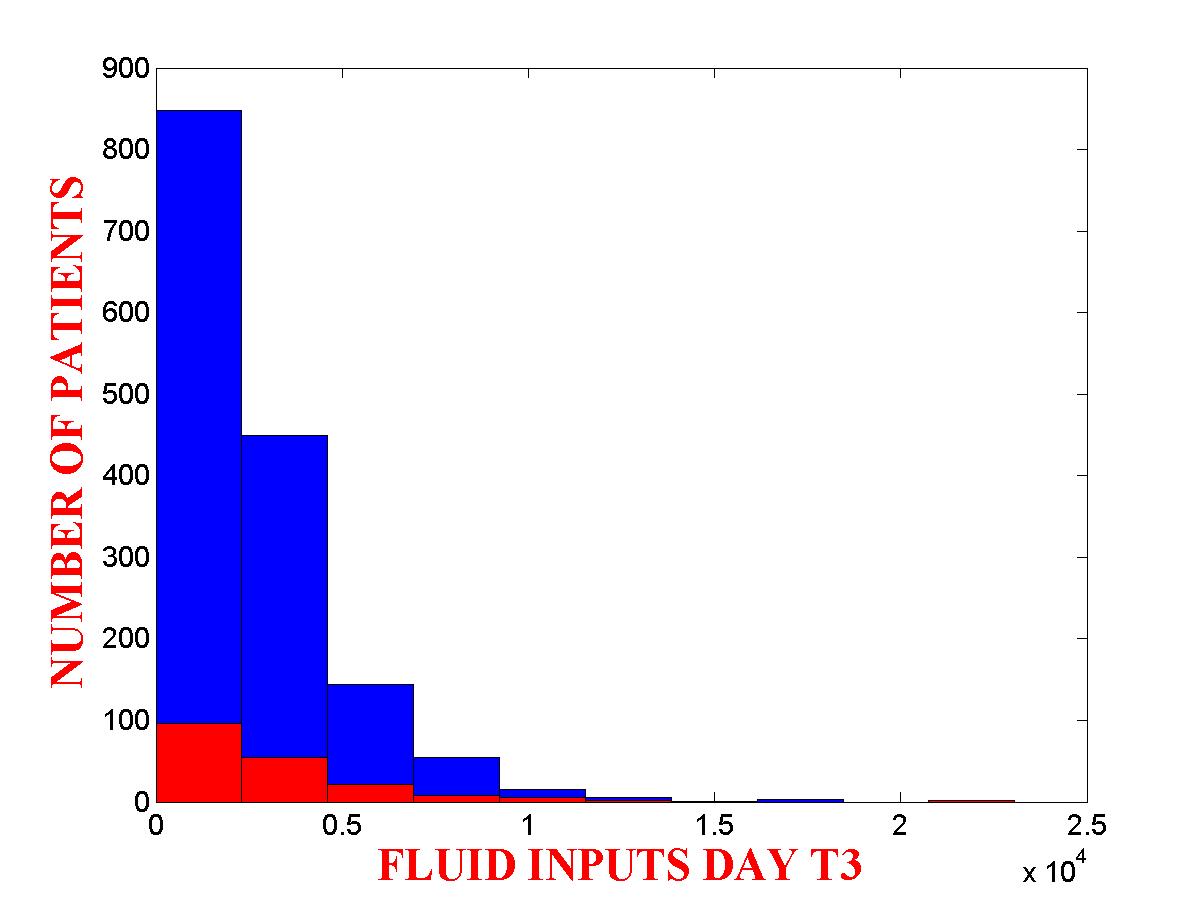}
\captionof{subfigure}[]{Fluids inputs day T3 is centered on 2.0 liters.}
\label{fig:hist_38}
\end{minipage}
\begin{minipage}{0.33\textheight}
\centering
\includegraphics[width=.9\linewidth]{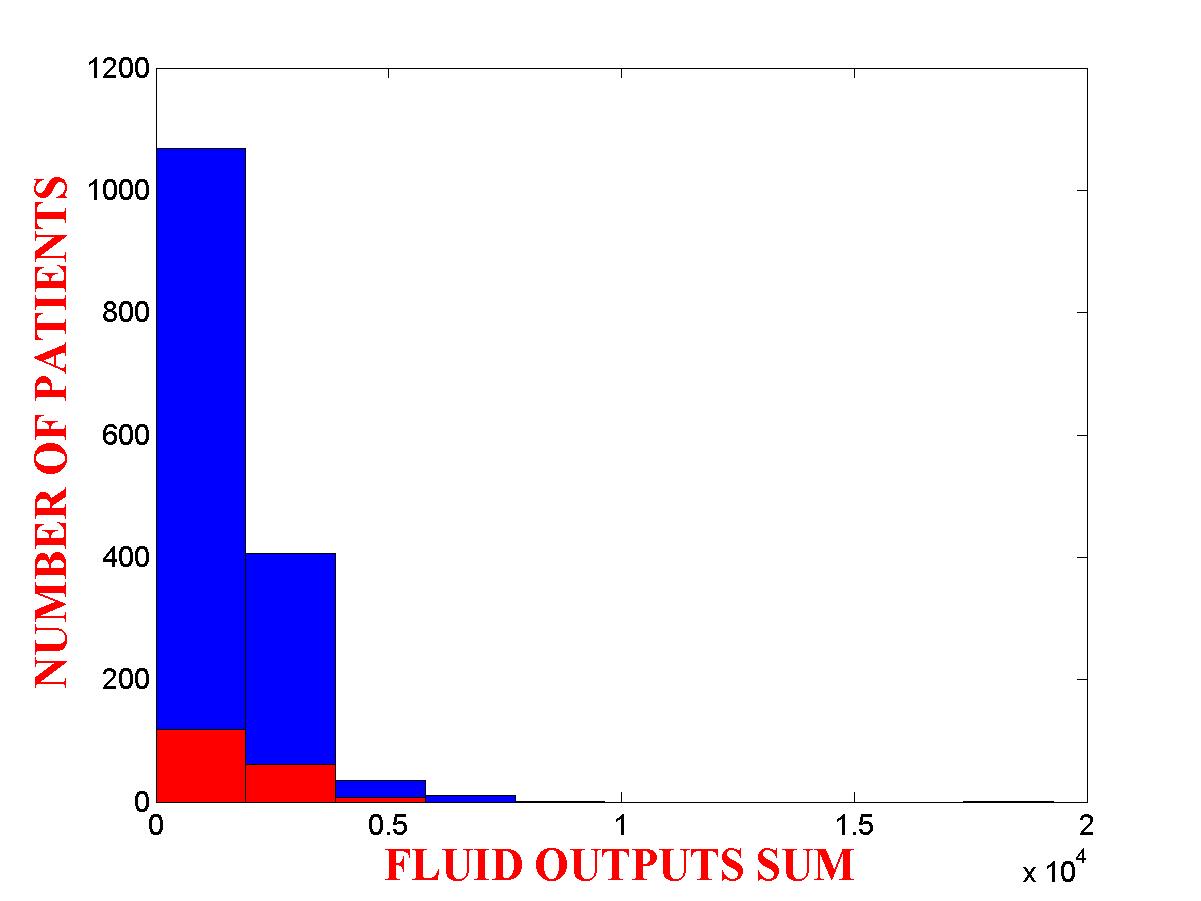}
\captionof{subfigure}[]{Fluids outputs sum is centered on 1.3 liters.}
\label{fig:hist_39}
\end{minipage}
\begin{minipage}{0.33\textheight}
\centering
\includegraphics[width=.9\linewidth]{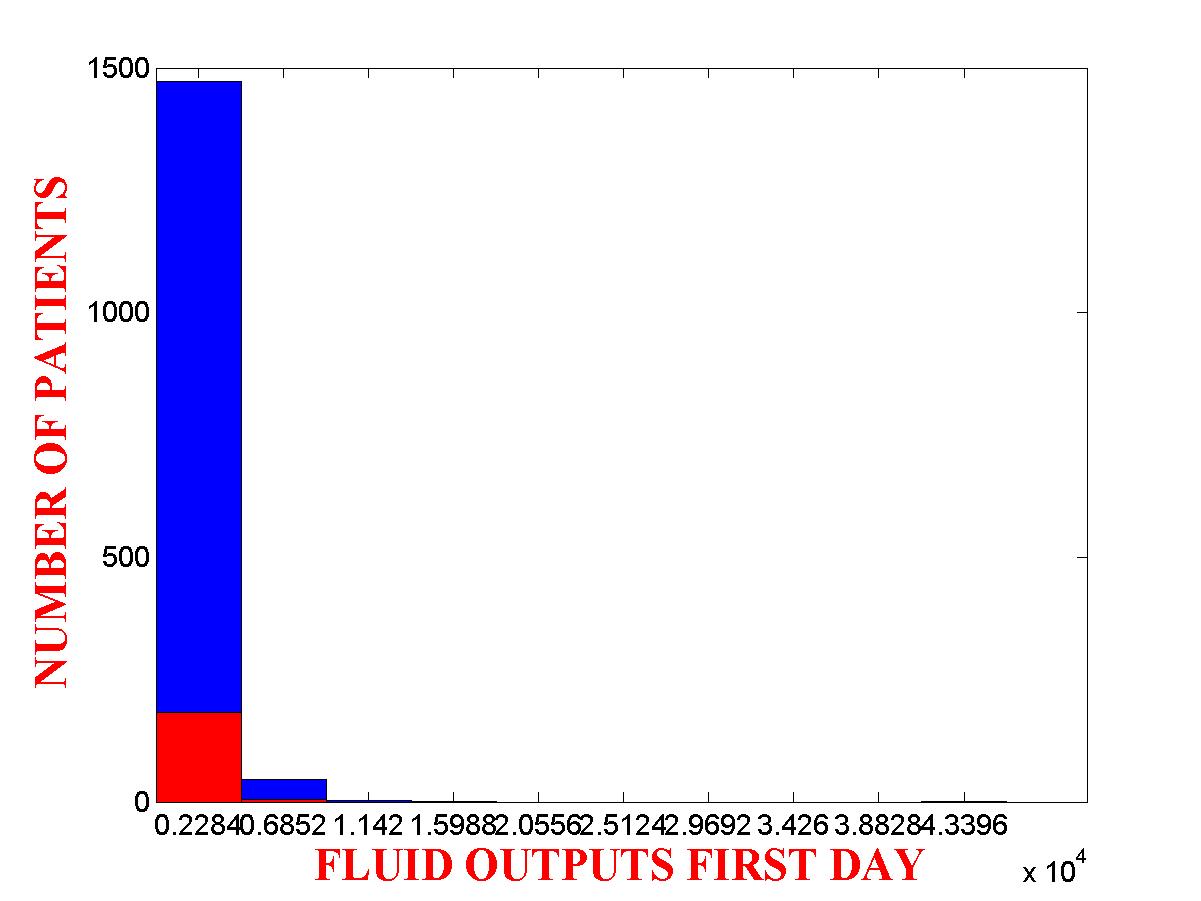}
\captionof{subfigure}[]{Fluids outputs first day is centered on 1.2 liters.}
\label{fig:hist_40}
\end{minipage}
\addtocounter{figure}{-1}
\captionof{figure}[Histograms of the Numeric Variables, part 4]{Histograms of the numeric variables, part 4}
\label{fig:histo_group_4}
\end{minipage}
}
\end{figure}
\begin{figure}
\centering
\rotatebox{90}{
\stepcounter{figure}
\setcounter{subfigure}{0}
\begin{minipage}{\textheight}
\begin{minipage}{0.33\textheight}
\centering
\includegraphics[width=.9\linewidth]{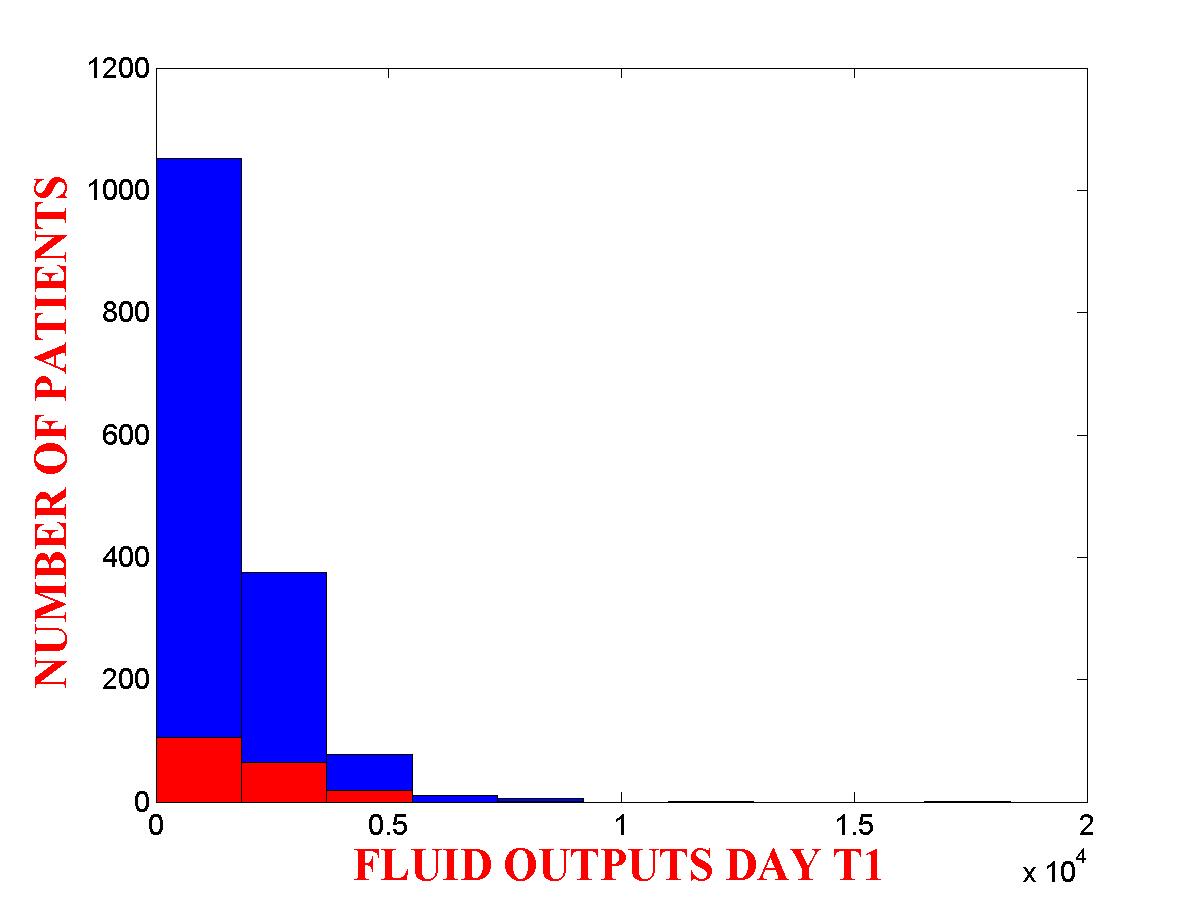}
\captionof{subfigure}[]{Fluids outputs day T1 is centered on 1.1 liters.}
\label{fig:hist_41}
\end{minipage}
\begin{minipage}{0.33\textheight}
\centering
\includegraphics[width=.9\linewidth]{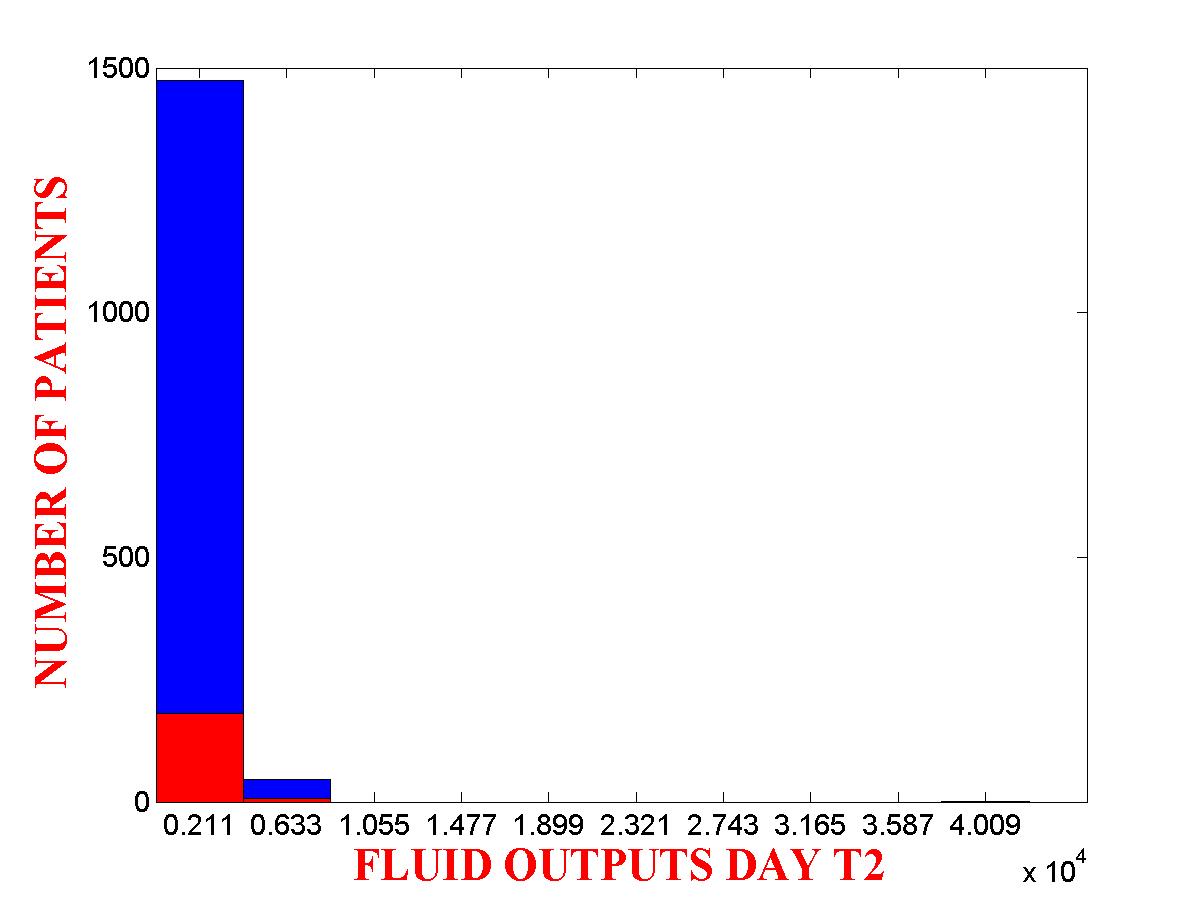}
\captionof{subfigure}[]{Fluids outputs day T2 is centered on 1.1 liters.}
\label{fig:hist_42}
\end{minipage}
\begin{minipage}{0.33\textheight}
\centering
\includegraphics[width=.9\linewidth]{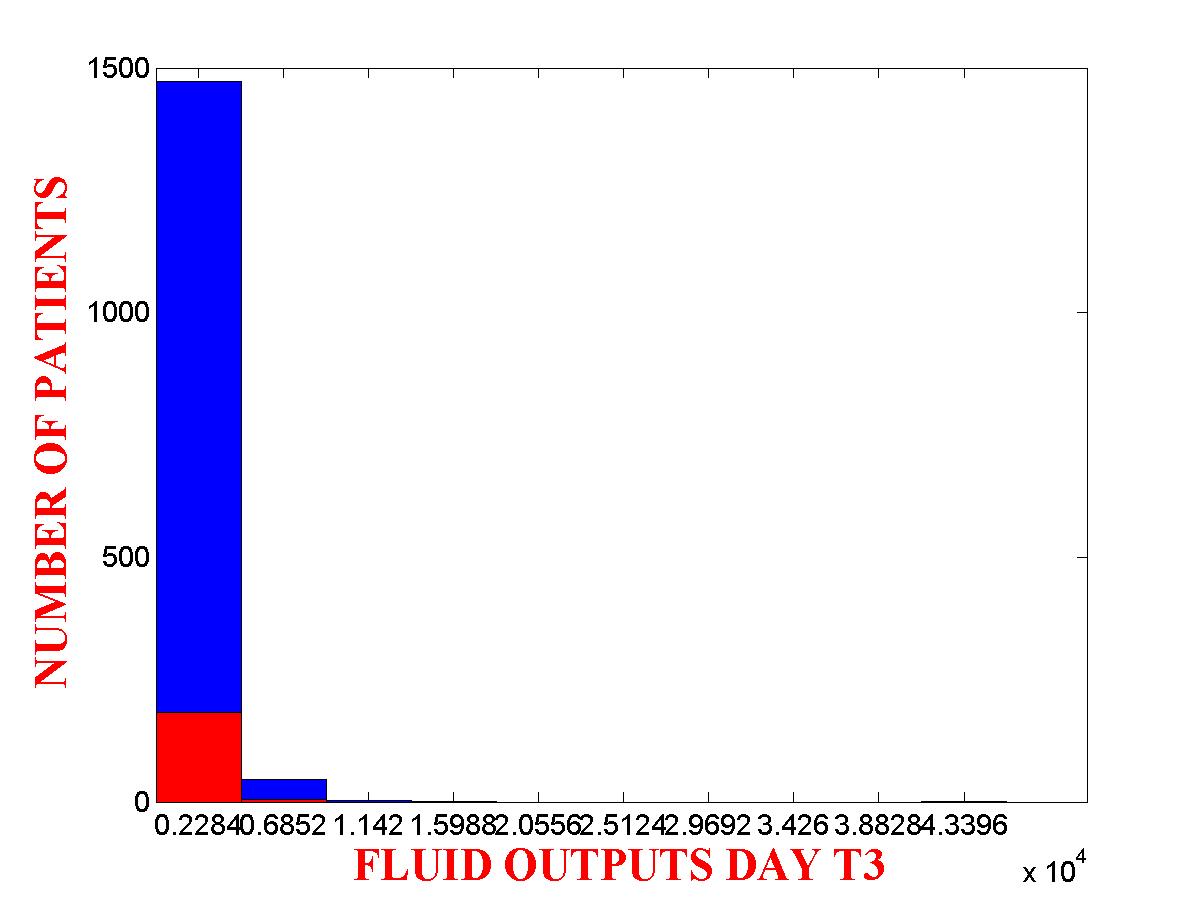}
\captionof{subfigure}[]{Fluids outputs day T3 is centered on 1.2 liters.}
\label{fig:hist_43}
\end{minipage}
\begin{minipage}{0.33\textheight}
\centering
\includegraphics[width=.9\linewidth]{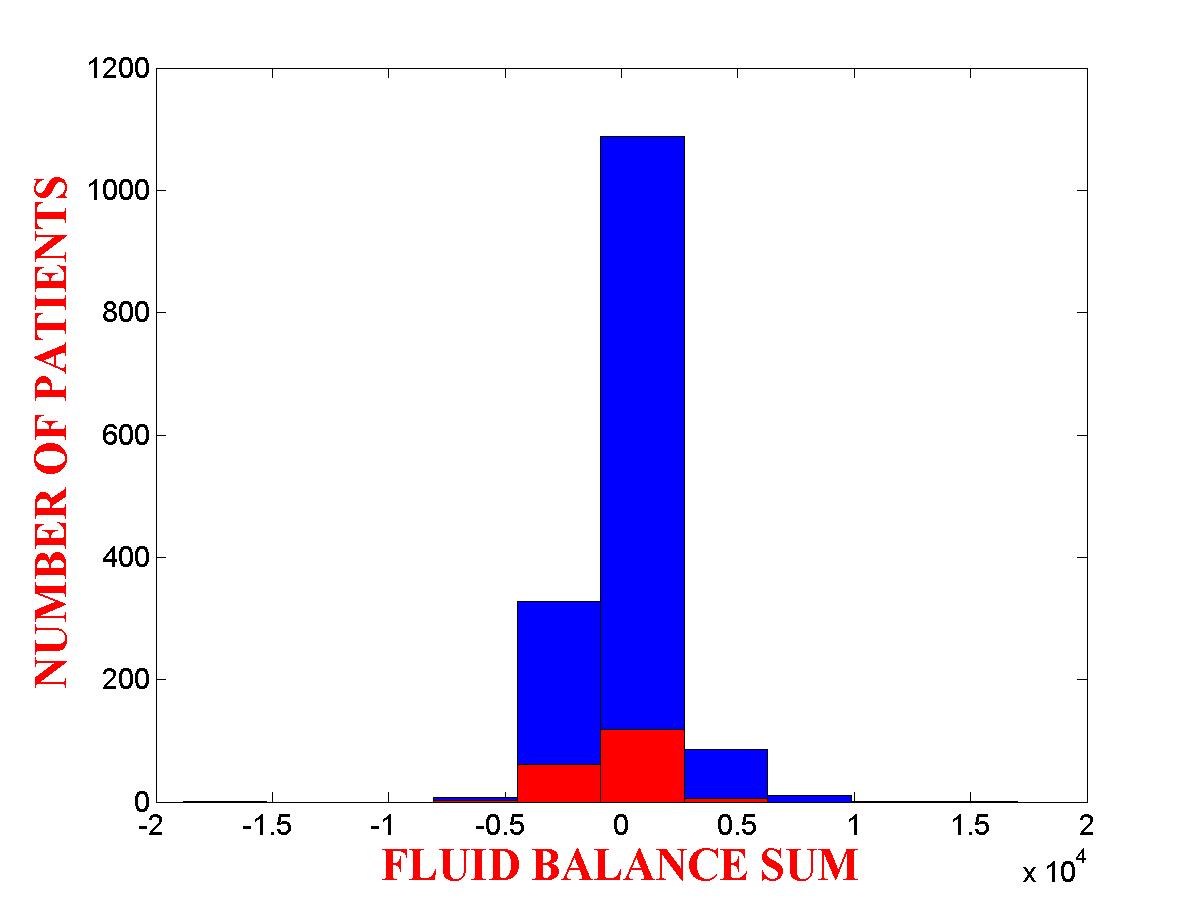}
\captionof{subfigure}[]{Fluids balance sum is centered on 0.05 liters.}
\label{fig:hist_44}
\end{minipage}
\begin{minipage}{0.33\textheight}
\centering
\includegraphics[width=.9\linewidth]{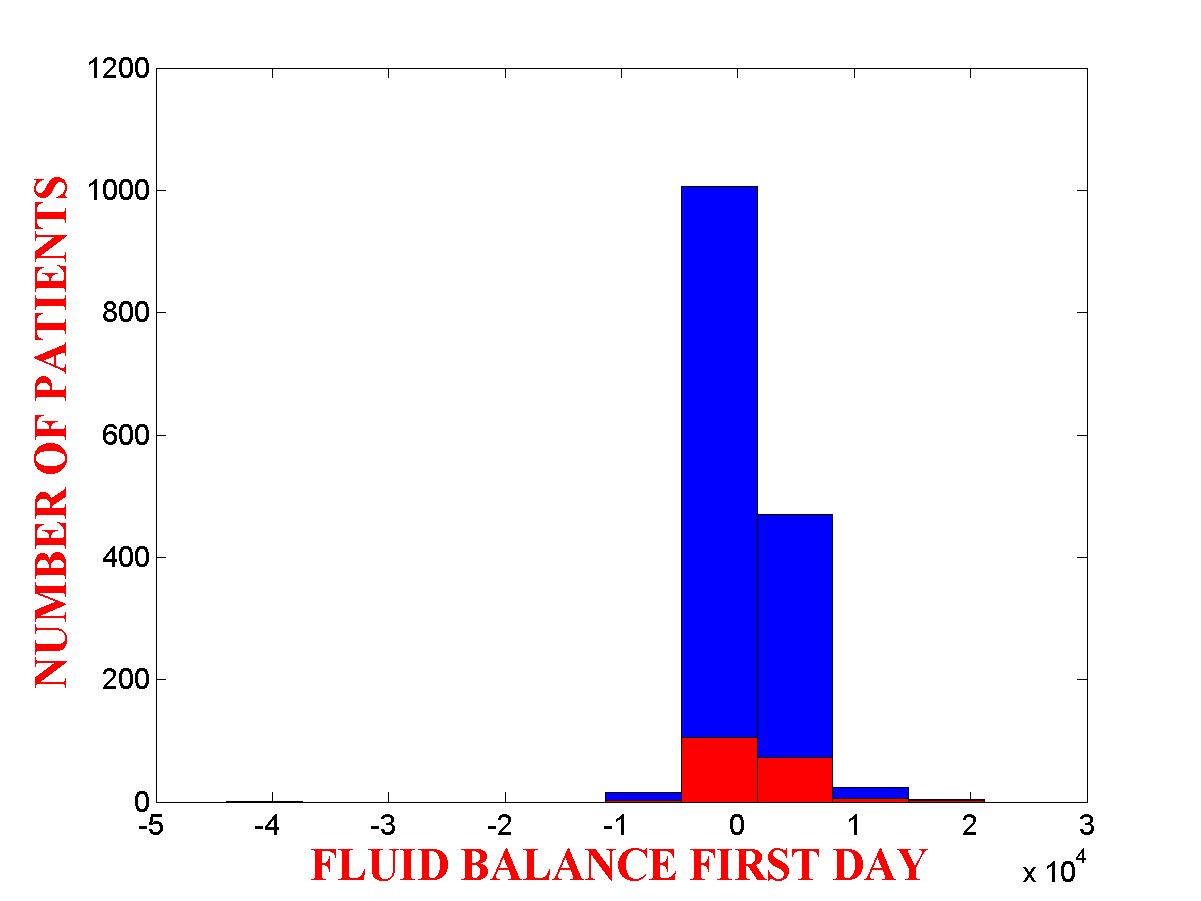}
\captionof{subfigure}[]{Fluids balance first day is centered on 0.67 liters.}
\label{fig:hist_45}
\end{minipage}
\begin{minipage}{0.33\textheight}
\centering
\includegraphics[width=.9\linewidth]{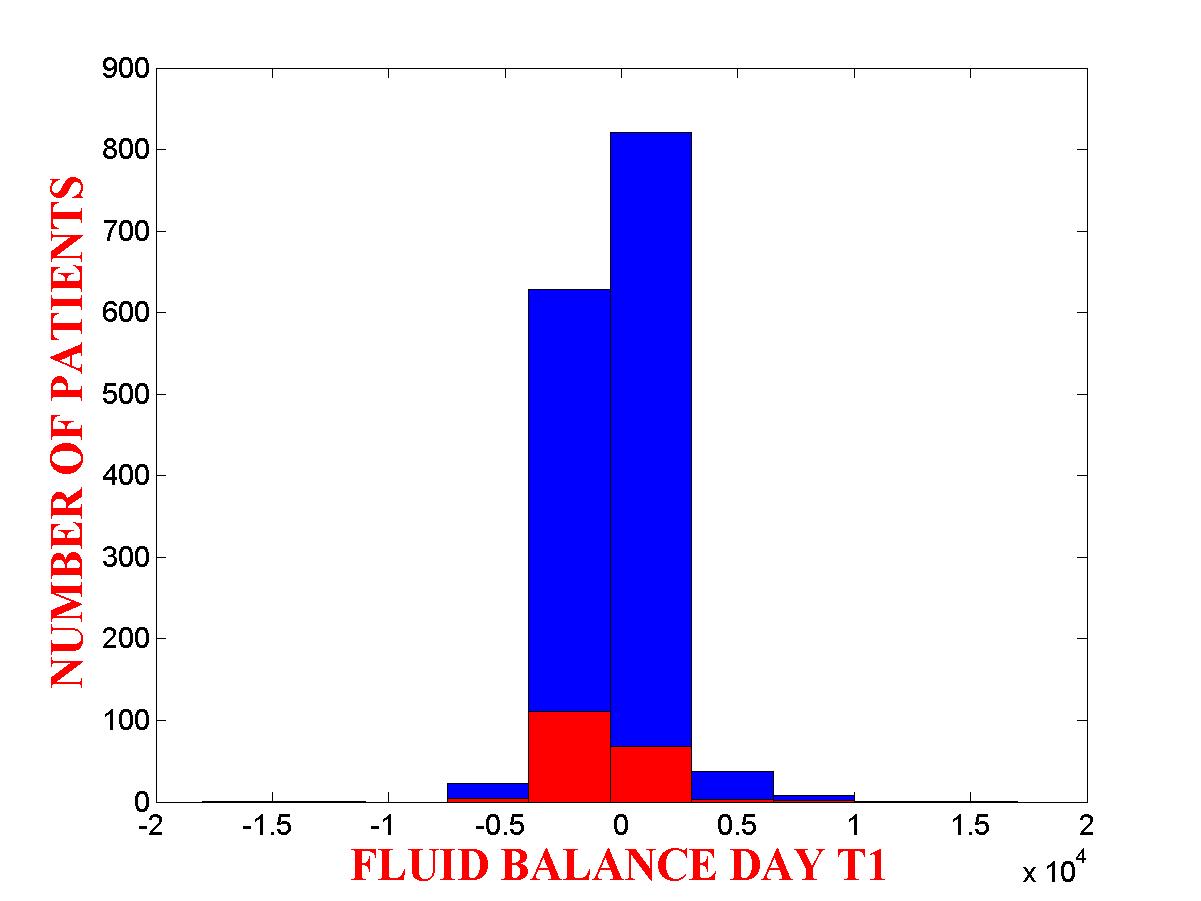}
\captionof{subfigure}[]{Fluids balance day T1 is centered on -0.25 liters.}
\label{fig:hist_46}
\end{minipage}
\addtocounter{figure}{-1}
\captionof{figure}[Histograms of the Numeric Variables, part 5]{Histograms of the numeric variables, part 5}
\label{fig:histo_group_5}
\end{minipage}
}
\end{figure}
\begin{figure}
\centering
\rotatebox{90}{
\stepcounter{figure}
\setcounter{subfigure}{0}
\begin{minipage}{\textheight}
\begin{minipage}{0.33\textheight}
\centering
\includegraphics[width=.9\linewidth]{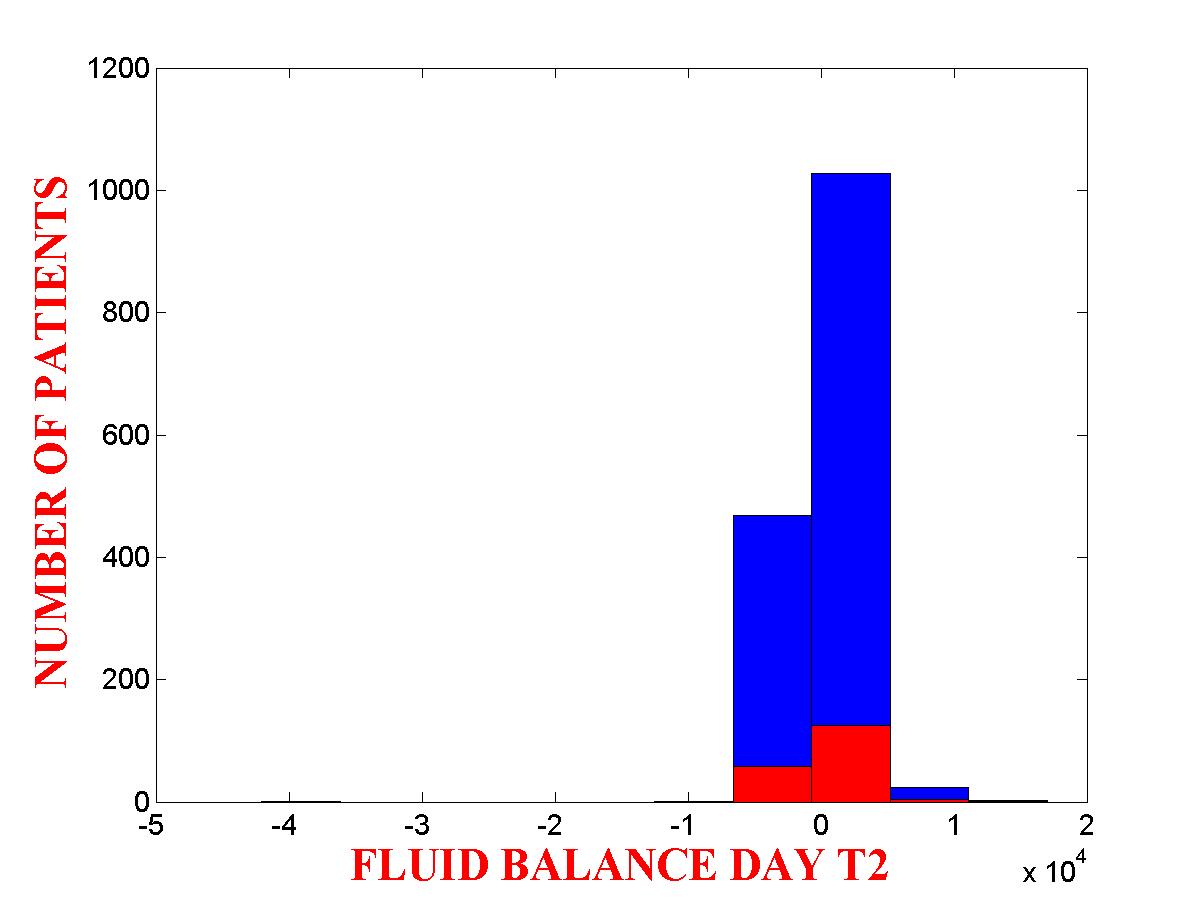}
\captionof{subfigure}[]{Fluids balance day T2 is centered on -0.002 liters.}
\label{fig:hist_47}
\end{minipage}
\begin{minipage}{0.33\textheight}
\centering
\includegraphics[width=.9\linewidth]{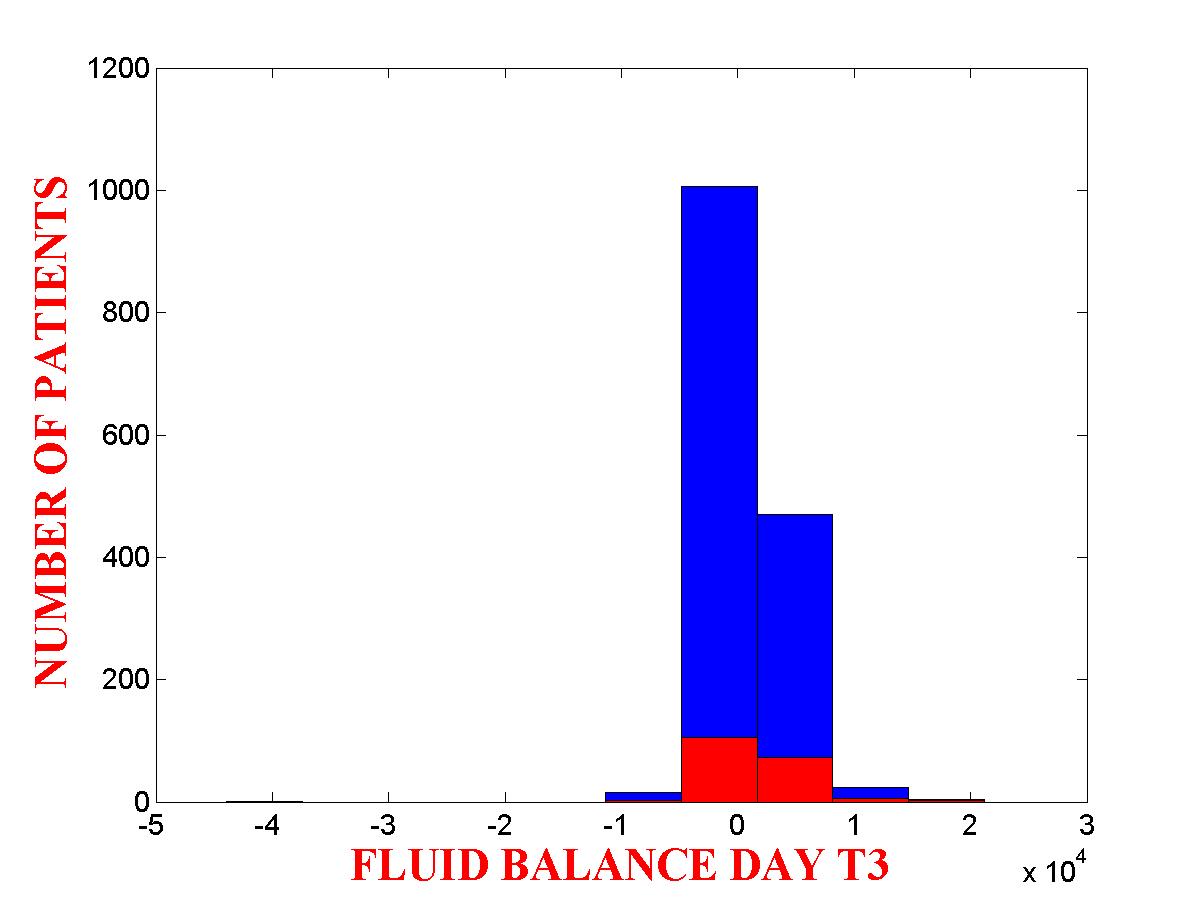}
\captionof{subfigure}[]{Fluids balance day T3 is centered on 0.67 liters.}
\label{fig:hist_48}
\end{minipage}
\begin{minipage}{0.33\textheight}
\centering
\includegraphics[width=.9\linewidth]{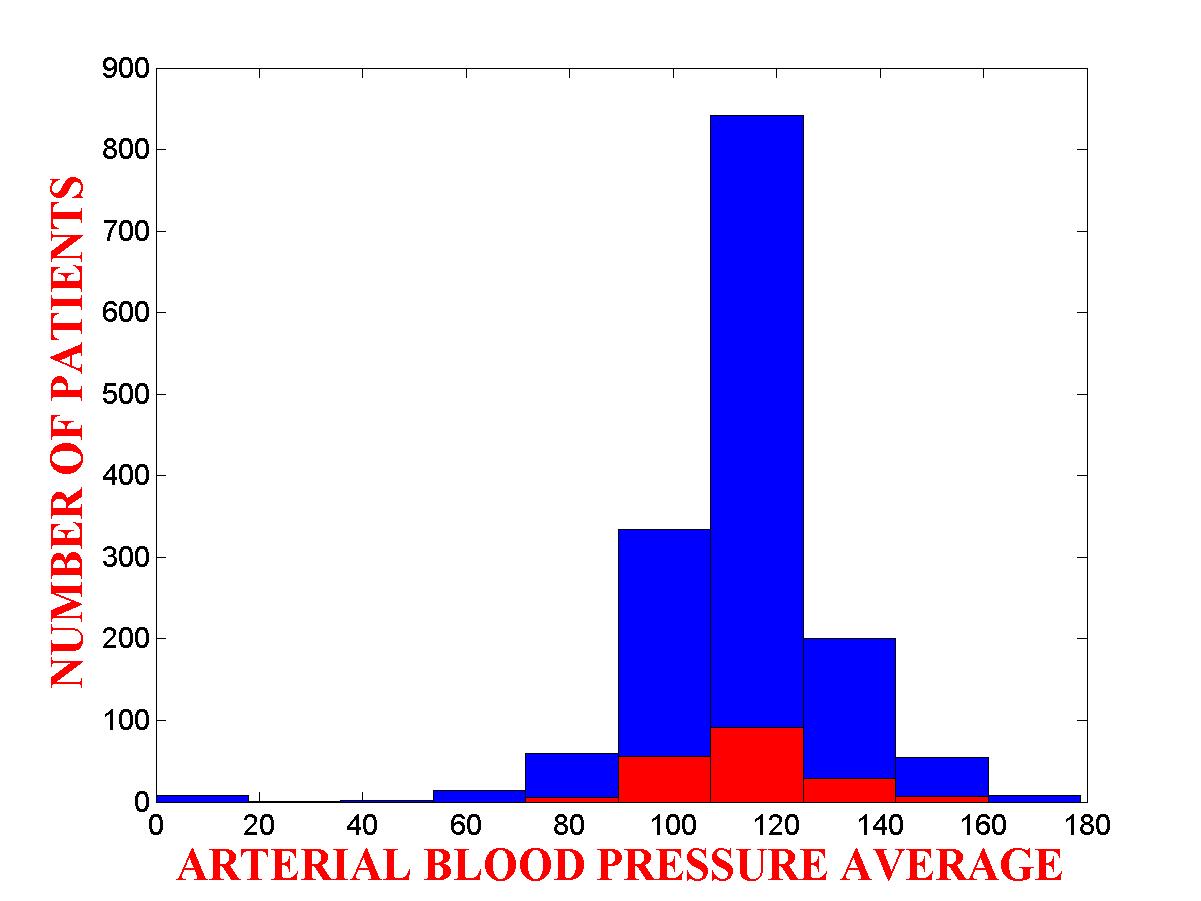}
\captionof{subfigure}[]{Arterial blood pressure average is centered on 114.9.}
\label{fig:hist_49}
\end{minipage}
\begin{minipage}{0.33\textheight}
\centering
\includegraphics[width=.9\linewidth]{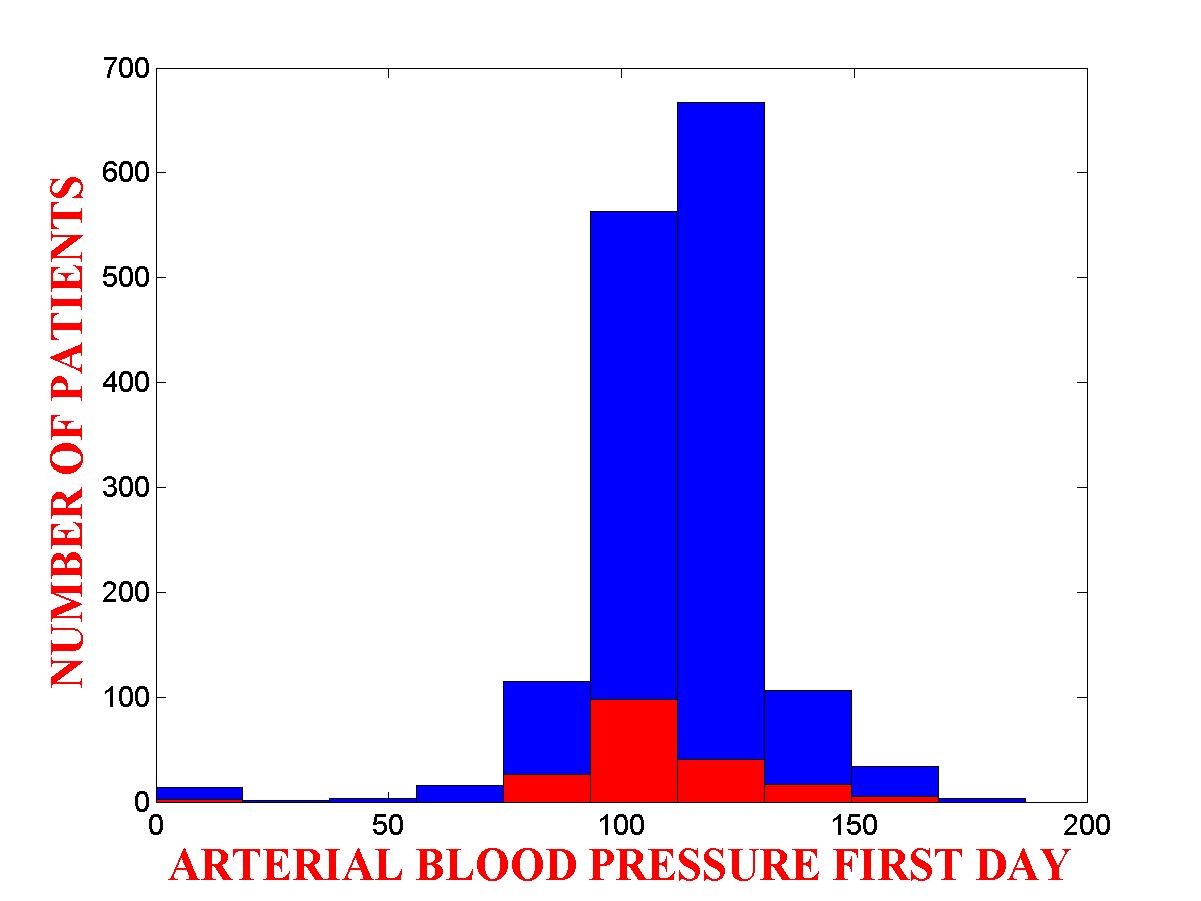}
\captionof{subfigure}[]{Arterial blood pressure first day is centered on 114.1.}
\label{fig:hist_50}
\end{minipage}
\begin{minipage}{0.33\textheight}
\centering
\includegraphics[width=.9\linewidth]{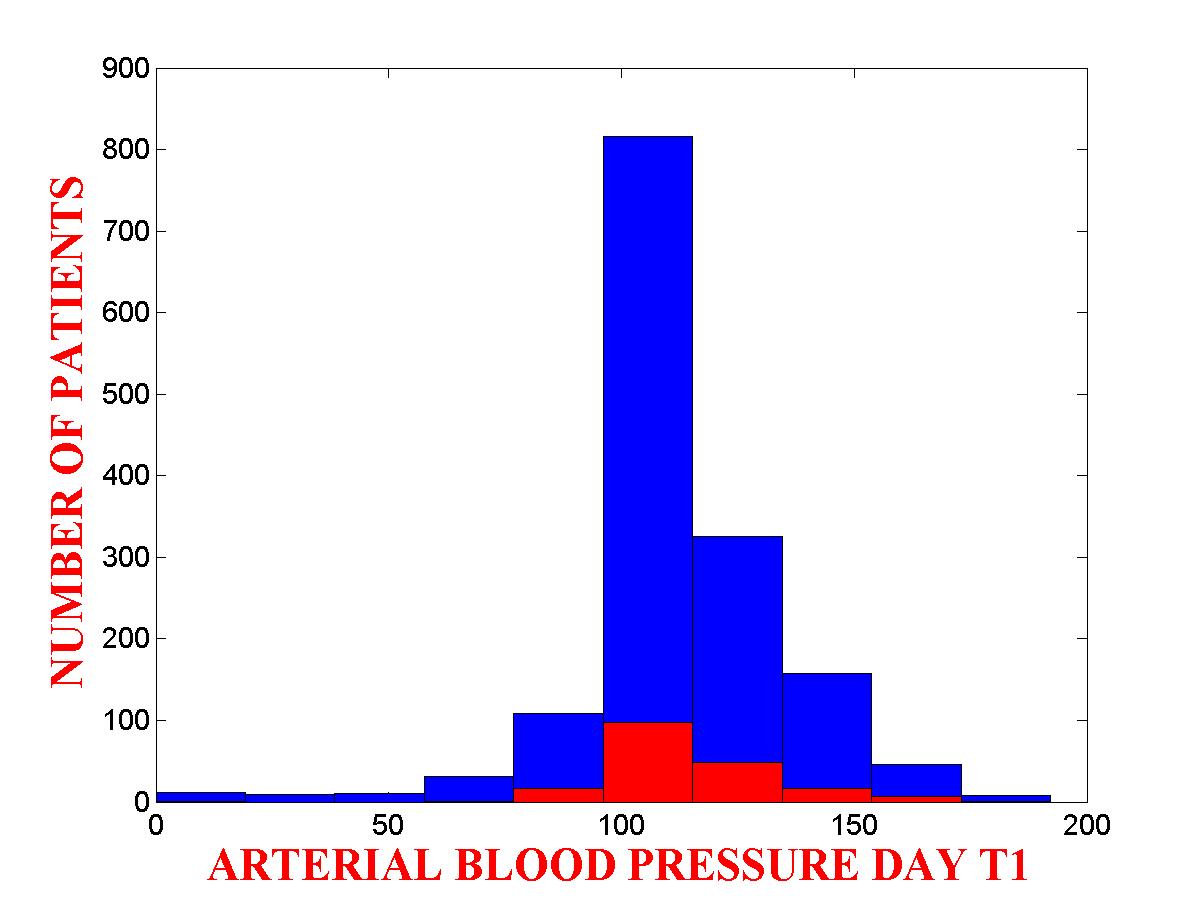}
\captionof{subfigure}[]{Arterial blood pressure day T1 is centered on 114.9.}
\label{fig:hist_51}
\end{minipage}
\begin{minipage}{0.33\textheight}
\centering
\includegraphics[width=.9\linewidth]{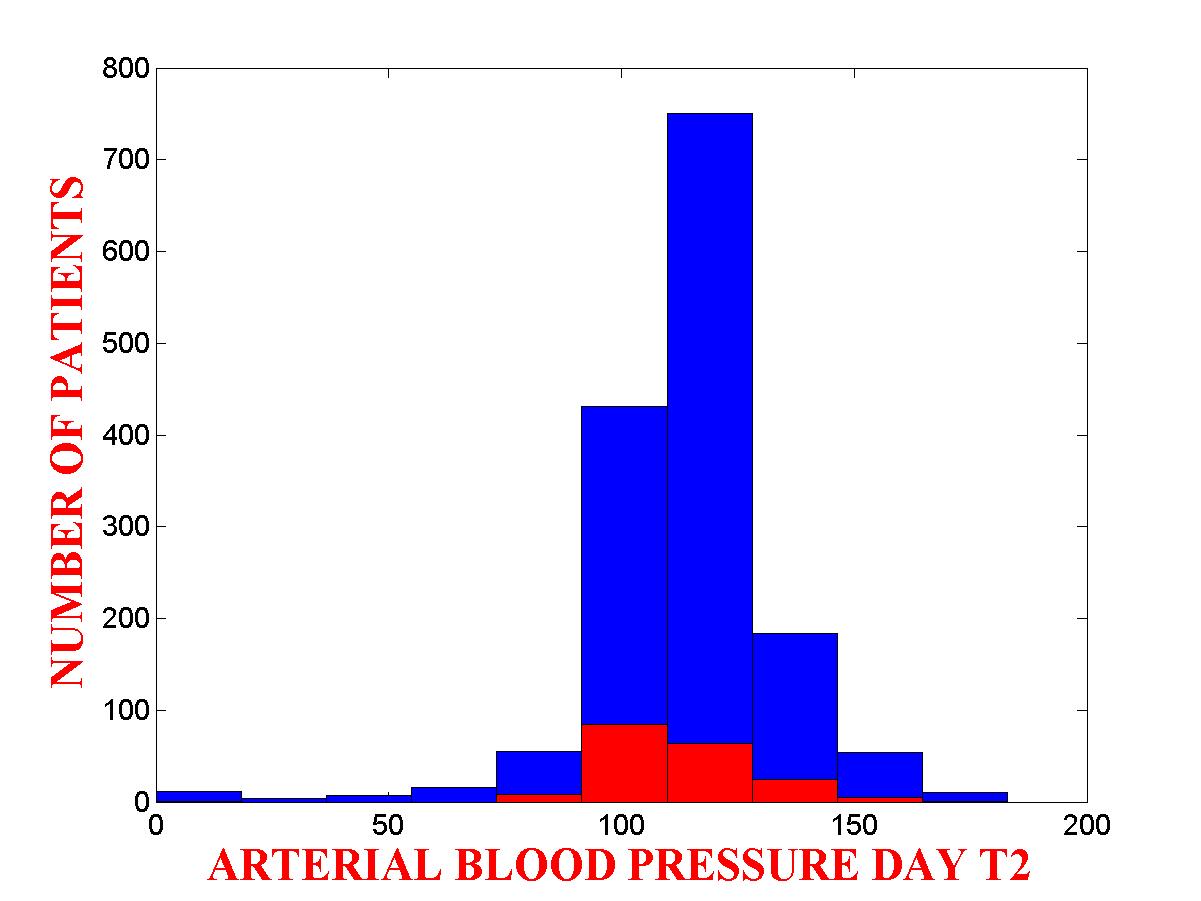}
\captionof{subfigure}[]{Arterial blood pressure day T2 is centered on 114.9.}
\label{fig:hist_52}
\end{minipage}
\addtocounter{figure}{-1}
\captionof{figure}[Histograms of the Numeric Variables, part 6]{Histograms of the numeric variables, part 6}
\label{fig:histo_group_6}
\end{minipage}
}
\end{figure}
\begin{figure}
\centering
\rotatebox{90}{
\stepcounter{figure}
\setcounter{subfigure}{0}
\begin{minipage}{\textheight}
\begin{minipage}{0.33\textheight}
\centering
\includegraphics[width=.9\linewidth]{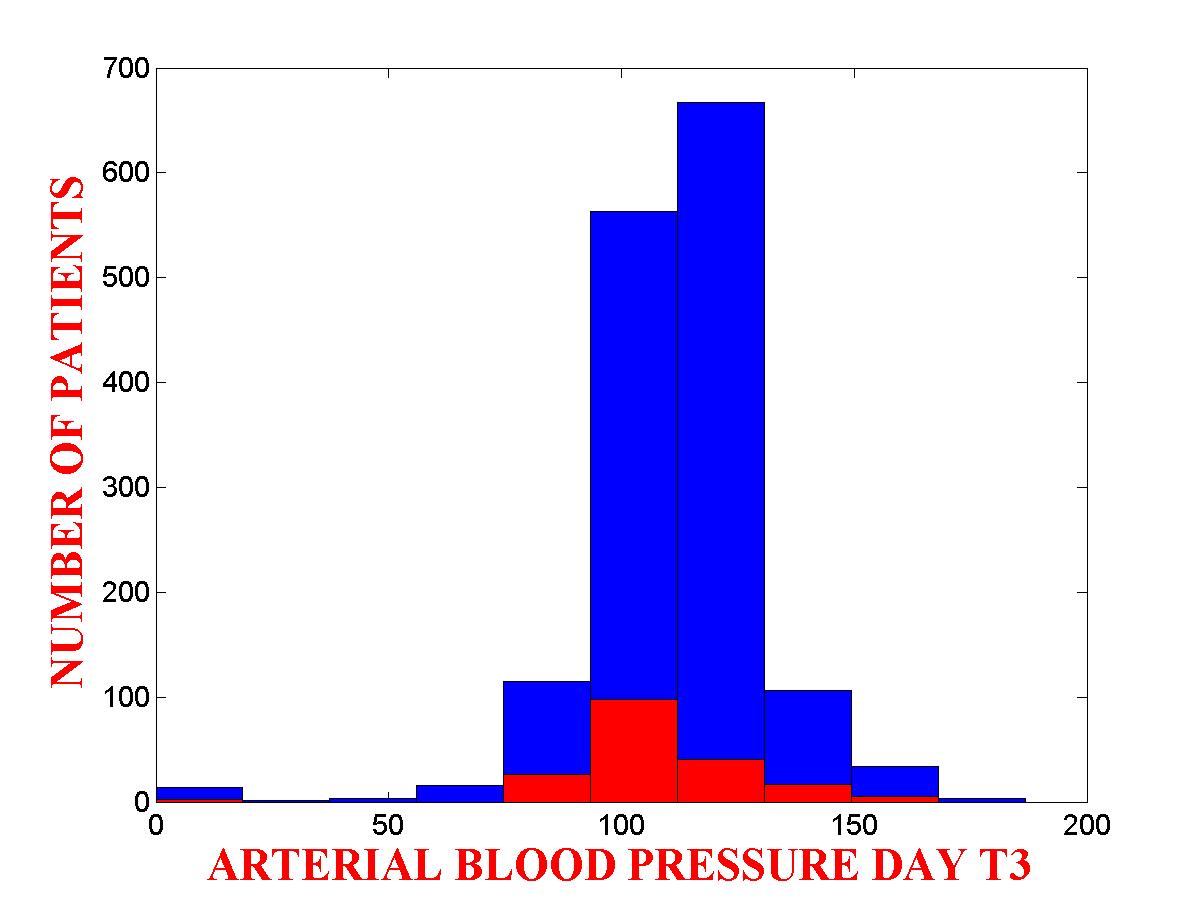}
\captionof{subfigure}[]{Arterial blood pressure day T3 is centered on 114.1.}
\label{fig:hist_53}
\end{minipage}
\begin{minipage}{0.33\textheight}
\centering
\includegraphics[width=.9\linewidth]{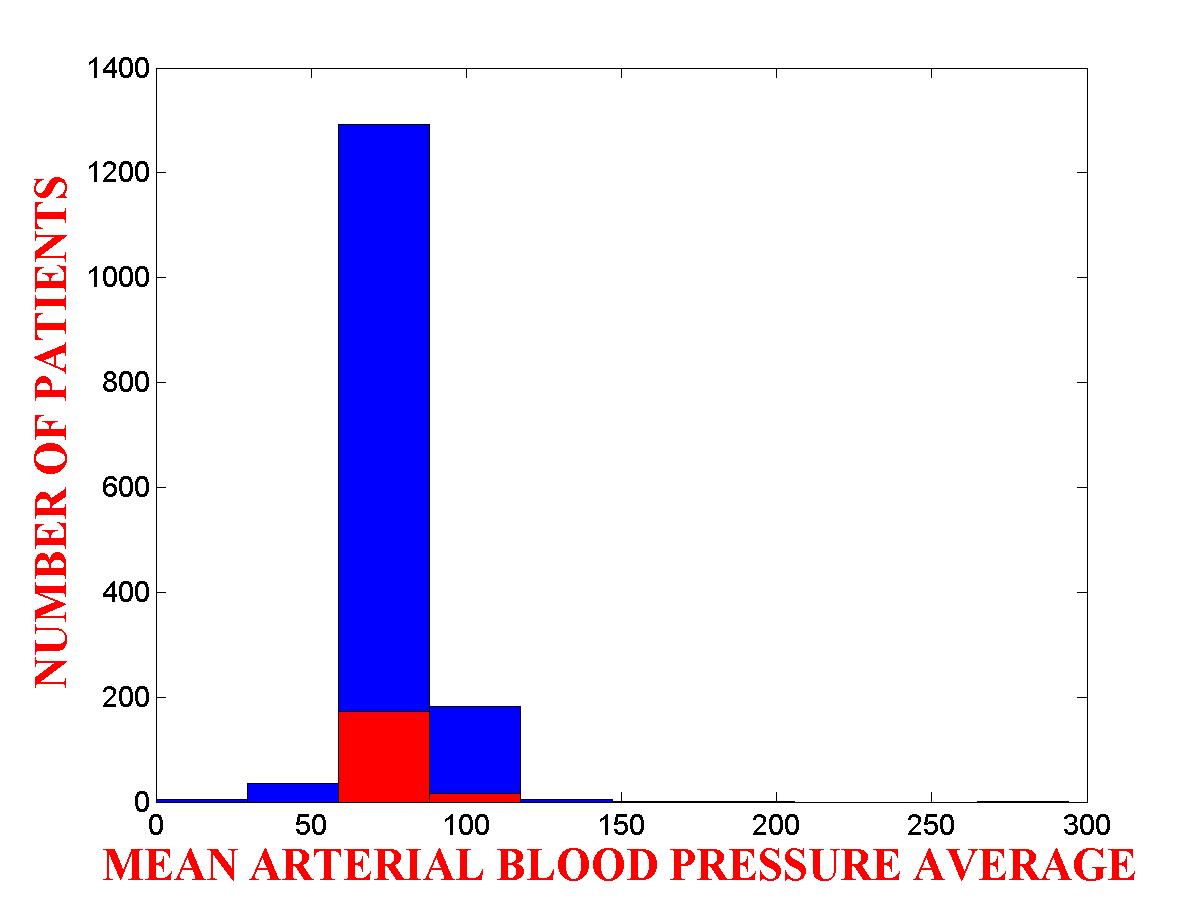}
\captionof{subfigure}[]{Mean arterial blood pressure average is centered on 79.2.}
\label{fig:hist_54}
\end{minipage}
\begin{minipage}{0.33\textheight}
\centering
\includegraphics[width=.9\linewidth]{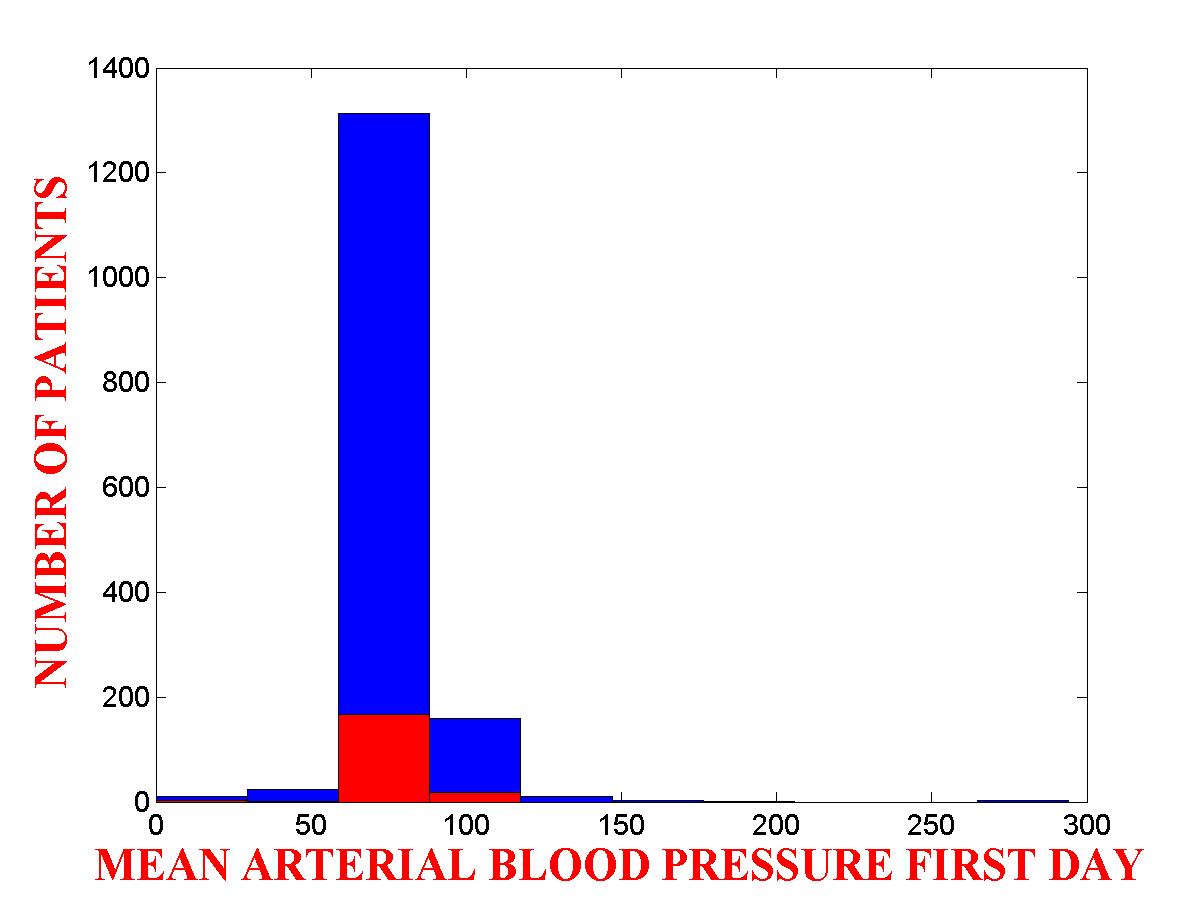}
\captionof{subfigure}[]{Mean arterial blood pressure first day is centered on 79.2.}
\label{fig:hist_55}
\end{minipage}
\begin{minipage}{0.33\textheight}
\centering
\includegraphics[width=.9\linewidth]{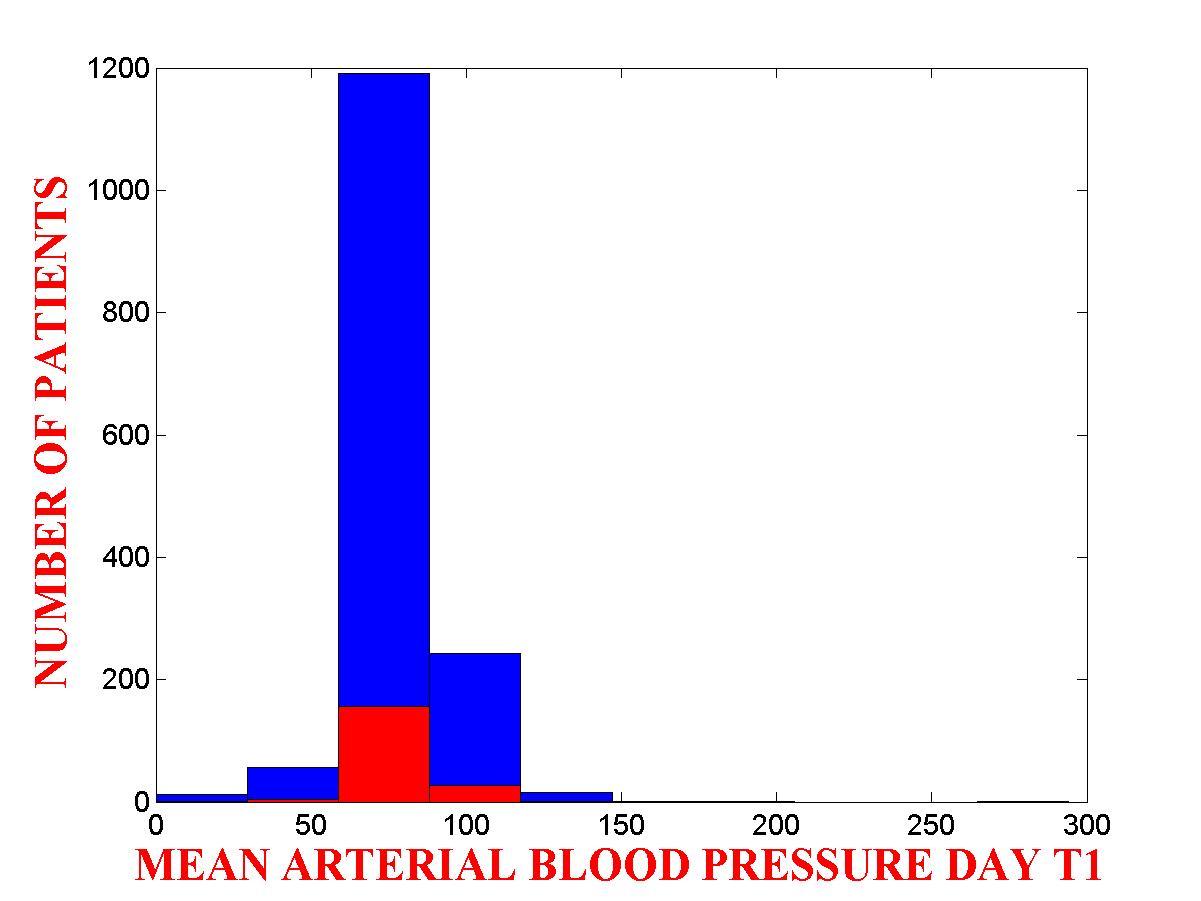}
\captionof{subfigure}[]{Mean arterial blood pressure day T1 is centered on 79.2.}
\label{fig:hist_56}
\end{minipage}
\begin{minipage}{0.33\textheight}
\centering
\includegraphics[width=.9\linewidth]{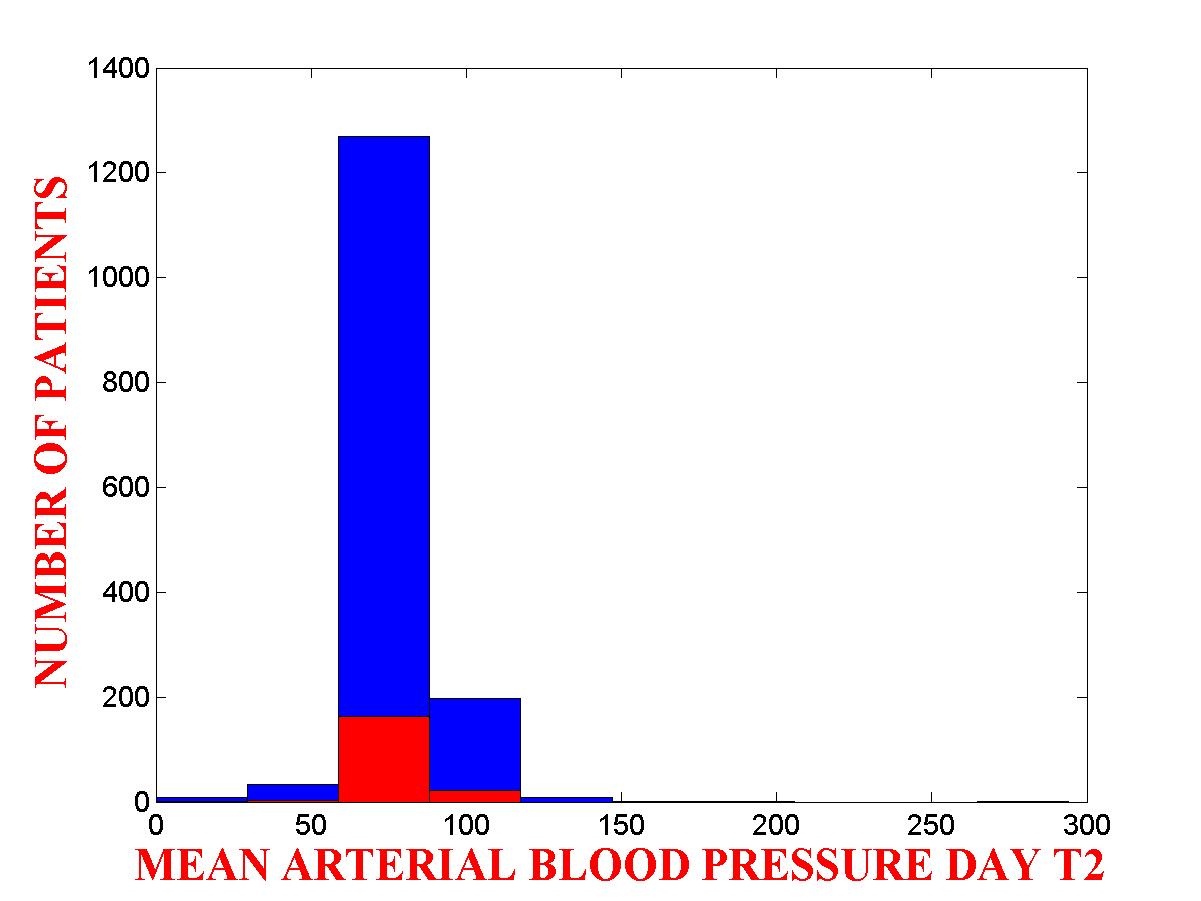}
\captionof{subfigure}[]{Mean arterial blood pressure day T2 is centered on 79.2.}
\label{fig:hist_57}
\end{minipage}
\begin{minipage}{0.33\textheight}
\centering
\includegraphics[width=.9\linewidth]{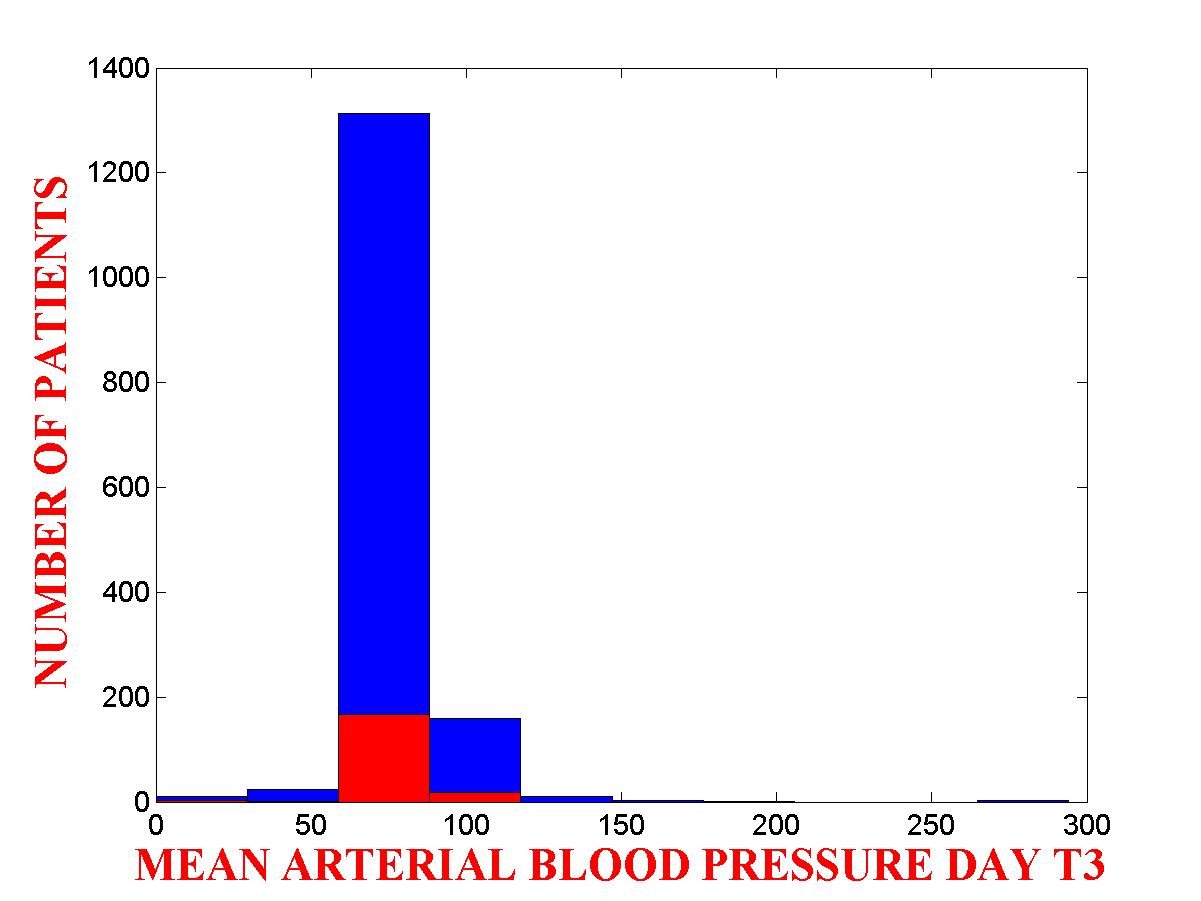}
\captionof{subfigure}[]{Mean arterial blood pressure day T3 is centered on 79.2.}
\label{fig:hist_58}
\end{minipage}
\addtocounter{figure}{-1}
\captionof{figure}[Histograms of the Numeric Variables, part 7]{Histograms of the numeric variables, part 7}
\label{fig:histo_group_7}
\end{minipage}
}
\end{figure}

\clearpage

\section{Timeline Values Discussion}
\label{apx:timesCorrs}
In this Section an overview of the correlations of the values for the variables with timelines at 8 time points will be provided. The final decisions on how to prepare timeline variables is in Chapter~\ref{cpt:datasetExtraction}.

Here the earlier data that have been collected which informed the final decisions are described. Follows the list of the studied time points:
\begin{description}
\item[Timepoint 1:] Diuretics average over $D^+$ patients (189 of 1,522 patients);
\item[Timepoint 2:] Diuretics average over $D^+$ patients as fraction of \los ;
\item[Timepoint 3:] First fluids balance minimum;
\item[Timepoint 4:] Second fluids balance minimum;
\item[Timepoint 5:] Stop of vasopressors average on 1,028 of 1,522 patients;
\item[Timepoint 6:] Stop of vasopressors average on 1,028 of 1,522 patients as fraction of \los ;
\item[Timepoint 7:] First blood pressure minimum;
\item[Timepoint 8:] Second blood pressure minimum.
\end{description}
These results showed that most of the defined time points were correlated and not introduced new information regarding diuretics to the dataset. So, it was decided to adopt the time points defined in Chapter~\ref{cpt:datasetExtraction}

\clearpage

\begin{table}[htbp]
\centering
\includegraphics[width=\textwidth]{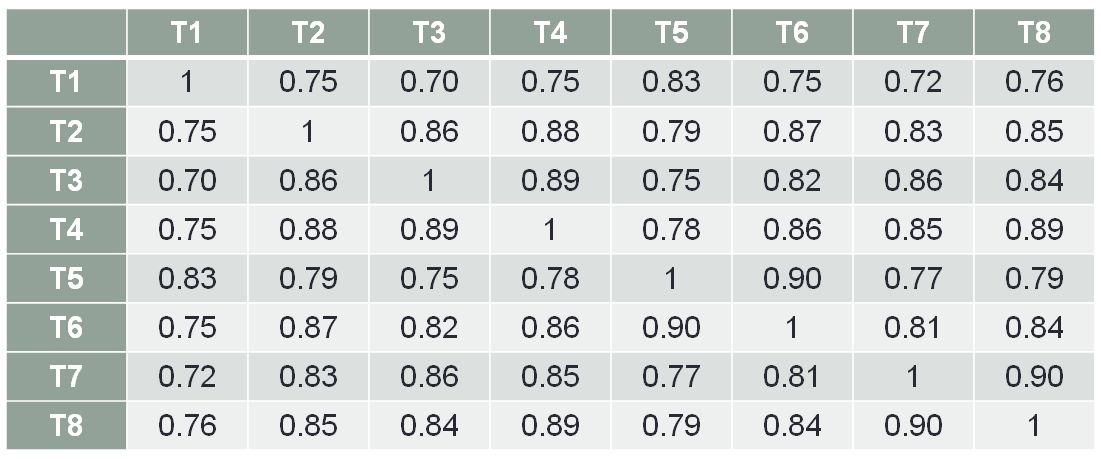}
\caption[Times Correlation for Saps]{The correlations between 8 time points for the saps variable.}
\label{fig:corr1}
\end{table}
\begin{table}[htbp]
\centering
\includegraphics[width=\textwidth]{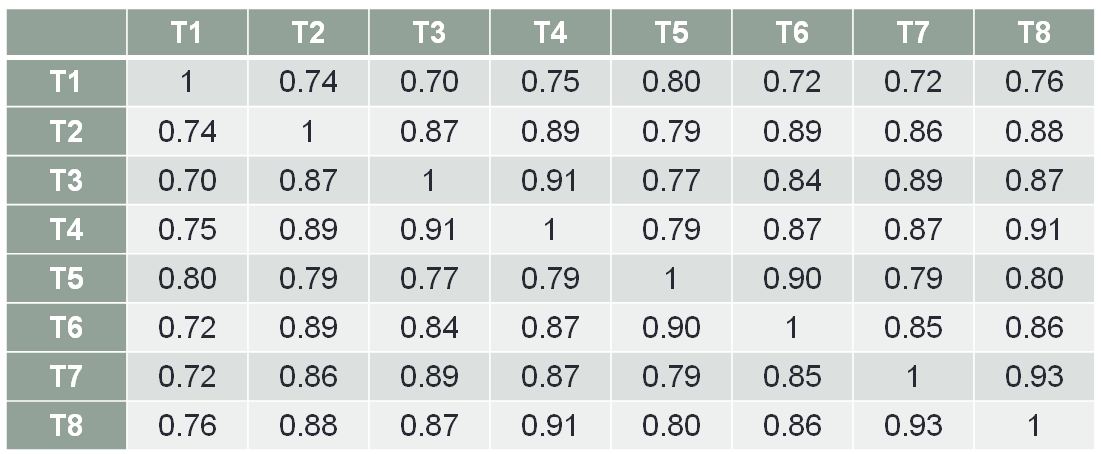}
\caption[Times Correlation for Sofa]{The correlations between 8 time points for the sofa variable.}
\label{fig:corr2}
\end{table}
\begin{table}[htbp]
\centering
\includegraphics[width=\textwidth]{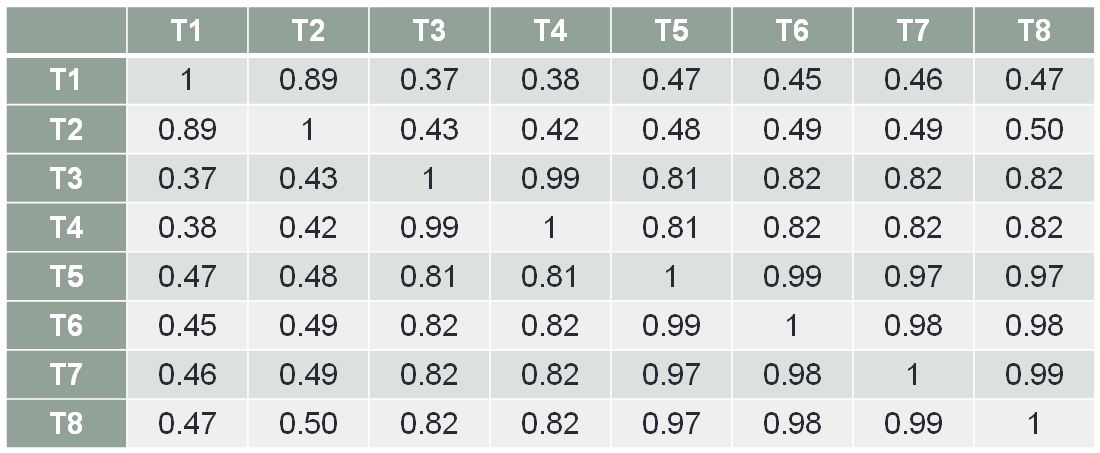}
\caption[Times Correlation for Creatinine]{The correlations between 8 time points for the creatinine variable.}
\label{fig:corr3}
\end{table}
\begin{table}[htbp]
\centering
\includegraphics[width=\textwidth]{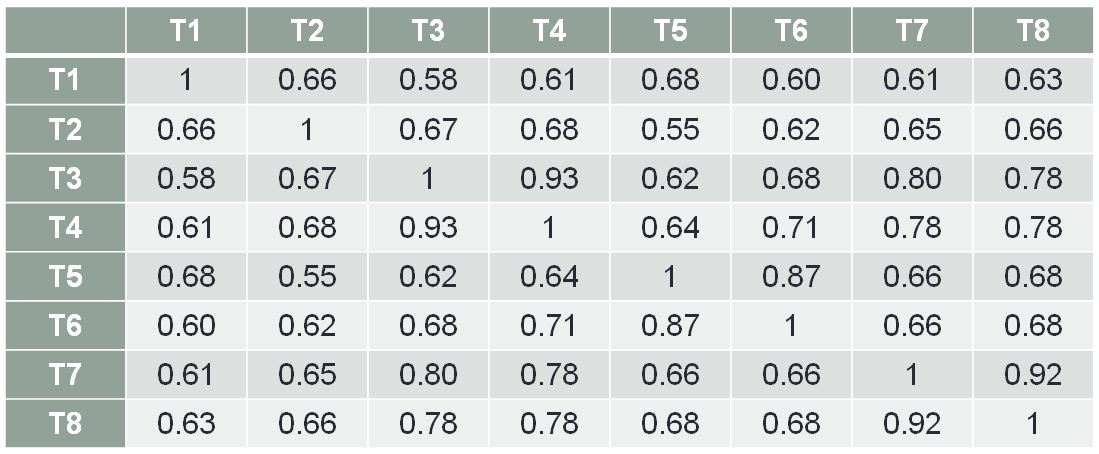}
\caption[Times Correlation for Fluids Inputs]{The correlations between 8 time points for the fluids inputs variable.}
\label{fig:corr4}
\end{table}
\begin{table}[htbp]
\centering
\includegraphics[width=\textwidth]{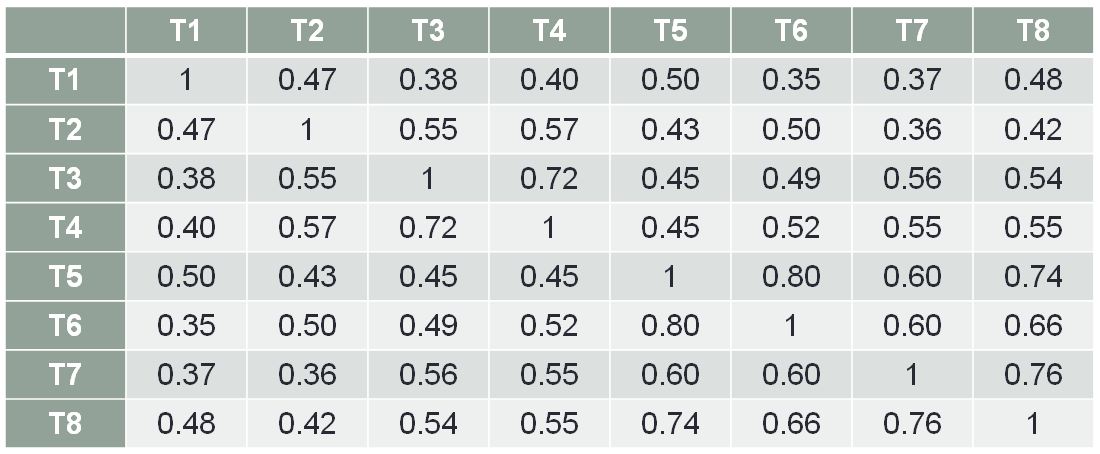}
\caption[Times Correlation for Fluids Outputs]{The correlations between 8 time points for the fluids outputs variable.}
\label{fig:corr5}
\end{table}
\begin{table}[htbp]
\centering
\includegraphics[width=\textwidth]{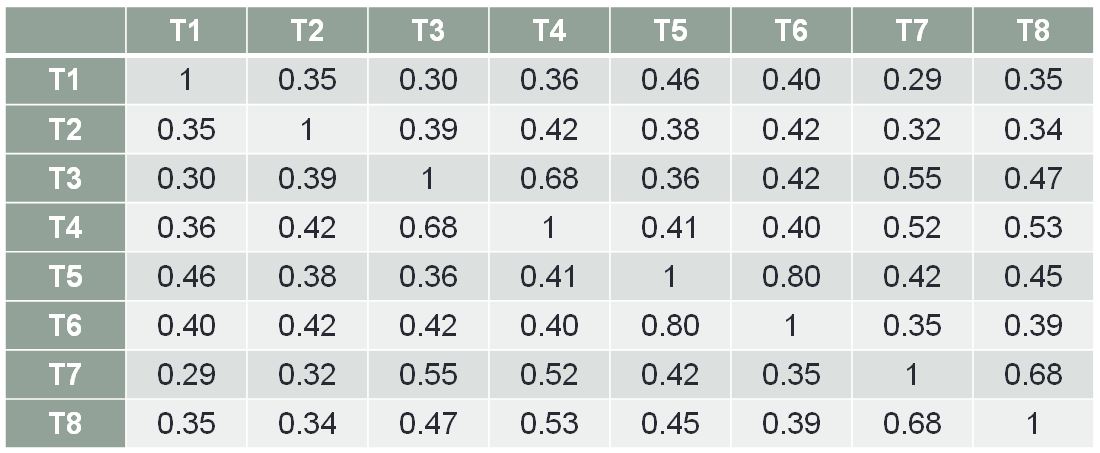}
\caption[Times Correlation for Fluids Balance]{The correlations between 8 time points for the fluids balance variable.}
\label{fig:corr6}
\end{table}
\begin{table}[htbp]
\centering
\includegraphics[width=\textwidth]{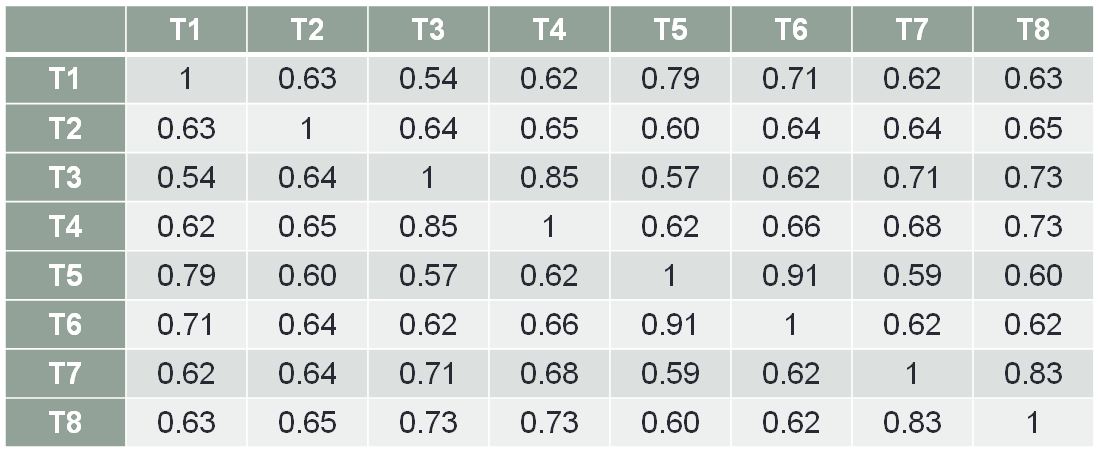}
\caption[Times Correlation for Vasopressors]{The correlations between 8 time points for the vasopressors amounts variable.}
\label{fig:corr7}
\end{table}
\begin{table}[htbp]
\centering
\includegraphics[width=\textwidth]{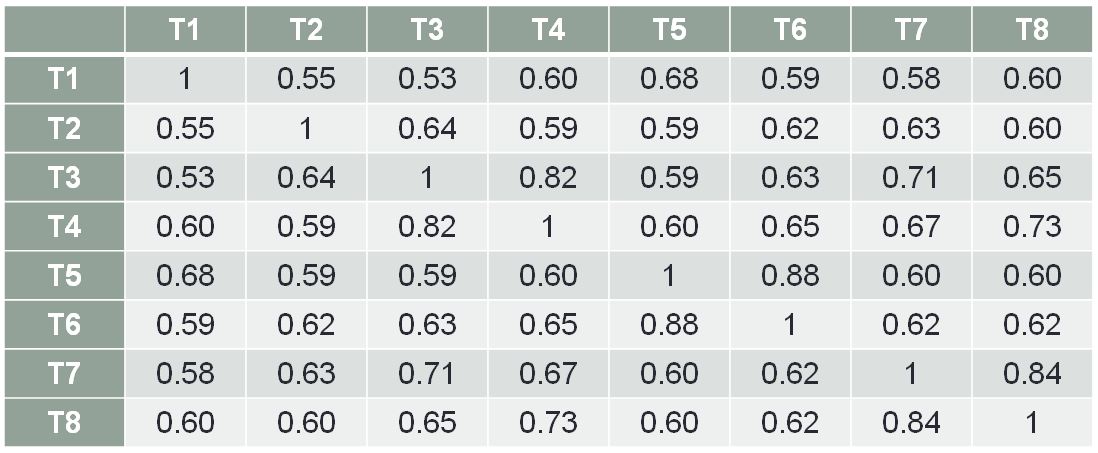}
\caption[Times Correlation for Blood Pressure]{The correlations between 8 time points for the blood pressure variable.}
\label{fig:corr8}
\end{table}

\clearpage

\section{Dataset Correlations}
\label{apx:dataCorrs}
Tables~\vref{fig:corr9} and~\vref{fig:corr10} show an overview of the correlations between the variables of the study at an earlier point at time. Table~\vref{tab:corrAbb} provide full description of the abbreviations. This shows that all the parameters are sensible and that the weight of the possible outliers present in the database has been mitigated. In red are the correletion values which are significant.
\begin{table}[htbp]
\centering
\includegraphics[width=\textwidth]{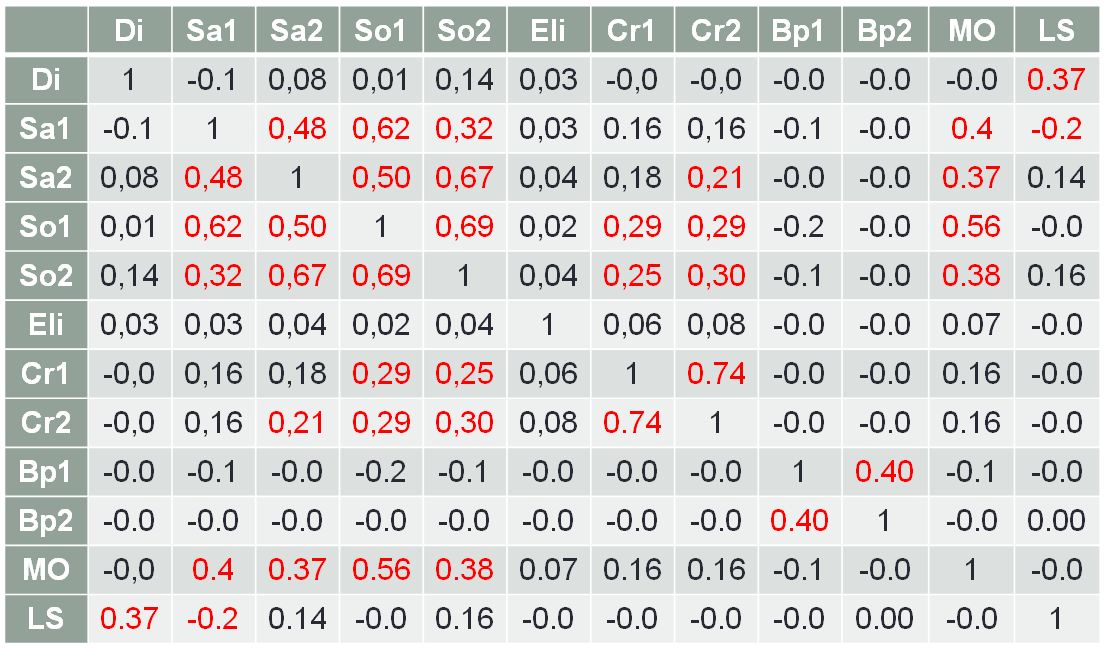}
\caption[Correlations in Automatic Dataset - First Part]{Correlations between a subset of the variables of the final dataset. In red the values with high correlation.}
\label{fig:corr9}
\end{table}
\begin{table}[htbp]
\centering
\includegraphics[width=\textwidth]{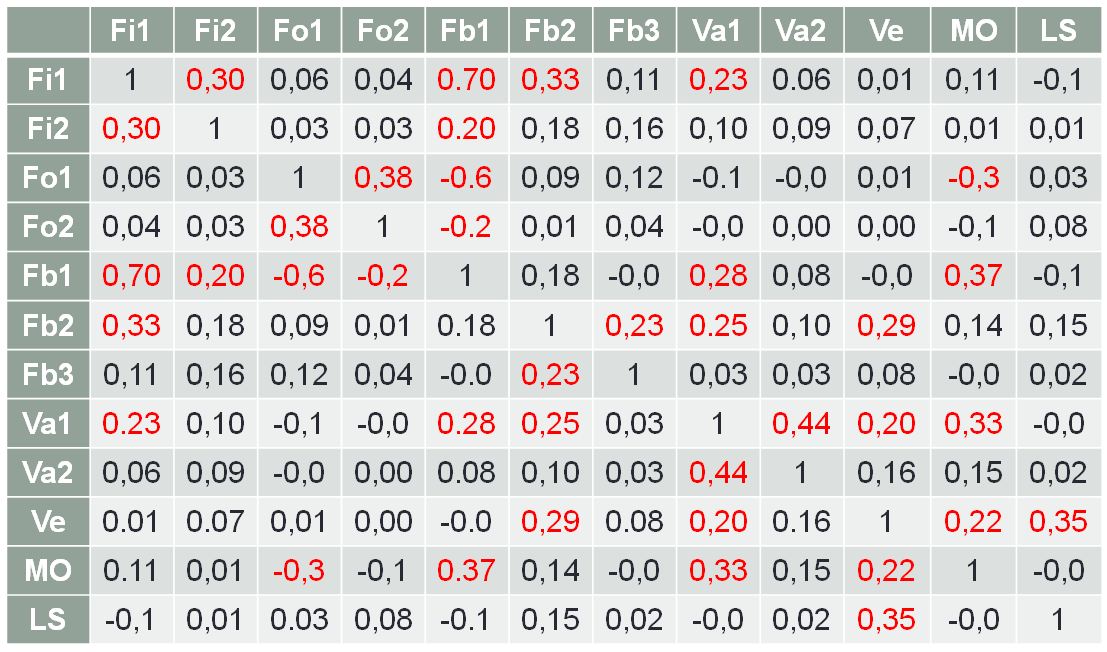}
\caption[Correlations in Automatic Dataset - Second Part]{Correlations between a subset of the variables of the final dataset. In red the values with high correlation.}
\label{fig:corr10}
\end{table}
\begin{table}
\centering
\begin{tabular}{|c|c|}
\hline {\bf Abbreviation} & Description\\
\hline Di & Abministration of Diuretics\\
\hline Sa1 & SAPS at the first day in ICU\\
\hline Sa2 & SAPS sum during the stay in ICU\\
\hline So1 & SOFA at the first day in ICU\\
\hline So2 & SOFA sum during the stay in ICU\\
\hline Eli & Elixahuser Score\\
\hline Cr1 & Creatinine at the first day in ICU\\
\hline Cr2 & Creatinine sum during the stay in ICU\\
\hline Bp1 & Arterial Blood Pressure at the first day in ICU\\
\hline Bp2 & Arterial Blood Pressure sum during the stay in ICU\\
\hline Fi1 & Fluids Inputs at the first day in ICU\\
\hline Fi2 & Fluids Inputs sum during the stay in ICU\\
\hline Fo1 & Fluids Outputs at the first day in ICU\\
\hline Fo2 & Fluids Outputs sum during the stay in ICU\\
\hline Fb1 & Balance $Input-Outputs$ sum during the stay in ICU\\
\hline Fb2 & Balance sum from \textit{TotalBalanceEvents}\\
\hline Fb3 & Balance at the first day in ICU from \textit{TotalBalanceEvents}\\
\hline Va1 & Amount of Vasopressors at the first day in ICU\\
\hline Va2 & Amount of Vasopressors sum during the stay in ICU\\
\hline Ve & Mechanical Ventilation\\
\hline MO & \Mort\\
\hline LS & \LOS\\
\hline
\end{tabular}
\caption[Abbrevation in the Correlations Tables]{Abbrevation used in the correlations tables.}
\label{tab:corrAbb}
\end{table}

\clearpage

\section{Experts Datasets}
\label{apx:expertsDatasets}
Except for the datasets described in Chapter~\ref{cpt:propensityAnalysis}, the \pa was performed on two more lists provided by medical experts\footnote{The work has been performed with the support of medical experts, including Dr. Leo Celi, MD Critical Care Physician - Boston, MA and John Danziger, MD Department of Medicine, Division of Nephrology, Beth Israel Deaconess Medical Center - Boston, MA.}. In this Section the results on this two datasets will be presented.

List of the variables chosen by the doctors follow:
\begin{itemize}
\item {\bf Experts list 1:} Chosen variables:
\begin{enumerate}
\item Age when admitted in the ICU ($x_2$)
\item Race (white vs not white) ($x_4$)
\item Elixhauser overall ($x_15$)
\item Elixhauser binary (selected 9 fields) ($x_16 \to x24$)
\item Creatinine mean of values during the first day ($x_26$)
\item Fluids inputs sum of values during the first day ($x_31$)
\item Fluids outputs sum of values during the first day ($x_36$)
\item Fluids balance sum of values during the first day ($x_41$)
\item Use of vasopressors in the ICU ($x_45$)
\item Mechanical ventilation in the ICU ($x_46$)
\item Arterial bp mean of values during the first day ($x_48$)
\end{enumerate}
\item {\bf Experts list 2:} Chosen variables:
\begin{enumerate}
\item Age when admitted in the ICU ($x_2$)
\item Race (white vs not white) ($x_4$)
\item Elixhauser overall ($x_15$)
\item Elixhauser binary (selected 9 fields) ($x_16 \to x24$)
\item Creatinine mean of values during the first day ($x_26$)
\item Fluids inputs sum of values during the first day ($x_31$)
\item Fluids outputs sum of values during the first day ($x_36$)
\item Fluids balance sum of values during the first day ($x_41$)
\item Use of vasopressors in the ICU ($x_45$)
\item Mechanical ventilation in the ICU ($x_46$)
\item Arterial bp mean of values during the first day ($x_48$)
\item Creatinine mean of values during day T1 ($x_27$)
\item Fluids inputs sum of values during day T1 ($x_32$)
\item Fluids outputs sum of values during day T1 ($x_37$)
\item Fluids balance sum of values during day T1 ($x_42$)
\end{enumerate}
\end{itemize}

In Figures~\vref{fig:F-test3-4} and~\vref{fig:F-test5-6} the improvements in the balance after subclassifiction with the list of variables provided above are shown:
\begin{figure}
\centering
\rotatebox{90}{
\stepcounter{figure}
\setcounter{subfigure}{0}
\begin{minipage}{\textheight}
\begin{minipage}{0.45\textheight}
\centering
\includegraphics[width=\textwidth]{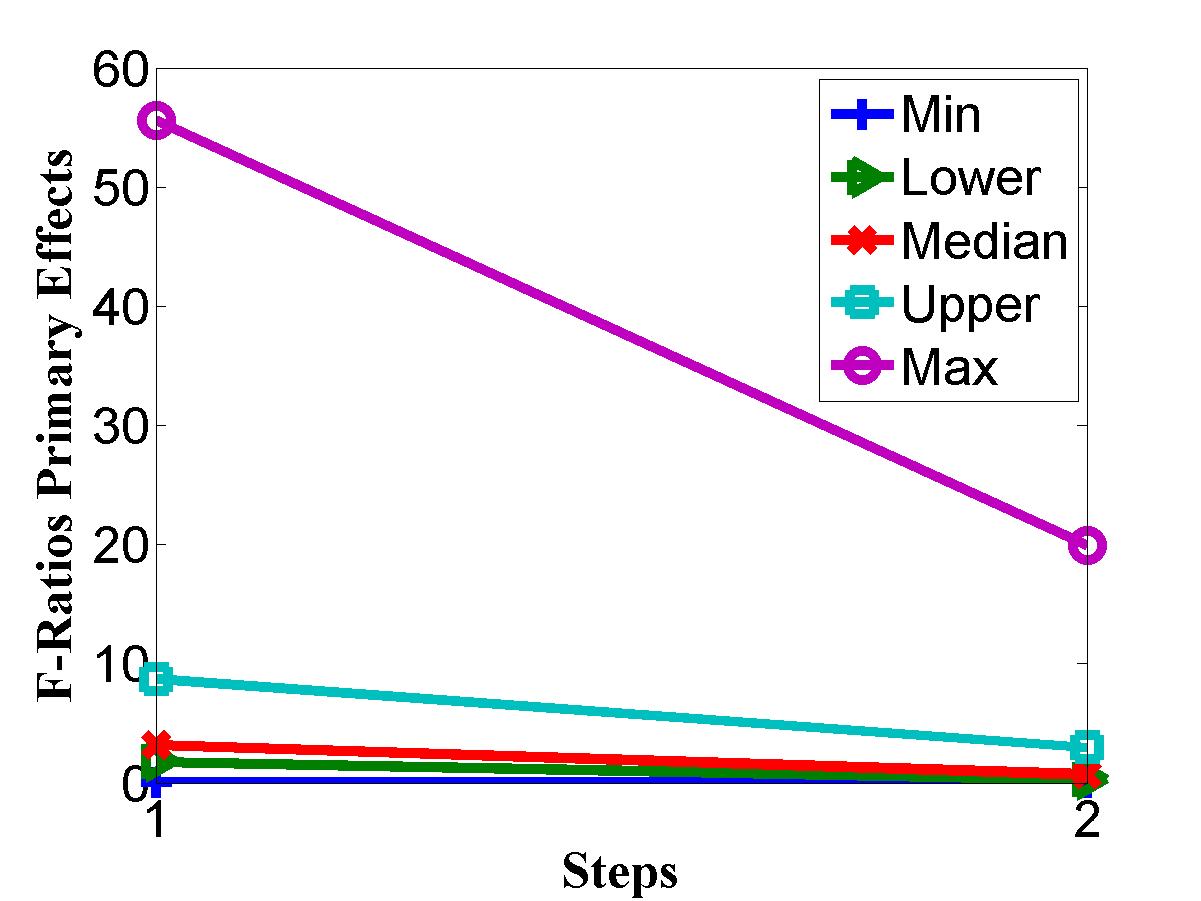}
\captionof{subfigure}[]{Primary Effects.}
\label{fig:F-test3}
\end{minipage}
\begin{minipage}{0.45\textheight}
\centering
\includegraphics[width=\textwidth]{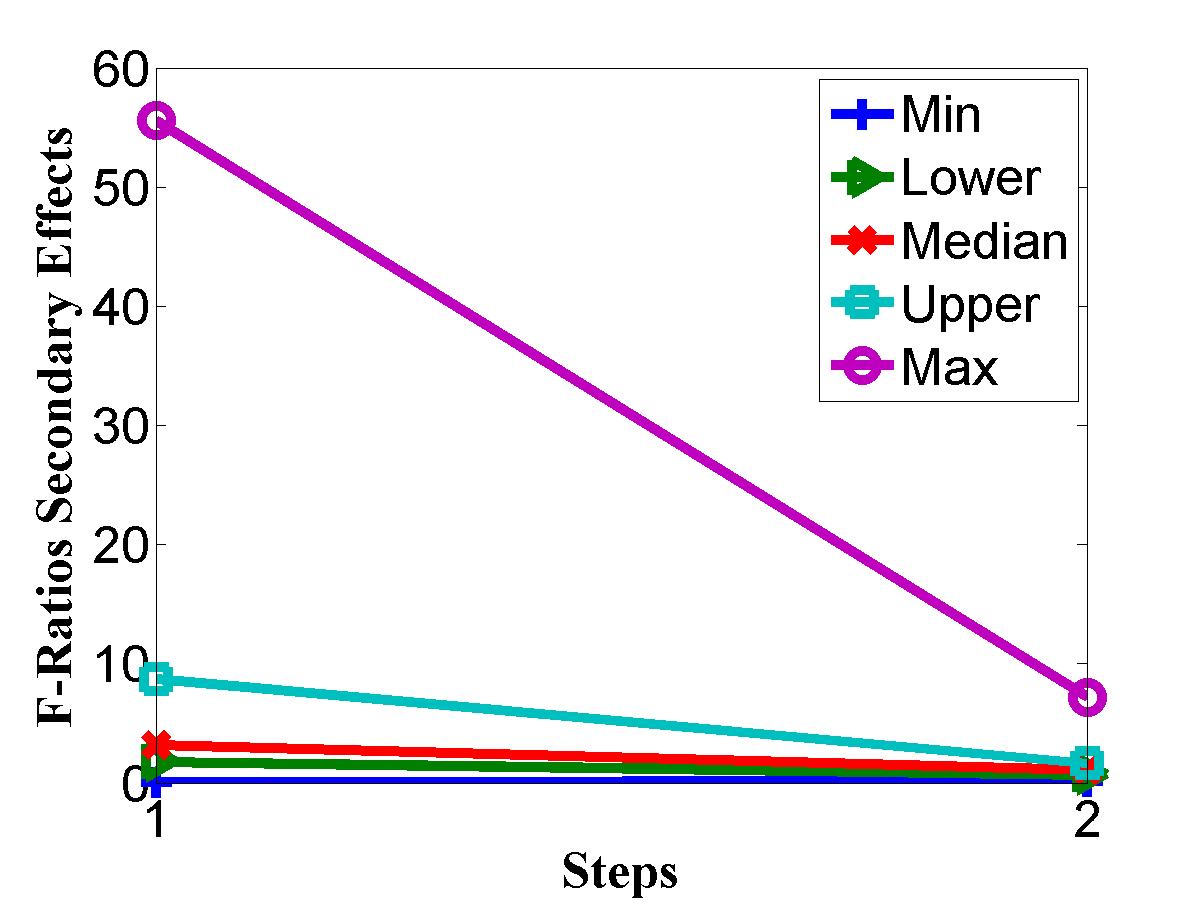}
\captionof{subfigure}[]{Secondary Effects}
\label{fig:F-test4}
\end{minipage}
\addtocounter{figure}{-1}
\captionof{figure}[F-Ratios on Experts List 1 Dataset]{The F-statistics on the experts list 1 dataset on the primary and secondary effects. The values in abscissa 1 refers to the balance on the original dataset, while on abscissa 2 there are the balance after the propensity score method.}
\label{fig:F-test3-4}
\end{minipage}
}
\end{figure}
\begin{figure}
\centering
\rotatebox{90}{
\stepcounter{figure}
\setcounter{subfigure}{0}
\begin{minipage}{\textheight}
\begin{minipage}{0.45\textheight}
\centering
\includegraphics[width=\textwidth]{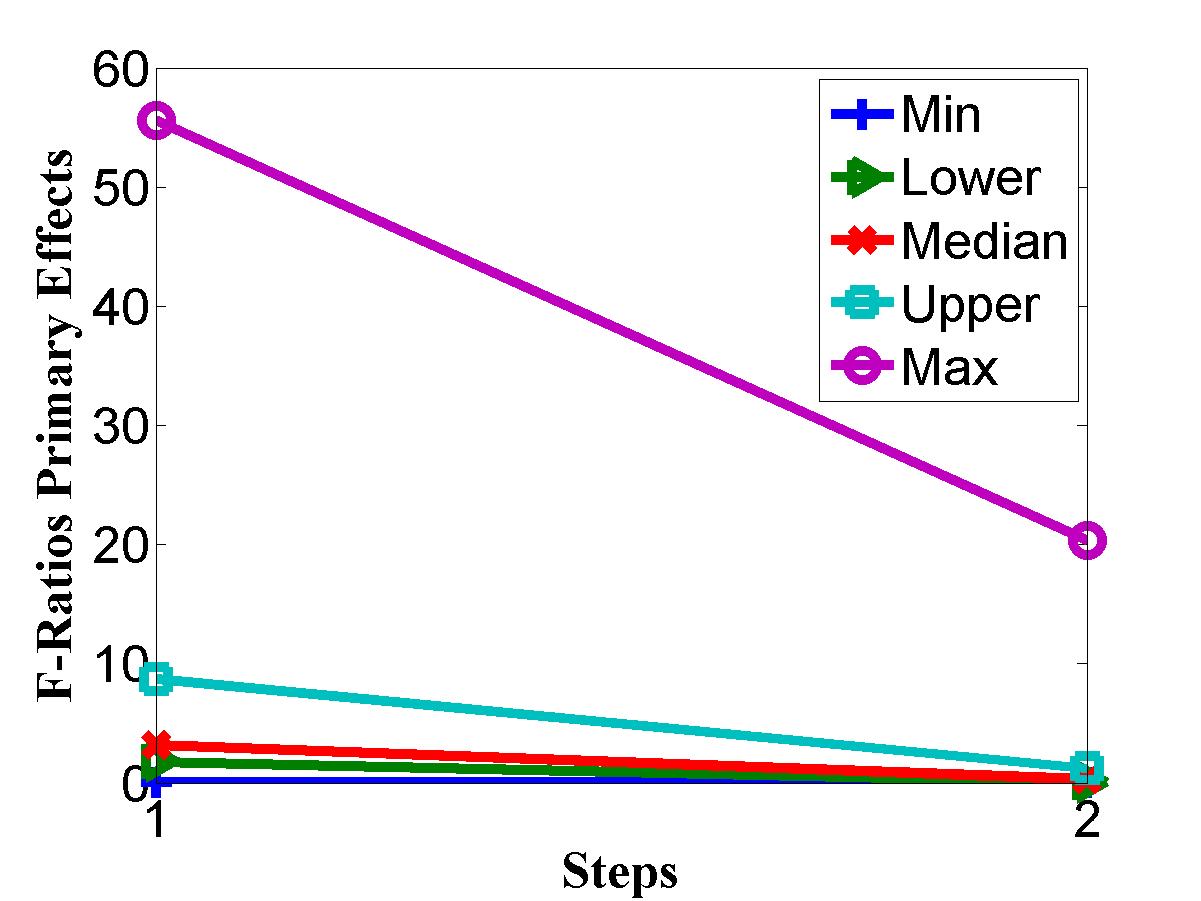}
\captionof{subfigure}[]{Primary Effects.}
\label{fig:F-test5}
\end{minipage}
\begin{minipage}{0.45\textheight}
\centering
\includegraphics[width=\textwidth]{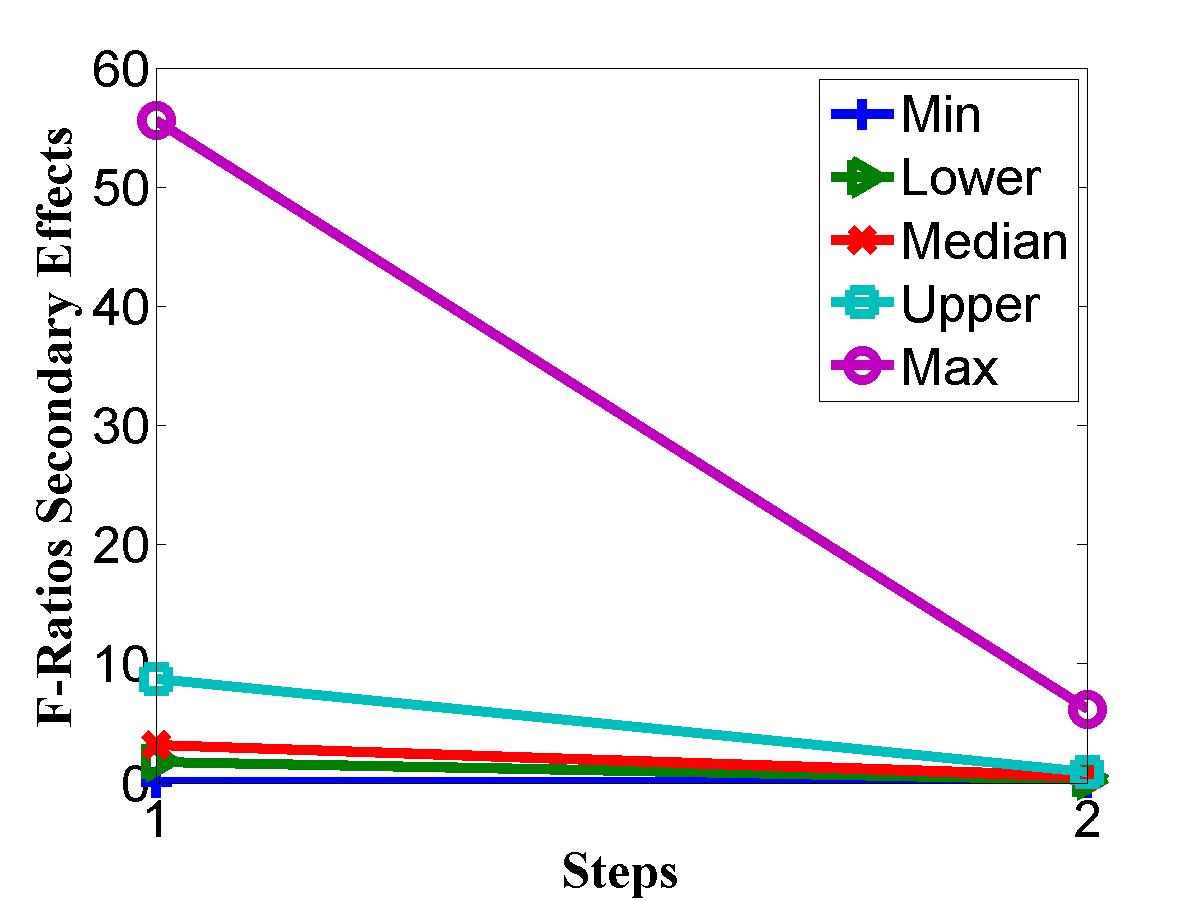}
\captionof{subfigure}[]{Secondary Effects}
\label{fig:F-test6}
\end{minipage}
\addtocounter{figure}{-1}
\captionof{figure}[F-Ratios on Experts List 1 Dataset]{The F-statistics on the experts list 2 dataset on the primary and secondary effects. The values in abscissa 1 refers to the balance on the original dataset, while on abscissa 2 there are the balance after the propensity score method.}
\label{fig:F-test5-6}
\end{minipage}
}
\end{figure}

Finally, a comparison of the results between these 2 new datasets is now provided:
\begin{itemize}
\item {\bf Experts list 1:} These \groups were created with the help of medical experts. They seems to be more balanced then the ones created with the stepwise method, because the propensity number calculated in this way was less predictive. The results for this dataset are shown in Table~\vref{tab:list1-dataset}.

\Group~1 is unbalanced and it should not be considered for the analysis. The length of stay in ICU appears to be longer for the patients who got diuretics in this case too, while the chances of survival are better for patients who did not get diuretics belonging to \group~3 and for patients who got diuretics belonging to \group~5.
\begin{table}
\centering
\begin{tabular}{|c|c|c|}
\hline {\bf \Group~1 $PS \in \left[ 0.00;0.02 \right] $} & Diuretics given & Diuretics not given\\
\hline Number of patients & 4 & 300\\
\hline Deaths & 0\% & 26\%\\
\hline Mean length of stay & 4 days & 2.3 days\\
\hline {\bf \Group~2 $PS \in \left[ 0.02;0.08 \right] $} & Diuretics given & Diuretics not given\\
\hline Number of patients & 18 & 286\\
\hline Deaths & 44\% & 26\%\\
\hline Mean length of stay & 12 days & 5 days\\
\hline {\bf \Group~3 $PS \in \left[ 0.08;0.13 \right] $} & Diuretics given & Diuretics not given\\
\hline Number of patients & 34 & 270\\
\hline Deaths & 41\% & 48\%\\
\hline Mean length of stay & 10.9 days & 7 days\\
\hline {\bf \Group~4 $PS \in \left[ 0.13;0.19 \right] $} & Diuretics given & Diuretics not given\\
\hline Number of patients & 46 & 258\\
\hline Deaths & 39\% & 43\%\\
\hline Mean length of stay & 12.4 days & 9.9 days\\
\hline {\bf \Group~5 $PS \in \left[ 0.19;0.65 \right] $} & Diuretics given & Diuretics not given\\
\hline Number of patients & 86 & 218\\
\hline Deaths & 34\% & 47\%\\
\hline Mean length of stay & 19.6 days & 8.2 days\\
\hline
\end{tabular}
\caption[Experts List 1 Dataset]{Results on experts list 1 dataset.}
\label{tab:list1-dataset}
\end{table}
\item {\bf Experts list 2:} These \groups were created with the help of medical experts too and they are similar to, even though not the same, the previous one. In fact in this case the accurancy of the model seems to be improved.
\begin{table}
\centering
\begin{tabular}{|c|c|c|}
\hline {\bf \Group~1 $PS \in \left[ 0.00;0.02 \right] $} & Diuretics given & Diuretics not given\\
\hline Number of patients & 2 & 302\\
\hline Deaths & 0\% & 29\%\\
\hline Mean length of stay & 3.5 days & 2.2 days\\
\hline {\bf \Group~2 $PS \in \left[ 0.02;0.06 \right] $} & Diuretics given & Diuretics not given\\
\hline Number of patients & 15 & 289\\
\hline Deaths & 26\% & 32\%\\
\hline Mean length of stay & 14.2 days & 5 days\\
\hline {\bf \Group~3 $PS \in \left[ 0.06;0.12 \right] $} & Diuretics given & Diuretics not given\\
\hline Number of patients & 28 & 276\\
\hline Deaths & 42\% & 45\%\\
\hline Mean length of stay & 12.4 days & 7.2 days\\
\hline {\bf \Group~4 $PS \in \left[ 0.12;0.19 \right] $} & Diuretics given & Diuretics not given\\
\hline Number of patients & 46 & 258\\
\hline Deaths & 32\% & 46\%\\
\hline Mean length of stay & 11.5 days & 8.9 days\\
\hline {\bf \Group~5 $PS \in \left[ 0.19;0.77 \right] $} & Diuretics given & Diuretics not given\\
\hline Number of patients & 97 & 207\\
\hline Deaths & 42\% & 37\%\\
\hline Mean length of stay & 18.2 days & 8.3 days\\
\hline
\end{tabular}
\caption[Experts List 2 Dataset]{Results on experts list 2 dataset.}
\label{tab:list2-dataset}
\end{table}
\end{itemize}

\clearpage \mbox{} \clearpage 
\chapter{Statistical Methods}
\label{apx:stats}

In this Appendix will be provided some details on the used statistical methods.

\section{Basic Stuff on Calculating a Propensity Score}
\label{apx:basic}
It is essential to realize that the outcome variable is not used in this step. 

\subsection{Using the Propensity Score}
\label{apx:usingPS}
There are different ways of using the propensity score even if, regardless of the technique, it is always calculated in the same way. However, once the propensity score is calculated, its application is different, and this will now be described. The following contents are drawn from\cite{PS-SURGEON}.

The three most common analytical techniques based on the propensity score are matching, stratification and regression adjustment. In the study described in Chapter~\ref{cpt:propensityAnalysis}, the stratification approach have been followed.

\begin{itemize}
\item\nibf{Matching by Propensity Scores:} Matching is a technique for adjusting baseline characteristics. Control subjects are matched with treatment subjects on important baseline characteristics, which need to be controlled for (potential confounders). However, an important drawback is the difficulty in finding close matches for all important confounders. The more confounders that require matching, the harder it is to find suitable patients in each group, with a corresponding reduction in sample size. As already noted, propensity scoring summarizes all measured confounders in a single score. So, using the propensity score requires the matching of only one (composite) factor and offers greater ease of use. This is one of the great advantages of this statistical technique.
\item\nibf{Stratification by Propensity Scores:} Stratification is another commonly used technique in non-randomized observational studies to control for measured differences in baseline characteristics. Patients are first grouped in strata determined by their propensity score and then treated and control patients in the same strata are compared directly. Similarly to matching, difficulty arises when the number of baseline characteristics increases.
\item\nibf{Regression Adjustment based on Propensity Scores:} Propensity scores can also be used in a regression adjustment. Recall, the propensity score is obtained by using a logistic regression model, with exposure to treatment as the dependent variable and all baseline characteristics as independent variables. In regression adjustment, the propensity score is used as the only confounding variable in association with the exposure to treatment (the primary predictor variable) to estimate the effect of the exposure on the outcome.
\end{itemize}

In the analysis that have been conducted in Chapter~\ref{cpt:propensityAnalysis}, the propensity score was used by including it out logistic regression models. However, it was \textit{NOT} the only confounding variable. Variables related to the \illness of the patient have added because was suspected that they were also effecting \mort or effect \los .

\clearpage

\section{Linear Regression}
\label{apx:regressions}
Linear regression is an approach to modelling the relationship between a scalar dependent variable  \textbf{y} and one or more explanatory variables denoted \textbf{X}.

The model of linear regression is:
\begin{equation}
\bar{Y_i} = \beta_0 + \bar{\beta_i} \cdot \bar{X_i} + \mu_i.
\end{equation}
where:\\
$i \in \left[ 1,n \right]$;\\
$\bar{Y_i}$ is the dependent variable;\\
$\bar{X_i}$ is the independent variable;\\
$\beta_0 + \bar{\beta_i} \cdot \bar{X_i}$ is the regression function;\\
$\beta_0$ is the y-intercept of the regression function;\\
$\bar{\beta_i}$ is the slope of the regression function;\\
$\mu_i$ is the statistical error.\\

In linear regression, data are modelled using linear predictor functions, and unknown model parameters are estimated from the data. Such models are called linear models. Like all forms of regression analysis, linear regression focuses on the conditional probability distribution of \textbf{y} given \textbf{X}.

\subsection{Generalized Linear Model}
\label{apx:regressionsGeneralized}
The generalized linear model generalizes linear regression by allowing the linear model to be related to the response variable via a \textbf{link function} and by allowing the magnitude of the variance of each measurement to be a function of its predicted value.

\subsection{Logistic Regression}
\label{apx:regressionsLogistic}
Logistic regression is a special case of generalized linear model with link function as logit function.
\begin{equation}
e(x)=  \frac{e(y)}{1 - e(y)} = \alpha + \beta ^Tf(x),
\end{equation}
It is a regression model applied in cases where the dependent variable \textbf{y} is a dichotomous attributable to the values ​​0 and 1.

\section{Medical Studies: \Pvalue{s} and Statistical Significance}
\label{apx:pVal}
Null Hypothesis:  The independent variable is responsible for random effects (rather than actual difference) in outcome. Whether the difference in outcome is just pure chance based on the effect of this variable. To answer, \Pvalue derivation considers independent variable's coefficient. \textbf{The only thing can be said analyzing the \Pvalue is that, when repeating the experiment, in the 97\% of the cases  a smaller difference between the groups than in the observed ones would be observed, while in the remaining 3\% the difference would be greater.}

In this Section the use of \Pvalue in this context will be described. The following contents are drawn from\cite{PVALUES}

\subsection{Null and Alternative Hypothesis}
\label{apx:nullHyp}
The statistical and probabilistic formalization of medical studies is based on the formulation of a hypothesis to be tested on the basis of collected data. The null hypothesis is that the studied treatment produces absence of effect to the patients or more generally that there is absence of difference between the two treated and untreated patients. The alternative to the null hypothesis (which defines what is expected to be true if the null hypothesis is false), that is that there is a difference between the two groups of patients.

In this context, the statistical and probabilistic formalization of medical studies aim at statistically defining if a given treatment is making or not the difference between the treated and untreated patients in the collected data. Once the data have been collected, their consistency with the null hypothesis will be measured. More precisely, will be determined which of the two hypotheses is statistically more plausible.

The \pvalue can be used to analyze the importance of a variable in a model. Being in fact the null hypothesis that the inclusion of a given variable $x_i$ in the model does not make a significant contribution, as discussed above a small \pvalue goes along with the rejection of this hypothesis. Hence, a small \pvalue , in the already discussed ranges, could indicate if a given variable $x_i$ is important or not for the outcome.

\subsection{Parametric and Non-Parametrics Hypothesis Tests}
\label{apx:tests}
A statistical test is parametric if assumes that the data has come from a type of probability distribution and makes inferences about the parameters of the distribution. In a non-parametric test, instead, the data are not assumed to come from a given distribution.

In the analysis made on the diuretics problem, mortality was studied with the Chi-squared test, a non-parametric test usable for binary (dichotomous) variables, while length of stay was studied with T test, a parametric test usable for continuous variables.

\subsection{Hypothesis Test with \Pvalue}
\label{apx:pvalueInfo}

\subsubsection{How a \Pvalue is calculated}
The \Pvalue is the probability of obtaining a test statistic at least as extreme as the one that was actually observed, assuming that the null hypothesis is true. Being $\bar{X}$ the expected value, $\bar{Z}$ the observed value and $H_0$ the null hypotesis, the \Pvalue is:
\begin{equation}
P(\bar{X}>\bar{Z} \mid H_0)
\end{equation}

\subsubsection{Interpretation of a \Pvalue}
\label{apx:interpretingPVAL}
In this context the \Pvalue is defined as the probability that quantifies the strength of evidence expressed by the observed data against the null hypothesis and in favor of the alternative one. In other words, the \Pvalue is a probability that expresses whether it is more plausible that the observed data come from the null hypothesis or the alternative.

A big \Pvalue, more than 0.05, defines that the results on the two compared groups of patients are likely following the same probability distributions and that it is more likely that any obtained difference in the results on these groups is caused by random effects rather then by actual differences. On the contrary, a small \Pvalue, less then 0.05, rejects the null hypothesis, and this means that the differences in the results between the treated and untreated patients are likely to be due to actual differences between the outcomes in the two groups.

\begin{figure}[htbp]
\centering
\includegraphics[width=\textwidth]{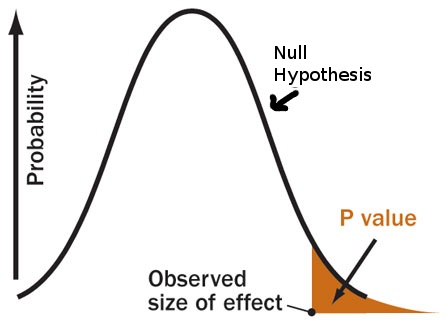}
\caption[\Pvalue and Null Hypothesis]{A \pvalue can be used to deduce the likelihood of the null hypothesis.}
\label{fig:pvalue_null}
\end{figure}

Figure~\vref{fig:pvalue_null} shows a graphical visualization of how a \Pvalue can indicate the probability of the null hypothesis.

The \pvalue is a probability, i.e. between 0 to 1. The levels of significance of the \pvalue follow:
\begin{itemize}
\item {\bf \pvalue $> 0.1$:} absence of evidence against the null hypothesis.
\item {\bf \pvalue $\in \left( 0.05;0.1 \right]$:} weak evidence against the null hypothesis.
\item {\bf \pvalue $\in \left( 0.01;0.05 \right]$:} moderate evidence against the null hypothesis.
\item {\bf \pvalue $\in \left[ 0.001;0.01 \right]$:} strong evidence against the null hypothesis.
\item {\bf \pvalue $< 0.001$:} very strong evidence against the null hypothesis.
\end{itemize}

\subsubsection{Clarification in the Interpretation of the \Pvalue}
\label{apx:clarificationPS}
The significance of the \pvalue is often misunderstood. A \pvalue equal to 0.03 expresses that there is a 3\% probability of observing a difference equal to the one observed in the data even if the means of the two populations are identical, that is even if the null hypothesis is true. One might be tempted to say that there is a probability of the 97\% that the observed difference reflects a real difference between the populations and a 3\% probability that the difference is due to chance. However, this conclusion would be incorrect: \textbf{the only thing you can say analyzing the \pvalue is that, by repeating the experiment would be observed in the 97\% of the cases a smaller difference between the groups than in the observed ones, while in the remaining 3\% the difference would be greater.}

Often the \pvalue is interpreted, wrongly, as the probability that the null hypothesis is true. It is necessary to clarify that a small \pvalue does not mean that the probability that the null hypothesis is true is lower, but only that it is more reasonable that the observed data were generated under the alternative hypothesis.

\nibf{Bottom Line:} \textbf{The only thing you can say analyzing the \Pvalue is that, when repeating the experiment, in the 97\% of the cases  a smaller difference between the groups than in the observed ones would be observed, while in the remaining 3\% the difference would be greater.}

\clearpage \mbox{} \clearpage 
\chapter{Machine Learning}
\label{apx:machineLearning}

The focus of this Appendix is to describe some methodologies used in the context of problems similar to the one analyzed in this work. Most of the contents are drawn from related papers.

In the first part of the Chapter, will be presented a series of concepts related to knowledge discoverty in clinical databases and further on discribed the methodologies used in similar works.

In the second part of the Chapter, will be provided a background in the context of machine learning and evolutionary computation. In particular the focus will be on describing the Genetic Programming (GP) methodology.

\section{Knowledge Discovery}
\label{apx:kd}
In this Section Knowledge discovery will be defined. The contents are drawn from\cite{KDD}.
 
Knowledge discovery is a concept that describes the process of automatically searching large volumes of data for patterns that can be considered knowledge about the data. It is often described as deriving knowledge from the input data. This complex topic can be categorized according to 1) what kind of data is searched and 2) in what form is the result of the search represented. Knowledge discovery developed out of the Data mining domain, and is closely related to it both in terms of methodology and terminology.

Knowledge discovery is the nontrivial extraction of implicit, previously unknown, and potentially useful information from data. Given a set of facts (data) $F$, a language $L$, and some measure of certainty $C$, a pattern is defined as a statement $S$ in $L$ that describes relationships among a subset $F_S$ of $F$ with a certainty $c$, such that $S$ is simpler (in some sense) than the enumeration of all facts in $F_S$. A pattern that is interesting (according to a user-imposed interest measure) and certain enough (again according to the user’s criteria) is called knowledge. The output of a program that monitors the set of facts in a database and produces patterns in this sense is discovered knowledge\cite{KDD}.
\begin{figure}[htbp]
\centering
\includegraphics[width=\textwidth]{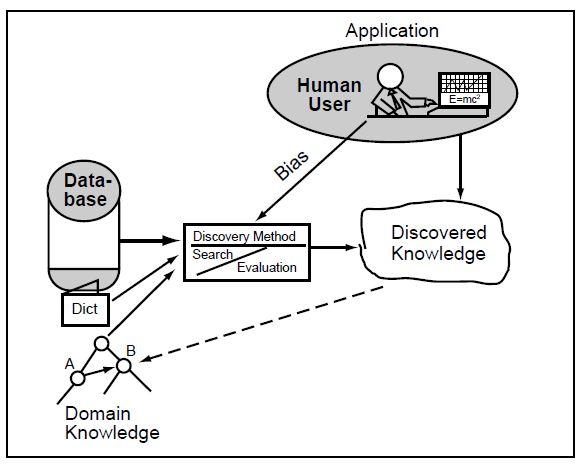}
\caption[Framework for Knowledge Discovery in Databases]{A Framework for Knowledge Discovery in Databases.}
\label{fig:kdd-framework}
\end{figure}
The most well-known branch of data mining is knowledge discovery, also known as Knowledge Discovery in Databases (KDD). Just as many other forms of knowledge discovery it creates abstractions of the input data. The knowledge obtained through the process may become additional data that can be used for further usage and discovery.

Although machine learning is the foundation for much of the work in this area, knowledge discovery in databases deals with issues relevant to several other fields, including database management, expert systems, statistical analysis, and scientific discovery.
\begin{itemize}
\item {\bf Database Management:} provides procedures for storing, accessing, and modifying the data. Typical operations include retrieval, update, or deletion of all tuples satisfying a specific condition, and maintaining user-specified integrity constraints. The ability to extract tuples satisfying a common condition is like discovery in its ability to produce interesting and useful statements (for example, “Bob and Dave sold fewer widgets this year than last”). These techniques, however, cannot by themselves determine what computations are worth trying, nor do they evaluate the quality of the derived patterns. Interesting discoveries uncovered by these data-manipulation tools result from the guidance of the user. However, the new generation of deductive and objectoriented database systems (Kim, Nicolas, and Nishio 1990) will provide improved capabilities for intelligent data analysis and discovery.
\item {\bf Expert Systems:} attempt to capture knowledge pertinent to a specific problem. Techniquesexist for helping to extract knowledge from experts. One such method is the induction of rules from expertgenerated examples of problem solutions. This method differs from discovery in databases in that the expert examples are usually of much higher quality than the data in databases, and they usually cover only the important cases, for a comparison between knowledge acquisition from an expert and induction from data). Furthermore, experts are available to confirm the validity and usefulness of the discovered patterns. As with database management tools, the autonomy of discovery is lacking in these methods.
\item {\bf Statistics:} slthough they provide a solid theoretical foundation for the problem of data analysis, a purely statistical approach is not enough. First, standard statistical methods are ill suited for the nominal and structured data types found in many databases. Second, statistics are totally data driven, precluding the use of available domain knowledge, an important issue that will be discussed later. Third, the results of statistical analysis can be overwhelming and difficult to interpret. Finally, statistical methods require the guidance of the user to specify where and how to analyze the data. However, some recent statistics-based techniques such as projection
pursuit (Huber 1985) and discovery of causal structure from data (Glymour et al. 1987; Geiger, Paz, and Pearl 1990) address some of these problems and are much closer to intelligent data analysis. That methods using domain knowledge is expected to be developed by the statistical community. Statistics should have a vital role in all discovery systems dealing with large amounts of data.
\item {\bf Scientific Discovery:} discovery in databases is significantly different from scientific discovery in that the former is less purposeful and less controlled. Scientific data come from experiments designed to eliminate the effects of all but a few parameters and to emphasize the variation of one or a few target parameters to be explained. However, typical business databases record a plethora of information about their subjects to meet a number of organizational goals. This richness (or confusion) both captures and hides from view underlying relationships in the data. Moreover, scientists can reformulate and rerun their experiments should they find that the initial design was inadequate. Database managers rarely have the luxury of redesigning their data fields and recollecting the data\cite{KDD}.
\end{itemize}

\subsection{Knowledge Discovery in Databases}
\label{apx:kddb}
This and next Sections describe the use of database in the context of knowledge discovery. The contents are drawn from\cite{DM-KD} and \cite{KDD}. 

Historically, the notion of finding useful patterns in data has been given a variety of names, including data mining, knowledge extraction, information discovery, information harvesting, data archaeology, and data pattern processing. The term data mining has mostly been used by statisticians, data analysts, and the management information systems (MIS) communities. It has also gained popularity in the database field. The phrase knowledge discovery in databases was coined at the first KDD workshop in 1989 (Piatetsky-Shapiro 1991) to emphasize that knowledge is the end product of a data-driven discovery. It has been popularized in the AI and machine-learning fields.

KDD refers to the overall process of discovering useful knowledge from data, and data mining refers to a particular step in this process. Data mining is the application of specific algorithms for extracting patterns from data. The distinction between the KDD process and the data-mining step (within the process) is a central point of this article. The additional steps in the KDD process, such as data preparation, data selection, data cleaning, incorporation of appropriate prior knowledge, and proper interpretation of the results of mining, are essential to ensure that useful knowledge is derived from the data. Blind application of data-mining methods (rightly criticized as data dredging in the statistical literature) can be a dangerous activity, easily leading to the discovery of meaningless and invalid patterns\cite{DM-KD}.

\begin{figure}[htbp]
\centering
\includegraphics[width=\textwidth]{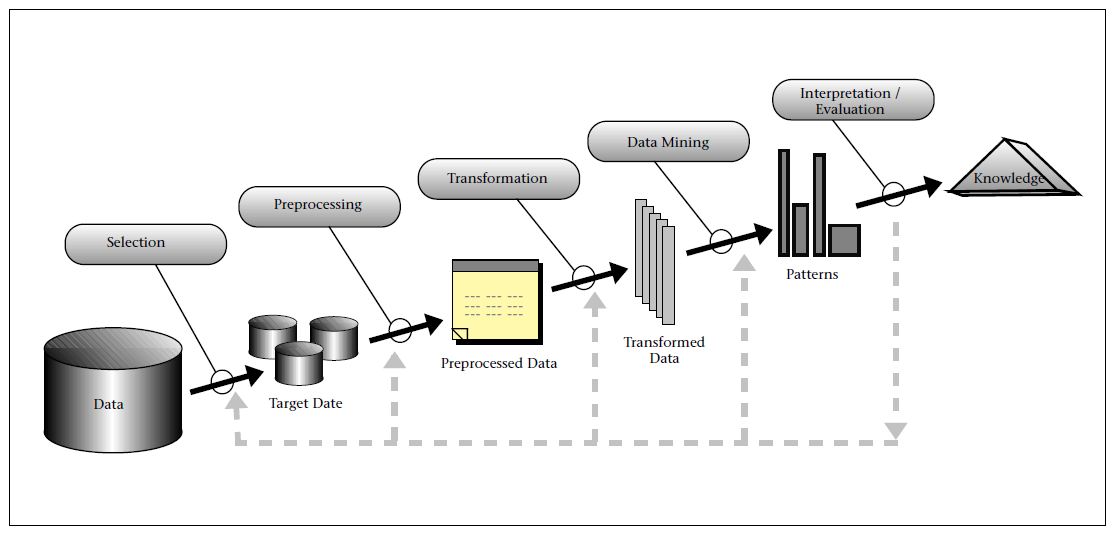}
\caption[Knowledge Discovery in Databases Process]{An Overview of the Steps That Compose the Knowledge Discovery in Databases Process.}
\label{fig:kdd-process}
\end{figure}

An example of Knowledge Discovery in Databases Process is shown in Figure~\vref{fig:kdd-process}.

\subsection{Complexity in Knowledge Discoverty}
\label{apx:complKD}
Discovery algorithms for large databases must deal with the issue of computational complexity. Algorithms with computational requirements that grow faster than a small polynomial in the number of records and fields are too inefficient for large databases.

Empirical methods are often overwhelmed by large quantities of data and potential patterns. The incorporation of domain knowledge can improve efficiency by narrowing the focus of the discovery process but at the risk of precluding unexpected but useful discovery. Data sampling is another way of attacking the problem of scale; it trades a degree of certainty for greater efficiency by limiting discovery to a subset of the database (see previous Section on uncertainty)\cite{KDD}.

\subsection{Clinical Decision Support Systems}
\label{apx:cdss}
This Section discuss the use of computer based techniques in clinical contexts. The contents are drawn from \cite{MDDS} and \cite{CDSS}.

Computerized Clinical decision support systems (CDSSs) are interactive decision support systems (DSS)\footnote{A decision support system (DSS) is a computer-based information system that supports business or organizational decision-making activities. DSSs serve the management, operations, and planning levels of an organization and help to make decisions, which may be rapidly changing and not easily specified in advance.} Computer Software, which are designed to assist physicians and other health professionals with decision making\footnote{Decision making can be regarded as the mental processes (cognitive process) resulting in the selection of a course of action among several alternative scenarios.} tasks, as determining diagnosis of patient data.

The goal of diagnosis is to place a nosologic\footnote{Nosology is a branch of medicine that deals with classification of diseases.} label on a process that manifests itself in a patient over time. However, diagnosis is a complex procedure more involved than producing a nosologic label for a set of patient descriptors. Efficient and ethical diagnostic evaluation requires a broad knowledge of people and of disease states. The nosologic labels used in diagnosis reflect the current level of scientific understanding of pathophysiology and disease, and may change over time without the patient or the patient’s illness per se changing.

The utility of making specific diagnoses lies in selection of effective therapies, in making accurate prognoses, and in providing detailed explanations. In some situations, it is not necessary to arrive at an exact diagnosis in order to fulfill one or more of these objectives. Treatment is often initiated before an esact diagnosis is made. Furthermore, the utility of making certain diagnoses is debatable. Labeling a patient as having “obesity” does not flatter the patient, and even worse, may cause the physician to do more harm than good.

In medical diagnostic reasoning, there are also cases where recognition from compiled knowledge does not pertain. Some cases present an overwhelming army of seemingly contradictory information; others present with common conditions in unexpected or unusual manners; some patients manifest rare findings or disorders. Unlike expert chess players who are no better than novices in reproducing random board positions from memory, medical experts have different modes of reasoning that can be invoked when simple pattern recognition based on experience fails. Medical diagnosticians in such settings attempt to reason from first principles, using their detailed knowledge of pathophysiologic processes, to construct scenarios under which an illness similar to the patient’s might occur\cite{MDDS}.

A CDSS allows to match characteristics of individual patients to a computerized knowledge base, and software algorithms generate patientspecific recommendations\cite{CDSS}.

There is currently widespread enthusiasm for introducing electronic medical records, computerized physician order entry systems, and CDSSs into hospitals and outpatient settings.

\section{Machine Learning}
\label{apx:MLDescr}
Learning is the process of knowledge acquisition in the absence of explicit programming. It can be seen as the process of construction of a program to run a job on the basis of information that do not provide an explicit description of the program itself.
\begin{figure}[htbp]
\centering
\includegraphics[width=\textwidth]{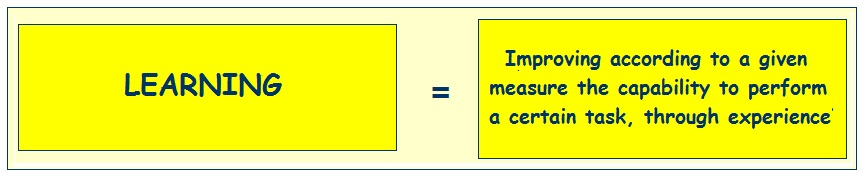}
\caption[Learning Process]{A computer program is said to learn from experience $E$ with respect to some class of tasks $T$ and performance measure $P$, if its performance at tasks in $T$, as measured by $P$, improves with experience $E$\cite{Learning-Mitchell}.}
\label{fig:ToLearn}
\end{figure}
Machine learning concerns with the design and development of algorithms that allow computers to evolve behaviors based on empirical data, such as from sensor data or databases. A learner can take advantage of examples (data) to capture characteristics of interest of their unknown underlying probability distribution. Data can be seen as examples that illustrate relations between observed variables. A major focus of machine learning research is to automatically learn to recognize complex patterns and make intelligent decisions based on data.

Tom Mitchell in \cite{ML-Mitchell} stated that
\begin{quote}
Machine Learning is a natural outgrowth of the intersection of Computer Science and Statistics. Could be said that the defining question of Computer Science is how machines that solve problems can be built, and which problems are inherently tractable/intractable? The question that largely defines Statistics is What can be inferred from data plus a set of modeling assumptions, with what reliability?

The defining question for Machine Learning builds on both, but it is a distinct question. Whereas Computer Science has focused primarily on how to manually program computers, Machine Learning focuses on the question of how to get computers to program themselves (from experience plus some initial structure).

Whereas Statistics has focused primarily on what conclusions can be inferred from data, Machine Learning incorporates additional questions about what computational architectures and algorithms can be used to most effectively capture, store, index, retrieve and merge these data, how multiple learning subtasks can be orchestrated in a larger system, and questions of computational tractability.
\end{quote}

Machine learning, knowledge discovery in databases (KDD) and data mining often employ the same methods and overlap strongly. Infact these fields work with similar basic assumptions: in machine learning, the performance is usually evaluated with respect to the ability to reproduce known knowledge, while in KDD the key task is the discovery of previously unknown knowledge. Evaluated with respect to known knowledge, an uninformed (unsupervised\footnote{Supervised learning is the machine learning task of inferring a function from supervised (labeled) training data. On the contrary unsupervised learning refers to the problem of trying to find hidden structure in unlabeled data.}) method will easily be outperformed by supervised methods, while in a typical KDD task, supervised methods cannot be used due to the unavailability of training data.

\subsection{Complexity of a Problem}
\label{apx:problemComplexity}
The gap between the development of hardware and software technology appears to be one of the biggest unsolved problems in Computer Science. Hardware speed and capabilty has inscreased exponentially during the last few years. Yet an adequate development of software production techniques does not correspond to such a quick and continuous improving in computer hardware performances.

Demand for computer code, more and more efficient and sophisticated, keeps growing in almost every field of industry, but the process of writing code still appears to be slow and obsolete: structured programming, object-oriented programming, and many other techniques allow, today, to write programs in a clean and friendly way, but still each single piece of code is handmade by a {\it 'craftsman'}, the programmer\cite{Vanneschi-Thesis}.

Hence the attempt to produce techniques that allow computers to learn.

In particular, machine learning methods are already the best methods available for developing particular types of software, in applications where:
\begin{itemize}
\item The application is too complex for people to manually design the algorithm. For example, software for sensor-base perception tasks, such as speech recognition and computer vision, fall into this category. All of us can easily label which photographs contain a picture of our mother, but none of us can write down an algorithm to perform this task. Here machine learning is the software development method of choice simply because it is relatively easy to collect labeled training data, and relatively ineffective to try writing down a successful algorithm.
\item The application requires that the software customize to its operational environment after it is fielded. One example of this is speech recognition systems that customize to the user who purchases the software. Machine learning here provides the mechanism for adaptation. Software applications that customize to users are growing rapidly - e.g., bookstores that customize to your purchasing preferences, or email readers that customize to your particular definition of spam. This machine learning niche within the software world is growing rapidly.
\end{itemize}
Viewed this way, machine learning methods play a key role in the world of computer science, within an important and growing niche. While there will remain software applications where machine learning may never be useful (e.g., to write matrix multiplication programs), the niche where it will be used is growing rapidly as applications grow in complexity, as the demand grows for self-customizing software, as computers gain access to more data, and as increasingly effective machine learning algorithms\cite{ML-Mitchell} are developed.

\subsection{Classification Problems}
\label{apx:ClassificationProblems}
Classification is the problem of identifying which of a set of categories (sub-populations) a new observation belongs, on the basis of a training set of data containing observations (or instances) whose category membership is known. The individual observations are analyzed into a set of quantifiable properties, known as various explanatory variables or features. These properties may variously be categorical (e.g. "A", "B", "AB" or "O", for blood type), ordinal (e.g. "large", "medium" or "small"), integer-valued (e.g. the number of occurrences of a particular word in an email) or real-valued (e.g. a measurement of blood pressure). Some algorithms work only in terms of discrete data and require that real-valued or integer-valued data be discretized into groups (e.g. less than 5, between 5 and 10, or greater than 10). An example would be assigning a given email into "spam" or "non-spam" classes or assigning a diagnosis to a given patient as described by observed characteristics of the patient (gender, blood pressure, presence or absence of certain symptoms, etc.).

An algorithm that implements classification, especially in a concrete implementation, is known as a classifier. The term {\it 'classifier'} sometimes also refers to the mathematical function, implemented by a classification algorithm, that maps input data to a category.

In the terminology of machine learning, classification is considered an instance of supervised learning\footnote{Learning where a training set of correctly-identified observations is available.}. The corresponding unsupervised procedure is known as clustering (or cluster analysis), and involves grouping data into categories based on some measure of inherent similarity (e.g. the distance between instances, considered as vectors in a multi-dimensional vector space).

Terminology across fields is quite varied. In statistics, where classification is often done with logistic regression\footnote{Logistic regression is a type of regression analysis used for predicting the outcome of a binary dependent variable (a variable which can take only two possible outcomes, e.g. "yes" vs. "no" or "success" vs. "failure") based on one or more predictor variables. Logistic regression attempts to model the probability of a "yes/success" outcome using a linear function of the predictors. Specifically, the log-odds of success (the logit of the probability) is fit to the predictors using linear regression.} or a similar procedure, the properties of observations are termed explanatory variables (or independent variables, regressors, etc.), and the categories to be predicted are known as outcomes, which are considered to be possible values of the dependent variable. In machine learning, the observations are often known as instances, the explanatory variables are termed features (grouped into a feature vector), and the possible categories to be predicted are classes. There is also some argument over whether classification methods that do not involve a statistical model can be considered {\it statistical}. Other fields may use different terminology: e.g. in community ecology, the term {\it 'classification'} normally refers to cluster analysis, i.e. a type of unsupervised learning, rather than the supervised learning described in this article.

\subsection{Evolutionary Computation}
\label{apx:evolutionaryComputation}
The contents of this Section are drawn from \cite{EVO-COMPUTATION}.

Evolution is any change across successive generations in the heritable characteristics of biological populations. Evolutionary processes give rise to diversity at every level of biological organisation, including species, individual organisms and molecules such as DNA and proteins.

Evolutionary computation uses iterative progress, such as growth or development in a population. This population is then selected in a guided random search using parallel processing to achieve the desired end. Such processes are often inspired by biological mechanisms of evolution.

The principle of evolution is the primary unifying concept of biology, linking every organism together in a historical chain of events. Every creature in the chain is the product of a series of {\it accidents} that have been sorted out thoroughly under selective pressure from the environment. Over many generations, random variation and natural selection shape the behaviors of individuals and species to fit the demands of their surroundings.

This fit can be quite extraordinary and compelling, a clear indication that evolution is creative. While evolution has no intrinsic purpose, it is merely the effect of physical laws acting on and within populations and species, it is capable of engineering solutions to the problems of sutvival that are unique to each individual's circumstance and, by any measure, quite ingenious\cite{EVO-COMPUTATION}.

The most important scientific theory on evolution of species is due to Charles Darwin. He established that all species of life have descended over time from common ancestors, and proposed the scientific theory that this branching pattern of evolution resulted from a process that he called natural selection\footnote{Natural selection is the gradual, nonrandom process by which biological traits become either more or less common in a population as a function of differential reproduction of their bearers.}.

Darwin identified a small set of essential elements to rule evolution by natural selection: reproduction of individuals, variation phenomena that effect the likelihood of survival of individuals, heredity of many of the parents' features by sons in reproduction and the presence of a finite amount resources causing competition for survival between individuals.

These simple features , reproduction, likelihood of survival, variation, heredity and competition are the bricks that build the simple model of evolution that inspired the machine learning technique known as evolutionary algorithms (EAs).

An EA uses some mechanisms inspired by biological evolution: reproduction, mutation, recombination, and selection. Candidate solutions to the optimization problem play the role of individuals in a population, and the fitness function\footnote{A fitness function is a particular type of objective function that is used to summarise, as a single figure of merit, how close a given design solution is to achieving the set aims.} determines the environment within which the solutions "live". Evolution of the population then takes place after the repeated application of the above operators.

During the years, many diffent kinds of EAs have been developed. The main feature characterizing the different paradigms of EAs is how the individuals are represented. In particular now will be described the generic algorithms, ancestor of genetic programming.

\subsubsection{Genetic Algorithms}
\label{apx:geneticA}
Genetic Algorithms (GA) were invented for the first time by Holland in 1970s, see \cite{Holland-Book}, and later on extended by Goldberg in his works, see \cite{Goldberg-Book}.

GA's key idea is to adapt the principles of evolution in the way of being able to implement these concepts in a computer so that they can be used to find the solution (or approximate it) for particular problems. In this context, these principles are called genetic operators. Given its nature inspired by natural evolution, many terms used are taken from biology and adapted for this use.

Genetic algorithms, in short, try to simulate the evolution of a species: defined an optimization problem\footnote{It refers to the selection of a best element from some set of available alternatives. In the simplest case, an optimization problem consists of maximizing or minimizing a real function by systematically choosing input values from within an allowed set and computing the value of the function}, starting from a random set of candidate solutions\footnote{A candidate solution is a member of a set of possible solutions to a given problem. A candidate solution does not have to be a likely or reasonable solution to the problem, it is simply in the set that satisfies all constraints.} of the problem, they attempt to improve their quality in an iterative way, applying the genetic operators.

At the beginning of the algorithm is generated a set of possible solutions to the problem, which is indicated with the term of population. Each solution present in the population is codified through a string of bits fixed length and takes the name of individual.

The quality of a solution is instead indicated by the term fitness, which is usually determined by a particular function that takes the name of fitness function.

After the initialization, the evolutionary process starts, which consists of updating at each iteration the set of hypotheses. Each iteration takes the name of generation and it is performed in two phases: the selection process and the variation process.

In the selection phase are calculated the fitness values of all individuals in the population and then is performed a probabilistic extraction on the population based on the values ​​of these fitness. The selected individuals are then used to form the population of the new generation.

In the variation phase are used one or more genetic operators on the individuals of the new population. Not all individuals undergo the process of variation and therefore some of them are replicated unchanged into the next population.

The process ends on the basis of a termination criterion: most commonly the process ends when at least one individual in the population has a satisfactory fitness or when a prefixed number of generation has occurred.

Follows a deeper discussion on how a genetic algorithm works:
\begin{itemize}
\item {\bf Crossover:} is applied to randomly paired strings with a probability denoted $p_c$. It produces two offsprings, usually different from their two parents and different from each other, but conteining some genetic material from each of their parents. The offsprings are then put into the new population. Many crossover algorithms have been developed for GAs. The most common one is called one point crossover. Its behaviour is shown in Figure~\vref{fig:Genetic Algorithms Crossover}. 
\begin{figure}[htbp]
\centering
\includegraphics[width=\textwidth]{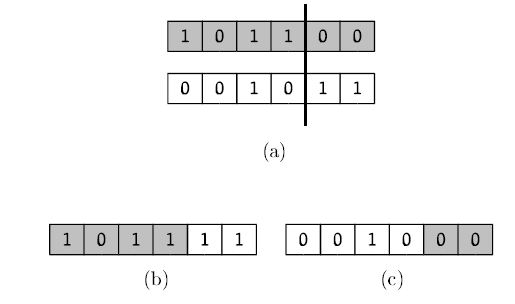}
\caption[Genetic Algorithms Crossover]{The GA crossover. Two (a) individuals are selected for crossover. The crossover point, in this case, is 4. In (b) and (c) are shown the results of the crossover process. The two individual are generated by combining the crossover fragment s of the parents.}
\label{fig:Genetic Algorithms Crossover}
\end{figure}
First of all, a number between 1 and $L-1$ is randomly generated using an uniform distribution, being $L$ the length of the individuals' string. This number, that in Figure~\vref{fig:Genetic Algorithms Crossover} is 4, becomes the crossover point. Each parent is then split at this crossover point into a crossover fragment and a remainder. For example, in the picture, the crossover fragment is 1011 and the remainder is 00. The crossover fragment of the first individual is then combined with the reminder of the second one and the crossover fragment of the second individual with the remainder of the first one. The resulting individual are then inserted in the new population.
\item {\bf Mutation:} modifies a sub-string of an individual, with a certain probability $p_m$, and the resulting individual is put into the new population. Many mutation algorithms have been developed. The most commonly used is called point mutation. Its behaviour is shown in Figure~\vref{fig:Genetic Algorithms Mutation}.
\begin{figure}[htbp]
\centering
\includegraphics[width=0.6\textwidth]{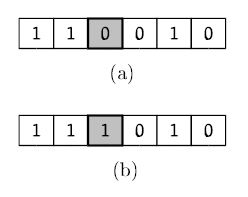}
\caption[Genetic Algorithms Mutation]{The GA mutation. In (a) are shown the individuals chosen for mutation with, in this case, mutation point 3. In (b) are shown the individuals resulting by mutation.}
\label{fig:Genetic Algorithms Mutation}
\end{figure}
Each position of the current string is chosen with distribution $p_m$ and the character contained in that positionis then replaced with another randomly chosen character.
\item {\bf The algorithm:} the pseudo-code is shown in Algorithm~\vref{GA Algorithms}. The algorithm takes as an input the dimension $n$ of the population, la probability of crossover $p_c$ and the probability of mutation $p_m$. The output is the best individual till the last population.
\begin{algorithm}
\caption[Genetic Algorithms]{Below is shown and example of pseudocode of a genetic algorithm}
\begin{algorithmic}
\REQUIRE $n \geq 0$
\STATE Generation of a random initial population $P$ of $n$ individuals
\WHILE{(Termination criterion is not verified)}
\STATE Computation of the fitness value for each individual in the population
\STATE Generation of an empy population $P'$
\WHILE{(Population $P' < n$)}
\STATE Perform the selection of a pair of individuals $x_1$ and $x_2$
\STATE Random extraction of a value r in the interval $[0,1]$
\IF{$r < p_c$}
\STATE Perform crossover on $x_1$ and $x_2$ optaining their sons $y_1$ and $y_2$
\ELSE
\STATE $y_1 \leftarrow	x_1$
\STATE $y_2 \leftarrow	x_2$
\ENDIF
\STATE Perform mutation on $y_1$ with probability $p_m$ for each bit
\STATE Perform mutation on $y_2$ with probability $p_m$ for each bit
\STATE $P' = P' \cup \{ y_1\} \cup \{ y_2\}$
\ENDWHILE
\STATE $P \leftarrow P'$
\ENDWHILE
\STATE {\bf Return} the best individual of $P'$
\end{algorithmic}
\label{GA Algorithms}
\end{algorithm}
\end{itemize}

\subsection{Genetic Programming}
\label{apx:geneticP}
A lot of the contents of this Section are drawn from \cite{Vanneschi-Thesis} and see it for a deeper discussion.

Genetic Programming (GP) is an evolutionary algorithm-based methodology inspired by biological evolution to find computer programs that perform a user-defined task. It is a specialization of genetic algorithms (GA) where each individual is a computer program. It is a machine learning technique used to optimize a population of computer programs according to a fitness landscape\footnote{In evolutionary optimization problems, fitness landscapes are evaluations of a fitness function for all candidate solutions.} determined by a program's ability to perform a given computational task.

The concept of GP was introduced by Koza in \cite{Koza1} and then refined in \cite{Koza2} and \cite{Koza3}. This tecnique aimed at overcoming the fixed length rapresentation of Genetic Algorithms's individuals. This limitetion, infact, is unnatural and contraining for a wide set of applications. For examples, fixed length strings do not readily support the hierarchical organization of tasks into subtasks typical of computer programs, they do not provide any convenient way of incorporation iteration and recursion and so on. But above all, GA rapresentation schemes do not have any dynamic variability: the initial selection of strings length limits in advance the number of internal states  of the system and limits what the system can learn\cite{Vanneschi-Thesis}.

GP, as originally defined by Koza, considers individuals as LISP-like\footnote{LISP is a family of computer programming languages. It is an expression-oriented language. Unlike most other languages, no distinction is made between "expressions" and "statements"; all code and data are written as expressions. When an expression is evaluated, it produces a value (in Common Lisp, possibly multiple values), which then can be embedded into other expressions. Each value can be any data type.} tree structures. These structures are perfectly capable of capturing all the fundamentals properties and features of modern programming languages. The tree-based GP is the oldest and the most commonly used rapresentation, althrough not the only one existing\footnote{In particular, in the last few years, a growing attention as been dedicated to linear and graph rapresentations.}\cite{Vanneschi-Thesis}.

\begin{figure}[htbp]
\centering
\includegraphics[width=0.6\textwidth]{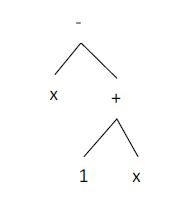}
\caption[Genetic Programming Tree-Like Individual]{An example of tree-like GP individual.}
\label{fig:GP-individual}
\end{figure}

An example of GP individuals is shown in Figure~\vref{fig:GP-individual}.

\subsubsection{GP Individuals}
\label{apx:GPI}
All the individuals are composed within two groups of symbols: the first one, $F$, composed by the function symbols $F = \{f_1, ..., f_n\}$ and the second one, $T$, composed by the terminal symbols $T = \{t_1, ..., t_n\}$. Every function in $F$ has a fixed number of arguments defined arity. Every element of the terminal symbols is a variable or a constant, defined according to the problem. For instance let's consider $F = \{+, -\}$ and $T = \{x, 1\}$, a possible LIPS-like individual could be $( + x ( - 1 + x ) )$.

The sets of symbols $F$ and $T$ should have the following properties:
\begin{itemize}
\item {\bf Closure:} Every function should be able to take as an argument every possible value of every element of $F$ and $T$.
\item {\bf Sufficiency:} The symbols in $F$ and $T$ should be able to define a solution for the problem. Often the two groups of symbols are not known a priori, but decided according to the problem.
\end{itemize}

\subsubsection{Initialization of the population}
\label{apx:GPInit}
Due to the complexity of the individuals, it is necessary to introduce particular initialization methods of the population. Koza in \cite{Koza3} proposed three methods: {\it grow}, {\it full} and {\it ramped half and half}. For each of those is necessary to specify the set $F$ of functional symbols, the set $T$ of terminals and a maximum allowed depth of the trees.

The {\it grow} method is performed as follows:
\begin{enumerate}
\item Random extraction of a function symbol $f_i$ from $F$ to be used as root of the tree;
\item Being $n$ the arity of $f_i$, if the current depth is lower then $d - 1$, randomly extract $n$ nodes from $F \cup T$ to be used as sons of $f_i$, otherwise the extraction is performed only on $T$;
\item Recursively repeat the procedure for all the $n$ extracted nodes.
\end{enumerate}
The {\it full} method uses the same procedure of the {\it grow} one, with the difference that extracts the nodes from $F$ when the depth is lower then $d$ instead that extracting them from $F \cup T$.

Both the methods, as observed by Koza in \cite{Koza3}, generate a population of trees a lot similar with each others. To avoid this, the  method {\it ramped half and half} has the purpose of preserving divertiry in the population. The idea is to divide the population in $d$ subgroups with the same dimension and assign to each of those a different maximum depth (between 1 and d). Afterwards half of each group is initialized with the {\it grow} method and half with the {\it full} one.

\subsubsection{Fitness Evaluation}
\label{apx:GPFitness}
Each program in the population is assigned a fitness value, rapresenting its ability to salve the problem. This values is calculated by means of same well defined explicit procedure. The two most commonly used measures in GP are the {\it raw fitness} and the {\it standardized fitness}.

The {\it raw fitness}, as defined by Koza, is "the measurement of fitness that is stated in the natural terminology of the problem itself". It is, therefore, the most natural way to calculate the ability of a program to solve a problem. For instance, if the task is to drive a robot to pick up the maximum number of objects, the {\it raw fitness} is the number of object picked up by the robot.

The {\it standardized fitness} restates the {\it raw fitness} so that a lower value is always better one. Problems where the {\it raw fitness} is used are also called {\it maximization problems}, instead problems where the {\it standardized fitness} is used are also called {\it minimization problems}.

\subsubsection{Genetic Operators}
\label{apx:GPOperators}
For each individual of the in the GP population, three possible actions can be choses: genetic operators can be applied to that individual, it can be copied into the new population as it is, or it can be discarded and replaced by a new individual.

Now will be discussed three genetic operators: {\it selection}, {\it crossover} and {\it mutation}.

The {\it selection} operator makes the decision of which of the three actions should be applied to the individual. Many possible algorithms have been developed for selection, of those three are the most common: {\it fitness proportional} (or {\it roulette wheel}) selection, {\it ranking} selection, {\it tournament} selection.

In the {\it fitness proportional} selection, being $N$ the number of individuals of the population $P$ and $\{f_1, ..., f_{N-1}\}$ their fitness values, each individual has the probability of beeing chosen $p_i = \frac{f_i}{\sum_{i=0}^{N-1} f_i}$. In practice, the probability of being selected is proportional to the value of the fitness.

In the {\it ranking} selection every individual is sorted according to their fitness values. Every individual is then associated with a function to be chosed according to their rank. This approach was proposed to mitigate the importance of high fitness values in the selection process.

In the {\it tournament} selection a number of individuals, called {\it tournament size}, is randomply selected and of those the one with best fitness is chosen.

\begin{figure}[htbp]
\centering
\includegraphics[width=\textwidth]{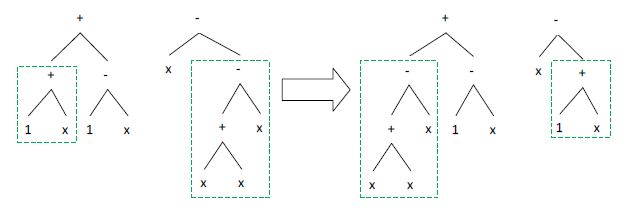}
\caption[Genetic Programming Crossover Operator]{An example of crossover in Genetic Programming. On the left are shown the parent individuals and on the right the sons. In green are pointed out the swaped subtrees.}
\label{fig:GP-crossover}
\end{figure}

\begin{figure}[htbp]
\centering
\includegraphics[width=\textwidth]{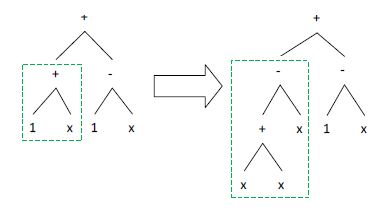}
\caption[Genetic Programming Mutation Operator]{An example of mutation in Genetic Programming. On the left is shown the original individual while on the right is the shown the tree after the mutation. In green is pointed out the mutated subtree.}
\label{fig:GP-mutation}
\end{figure}

The {\it crossover} operator in GP, as in GA, genetates two individuals, $y_1$ and $y_2$, from two parents $x_1$ e $x_2$. This operator selects a subtree of the individual $x_1$ and one of $x_2$ and swaps them, generating in this way two new individuals with genetic material from both the two parent individuals. In Figure~\vref{fig:GP-crossover} is shown an example of this process.
The {\it mutation} operator choses a subtree of an individual and replace it with a randomply generated one. The depth of the new subtree is limited to the maximum depth of the whole tree. An example of mutation is shown in Figure~\vref{fig:GP-mutation}.
Differently from the GA operators, the operators defined in GP are very destructive as they modify the individuals a lot. For this reason, less destructive variants exist. The following techniques aim at this:
\begin{itemize}
\item {\bf Steady State:} in this case, after variation, one or two individuals are directly merged into the new population. After the new individuals have been inserted into the population, the new individual are already taken in count for selection. In this way, the steady state works as the variation takes place just one time per generation.
\item {\bf Automatically Define Fuction:} some subtrees are considered unchangeble atomic objects. They can be defined ad hoc for the application.
\end{itemize}

\clearpage

\subsubsection{GP Algorithm}
\label{apx:GPAlg}
The pseudo-code for the GP algorithm is the same as the one shown in Algorithm~\vref{GA Algorithms} for GA.

In synthesis, the GP paradigm breeds computer programs to solve problems by executing the followind steps:
\begin{enumerate}
\item Generate an initial population of computer programs (or individuals);
\item Iteratively perform the following steps untill the termination criterion has been satisfied:
\begin{enumerate}
\item Execute each program in the population and assign it a fitness value according to how well it solves the problem;
\item Create a new population by applyng the following operations:
\begin{enumerate}
\item Probabilistically select a set of computer programs to be reproduced, on the basics on thier fitness ({\it selection});
\item Copy same of the selected individuals, without modifying them, into the new population ({\it reproduction});
\item Create new computer programs by genetically recombining randomply chosed ({\it crossover}) parts of two selected individuals;
\item Create new computer programs substituting ({\it mutation}) randomply chosed parts of some selected individuals with new randomly generated ones;
\end{enumerate}
\end{enumerate}
\item The best computer program appeared in each generation is designed as the result of the GP process at that generation. This result may be a soluction (or an approximate solution) to the problem\cite{Vanneschi-Thesis}.
\end{enumerate}

\clearpage \mbox{} \clearpage 
\thispagestyle{empty} 

\end{document}